\documentclass{scthesis}
\usepackage{epsf}
\usepackage{graphics}
\def\be{\begin{eqnarray}}
\def\e{\end{eqnarray}}
\def\pp{\partial}

\def\ov{\over}
\def\k{\kappa}
\def\p{\perp}
\def\s{\sigma}
\def\ep{\epsilon}

\def\st{\stackrel}
\def\lh{\leftrightarrow}

\begin{document}

\submissionmonth{MAY}
\submissionyear{2001}
\author{\bf {ASMITA MUKHOPADHYAY}\\ SAHA INSTITUTE OF NUCLEAR PHYSICS \\
KOLKATA}
\title{POWER CORRECTIONS IN DEEP INELASTIC SCATTERING AND NON-PERTURBATIVE
LIGHT FRONT DYNAMICS}
\maketitle

%%%%%%%%%%%%%%%%%%%%%%%%%%%%%%%%%%%%%%%%%%%%%%%%%%%%%%%%%%%%%%%%%%%%%%%

\begin{acknowledgements}
I would like to express my deep gratitude to my supervisor Prof. 
A. Harindranath for
giving me  all the inspiration and guidance to do the research work
contained in this thesis. I am grateful to him for prompt help
whenever I needed it. I would like to thank him specially for careful
reading of all the manuscripts. 

I would also like to thank my collaborators, Prof. J. P. Vary, Prof. H. C. Pauli,
R. Ratabole, R. Kundu, D. Chakrabarti and S. Bhattacharya. I am also
grateful to Prof. P. Mitra for many illuminating discussions during my
research.   

I am indebted to  all the members of Theory Group of
SINP for their  help and cooperation during my research work.
Finally, I would like to thank all my family members for their inspiration and
support. 

\vskip 2.5cm
\noindent Kolkata, \hspace{8.6cm} Asmita Mukhopadhyay\\
\noindent May, 2001 \hspace{8.3cm} Theory Group, SINP
\end{acknowledgements}
%\include{list}

%
% Make the Table of Contents, List of Figures, and List of Tables.
%

\tableofcontents
\listoffigures
%\listoftables

%
% The following is a list of chapters.  Each is brought in from a
% separate file using the \include{} command.
%

%\documentstyle[aps,preprint]{revtex}
%\newcommand{\be}{\begin{eqnarray}}
%\newcommand{\e}{\end{eqnarray}}
%\newcommand{\pp}{\partial}
%\setcounter{page}{1}
%\begin{document}
%\newcommand{\kw}{\kappa}
%\setcounter{section}{0}
%\renewcommand{\thesection}{1.\arabic{section}}
%\setcounter{equation}{0}
%\renewcommand{\theequation}{1.\arabic{equation}}
%{\flushleft\huge\bf {Chapter 1}}
\chapter{Introduction}
%\vskip .2in
%%%%%%%%%%%%%%%%%%%%%%%%%%%
\section{Motivation}\label{mot}
%%%%%%%%%%%%%%%%%%%%%%%%%%%%
Lepton-hadron deep inelastic scattering (DIS) is a very important tool to 
unravel the structure of
hadrons. To the lowest order, this scattering takes place via the exchange 
of a virtual photon.
When a very low mass virtual photon ($Q^2<1 Gev^2$) scatters off a proton,
the photon `sees' only the total charge and magnetic moment of the proton
and the proton appears to be a pointlike object. However, a higher virtual mass photon
resolves the fine structure of the proton charge distributions and sees its
elementary constituents.  
The cross section of such scattering can be expressed in terms of several
functions which depend on two variables, $Q^2$ (momentum transfer square
in the process) and $\nu$ (energy of the virtual photon). These functions, to
start with, are unknown and they depend on the hadron structure. They are 
called structure functions. 
Early SLAC experimental results showed that the 
structure functions show a phenomenon called scaling, which means that they
depend only on the ratio of $Q^2$ and $\nu$, and do not depend
on them separately. In the pre-QCD era, Bjorken used current algebra
techniques to explain such behavior of structure functions. Later, Feynman
proposed parton model idea according to which, the proton consists of
pointlike, massless, noninteracting particles called partons and the
scattering takes place incoherently from the partons. The parton model
successfully explained scaling of the structure functions. However,
substantial scaling violations were observed in later SLAC experiments and
the structure functions were found to evolve with $Q^2$. This violation of
scaling occurs because of the interactions between the constituents of the
proton, which are now known as the quarks and gluons. Their interaction is
governed by QCD (quantum chromodynamics).

The $Q^2$ dependence of the structure functions can be logarithmic or there
may be power dependence. The contributions to the structure functions which
are  proportional to some power of ${1\over Q}$ are commonly called `higher
twist' contributions. The most widely used formal approach to DIS structure
functions is based on Wilson's operator product expansion (OPE). There
exists a more formal definition of twist in terms of OPE, which we shall
discuss shortly. DIS is a high energy process, and $Q^2$ is very large. This
means that higher and higher twist contributions are suppressed compared to
the leading twist contributions. However, recent experiments indicate that
the higher twist effects play a very important role in the SLAC kinematical
range. Understanding the higher twist effects requires non-trivial
non-perturbative information of the structure of the hadron. So far, a clear
and physical picture of the higher twist contribution is lacking. To
understand these effects, there is an urgent need to develop a
non-perturbative technique to analyze DIS. For this, one needs theoretical
tools which are based on physical intuitions and at the same time employs
well-defined field theoretic calculational procedure.

The usual field theoretical formulation is Lagrangian formulation, in which
one starts from the local Lagrangian, gets the equations of motion and
then quantizes the system by imposing suitable commutation or
anti-commutation relations between the fields and their conjugate momenta on 
a space-like hypersurface. It is a manifestly covariant formalism. However,
for interacting theory, the equations of motion involve non-trivial
interacting fields and can only be solved approximately, using 
 perturbative techniques. Perturbative methods cannot be applied
to low energy QCD bound state problems since the coupling constant takes a
large value in this energy range. Another point is that, in the covariant
Lagrangian formulation, the intuitive picture of quantum mechanics is lost.
There exists another formulation known as Hamiltonian formulation of quantum field theory
in which one starts with the Hamiltonian, obtains the Hamiltonian equations
of motion and sets the quantization conditions on a space-like hypersurface.
The Hamiltonian gives the time evolution of the system. Though it is physically
more intuitive, there are some disadvantages of Hamiltonian formulation. It
is manifestly non-covariant since one has to choose a time axis and perturbation theory becomes more complicated.
The latter is due to the fact that here one has to do old-fashioned time
ordered perturbation theory and instead of Feynman diagrams, one has to
calculate a larger number of time-ordered diagrams at each order. It has
been shown that, light-front quantization (where the surface of quantization
is tangential to the light cone) makes Hamiltonian formulation simpler. This
is due to the fact that. the diagrams which produce particles from the vacuum
are absent here (this point will be clarified in chapter 3). So one 
encounters lesser number of time ordered diagrams. In
what follows, we have used Hamiltonian light-front QCD framework
(Hamiltonian formulation of QCD where the quantization surface is tangential
to the light-cone) to analyze DIS structure functions.

In order to get an intuitive picture of DIS, it is necessary to keep close
contact with Feynman's parton model. As mentioned before, partons were
originally introduced as collinear, massless, non-interacting constituents
of the proton. It is necessary to generalize the parton ideas to introduce
field theoretic partons which are massive, non-collinear and their
interaction is governed by QCD. This goal is achieved in light-front
Hamiltonian QCD analysis. 
This is
essentially a non-perturbative method which allows us to explore the
non-perturbative higher twist effects explicitly. In this approach, instead
of calculating the moments of the structure functions as one does in the
conventional OPE formalism, one directly calculates the structure functions
themselves which involve hadron matrix elements of light-front bilocal operators.
The structure functions can be calculated once the light-front wave
functions are known. However, perturbative calculations are also possible,
 and both perturbative and non-perturbative aspects can be
investigated within the same framework. 

In what follows, we have used light-front Hamiltonian QCD framework 
 to study the twist four part of the longitudinal
 structure function $F_L$ in unpolarized DIS and the transverse polarized structure
function $g_T$ in polarized DIS, which is a twist three contribution. Both
of these structure functions involve non-trivial operators which brings in
non-perturbative dynamics. It is well known that the evolution of $F_L$ and
$g_T$ with the scale, $Q^2$, are highly complicated. We have shown that the reason of this
complexity is the fact that $F_L$ is related to the light-front QCD
Hamiltonian density and $g_T$ is related to the transverse spin operator,
both of which are complicated dynamical operators (they change the surface
of quantization, see chapter 3). As a result, both these structure
functions acquire off-diagonal contributions from the matrix elements, in
contrast to the other two electroproduction structure functions, $F_2$ and
$g_1$, which involve diagonal matrix elements only. Another interesting
point is, the operators involved acquire divergences in perturbation theory.
The experimentally measured structure functions, of course, are finite
quantities. Therefore, one must know how to remove these divergences
(renormalization). We have addressed the issue of renormalization of these
composite operators which, in light-front theory is highly non-trivial. As
we mentioned before, a very interesting point is that, both perturbative and
non-perturbative issues can be address in our approach. In
the non-perturbative context, we have analyzed twist-four longitudinal
structure function for a meson in $1+1$ dimensional QCD and for a
positronium in $3+1$ dimensional QED in weak coupling limit.

Earlier, it has been shown that the integral of the structure function 
$F_2$ is related to the light-front longitudinal momentum density and the
integral of the polarized structure function $g_1$ is related to the
helicity. In this work, we have shown that the integral of $F_L$ is directly related to the
light-front Hamiltonian density. Also the integral of the transverse
polarized structure function $g_T$ is related to the transverse spin
operator in light-front QCD. Thus our approach gives a unified picture of
DIS structure functions and connects them to the light-front Poincare
generators. We emphasize  that such connections arise because our approach
is different from the usual one and we get many new and interesting results.   
\vskip .2in
%%%%%%%%%%%%%%%%%%%%%%%%%%%%%%%%%%%%%%%%%%
\section{Plan of the Thesis}\label{plan}
%%%%%%%%%%%%%%%%%%%%%%%%%%%%%%%%%%%%%%%%%%%%
The usual approach to analyze DIS structure functions is based on OPE. This
 mainly concentrates on the
evolution of the structure functions in perturbative QCD. In these methods,
the moments of the structure functions automatically arise. We briefly
discuss the OPE method to analyze DIS structure functions in chapter 2. In
this chapter, we  also explain the term `twist' in the context of OPE. 
It is also important to understand the origin and nature of the higher twist
contributions. We discuss how these higher twist effects are
addressed in the covariant formalism. For the sake of simplicity, we briefly
outline the procedure and omit the mathematical details. 
The effect of the target mass gives a higher twist contribution, which, in
the free theory assumption produces $\xi$ scaling. The dynamical higher
twist effects are highly complicated. We point out that 
in the QCD improved parton model approach, these contributions are expressed
as the Fourier transform of the target matrix elements of various 
bilocal operators. However, these matrix elements  cannot be further
investigated because of the non-perturbative hadronic state.

We follow an alternative approach based on light-front
Hamiltonian QCD. The light-front formulation of quantum field theory has
many strikingly different features compared to the usual equal time
formulation. Some of these features makes it more suitable for the analysis
of DIS process. It is important to remember that the parton model was
formulated in the infinite momentum frame, which, in some sense is
equivalent to light-front formulation. In chapter 3, we discuss the basic
features of light-front field theory and light-front QCD.

In chapter 4, we outline our approach. Bjorken-Johnson-Low (high energy)
expansion of the Compton scattering amplitude gives the expressions of the
structure functions as the Fourier transform of the matrix elements of
light-front bilocal operators. We work in the light-front gauge. In this method, one gets the structure
functions themselves and not their moments. By expanding the target state in
Fock space and using truncation, the structure functions are expressed in
terms of light cone wave functions. Splitting functions 
are obtained in perturbation theory simply by replacing the target by a
dressed parton. In this chapter, we review the works done so far in
this approach.  We show that this method  not only reproduces the well known results of the
covariant formalism, like the $Q^2$ evolution of the twist two structure
function $F_2$, but also addresses some important issues, like the effect of
quark mass in $g_T$ and the helicity sum rule of the proton, in a more
transparent way.  

The actual work done in this research program has been discussed in detail
in chapters 5, 6, 7 and 8. These works are concerned with the higher twist
effects $F_L$ and $g_T$. 
All of these results are new and interesting, both from theoretical and
phenomenological point of view. Since our approach is different from the
usual one, such new outcomes are not unexpected. Also, our approach gives
better insights into problems that are not easy to understand in the
commonly used procedure.   

In chapter 5, we discuss a new sum rule that we
have obtained which connects the twist four part of $F_L$ to the fermionic
part of light-front QCD Hamiltonian density. Introducing a
gluonic part of $F_L$, we show that the sum rule relates the total $F_L$ to
the target mass. To our knowledge, this is the first sum rule for a higher
twist structure function. In this chapter, we also perform one loop renormalization
of $F_L$ in light-front Hamiltonian perturbation theory. Replacing the
target by a dressed quark, we show that the twist four part of $F_L$ is
directly related to the quark mass shift in perturbation theory.

In chapter 6, we continue our investigation of $F_L$. In the non-perturbative 
context, we verify that the sum rule is obeyed for a meson in $1+1$ dimensional 
QCD because of t' Hooft equation. An analysis for a bound state in QCD in $3+1$
dimension is very complicated and there one has to use the recently
developed similarity renormalization technique. In order to understand the
calculational procedure, we  consider a simpler case, namely a
positronium-like bound state in $3+1$ dimensional QED in the weak coupling
limit. We explicitly verify the sum rule in this case.

In chapters 7 and 8, we investigate another most challenging problem,
namely that of the spin operator for a composite system in an arbitrary
reference frame in relativistic quantum field theory. We
 show that, this issue is very important in the context of polarized
scattering, since the transverse polarized structure function $g_T$ is
related to a part of the transverse spin operator. It is also known that
$g_1$ is related to the helicity operator. In chapter 7, we 
discuss the problems associated with the definitions of the relativistic
spin operators in equal-time theory and contrast with the light-front
case. The light-front theory is more suitable for this purpose because boost
is kinematical. 
We  discuss the light-front spin operators for the massive and
massless cases separately. The transverse spin operator contains the
transverse rotation operator which is interaction dependent. We 
construct the transverse rotation operator  in light-front QCD. In this
chapter, we  also show that though the light-front transverse spin
operators cannot be separated into orbital and spin parts, there exists a
decomposition of it which is physically interesting. 

In chapter 8, We explore the physical significances of this
decomposition and compare and contrast it with the decomposition of the
helicity operator into orbital and spin parts. We  also perform one
loop renormalization of the full transverse spin operator.  Our results show 
that only one counterterm
is needed and that has the same form as the linear mass counterterm in light-front
Hamiltonian perturbation theory. We  explicitly show that in an arbitrary
reference frame, all the center of mass momentum dependence get canceled in
the matrix element of the transverse spin operator. We emphasize that in the
equal time case,  this cancellation is almost impossible to prove because of
the dynamical boost operators. 

Summary and conclusions are given in chapter 9. References are given at the
end of each chapter. In order to clarify the
notations and also the intermediate steps of calculations, we have provided
several appendices.

\chapter{Operator Product Expansion Approach to DIS}

The standard approach to DIS structure functions is based on Wilson's
Operator Product Expansion (OPE). In this chapter, we introduce the idea of
OPE and also the definition of twist of an operator. Then we describe
briefly the OPE method of analyzing DIS. 
\vskip .2in 
%%%%%%%%%%%%%%%%%%%%%%%%%%%%%%%%%%%%%%%%%%%%%%%%%%%%%%
\section{`Twist' of an Operator}\label{c2twist}
%%%%%%%%%%%%%%%%%%%%%%%%%%%%%%%%%%%%%%%%%%%%%%%%%%%%%%
Products of fields at the same space-time point are called 
composite operators. The composite operator is not well-defined. The
easiest way to see this is to calculate the vacuum expectation value of the
propagator of free scalar field, $\Delta(x-y)$. In coordinate space, one
clearly sees that $\Delta(x-y)$ diverges as $x \rightarrow y$. For free
field theory, one then subtracts out the singularity by taking the normal
ordered product. However, for interacting fields, this cannot be generalized
in a straightforward manner. Wilson introduced the idea of operator product
expansion (OPE) in which the singularities of operator products are
expressed as a sum of non-singular operators with singular c-number
coefficients. There are two types of OPE \cite{c2cheng}:

\subsection{Short Distance Expansion}

The product of two operators at short distance $(x-y)^\mu \rightarrow 0$ is
given by,
\be
A(x)B(y) \approx \sum_i C_i(x-y) O_i( {1\over 2}(x+y));~~~~~~(x-y)^\mu
\rightarrow 0
\e
where $A, B$ and $O_i$ are local operators. $C_i$ are singular c number
functions which are called Wilson coefficients. 
The short distance behavior of the Wilson coefficients is expected to be
given by naive dimensional counting. The most dominant term in the expansion
at $x\rightarrow 0$ will be the term for which ${x}^{d_i - d_A-d_B}$ is
most singular, where
$d_i, d_A$ and $d_B$ are the mass dimensions of the operators. Higher is the
dimension of the operator $O_i$, less singular is the Wilson coefficient.

\subsection{Light Cone Expansion}

The product of two operators at short light cone distance $x^2 \rightarrow
0$ is given by,
\be
A({x\over 2})B(-{x\over 2}) \approx \sum_i C_i(x) O_i ({x\over 2}, -{x\over
2}),~~~~~~~~~~~x^2 \approx 0.
\e
$O_i(x, y) $ are regular bilocal operators which can be expanded in a Taylor
series,
\be
O_i( {x\over 2}, -{x\over 2}) = \sum_j x^{\mu_1} x^{\mu_2}....x^{\mu_j}
O^{j,i}_{\mu_1 ... \mu_j}(0).
\e 
So the OPE on the light cone becomes, in terms of local operators,
\be
A({x\over 2}) B(-{x\over 2}) \approx \sum_{j,i} C_i^{(j)}(x^2) x^{\mu_1} x^{\mu_2}....x^{\mu_j}
O^{j,i}_{\mu_1 ... \mu_j}(0)   
\e
where $j$ is the maximum spin of the operators $O^{j,i}_{\mu_1 ...
\mu_j}(0)$. The light cone behavior of the Wilson coefficients is given by
naive dimensional counting, and the most dominant singularity is for the
term which has the lowest value of $(d_{j,i} - j)$, i.e. the dimension of $O^{j,i}_{\mu_1 ...
\mu_j}(0)$ minus the spin of $O^{j,i}_{\mu_1 ...\mu_j}(0)$. The {\it twist}
of an operator is defined as
\be
\tau = dimension - spin.
\e

\subsection{Spin of an Operator}

The representations of the homogeneous Lorentz group are labeled by the
values of two positive integers and/or half integers, $A$, $B$.  
An operator that transforms like the $(A, B)$ representation has components
that rotate like objects of spin $j$, with
\be
j=A+B, A+B-1, ....\mid A-B\mid.
\e
A traceless symmetric tensor of rank $2A$ transforms like the $(A, A)$
representation and has components of spin $2A, 2A-1, ...0$ \cite{c2wein}. 
In the definition of twist, one takes into account the maximum spin.

Deep inelastic scattering is a light cone dominated process, which means
that it gets major contributions near the light cone and the light cone
expansion is applied here. 

In the next section, we shall describe briefly the conventional approach
based on OPE to deep inelastic scattering.

%%%%%%%%%%%%%%%%%%%%%%%%%%%%%%%%%%%%%%%%%%%%%%%%%%%%%%
\section{A Brief Overview of Deep Inelastic Scattering in the Conventional
Approach}\label{c2over}
%%%%%%%%%%%%%%%%%%%%%%%%%%%%%%%%%%%%%%%%%%%%%%%%%%%%%%%%    

In this section, we describe the approach based on OPE to polarized and
unpolarized deep inelastic scattering. Because of the complexities of the expressions, we
shall avoid the explicit details but only outline the procedure.

We begin with a brief review of the basic ingredients of  
lepton-nucleon deep inelastic scattering (DIS):
\begin{equation}
	e(k) + h(P) \longrightarrow e(k^\prime)  + X(P+q) \, ,
\end{equation}
where we have specified the four momenta of the particles explicitly 
and $q=k-k^\prime$
is the momentum transfer in the process through the virtual photon. 
The inclusive cross section for the above scattering process is given by 
\begin{equation}
	{d \sigma \over d\Omega dE'} = {1\over 2M}{\alpha^2 \over q^4}
		{E'\over E} L_{\mu \nu} W^{\mu \nu}.
\end{equation}
Here $E$ ($E'$) is the energy of the incoming (outgoing) lepton,
$L_{\mu \nu}$ is the leptonic tensor,
\begin{eqnarray}
	L_{\mu \nu} &=& {1\over 2}\sum_{s'}[\overline{u}(k,s)\gamma_\mu 
		u(k',s') \overline{u}(k',s')\gamma_\nu u(k,s)] \nonumber \\
	&=& 2(k'_\mu k_\nu + k'_\nu k_\mu) - 2 g_{\mu \nu} k \cdot k'
		- 2i \epsilon_{\mu \nu \rho \sigma} q^\rho s^\sigma .
\end{eqnarray}
$W^{\mu \nu}$ is the hadronic tensor which contains all the hadronic 
dynamics involved in DIS process, 
\begin{equation}
	W^{\mu \nu} = {1\over 4\pi} \int d^4 \xi~ e^{iq \cdot \xi} 
		\langle PS |[J^\mu(\xi), J^\nu(0)]|PS \rangle  \, ,
\label{c2wmunu}
\end{equation}
where $P$ and $S$ are the target four-momentum and polarization 
vector respectively ($P^2=M^2, S^2=-M^2, S\cdot P=0$), $q$ is 
the virtual-photon four momentum, and $J^\mu(x)=\sum_\alpha 
e_\alpha \overline{\psi}_\alpha(x) \gamma^\mu \psi_\alpha(x)$ the 
electromagnetic current with quark field $\psi_\alpha (x)$ carrying
the flavor index $\alpha$ and the charge $e_\alpha$. 

The above hadronic tensor can be decomposed into independent
Lorentz invariant scalar functions:
\begin{eqnarray}
	W^{\mu \nu} &=&\Big(-g^{\mu \nu} + {q^\mu q^\nu 
		\over q^2} \Big) W_1(x,Q^2) + \Big(P^\mu - {\nu 
		\over q^2} q^\mu\Big)\Big(P^\nu -{\nu \over q^2} 
		q^\nu\Big)W_2(x,Q^2) \nonumber \\
	& & - i \epsilon^{\mu \nu \lambda \sigma}q_\lambda \Big[
		S_\sigma W_3(x,Q^2)+ P_\sigma S\cdot q W_4(x,Q^2) 
		\Big] \nonumber \\
	&=&  \Big( g^{\mu \nu} - {q^\mu q^\nu \over q^2} \Big)\Big({1\over 2}
		F_L (x,Q^2) -{M^2\over \nu}F_2(x,Q^2)\Big)\, 
		+ \Big [P^\mu P^\nu - {\nu \over q^2} \Big(
		P^\mu q^\nu + P^\nu q^\mu \Big)  \nonumber \\
	& & \quad\quad\quad\quad + g^{\mu \nu} {\nu^2 
		\over q^2} \Big] {F_2(x,Q^2) \over \nu}
		- i \epsilon^{\mu \nu \lambda \sigma}{q_\lambda \over \nu}
		\Big[ S_{\sigma L} g_1(x,Q^2) + S_{\sigma T} g_T
		(x,Q^2) \Big] \, . \label{c2wfun}
\end{eqnarray}
The dimensionless functions
\begin{eqnarray}
F_L(x,Q^2)&=& 2 \Big [-W_1 + \big [ M^2 -{(P.q)^2 \over q^2}\big ] W_2
\Big ]\\
F_2(x,Q^2) &=& \nu W_2(x,Q^2)
\e
are the unpolarized structure functions measured from the unpolarized target
and
\be
g_1(x,Q^2) &=&\nu \Big[ W_3(x,Q^2) + \nu W_4 (x,Q^2) \Big]~~~\\	
g_T(x,Q^2) &=& g_1(x,Q^2) + g_2(x,Q^2) = \nu W_3(x,Q^2)~~, 
\end{eqnarray}
 where $g_1(x,Q^2)$ and $g_T(x,Q^2)$ are 
known as the longitudinal
and transverse polarized structure functions respectively. Here $x = { Q^2
\over {2 \nu}}$ is the Bjorken variable, $\nu = P \cdot q$ is the energy
transfer and $Q^2 = -q^2$.  
The longitudinal and transverse polarization vector 
components are given by
\begin{equation}
	S_{\mu L} = S_\mu - S_{\mu T}~~, ~~~~~
	S_{\mu T} = S_\mu - P_\mu {S \cdot q \over \nu} \, .
\end{equation}

The hadronic tensor is related to the forward 
virtual-photon hadron Compton scattering amplitude as 
\begin{equation}
	W^{\mu \nu} = {1\over 2\pi}{\rm Im} T^{\mu \nu} ,\label{c2comp}
\end{equation}
where $T^{\mu \nu} $ is given by
\be
T^{\mu \nu} = &&i \int d^4 \xi~ e^{iq \cdot \xi} 
		\langle PS |T(J^\mu(\xi), J^\nu(0))|PS \rangle \nonumber\\
            =&&\Big(-g^{\mu \nu} + {q^\mu q^\nu 
		\over q^2} \Big) T_1(x,Q^2) + \Big(p^\mu - {\nu 
		\over q^2} q^\mu\Big)\Big(p^\nu -{\nu \over q^2} 
		q^\nu\Big)T_2(x,Q^2) \nonumber \\
	& & ~~~~~~~~~~~~~ - i \epsilon^{\mu \nu \lambda \sigma}q_\lambda 
		\Big[ S_\sigma T_3(x,Q^2)+ P_\sigma S\cdot q T_4(x,Q^2) 
		\Big ] \, .
\end{eqnarray}
It can be shown that the most dominant contribution to $W^{\mu \nu}$ comes
from the region near the light cone, $\xi^2 \approx 0$, at high value of
$Q^2$. So one can expand the products of currents near the light cone. For
the sake of clarity and simplicity we suppress the Lorentz indices and write
\be
{T(\nu,Q^2)}_{\xi^2 \rightarrow 0}&&= \sum_{i,n} \int d^4 \xi
e^{iq\xi}C_1^n(\xi^2) \xi_{\mu_1}...\xi_{\mu_n}\langle P \mid O_i^{\mu_1
...\mu_n} \mid P \rangle \nonumber\\&&
=\sum_{i,n} 2q_{\mu_1}...2q_{\mu_n}{\pp^n\over {\pp(iq^2)^n}}
 \int d^4 \xi
e^{iq\xi}C_1^n(\xi^2) \langle P \mid O_i^{\mu_1
...\mu_n} \mid P \rangle.
\e
Here, in the last step we have replaced $\xi^\mu$s by ${\pp\over \pp
q_\mu}$s and used the relation
\be
{\pp\over \pp q_{\mu_1}}...{\pp\over \pp q_{\mu_n}}= 
2q_{\mu_1}...2q_{\mu_n}{\pp^n\over {\pp(iq^2)^n}}+{\rm trace~ terms}.
\e
The trace terms are neglected since they are suppressed.
So we get
\be
T(\nu,Q^2)=\sum_{i,n} ({Q^2\over 2})^{-n}q_{\mu_1}...q_{\mu_n} C^n_i(Q^2)
\langle P \mid O_i^{\mu_1...\mu_n}(0) \mid P \rangle.
\e
Here $C_i^n(Q^2)$ are (Fourier transform of) the Wilson coefficients and 
$O_i^{\mu_1...\mu_n}(0)$ are local operators. According to their most
general Lorentz structure, the spin averaged matrix elements of the
operators $O_i$ can be written as \cite{c2reya} 
\be
\langle P \mid O_i^{\mu_1...\mu_n}(0) \mid P
\rangle=A^n_i(p^{\mu_1}...p^{\mu_n}-m^2 g^{\mu_1 \mu_2}p^{\mu_3}...).
\e
The terms proportional to $g^{\mu_1 \mu_2}$ are trace terms which are
subtracted to project out operators of definite spin. We then get
\be
T(\nu, Q^2)=\sum_{i,n}C_i^n(Q^2)x^{-n}A_i^n+O[x^{-n+2}{m^2\over Q^2}].
\label{c2tmn}
\e
The neglected terms are either contributions from the trace terms (which are
called target mass effects) or they arise from higher twist operators (in
this case, the mass scale $m^2$ is not necessarily the target mass). 

The Wilson coefficients $C_i^n(Q^2)$ depend on the renormalization scale
$\mu$ and the coupling $g(\mu)$. They obey the renormalization group
equation
\be
(\mu {\pp\over \pp \mu}+\beta{\pp\over \pp g}-\gamma_{O_i^n})C_i^n({Q^2\over
\mu^2},g(\mu))=0
\e
where $\beta$ is the QCD beta function and $\gamma_{O_i^n}$ is the first
non-trivial (one loop) order anomalous 
dimension of the operator $O_i^{\mu_1...\mu_n}$ \cite{c2cheng}. The leading $Q^2$
dependence to all orders in $\alpha_s$ can be obtained from the solution of
the RG equation.
One then uses optical theorem to connect $T^{\mu \nu}$ to $W^{\mu \nu}$.
This is not straightforward, since the physical region for DIS is $0 \le x
\le 1$. But from Eq. (\ref{c2tmn}) one can see that $T^{\mu \nu}$ diverges in
the physical region. This difficulty is overcome by analytically continuing
$T^{\mu \nu}$ to complex $x$ and get the result corresponding to the
physical region as a limit of the analytic function in the unphysical
region. In this way one automatically gets the $x$ moments of the structure
functions. We define $T(x, Q^2)$ as a function of complex $x$ which is
analytic as $\mid x \mid \rightarrow \infty$ and has a cut from $-1$ to
$+1$. The coefficient of $x^{-n}$ in Eq. (\ref{c2tmn}) is isolated by taking
the $x^{n-1}$ moment and integrating along a closed contour in the complex $x$
plane such that the cut lies inside the contour. We then get
\be
{1\over {2\pi i}}\int_C dx x^{n-1} T(x, Q^2)=\sum_i C_i^n(Q^2)A_i^n.
\e
This is called Mellin transform. Shrinking the contour to the physical cut,
and also using optical theorem, we get
\be
4\int_0^1 dx x^{n-1} W(x,Q^2) =\sum_i C_i^n(Q^2)A_i^n.
\e
This equation uniquely predicts the $Q^2$ evolution of the DIS structure
functions $W(x, Q^2)$ provided we know the non-perturbative part, $A_i^n$.
These depend on the target hadron state and are related to the
experimentally measured parton distribution functions. One takes the ratio
of moments at different $Q^2$ to get rid of the $A^n_i$'s and compare
moments at various scales of $Q^2$.
The moments can be inverted by what is called an inverse Mellin
transformation to get the structure functions. 

So far, we considered only lowest twist contributions. 
Higher twist effects are suppressed by powers of
${1\over Q^2}$ compared to the lowest twist effects in DIS. They may become
important at lower value of $Q^2$ where perturbative QCD is no longer
applicable since the coupling constant becomes large. In this kinematical
region, the effect of target mass also becomes important. Correction terms
that appear when the target mass effects are included involve powers of
${1\over Q^2}$ and are often referred to as kinematical higher twist
effects, whereas terms coming from higher twist operators are called
dynamical higher twist effects.

Operators of higher dimensions
contribute at higher twist. 
The detailed analysis of DIS using OPE keeping only the twist two
contributions can be found in textbooks \cite{c2cheng,c2wil}. The analysis
involving the higher twist operators is very complicated because there are a
large number of operators. 

The target mass enters in the structure functions in the OPE method through
the subtraction of the trace terms. The
operators in OPE are made traceless to project out operators of definite
spin. The trace terms are proportional to ${M^2\over
Q^2}$ which are subtracted. These are higher twist effects and in the leading 
twist analysis they are neglected. Target mass effects have been analyzed by
Natchman by introducing a variable $\xi$ defined as,
\be
\xi={2x\over {1+(1+{4M^2x^2\over Q^2})}^{1\over 2}}.
\label{c2xi}
\e     
The basic assumption in this case is that the effective coupling constant is
small such that free field OPE is a good approximation. Instead of scaling
in terms of the Bjorken variable $x$, the structure functions scale in terms
of the variable $\xi$. In other words, if the structure functions $F(x,
Q^2)$ are expressed in terms of the variables $\xi$, such that,
\be
{\tilde F}=({\pp \xi\over \pp x})^{-1} F,
\e  
the ${\tilde F}(\xi, Q^2)$ is independent of the scale $Q^2$ in free theory. $\xi$ scaling
was derived by Natchman \cite{c2natch} by expanding the products of currents in terms of
operators of definite spin. He showed that, the light cone expansion of the
product of currents in Mincowski space corresponds to the expansion 
 in terms of $O_4$ spherical harmonics. These are orthogonal Gegenbauer 
polynomials on the interval $-iQ \le \nu \le iQ$. The orthogonality relations among the
spherical harmonics project out operators of definite spin $n$. 
An alternative way of deriving $\xi$ scaling is given in  \cite{c2gorgi}. 
 Violations of $\xi$ scaling occurs due to interactions.

$\xi$ scaling can be interpreted in the language of the parton model in the
following way. We take the initial proton four-momentum as
$(p^0,0,0,p^3)$ and the virtual photon four-momentum as $(q^0,0,0,q^3)$.
Then taking both the initial and the final state of the struck quark to be
on the mass shell, it
can be shown that the  expression for $\xi$ given by Eq. (\ref{c2xi}) can be written as
\be
\xi={p_I^0+p_I^3\over {p^0+p^3}}
\e
where $p_I$ is the initial momentum of the struck quark and we have assumed
that $p_T^2=0$, where $p_T$ is the part of $p_I$ transverse to $p$ and $q$. 

As we mentioned before, higher twist contributions to DIS also come from the 
higher twist operators. These dynamical higher twist effects involve
nontrivial quark-gluon interaction. There are two approaches to analyze the higher twist
contributions:

(1) OPE is used to relate the $O({1\over Q^2})$ corrections to the target
matrix elements of certain classes of twist four local operators
\cite{c2soldate}.

(2) The other method is a graphical method to express $O({1\over Q^2})$
corrections  in terms of certain quark-gluon-target forward scattering
amplitudes \cite{c2EFP}. 

Both of these methods are very complicated. The starting point
is the assumption that the structure function can be factorized into two
parts: the coefficient functions $C_i$ which describes the short distance
part of the interaction and can be calculated in perturbation theory, and
the parton correlation functions $f_i$ which contain non-perturbative
information about the distribution of partons inside the struck hadron. 

OPE deals with the moments of the structure functions. In the QCD improved
parton model analysis one gets the parton correlation functions as the
Fourier transform of the matrix elements of light-front bilocal currents.
One cannot explore these
expressions further because of the non-trivial hadronic matrix elements.
Here
one has to parametrize the nonperturbative information and it is only the
$Q^2$ evolution which is calculated perturbatively.
We have followed a recently developed alternative approach based on
light-front Hamiltonian technique. 
It addresses the structure functions directly instead of its moments, using
the Bjorken-Johnson-Low (high energy) expansion of the Compton scattering
amplitude. More important is the point that our approach gives a complete
description of the structure functions because it has the potential of 
incorporating the non-perturbative contents of the structure functions.
It is possible to investigate the structure functions for a bound state
target by solving the light-front QCD bound state equation. At present,
active research is going on for solving light-front QCD bound state equation
using a technique called similarity renormalization (see chapter 3 for
references). Another advantage of our method is that the target and quark
mass effects are automatically included and one does not have to introduce
them by hand. The intrinsic picture of parton model is there but the partons
are massive, non-collinear and interacting.

%%%%%%%%%%%%%%%%%%%%%%%%%%%%%%%%%%%%%%%%%%%%%%%%%%%%%%%%%%%%%%%%%%%

\chapter{Basic Features of Light-Front Field Theory}
In this chapter, we introduce the main features of light-front field theory
and light-front QCD.

The light-front formulation of a dynamical system was originally introduced
by Dirac in 1949 \cite{c3dir}. He called it `Front Form'. Simply speaking, light-front
dynamics is the description of the evolution of a relativistic system along
a light-front direction \cite{c3chi}. 

In the usual (equal-time) or instant formulation of field theory, the initial
conditions are fixed on a spacelike hypersurface, $x^0 = 0$, which in the
quantum theory becomes the surface of quantization. The time evolution of
the system is given by the equal time Hamiltonian. In the front form, the
initial conditions are fixed on the light like hypersurface, $x^+ = x^0+x^3
= 0$, which is the surface of quantization. It is a surface tangential to
the light cone. The light-front time is $x^+$ and the $x^+$ evolution of the
system is governed by the light-front Hamiltonian, $P^-$. The light-front
coordinates are,
\be
x^{\pm} = x^0 \pm x^3,~~~~~~~~~~~~~~x^\perp = ( x^1, x^2 ).
\e
Light-front time derivative is, $\pp^-=2{\pp \over {\pp x^+}}$, the space
derivatives are, $\pp^+=2{\pp \over {\pp x^-}}$ and $\pp^\perp=-
{\pp \over {\pp x^\perp}}$. The metric and other useful relations are given
in Appendix A.
For a system of particles moving with relativistic speed, the particle world
lines cluster near the light cone, in contrast to the non-relativistic
situation where the particles move with a velocity much less than the
velocity of light $c$ and the world lines cluster near the time axis. If we
choose $x^+$ as the new time direction, the structure of the time evolution
of a relativistic system may look similar to a non-relativistic system in
the ordinary space time coordinate system. As a result, relativistic
dynamics is expected to become simple  and more transparent \cite{c3ped}. 

It may be mentioned here that, there exists another formulation of field
theory, known as the infinite momentum frame formulation \cite{c3suss}, where the
equal-time results are boosted to an infinite momentum frame.
Although the Lorentz transformation is highly singular, the singularity
cancels in the physical objects like Poincare generators. It can be shown
that such a transformation basically produces the coordinate change,
\be
x^{\pm}=x^0 \pm x^3, ~~~~~~x^\perp=(x^1,x^2) 
\e
which are the same as in light-front field theory. However, in light-front
formulation, we start with these coordinates and no infinite boost is involved.
One works in an arbitrary reference frame in contrast to the infinite
momentum frame formalism where one is restricted to one particular reference
frame.

Deep inelastic lepton-hadron scattering is a highly relativistic process.
As described earlier, Feynman introduced the parton model in the infinite
momentum frame which successfully described scaling of the structure
functions. Light-front formulation of QCD gives an intuitive picture of
DIS by keeping close contact with the parton ideas, however, the partons
here are field theoretic partons and one obtains the expected scaling
violations and many other interesting results.   
%%%%%%%%%%%%%%%%%%%%%%%%%%%%%%%%%%%%%%%%%%%%%%%%%%%%%%%%%%%%%%%%%%%%
\section{Some Special Features of Light-Front Field Theory}\label{c3sf}
%%%%%%%%%%%%%%%%%%%%%%%%%%%%%%%%%%%%%%%%%%%%%%%%%%%%%%%%%%%%%%
\subsection{Dispersion Relation}
 Using the metric tensor given in Appendix
A, we can
see that the dot product of two 4-vectors is given in terms of light-front
components as,
\be
a \cdot b = a^\mu b_\mu = {1\over 2}a^+b^-+{1\over 2}a^-b^+- a^\perp
\cdot b^\perp.
\e
So the dispersion relation for an on-mass-shell particle is given by,
\be
k^-= { (k^\perp)^2+m^2\over k^+}
\e
where $k^-$ is the light-front energy conjugate to light-front time $x^+$ 
and $k^+$ is the longitudinal
momentum. The above dispersion relation is remarkable in the following
aspects:

(i) Even though it is a relativistic dispersion relation, there is no square
root factor involved. This is helpful if one tries to solve eigenvalue
equations like, $H\mid \psi \rangle = E \mid \psi \rangle$. 

(ii) The dependence of the energy $k^-$ on the transverse momentum
$k^\perp$ is just like in the non-relativistic dispersion relation.

(iii) From the dispersion relation, we see that for an on-mass-shell
particle, for $k^+$ positive (negative), $k^-$ is also positive (negative). It so
happens that the particles with negative energy which must have negative
longitudinal momentum can be mapped into antiparticles with positive energy
and positive longitudinal momentum. As a result, we always have $k^+ \ge 0$.
This fact has several interesting consequences, particularly in the context
of the vacuum structure in light-front theory, which we shall discuss
shortly.

(iv) The dependence of energy on $k^\perp$ and $k^+$ is multiplicative and
large energy can result from large $k^\perp$ and/or small $k^+$. This has
drastic consequence for the renormalization aspects of light-front field
theories \cite{c3wil}.
%%%%%%%%%%%%%%%%%%%%%%%%%%%%%%%%%%%%%%%%%%%%%%%%%%%%%%%%%%%%%%%%% 
\subsection{Poincare Generators in Light-front Formulation}
A relativistic dynamical system must be Lorentz invariant. The inhomogeneous
Lorentz transformation is generated by ten operators, Hamiltonian, three
momenta, three boosts and three angular momenta. These generators satisfy
Poincare algebra, which is given in terms of 4-momenta $P^\mu$ and generalized
angular momentum $M^{\mu \nu}$ as,
\be
[P^\mu, P^\nu]=0,~~~[M^{\mu \nu}, P^\rho]= i(-g^{\mu \rho} P^\nu +g^{\nu \rho} 
P^\mu),
\e
\be
[M^{\mu \nu}, M^{\rho \sigma}]=i(-g^{\mu \rho} M^{\nu \sigma}+g^{\nu \rho} 
M^{\mu \sigma}-g^{\mu \sigma} M^{\rho \nu}+g^{\nu \sigma} M^{\rho \mu}).
\e
In the usual equal time or instant form of dynamics, 
the rotation and boost operators are given by, $M_{ij}=\epsilon_{ijk}J^k$
and $M^{0i}=K^i$ respectively. Four out of the ten generators, namely,
the Hamiltonian and the three boost generators, are
dynamical (depend on the dynamics), the remaining six are kinematical (do not
depend on the dynamics). The action of the dynamical generators change the
surface of quantization whereas the action of the kinematical generators
do not. In field theory, the ten Poincare generators
are constructed from the symmetric energy momentum density $\theta^{\mu \nu}$:
\be
P^\mu=\int d^3x  \theta^{0\mu},
\e
\be 
M^{\mu \nu}=\int d^3x [x^\mu \theta^{0\nu}-x^\nu \theta^{0\mu}].
\e
In light-front field theory, the ten Poincare generators are obtained in the
same way, from the energy momentum density;
\be
P^\mu={1\over 2}\int dx^- d^2x^\perp \theta^{+\mu},
\e
\be 
M^{\mu \nu}={1\over 2}\int dx^- d^2x^\perp [x^\mu \theta^{+\nu}-x^\nu 
\theta^{+\mu}].
\e
Here, one has to remember that $x^+$ is the light-front time.   
$P^-$ is the light-front Hamiltonian, this produces the $x^+$ evolution of a
system. $P^+$ is the longitudinal momentum and $P^\perp$ are the transverse momenta.
$M^{+-}=2K^3$ and $M^{+i}=E^i$ are light-front boosts. $M^{12}=J^3$ and 
$M^{-i}=F^i$ are the three rotation operators.  

A very important aspect of the light-front
theory is that, here, the boost operators are
kinematical. Moreover, the longitudinal boost generator $K^3$ produces just a scale
transformation and the transverse boosts behave like the non-relativistic Galilean
boosts \cite{c3ped}. On the other hand, the transverse rotation operators, $F^i$ are
dynamical which means that they change the surface of quantization, $x^+=0$. 
So the kinematical subgroup of the Poincare group is enlarged in
the light-front theory and instead of four dynamical generators we have
only three, which are the light-front Hamiltonian and the two transverse
rotation generators. Because of this different nature of the Poincare
generators, certain problems are easier to tackle
in the light-front formulation of field theory. For example, in any
practical calculation using the Fock space expansion of the state, one has
to do particle number truncation (called Tamm-Dancoff truncation). In the
equal time theory, such a truncation has a major difficulty since this
violates boost invariance. In the light-front theory, on the other hand,
such a truncation violates rotational invariance. It is argued that the
restoration of rotational invariance is easier than the restoration of boost
invariance because the rotational group is compact in contrast to the boosts
which are non-compact \cite{c3ji}. This makes the light-front Hamiltonian technique more
advantageous than the equal time Hamiltonian techniques. Because of this, 
the highly complex QCD bound states may be represented by a few boost
invariant multi-particle wave functions in the Fock space expansion, as we
shall see later in this section.
%%%%%%%%%%%%%%%%%%%%%%%%%%%%%%%%%%%%%%%%%%%%%%%%%%%%%%%%%%%%%%%%%%%%%               
\subsection{Simplicity of Light-front Vacuum}
 One of the most remarkable
features of the light-front field theory is the so called triviality of the
vacuum. In the equal time formulation of QCD, vacuum is highly non-trivial
and is the source of various non-perturbative effects like chiral symmetry
breaking, confinement etc. In light-front theory, a particle's momentum is
separated into two components, longitudinal component which is along the
light-front time direction $x^+$ and transverse components which are
perpendicular to $x^+$ direction.   
 As we said before, in the light-front theory,
longitudinal momenta $k^+ \ge 0$. Now, vacuum is an eigenstate of the longitudinal
momentum, $P^+ \mid 0\rangle = 0$. So the vacuum of a light-front theory 
can only have particles with zero longitudinal momenta, since the total $P^+$ is given by,
 $P^+ = \sum_i k_i^+$, where $k_i^+$ are the longitudinal momenta of
individual particles in the state. At this point, the vacuum is still nontrivial.
To construct hadronic bound states in the Fock space of quarks and gluons,
one needs to simplify the vacuum. 
If we impose a small longitudinal momentum cutoff $\epsilon$ in the
theory, light-front vacuum becomes trivial, i.e. it does not contain any
particle. It is important to understand that, this trivial vacuum is  the eigenstate of the full interacting
Hamiltonian. In the equal time case, there is no such restriction on the
longitudinal momenta $k^3$ which can be positive, zero or negative. So even
if we impose a cutoff $k^3 \ge \epsilon$, the total longitudinal momentum
$P^3 = \sum_i k_i^3$, can have zero eigenvalue (vacuum) for non-zero
$k^3_i$ and we do not get a trivial vacuum devoid of any particle. Because of
this triviality of the vacuum in light-front theory, Hamiltonian perturbation theory
becomes simpler since we no longer have the graphs that generate particles
from the vacuum (Z graphs and vacuum diagrams). All such diagrams require
negative $k_i^+$ which is not possible. In the case of 
QCD, the effects associated with the non-trivial vacuum in the equal time
formulation are expected to be recovered while removing the cutoff dependence
from the theory, i.e. by the process of renormalization which produces new
interactions in the theory. The wave function of a relativistic bound state
can be expressed as an ordinary Fock space expansion using the trivial
vacuum,
\be
\mid \psi \rangle=f(a^\dagger)\mid 0 \rangle
\e
For QCD, $f(a^\dagger)$ should consist of quark, antiquark and gluon
creation operators with non-zero longitudinal momenta and it must be a color singlet
operator.    
%%%%%%%%%%%%%%%%%%%%%%%%%%%%%%%%%%%%%%%%%%%%%%%%%%%%%%%%%
\section{Light Front Bound State Equation}\label{c3bseq}
%%%%%%%%%%%%%%%%%%%%%%%%%%%%%%%%%%%%%%%%%%%%%%%%%%%%%%%%
Due to the simplicity of the light-front vacuum,
 the hadronic bound states can be expanded in terms of the Fock space 
components. Explicitly, a hadronic bound state of momentum $P$ and helicity
$\lambda$ can be written as \cite{c3brod},
\be
\mid P^+, P^\perp, \lambda \rangle = \sum_{n, \lambda_i} \int' dx^i
d^2\kappa^\perp_i \mid n, x_iP^+, x_i P^\perp + \kappa^\perp_i, \lambda_i
\rangle \Phi^\lambda_n ( x_i, \kappa^\perp_i, \lambda_i ).
\e
Here $n$ represents $n$ constituents in the Fock state $\mid n, x_iP^+,
x_iP^\perp +\kappa^\perp_i, \lambda_i \rangle$, $\lambda_i$ is the helicity
of the $i$th constituent, $\int'$ denotes integration over the space,
\be
\sum_i x_i = 1, ~~~~~~~~~~~~~~~~\sum_i \kappa^\perp_i = 0,
\e
where $x_i$ is the fraction of the longitudinal momentum carried by the
$i$th constituent, $\kappa^\perp_i$ is its transverse momentum relative to
the center of mass frame,
\be
x_i = {p^+_i\over P^+}, ~~~~~~~~~\kappa^\perp_i = p_i^\perp -x_i P^\perp.
\e
Here $p_i^+$ and $p_i^\perp$ being the longitudinal and transverse momenta
of the $i$th constituent, $\Phi^\lambda_n ( x_i, \kappa^\perp_i, \lambda_i
)$ is the amplitude of the Fock state $\mid n, x_iP^+, x_i P^\perp + \kappa^\perp_i, \lambda_i
\rangle $ and satisfies the normalization condition,
\be
\sum_{n, \lambda_i} \int' dx_i d^2\kappa^\perp_i {\mid \Phi_n^\lambda (
x_i, \kappa^\perp_i, \lambda_i) \mid}^2 = 1.
\e
The states obey the relativistic Schroedinger-like equation,
\be
P^- \mid P, \lambda \rangle = { (P^\perp)^2 + M^2 \over P^+} \mid P, \lambda
\rangle.
\e
$P^-$ is the light-front Hamiltonian. One remarkable property of the light
front is that boost does not involve dynamics, so if one can solve the bound
state equation in the rest frame, one can understand the particle structure in
any frame. Solving the bound state equation for the full Fock space is, of
course, impossible, but one can restrict oneself to a few particle sector 
(Tamm-Dancoff truncation) \cite{c3per} and get useful information of the theory.   
It is important to mention here that Tamm-Dancoff truncation in the usual
equal time theory encounters a major problem since it violates boost
invariance. But in the light-front theory, since boost is kinematical, it
breaks only rotational invariance. It is argued that the restoration of
rotational invariance is easier than the restoration of boost invariance
since rotational group is compact whereas the boosts are
non-compact \cite{c3ji}.

The feasibility of describing hadronic bound states in terms of Fock states
gives a hope of reconciling constituent quark model, which has been so
successful in describing hadron spectra, with QCD. We shall see in the later
chapters, how this particularly helps us to investigate DIS
structure functions. 

%%%%%%%%%%%%%%%%%%%%%%%%%%%%%%%%%%%%%%%%%%%%%%%%%%%%%%
\section{Light-Front QCD}\label{c3lfqcd}
%%%%%%%%%%%%%%%%%%%%%%%%%%%%%%%%%%%%%%%%%%%%%%%%%%%%%%%

QCD (Quantum Chromodynamics) has now been taken as the fundamental theory of
strong interactions. 
Light-front QCD, which means the theory quantized on the surface $x^+=0$, is
basically used to explore non-perturbative effects of QCD and is based on
Hamiltonian theory. The technique was first developed by Kogut and Soper
\cite{c3sop} for QED. Because of the triviality of the vacuum and kinematical
boost, hadronic bound states can be studied in a truncated Fock space of
quarks and gluons. The low energy hadron structures like the parton
distribution functions, fragmentation functions and various hadronic form
factors can be addressed in this framework. Perturbative calculations are also
possible using the old fashioned light-front Hamiltonian perturbation
theory, which, as described before, is simpler than the equal-time
Hamiltonian perturbation theory because of less number of graphs at each
order.

Before the advent of QCD, Feynman proposed the parton
model in the infinite momentum frame. Since the infinite momentum frame
formulation is equivalent to the light-front formulation, the natural
framework to investigate parton phenomena in high energy scattering process
is the formulation of QCD on the light-front in light-front gauge. Tomboulis
\cite{c3tom}
first proposed Yang-Mills theory on the light-front in 1973. A systematic
formulation of light-front QCD was given by Casher \cite{c3cas} and Bardeen et al
\cite{c3bar}. 
Thorn \cite{c3thorn} studied various aspects of light-front QCD including
asymptotic freedom for pure Yang Mills theory.
Leapage and Brodsky \cite{c3leap} applied it to the study of
exclusive processes in the early eighties. However, only very recently 
there has been a lot of interesting investigation and development in light-
front QCD \cite{c3light}. In particular, active research is 
going on in the area of recently developed similarity renormalization 
\cite{c3similarity}. 

The QCD Lagrangian is given by
\be
{\cal L} = -{1\over 2} Tr [F^{\mu \nu}F_{\mu \nu}] + {\overline {\psi}} (
i\gamma_\mu D^\mu - m) \psi
\e
where $ F^{\mu \nu}= \partial^\mu A^\nu - \partial^\nu A^\mu -ig[ A^\mu,
A^\nu]$, $ A^\mu = \sum_a A^\mu_a T^a$ is the gluon field color matrix and
$T^a$ are the generators of the $SU(3)$  color group. The field variable
$\psi$ describes quarks with three colors and $N_f$ flavors, $D^\mu =
\partial^\mu-igA^\mu$ is the covariant derivative, $m$ is the quark mass
(matrix). The equations of motion are :
\be
\pp_\mu F^{\mu \nu a}+gF^{abc} A^b_\mu F^{\mu \nu c}+g {\bar \psi} \gamma^\nu
T^a \psi = 0,
\e
\be
(i\gamma_\mu \pp^\mu -m+g \gamma_\mu A^\mu)\psi=0.
\e 

{\bf Dynamical and Constrained Fields}:
The fermion spinor in light-front coordinates can be split into $ \psi =
\psi^+ + \psi^-$, $ \psi^\pm = \Lambda^\pm \psi$ where $\Lambda^\pm =
{1\over 2} \gamma^0 \gamma^\pm$ (see Appendix A). It turns out that, the fields $\psi^-$ and
$A^-$ ( minus component of $A^\mu$) are both constrained, the equations of
constraints are given by, in the light-front gauge $A^+ = 0$,
\be
i \partial^+ \psi^- &=& \big [ \alpha^\perp \cdot ( i \partial^\perp + g
A^\perp) + \gamma^0 m \big ] \psi^+,   \\
{ 1 \over 2} \partial^+ A^{-a} &=& \partial^i A^{ia} + g f^{abc} { 1 \over
\partial^+}(A^{ib} \partial^+A^{ic})
 + 2 g { 1 \over \partial^+} \Big (
\xi^\dagger T^a \xi \Big ), 
\e
where $\pp^+= {\pp\over {\pp x^-}}$.
It is a unique feature of light-front formulation of any field theory that
some of the fields which are dynamical in the ET (equal time) formulation become
constrained here, as a result the equations of motion of the dynamical fields
$\psi^+$ and $A^\perp$ are first order in light-front time. In order to
solve the constraint equations, it is needed to define the operator ${1\over
\pp^+}$. Assuming antisymmetric boundary condition, it is defined as
\cite{c3two},
\be
{1\over \pp^+} f(x^-) = {1\over 4} \int_{-\infty}^{+\infty} dy^- \epsilon
(x^- - y^-) f(y^-).
\e     
Choosing the antisymmetric boundary condition completely fixes the residual
gauge freedom.

The Hamiltonian can be obtained from the Lagrangian. Using the constraint
equations, the  constrained field can be removed from the Hamiltonian and it
can be written entirely in terms of dynamical variables in light-front
gauge. The light-front QCD Hamiltonian in terms of the dynamical variables
is given explicitly in Appendix C. There is another method of constructing
light-front QCD Hamiltonian which is based on light-front power counting
\cite{c3wil}, we shall not discuss it here. 

{\bf Quantization}:
In order to quantize such a constrained system, one has to follow Dirac
procedure \cite{c3sti} or the phase space quantization method
\cite{c3light,c3two}. 
The non-vanishing commutators
between the dynamical fields are given by,
\be
{\{ \psi^+(x), \psi^{\dagger +}(y)\}}_{x^+ = y^+} = \Lambda^+ \delta^3(x-y),\\
{[ A^i_a (x), A^j_b (y)]}_{x^+=y^+} = -i \delta_{ab} \delta^{ij} {1\over 4}
\epsilon(x^- - y^-) \delta^2 (x^\perp - y^\perp),
\e
where $\delta^3(x-y) = \delta(x^- -y^-) \delta^2 ( x^\perp - y^\perp)$. 
 It is to be noted that these commutators are evaluated in equal light-front
time, $x^+ = y^+$. The ET (equal-time) commutators between two field variables must
vanish because of causality since the separation is spacelike, however, the light-front commutators are
non-vanishing because here the two fields are separated by a light like
distance, and a signal traveling with the velocity of light can still 
communicate between them.

{\bf Non-locality}: Another interesting point to note is that, the above
commutators and anticommutators are non-local in the longitudinal $(x^-)$
direction. This non-locality  creeps in since in light
front coordinates, $x^2=x^+x^- - (x^\perp)^2$, and $x^+ = 0$ still allows
non-local $x^-$. This non-locality in the longitudinal direction is also
seen in the definition of ${1\over \pp^+}$ and has far reaching
consequences \cite{c3jac}.

{\bf Two Component Formulation}: By eliminating the constrained fields completely in light-front gauge, tht
light-front QCD Hamiltonian can be written completely in terms of the
dynamical field components, $\psi^+$ and $A^\perp$. A very interesting
point is that, in a particular representation of the gamma matrices, (see
Appendix A) the
four component fermion field can be reduced to a two component field, i. e.
the dynamical field $\psi^+$ can be written as \cite{c3two},
\be
 \psi^+ = 
\left [ \begin{array}{c} \xi \\
                       0 \end{array} \right ].
\e
where $\xi$ is the two component fermion field.
This is one of the major advantages of the light-front theory that simplify
the relativistic fermionic structure.
The two component fermionic field satisfy the anticommutaion relation,
\be
{\{ \xi, \xi^\dagger\}}_{x^+=y^+} = \delta^3(x-y).
\e 
In the two component formalism, the Fock space expansion of the dynamical
fields become,
\be
A^i(x)=\sum_{\lambda} \int {dq^+ d^2q^\perp\over {2(2\pi)^3 q^+}}[
\epsilon^i_\lambda a(q, \lambda)e^{iqx}+h.c.],
\e
\be
\xi(x)=\sum_{\lambda} \chi_\lambda  \int {dp^+ d^2p^\perp\over {2(2\pi)^3
{\sqrt p^+}}} [b(p,\lambda)e^{iqx}+d^\dagger(p,\lambda)e^{-ipx}],
\e
where $q^-={(q^\perp)^2\over q^+}$ and $p^-={{(p^\perp)^2+m^2}\over p^+}$.
$\lambda$ can be $1$ or ${-1}$ for the gluon and ${1\over 2}$ or $-{1\over
2}$ for the quark. The gluon polarization vectors are,
\be
\epsilon^i_1={1\over {\sqrt 2}} (1, i),~~~~
\epsilon^i_{-1}={1\over {\sqrt 2}} (1, -i).
\e
The quark two-component spinors are
\be
\chi_{1\over 2}=(1,0),~~~~~~~~\chi_{-{1\over 2}}=(0,1).
\e
\be
[a(q,\lambda),a^\dagger(q',\lambda')]=2(2 \pi)^3q^+\delta^3(q-q')
\delta_{\lambda \lambda'},\nonumber\\
\{b(p,\lambda),b^\dagger(p',\lambda')\}=
\{d(p,\lambda),d^\dagger(p',\lambda')\}= 2(2 \pi)^3p^+\delta^3(q-q').
\e
As described before, the constrained fields are eliminated from the theory
using the constraint equations and everything can now be written in terms of
the two-component dynamical fields, $\xi$ and $A^\perp$. This makes
computations easier.

{\bf Power Counting and Renormalization}: 
Power counting in light-front theory is different from the usual equal time
case. In order to understand this, let us look into the dispersion relation,
\be
k^-={{(k^\perp)^2+m^2}\over k^+}
\e
which means that $k^\perp$ scales as $m$ and the product of $k^+k^-$ scales
as $m^2$, but $k^+$ and $k^-$ individually can have different scaling
behavior. Among the light-front coordinates, only the
transverse directions $x^\perp$ carry inverse mass dimensions, while the
longitudinal coordinate $x^-$ does not carry any mass dimension \cite{c3wil}.   
 This makes the
crucial difference from the instant form where all directions carry the
same mass dimension.  

In perturbative calculations using field theory, one
often encounters divergences.  In light-front field theory, one has to treat
transverse and longitudinal directions separately in determining the
superficial degree of divergence of a divergent integral by power counting.
Also since light-front theory is manifestly non-covariant and gauge fixed, one   
gets new types of divergences. 
However, the physical quantities themselves
are finite. One has to regulate these divergences and remove them, this
process is called renormalization.  
Renormalization in light-front field
theory is completely different from the usual instant form of field theory.
 One has to add non-local non-covariant
counterterms to restore all the invariances in the renormalized theory.
Various investigations regarding renormalization of light-front theories are
in progress. Recently a special renormalization scheme, called similarity
renormalization has been proposed \cite{c3gla} which is suitable for the bound state
studies on the light-front. In this work, we have used ultraviolet 
transverse momentum
cutoff and a small infrared longitudinal momentum cutoff.      
%%%%%%%%%%%%%%%%%%%%%%%%%%%%%%%%%%%%%%%%%%%%%%%%%%%%%%%%

\chapter{Light-Front Hamiltonian QCD Approach to DIS}
A large number of recent experiments on polarized and unpolarized deep
inelastic scattering are providing valuable information on the so called
`higher twist' or power suppressed contributions to DIS. A theoretical
understanding of these effects require non-perturbative information on the
structure of the hadron. There is an
urgent need to develop a non-perturbative approach to DIS which is
preferably  based on physical intuitions but at the same time employs well
defined field theoretical calculational procedure. This goal is achieved in
light-front Hamiltonian QCD framework. We follow a recently
developed method \cite{c4nonpert} of calculating the structure functions which is a
combination of coordinate space approach based on light-front current
algebraic techniques and momentum space approach based on Fock space
expansion method in light-front theory in Hamiltonian QCD.

The most intuitive approach to DIS is the parton model \cite{c4par} proposed by Feynman. 
Partons were originally introduced as collinear, massless,
non-interacting pointlike constituents of the proton. However, the QCD governed
 interacting partons need not be collinear and massless. Can one
generalize this concept to introduce field theoretic partons, which are
non-collinear and massive, but still on-mass-shell objects in an interacting
field theory? The answer is yes, and this is achieved in light-front
Hamiltonian QCD by introducing many body (or multi-parton) wave functions. 

In order to introduce these many body wave functions in the description of DIS
structure functions, we use Bjorken-Johnson-Low (BJL) expansion of the
scattering amplitude \cite{c4jack}, 
which is essentially a non-perturbative approach where the expansion
parameter is the inverse of the light-front energy of the probe (virtual
photon). The leading term in this expansion in the high energy limit
contains the matrix element of the equal light-front time current
commutator which can be evaluated, knowing the light-front
commutators/anticommutators between the field variables. Using optical
theorem (see Chapter 2)    
one  obtains the expressions of the structure functions as the
Fourier transform of the matrix elements of light-front bilocal vector and
axial vector currents. The bilocality is only in the longitudinal
(minus) direction and the matrix element is between the target hadron state.
In the next step, one uses the Fock space expansion of the target state
using Tamm-Dancoff truncation. This introduces multi-parton wave functions in
the expression of the structure functions. The structure functions, then,
can be evaluated once these wave functions renormalized at scale $Q$ are
known. At present major efforts are under way to calculate these wave
functions. It is important to emphasize here that, such an approach is
possible because of certain special properties of light-front formulation,
in particular, the fact that light-front boost is kinematical (see chapter
3). The advantage
of this technique is that both perturbative and non-perturbative issues can
be addressed within the same framework. So, a unified description of 
perturbative and non-perturbative QCD underlying the deep inelastic
structure functions can be realized. 

In this section, we briefly outline our approach and summarize the previous
works done using this formalism.

%%%%%%%%%%%%%%%%%%%%%%%%%%%%%%%%%%%%%%%%%%%%%%%%%%
\section{Structure Functions in the BJL Limit}\label{c4bjl}
%%%%%%%%%%%%%%%%%%%%%%%%%%%%%%%%%%%%%%%%%%%%%%%%%%
The hadronic tensor $W^{\mu \nu}$ is given in terms of 
hadronic matrix elements
of the current commutator as in chapter 2. However, it is not
equal-time or equal-$x^+$ commutator. 
The DIS structure functions are related to the equal $x^+$ current
commutators through the BJL (high energy) limit of the forward virtual
photon-hadron Compton scattering amplitude $T^{\mu \nu}$ \cite{c4jack}:
\begin{equation} \label{c4lfccl}
	T^{\mu \nu} \stackrel{{\rm large}~q^-}{=} - {1\over q^-}
		\int d\xi^- d^2 \xi_\bot e^{iq\cdot \xi}
	  \langle PS | [J^\mu(\xi), J^\nu(0)]_{\xi^+=0}| PS\rangle  + ... 
\end{equation}

where $q^-$ is the light-front energy of the virtual photon,
and $(\xi^+,\xi^-,\xi^i)$ are the light-front space-time 
coordinates. 
For large $Q^2$ and large $\nu$ limits in DIS, without any  
loss of generality we can always select a Lorentz frame 
such that the light-front energy $q^-$ of the virtual photon 
becomes very large. Explicitly, in terms of light-front variables, we can
choose $q^+$ to be negative and finite for the virtual photon (since the
photon is virtual, longitudinal momentum can be negative). 
Also, keeping $q^i$ to be finite, one can get large space-like $q^2$ 
($Q^2 \rightarrow \infty$) by taking $q^-\rightarrow \infty$. In this case, 
$x=-{q^+\over P^+}$ is positive and finite, since $q^+$ is negative while both 
$q^+$ and $P^+$ are finite. So by taking the limit $q^- \rightarrow \infty$,
we do not go beyond the physical region of DIS. Also, in the large $Q^2$
region, we take only the first term in the above expansion since the higher
and higher terms are suppressed. 

The structure functions are related to the equal light-front time current
commutator by using optical theorem and dispersion relation (see chapter 2).
The light-front current commutator can be calculated exactly from
light-front QCD in the light-front gauge. 

It is worthwhile to point out that there exists an equal-time BJL expansion
of the scattering amplitude where the expansion parameter is ${1\over q^0}$.
But taking $q^0 \rightarrow \infty$ gives timelike $q^2$ which is unphysical
for DIS. In order to overcome this difficulty, one has to go over to the
complex $q^0$ plane and take $iq^0 \rightarrow \infty$. This increases the
complexity. Another important point is, equal-time BJL expansion gives an
infinite set of relations connecting the moments of the structure functions
to the corresponding term in the BJL expansion \cite{c4jack}. So in order to get the
structure functions, one has to invert these equations.  Light-cone BJL
expansion instead gives the structure functions themselves in terms of the
light-cone current commutators.

The various expressions of the DIS structure functions that are obtained in
the BJL limit and inserting the current commutator are given below:
\begin{eqnarray}
	{F_2(x,Q^2)\over x} 
	&=& {1\over 4\pi P^+} \int d\eta e^{-i\eta x} \sum_\alpha 
		e^2_\alpha \langle PS| \overline{\psi}_\alpha (\xi^-) 
		\gamma^+ \psi_\alpha (0) \nonumber\\&&~~~~~~~~~~~~
                          - \overline{\psi}_\alpha (0)
		\gamma^+ \psi_\alpha (\xi^-) |PS \rangle  \label{c4f2+} \\
	&=& {1\over 4\pi P^i_\bot} \int d\eta e^{-i\eta x} \sum_\alpha
                e^2_\alpha \langle PS | \overline{\psi}_\alpha (\xi^-) 
		\gamma_\bot^i \psi_\alpha (0) \nonumber\\&&~~~~~~~~~~~~~~
                      - \overline{\psi}_\alpha (0)
		\gamma_\bot^i \psi_\alpha(\xi^-) |PS \rangle \, . \label{c4f2i}
\end{eqnarray}
\begin{eqnarray}
	F_L(x,Q^2) 
	&=& {P^+\over 4\pi } \Bigg({2x\over Q}\Bigg)^2 \int d\eta 
		e^{-i\eta x} \sum_\alpha e^2_\alpha \langle PS| 
		\overline{\psi}_\alpha (\xi^-) \nonumber \\
	& & ~~~~~~~~~~~~~~~~~~~~~~~~~  \times 
		\Big(\gamma^--{P_\perp^2 \over (P^+)^2}
                       \gamma^+ \Big)\psi_\alpha 
		(0) - h.c. |PS \rangle \, .  \label{c4fl}
%	&=& {1\over 4\pi P^+} \Bigg({2x\over Q}\Bigg)^2 
%		\int d\eta e^{-i\eta x} \sum_\alpha 
%		e^2_\alpha \langle PS| \overline{\psi}_\alpha (\xi^-) 
%		\Big(\gamma^--{P^-\over P_\bot^i}\gamma^i_\bot \Big)
%		\psi_\alpha (0) - h.c. |PS \rangle \, ,
\end{eqnarray}
Eq. (\ref{c4fl}) may be reduced to the same expression obtained 
by the collinear expansion in the Feynman diagrammatic method up to the 
order twist-four \cite{c4EPF}. But it is obtained directly here in the 
leading order in the $1/q^-$ expansion without involving the concept 
of twist expansion.
The polarized structure functions come out to be, 
\begin{eqnarray}
	g_1(x,Q^2) 
	&=& {1\over 8 \pi S^+} \int d\eta e^{-i\eta x} \sum_\alpha
		e^2_\alpha \langle P S| \overline{\psi}_\alpha (\xi^-)
		\gamma^+ \gamma_5 \psi_\alpha(0) + h.c. |PS \rangle , \label{c4g1}\\
	g_T(x,Q^2) 
	&=& {1\over 8\pi S^i_T} \int d\eta 
		e^{-i\eta x} \sum_\alpha e^2_\alpha \langle PS|\overline{
		\psi}_\alpha(\xi^-) \Big(\gamma^i -{P^i\over P^+}
		\gamma^+ \Big)\gamma_5 \psi_\alpha(0)
                  \nonumber\\&&~~~~~~~~~~~~~~+~ h.c. |PS \rangle 
		\, . \label{c4gt}
\end{eqnarray}
We refer to \cite{c4nonpert} for a detailed derivation of the above results. In
our work, these are the starting expressions. The detailed derivation of the
structure function $F_L$ taking into account the flavor structure is given
in the next chapter.  
The above results are derived without recourse to perturbation theory,
and also without the use of the concept of collinear and massless partons.
Also, they are not derived using the twist expansion and these are the
leading terms in the ${1\over q^-}$ expansion and not the leading terms in
terms of twist, the target here is in
arbitrary Lorentz frame. 
In the rest frame, these expressions reduce to those obtained by Jaffe and
Ji in the impulse approximation \cite{c4jaji} and also by Efremov {\it et.
al.} in
QCD field theoretic model \cite{c4efre}.
Also, the above expressions are valid only in the
light-front gauge, $A^+=0$, otherwise the bilocal expressions should involve
a path ordered exponential to ensure gauge invariance. Since the bilocality
is only in the longitudinal direction, the exponential factor is unity.

In the above expressions, the bilocal operators arise from the light-front
current commutator. It can be seen from Eq. (\ref{c4f2+}) and (\ref{c4f2i}) that the plus
(involving $\gamma^+$) and perp (involving $\gamma^\perp$) components of the
bilocal current are related to the same structure function $F_2$, whereas
the minus component (involving $\gamma^-$) is related to the longitudinal
structure function $F_L$. Explicit calculation shows that the minus component
involves the constrained field, $\psi^-$, which, when eliminated using
constraint equation, gives rise to interaction dependent terms in the
operator. This interaction dependence is absent in the plus component.
Similarly in the polarized structure function $g_1$, the operator involved
is the plus component  of the axial vector current
which does not depend upon interactions whereas in transverse polarized
structure function $g_T$, the perp component is involved, which contains
$\psi^-$, elimination of which brings in interaction dependence. 
The evolution of the structure functions $F_2$ and $g_1$ with $Q^2$ are
generated entirely from the target state in the leading log approximation. 
In these cases, all the
complexities are buried in the target hadronic state and the operators
involved have simple structure. The situation is completely different for
the other two structure functions, namely $F_L$ and $g_T$, where complexity
is already there in the operator structure. The interaction dependence of
the operator gives off-diagonal contributions to the hadron matrix elements. 

The evaluation of the hadronic matrix elements is straightforward in this
approach and this clarifies the physical picture of DIS. In this framework,
we do not use dimensional regularization and the presence of $\gamma_5$ does
not produce any problem. Also, in contrast to the OPE formalism, the quark 
mass effects can be included without any difficulty. In this approach, we
deal with probability amplitudes rather than probability densities and
interference effects are easy to handle.    
 
In the next section,
 we briefly review the previous works done in this approach.

%%%%%%%%%%%%%%%%%%%%%%%%%%%%%%%%%%%%%%%%%%%%%%%%%%%%%%%%%%%%%%%%%%%
\section{Review of Previous Works Done in This Approach}\label{c4rev}
    
%%%%%%%%%%%%%%%%%%%%%%%%%%%%%%%%%%%%%%%%%%%%%%%%%%%%%%%%%%%%%%%%%%%%%
\subsection{$F_2$ From Transverse Component of Bilocal Current}
%%%%%%%%%%%%%%%%%%%%%%%%%%%%%%%%%%%%%%%%%%%%%%%%%%%%%%%%%%%%%%%%%%%%%
Earlier in this section, we have written $F_2$ in terms of both plus and
perp components of bilocal current. That the perp component also gives the
same $F_2$ structure function is a new result and comes out directly from
our approach based on BJL limit and light-front Hamiltonian QCD. This is
different from \cite{c4jaffe}, according to which, the plus component is
twist two and perp component is twist three. So the perp component should
not have partonic interpretation. Explicit calculation shows that the
perp component involves the constrained field $\psi^-$. When it is
eliminated using the constraint equation, the expression for $F_2$ becomes,
\be
{F_2(x)\over x}={1\over {8 \pi}} {P^+\over P^1} \int dy^- e^{-{i\over 2}P^+
y^-x}\langle P\mid \xi^\dagger (y) [O_m+O_{k^\perp}+O_g] \xi(0)\mid P \rangle
+h.c.
\e
h.c. is the Hermitian conjugate and,
\be
O_m=im{1\over {i \pp^+}}\sigma^2,\nonumber\\
O_{k^\perp}={1\over {i\pp^+}}[i\pp^1-\sigma^3 \pp^2],\nonumber\\
O_g=g{1\over {i\pp^+}} [A^1+i\sigma^3A^2].
\e
There is an interaction term $O_g$ in the operator. 
However, contribution from this term to the matrix
element vanishes and one gets the same structure function $F_2$ as from the
plus component and it carries the same parton interpretation \cite{c4matrix}. 

%%%%%%%%%%%%%%%%%%%%%%%%%%%%%%%%%%%%%%%%%%%%%%%%%%%%%%%%%%%%%%%%%%%
\subsection{Factorization of Hard and Soft Dynamics}
%%%%%%%%%%%%%%%%%%%%%%%%%%%%%%%%%%%%%%%%%%%%%%%%%%%%%%%%%%%%%%%%%%%

Any general cross section in hadron physics contains both short distance and
long distance behavior and hence is not accessible to perturbative QCD.
Factorization theorem allows one to separate the  two behaviors in a
systematic way. According to factorization theorem, the cross section can be
separated into two parts, the non-perturbative soft dynamics factors out of
the hard part \cite{c4fac}. So the hard (short distance) part of the cross section is calculated using
perturbative QCD and the soft (long distance) part is given in terms of a
set of operator matrix elements which have to be calculated
non-perturbatively using some non-perturbative techniques like lattice gauge
theory. This factorization theorem has not been proved for higher twist
contributions to the cross section. Recently, a factorization scheme is
presented in our approach based on light-front Hamiltonian QCD and Fock
space expansion \cite{c4nonpert}. The light-front Hamiltonian is separated into three parts,
depending on the hard and soft transverse momenta $k^\perp$,
\be
P^- = (P^-)^h + (P^-)^m + (P^-)^l
\e
where the soft part $(P^-)^l$ contains only those momenta for which
$(k^\perp)^2<\mu^2$ and hard part $(P^-)^h$ contains $Q^2>(k^\perp)^2>\mu^2$
where $\mu^2$ is some intermediate scale separating the hard and soft
dynamics. This is called factorization scale. $Q^2$ is the scale of the
virtual photon. The low energy effective Hamiltonian  can, in principle, be
obtained by integrating out all modes with $(k^\perp)^2>\mu^2$ from the
canonical QCD Hamiltonian. The mixed part contains interaction and mixes
the hard and soft partons. The structure function can be written as,
\be
F_i(x,Q^2) \approx \int d \eta exp^{-i \eta x} \sum_\alpha e_\alpha^2 \langle
P S \mid {\bar \psi_\alpha}(\xi^-) \Gamma_i \psi_\alpha(0) \pm h.c.\mid P S
\rangle
\e
where $\Gamma_i$ denote the gamma matrices. The target bound state can be
expressed as, 
\be
\mid P S \rangle =U_h \mid P S, \mu^2\rangle
\e
where $U_h=T^+ exp({-{i\over 2} \int^{-\infty}_0 dx^+ [(P^-)^h+ (P^-)^m]})$
and,
\be
(P^-)^l\mid P S, \mu^2 \rangle = {{(P^\perp)^2 +M^2}\over P^+} \mid P S,
\mu^2 \rangle.
\e
Then one can write, 
\be
F_i(x,Q^2) \approx &&\int d\eta exp^{-i \eta x} \sum_\alpha e_\alpha^2
\sum_{n_1, n_2}\langle
P S, \mu^2 \mid n_1 \rangle \nonumber\\&&~
\langle n_2 \mid P S, \mu^2 \rangle \langle n_1
\mid U_h^{-1}[{\bar \psi_\alpha}(\xi^-) \Gamma_i \psi_\alpha(0) \pm h.c.]U_h 
\mid n_2 \rangle
\e
where $\mid n_1 \rangle, \mid n_2 \rangle$ are complete sets of quark and
gluon Fock states with momenta $k_i^2 \le \mu^2$. The hard contribution is
described by the matrix element,
$\langle n_1
\mid U_h^{-1}[{\bar \psi_\alpha}(\xi^-) \Gamma_i \psi_\alpha(0) \pm h.c.]U_h 
\mid n_2 \rangle  $ which is evaluated in light-front Hamiltonian
perturbation theory. The soft contribution is characterized by an overlap of
the multiparton wave functions, $\langle
P S, \mu^2 \mid n_1 \rangle \langle n_2 \mid P S, \mu^2 \rangle \langle n_1
\mid $. For $F_2$ and $g_1$,    $\mid n_1 \rangle= \mid n_2 \rangle= \mid n \rangle $,
only one parton actively takes part in the scattering process, the remaining
ones are spectators. Such a factorization scheme is valid for the $F_2$ and
$g_1$ structure functions. The structure function can be written as,
\be
F_i(x, Q^2) \approx \sum_\alpha e_\alpha^2 \int_x^1 dy P_{p p'} (y, x, {Q^2
\over \mu^2}) q_{\alpha i}(y, \mu^2)
\e
where $P_{p p'} (y, x, {Q^2\over \mu^2})$ are the leading hard contributions
to the structure functions, called splitting functions, and $q_{\alpha i}
(y, \mu^2) $ is called the distribution function which is to be
calculated by non-perturbative light-front QCD approaches to hadronic bound
states. Thus, a unified treatment of both perturbative and non-perturbative
aspects of DIS is possible in this framework.

%%%%%%%%%%%%%%%%%%%%%%%%%%%%%%%%%%%%%%%%%%%%%%%%%%%%%%%%%%%%%%%%%%%%
\subsection{$F_2$ Structure Function, $Q^2$ Evolution and Sum Rule}
%%%%%%%%%%%%%%%%%%%%%%%%%%%%%%%%%%%%%%%%%%%%%%%%%%%%%%%%%%%%%%%%%%%%
The factorization scheme shows that the dressed parton structure functions
are important, since they give informations on the splitting functions and
scaling violations. The $F_2$ structure function has been calculated for
dressed quark and gluon target states \cite{c4rad}. The state is expanded in Fock space
for a dressed quark in terms of bare states of quark and  quark plus gluon
(upto $O(g^2)$) as,
\be
\mid P, \sigma \rangle &=& \phi_1 b^\dagger(P,\sigma) \mid 0 \rangle
\nonumber \\  
&& + \sum_{\sigma_1,\lambda_2} \int 
{dk_1^+ d^2k_1^\perp \over \sqrt{2 (2 \pi)^3 k_1^+}}  
\int 
{dk_2^+ d^2k_2^\perp \over \sqrt{2 (2 \pi)^3 k_2^+}}  
\sqrt{2 (2 \pi)^3 P^+} \delta^3(P-k_1-k_2) \nonumber \\
&& ~~~~~\phi_2(P,\sigma \mid k_1, \sigma_1; k_2 , \lambda_2) b^\dagger(k_1,
\sigma_1) a^\dagger(k_2, \lambda_2) \mid 0 \rangle 
\e
where $\sigma$ is the helicity, $\phi_2$ is the probability amplitude to
find a bare quark of momentum $k_1$, helicity $\sigma_1$ and a bare gluon of
momentum $k_2$ and helicity $\lambda_2$ in the dressed quark. 

Explicit evaluation leads to, upto $O(\alpha_s)$,
\be
{F_2 (x, Q^2)\over x} = \delta(1-x)+ {\alpha_s \over {2 \pi}} C_f ln
{Q^2\over \mu^2} [{{1+x^2}\over {(1-x)_+}}+{3\over 2} \delta (1-x)].
\e
Here we have taken the state to be normalized to unity.
The first term in RHS is the parton model result and the next term is the
QCD correction. $ln{Q^2\over \mu^2}$ in the second term arises from the
transverse momentum integration of the field theoretic partons and it gives
rise to scaling violations. The collinear $(x=1)$ singularity is canceled
between the real and virtual gluon emission, which is indicated by the plus
prescription in the above expression. The Altarelli-Parisi splitting
function can be obtained from this expression:
\be
P_{qq}=  C_f
{{1+x^2}\over {(1-x)_+}}+{3\over 2} \delta (1-x).
\e
It is known that the $F_2$ structure function gives the longitudinal
momentum sum rule. It has been shown that the integral of $F_2$ over $x$ is
related to the hadronic matrix element of the fermionic part of the
light-front QCD longitudinal momentum density $\theta^{++}$. Thus, it gives
the total fraction of the hadronic longitudinal momenta carried by the quarks
and antiquarks. The structure function for the gluons $F_2^g(x, Q^2)$ has
been calculated and it has been shown explicitly that the longitudinal
momentum sum rule is obeyed,
\be
\int_0^1 dx (F_2^q(x) + F_2^g(x))=1.
\e
This sum rule is verified upto $O(\alpha_s)$ in light-front Hamiltonian
perturbation theory. 

In order to understand the factorization of hard and soft dynamics in this
picture, $F_2$ has been calculated for a meson-like bound state by expanding
the target state in Fock space \cite{c4rad}. Upto $O(\alpha_s)$, one considers 
the bare
states  consisting of a quark-antiquark pair and quark-antiquark-gluon. 
$Q^2$ evolution
of $F_2$ has been calculated and it has been shown that it factorizes into
soft ($0 \le k^\perp \le \mu)$ and hard $(\mu \le k^\perp \le \Lambda)$,
where $\Lambda$ is the cutoff on transverse momenta.

%%%%%%%%%%%%%%%%%%%%%%%%%%%%%%%%%%%%%%%%%%%%%%%%%%%%%%%%%%%%%%%%%%%
\subsection{Polarized Structure Function: $g_1(x,Q^2)$ and Helicity Sum Rule}
%%%%%%%%%%%%%%%%%%%%%%%%%%%%%%%%%%%%%%%%%%%%%%%%%%%%%%%%%%%%%%%%%%%%%

The first moment of the polarized structure function $g_1(x,Q^2)$ gives a
measure of the total intrinsic helicity contributions of quark and antiquark
in the nucleon.  The EMC data in the late 80's produced a lot of
interest in
polarized structure function by confirming that Ellis-Jaffe sum rule,
according to which about 50-60 percent of the nucleon helicity should be 
carried by
quarks and antiquarks, is violated. The EMC data showed that only a very 
small fraction
of the nucleon helicity is carried by the charged constituents. This gave
rise to a lot of confusion in understanding the spin contribution of the
nucleon and this is what is known as the spin crisis. Later, it has been
shown by calculating the triangle graph anomaly that the gluon coming 
from the anomaly graph interacts with the virtual photon. This leads to a 
small quark/antiquark helicity contribution because of certain cancellations
\cite{c4ano}.
However, in the total helicity of the nucleon, in addition to the quark,
antiquark and gluon intrinsic helicity, one has to take into account the
orbital angular momenta of the constituents. In fact, the orbital angular
momentum of the constituents play a very important role in the nucleon
helicity.  The role of orbital helicity has been investigated recently in
our approach \cite{c4or}. A detailed analysis of light-front helicity operator is needed
for this purpose. Light-front helicity operator $J_3$ (in the frame
$P^\perp=0$, $J^3$ is the helicity operator; see chapter 7) can be written in
terms of manifestly symmetric gauge invariant energy-momentum tensor
$\theta^{\mu \nu}$ in QCD as, 
\be
J^3={1\over 2} \int dx^- d^2x^\perp [x^1 \theta^{+2}-x^2 \theta^{+1}].
\e
Recently it has been shown explicitly \cite{c4or} that $J_3$ constructed in light-front
gauge $A^+=0$, after eliminating the constrained variables $\psi^-$ and
$A^-$ using constraint equations, is independent of interaction and its form
agrees with the naive canonical form, provided we assume that the fields
vanish at the boundary. $J^3$ thus constructed can be separated into four
parts,
\be
J^3=J^3_{fi}+J^3_{fo}+J^3_{go} + J^3_{gi}
\e
where $J^3_{fi}$ is the fermion intrinsic part, $J^3_{fo}$ is the
fermion orbital part, $J^3_{go}$ is the gluon orbital part and $J^3_{gi}$
is the gluon intrinsic part. It has been shown that $J^3_{fi}$ is directly
related to the polarized structure function $g_1(x,Q^2)$ measured in
longitudinally polarized DIS. The helicity sum
rule can be written as
\be
\langle P S \mid J^3 \mid P S \rangle =\langle P S \mid J^3_{fi}+J^3_{fo}+
J^3_{go} + J^3_{gi}\mid P S \rangle = \pm {1\over 2},
\e
that is, the sum of the orbital and intrinsic helicities of the constituents
gives the nucleon helicity. This has been verified in perturbation theory
for a dressed quark a and dressed gluon explicitly.       
%%%%%%%%%%%%%%%%%%%%%%%%%%%%%%%%%%%%%%%%%%%%%%%%%%%%%%%%%%%%%%%%%%
\subsection{Polarized Structure Function: $g_2(x,Q^2)$ and the Violation of
Wandzura Wilczek Relation in pQCD}
%%%%%%%%%%%%%%%%%%%%%%%%%%%%%%%%%%%%%%%%%%%%%%%%%%%%%%%%%%%%%%%%%%
The polarized structure function $g_2(x,Q^2)$ is the least well known
structure function in DIS. The contribution of $g_2(x,Q^2)$ to the cross
section is suppressed by a factor of ${1\over Q}$ compared to the leading
contribution in the longitudinally polarized scattering. So it is commonly
known as higher twist effect. Wandzura and Wilczek showed that \cite{c4wand}  
$g_2(x,Q^2)$ can be separated into two parts, 
\be
g_2(x,Q^2)=g_2^{ww}(x,Q^2)+{\bar g_2}(x,Q^2)
\e
where ${\bar g_2}(x,Q^2)$ is a twist three piece and the twist two part is
related to $g_1(x,Q^2)$ through the Wandzura-Wilczek relation:
\be
g_2^{ww}(x,Q^2)=-g_1(x,Q^2) + \int_x^1 {dy\over y}g_1(x,Q^2).
\e
The twist three part was negligible in their model calculation and the quark
mass was neglected. Later, there has been a lot of interest in $g_2(x,Q^2)$.
The effect of the quark mass has been investigated \cite{c4alt} and it was
 found that the twist three part is a direct quark-gluon interaction effect
and plays a very important role \cite{c4shur}. The importance of quark mass in
$g_2(x,Q^2)$ has been investigated recently in the light-front QCD approach
\cite{c4wan}. The pure transversely polarized structure function is defined as,
\be
g_T(x,Q^2)=g_1(x,Q^2)+g_2(x,Q^2).
\e
The expression for $g_T$ in terms of the hadronic matrix element of
light-front bilocal current has been given earlier. After one eliminates the
constrained fields using the constraint equations, the expression of $g_T$
becomes,
\be
g_T(x,Q^2)={1\over 8\pi S^i_T} \int d\eta 
		e^{-i\eta x}  \langle PS| (O_m+O_{k_\perp}+O_g)+h.c. 
                |PS \rangle 
\e
where
\be
O_m=m\psi_+^\dagger (\xi^-)Q^2\gamma_\perp ({1\over {i\pp^+}})\gamma_5
\psi_+(0),\\
O_{k_\perp}=-\psi_+^\dagger(\xi^-)Q^2 (\gamma_\perp {1\over
{\pp^-}}\gamma^\perp \cdot \pp^\perp +2{P_\perp\over P^+})\gamma_5
\psi_+(0),\\	    
O_g=g\psi_+^\dagger(\xi^-)Q^2(\gamma^\perp \cdot A^\perp(\xi^-){1\over
{i\pp^+}}\gamma_\perp \cdot \gamma_5 \psi_+(0).
\e
$g_T$ has been calculated for a transversely polarized quark state dressed
with one gluon. This state can be expressed in terms of helicity states as,
\be
\mid k^+,k^\perp,s^1 \rangle = {1\over {\sqrt 2}} (\mid
k^+,k^\perp,\uparrow \rangle \pm \mid k^+,k^\perp,\downarrow \rangle)
\e
with $s^1=\pm m_R$, and $m_R$ is the renormalized quark mass. For this state,
\be
g_T(x,Q^2)={e_q^2\over 2}\Big \{ \delta(1-x)+{\alpha_s\over {2\pi}} C_f
ln{Q^2\over \mu^2}\Big [{(1+2x-x^2)\over {(1-x)_+}}\Big ] \Big \}
\e
which is independent of quark mass. However, it is important to start from
massive theory and renormalize the quark mass in light-front Hamiltonian
perturbation theory. However, starting from a massless theory one gets wrong
result for $g_T$. The structure function $g_1$ can also be calculated for a
dressed quark in the same method and it can be verified that 
Wandzura-Wilczek relation is not obeyed in perturbative QCD. It is also easy
to verify that the Burkhardt-Cottingham sum rule is obeyed, that is,
\be
\int_0^1 dx g_2(x, Q^2)=0.
\e

In the next few chapters, we discuss the works that we have done in this
research program using this framework.

%%%%%%%%%%%%%%%%%%%%%%%%%%%%%%%%%%%%%%%%%%%

\chapter{Twist Four Longitudinal Structure Function in Light-Front QCD}
In chapter 4, we have outlined our approach based on light-front
Hamiltonian QCD to DIS. In this chapter and in the next three chapters we
shall discuss the problems that we have investigated in this research program. 

In this chapter, we shall analyze twist four longitudinal structure
function.
%%%%%%%%%%%%%%%%%%%%%%%%%%%%%%%%%%%%%%%%%%%%%%%%%%%%%%%%%%%%%%%%%
\section{Motivation for Studying Twist Four $F_L$}
%%%%%%%%%%%%%%%%%%%%%%%%%%%%%%%%%%%%%%%%%%%%%%%%%%%%%%%%%%%%%%%%%%

The higher twist contributions or power corrections to DIS structure
functions play a very important role in the SLAC kinematical range
\cite{c5dasu}. The longitudinal structure function $F_L$ involves coefficient functions
which are zero at zeroth order and the lowest twist contributions in this
case involves calculations to the same order in $\alpha_s$ as the NLO results
for $F_2$ \cite{c5amanda}. Thus, twist four contributions to $F_L$ are extremely important.
These are the first non-perturbative contributions to $F_L$.
 The higher
twist contributions to the unpolarized structure functions $F_2$ and $F_L$
have been analyzed previously in operator product expansion method \cite{c5js} 
and also
in the Feynman diagram approach \cite{c5efp}. Qiu has given an alternate method  based
on special propagators utilizing some unique features of light-front
coordinates \cite{c5qiu}. At twist four level, in the OPE analysis, there appears a
proliferation of operators. The operators are not all independent and they 
are related through the equation of motion and the analysis is highly complex.
On the other hand, in the Feynman diagram approach, it has been shown that
one can make contact with light-front current algebra analysis, namely, twist
four part of $F_L$ can be expressed as the Fourier transform of the hadronic
matrix element of the minus component of light-front bilocal current. 
It is important to go beyond phenomenological parametrizations for a proper
understanding of the higher twist effects.
However, even after many years of investigations, an intuitive physical
picture of the interaction dependence of twist four longitudinal structure
function is lacking. In our analysis  we get an intuitive picture of $F_L$.

Another important problem of current interest is the perturbative aspect of
the twist four matrix element. Simple power 
counting indicates that in the bare theory the
twist four matrix element contains quadratic divergences. Understanding the
origin and nature of these divergences will be quite helpful in finding
procedures to remove them (the process of renormalization).     
     
%%%%%%%%%%%%%%%%%%%%%%%%%%%%%%%%%%%%%%%%%%%%%%%%%%%%%%%%%%%%%%%%%
\section{Plan of This Chapter}
%%%%%%%%%%%%%%%%%%%%%%%%%%%%%%%%%%%%%%%%%%%%%%%%%%%%%%%%%%%%%%%%%%%%

In this chapter, we show that using light-front Hamiltonian QCD
framework, one can resolve outstanding issues associated with the twist four
contribution to the longitudinal structure function. 
Our starting point is the Bjorken-Johnson-Low (BJL)
expansion for the forward virtual photon-hadron Compton scattering
amplitude. This leads us to the commutator of currents which we present in
detail for arbitrary flavors in SU(3). Next we consider the specific case of
electromagnetic currents and arrive at expressions for the twist two part of
$F_2$ and twist four part of $F_L$ in terms of specific flavor 
dependent form factors.
We identify the integral of ${F_L(x) \over x}$ with
the fermionic part of the light-front QCD Hamiltonian density. The
consideration of mixing in the flavor singlet channel
leads us to the definition of the twist four longitudinal gluon
structure function and then we find a sum rule, free from radiative
corrections. 

The sum rule which the physical structure function has to satisfy involves the
physical mass of the hadron which is a finite quantity. A theoretical
evaluation of the sum rule which starts with the bare theory, on the other 
hand, will contain various divergences  
depending on the regulator
employed. In order to compare with the physical answer resulting from the
measurement, we need to renormalize the result by adding counterterms. For
the dressed parton target, for example, these counterterms are dictated
by
mass counterterms in the light-front Hamiltonian perturbation theory. For a
dressed gluon target, calculations show that quadratic
divergences are generated and one does not automatically get the result
expected for a massless target. The divergence generated is shown to be
directly related to the gluon mass shift in old fashioned perturbation
theory. To a given order in perturbation theory, counterterms have to be
added to the calculated structure function. The precise selection of
counterterms is dictated entirely
 by the regularization and renormalization of the
light-front QCD Hamiltonian. The choice of counterterms in the
Hamiltonian, in turn, determines the counterterms to be added to the
longitudinal structure function which results in a theoretical prediction of
the physical longitudinal structure function. We recall that in Hamiltonian
perturbation theory we cannot 
automatically generate a massless gluon
by the use of dimensional regularization. The point we emphasize is that the
twist four longitudinal structure function is one to one related to the
Hamiltonian density and that there is no arbitrary freedom in this
relationship.

To understand the nature of quadratic divergences, we evaluate
the twist four longitudinal structure functions for quark and
gluon target each dressed through lowest order in perturbation theory. 
The sum rule allows us to relate these
divergences to quark and gluon mass corrections in QCD in
time-ordered light-front perturbation theory.  Finally we discuss
the relevance of our results for the problem of the partitioning
 of hadron masses
in QCD.

The plan of this chapter is as follows. In Sec. 5.3 we derive the expressions for
the twist two structure function $F_2$ and the twist four longitudinal
structure function
$F_L$ using the BJL expansion and equal time ($x^+$) current algebra and
also taking into account the flavor structure.
The flavor singlet expressions are given in Sec. 5.4. The sum
rule for $F_L$ is derived in Sec. 5.5. In Sec. 5.6 we 
evaluate $F_L$ for a massless quark dressed through lowest order
in perturbation theory
and explicitly verify the sum rule. Then we evaluate $F_L$ in
perturbation theory for dressed quark (massive) 
 and dressed gluon target. In Sec. 5.7 we discuss the
issue of the breakup of hadron mass in QCD in the context of our sum rule.
Finally, in Sec. 5.8, we discuss the implication of our results. Our
notations and conventions have been clarified in Appendix B. We also
refer to our original papers, \cite{c5fl1,c5fl2}.
%%%%%%%%%%%%%%%%%%%%%%%%%%%%%%
\section{Longitudinal Structure Function: Flavor Structure}
%%%%%%%%%%%%%%%%%%%%%%%%%%%%%%
%%%%%%%%%%%%%%%%%%%%%%%%%%%%%%%%%%%%%%%%%%%%%%%%%%%%%%%%%%%%%%%%
%\section{Flavor structure}
%%%%%%%%%%%%%%%%%%%%%%%%%%%%%%%%%%%%%%%%%%%%%%%%%%%%%%%%%%%%%%%
%The electromagnetic current 
%\begin{equation}
%J^{\mu}(x) = J^\mu_3(x) + {1 \over \sqrt{3}} J^\mu_8(x).
%\end{equation}
In this section we present the expressions for structure functions for
arbitrary flavors in SU(3) which follow from the use of the
Bjorken-Johnson-Low
expansion and light-front current algebra. 
In terms of the flavor current 
$J_a^\mu(x) = {\overline \psi}(x) \gamma^\mu {\lambda_a \over 2} \psi(x)$,
the hadron tensor relevant for deep inelastic scattering is given by 
\begin{equation}
	W^{\mu \nu}_{ab} = {1\over 4\pi} \int d^4 \xi~ e^{iq \cdot \xi} 
		\langle P |[J^\mu_{a}(\xi), J^\nu_{b}(0)]|P \rangle.
\label{c5wmn}
\end{equation}
The forward virtual photon-hadron Compton scattering amplitude is given by
\begin{eqnarray}	
	T^{\mu \nu}_{ab} &=& i \int d^4\xi e^{iq\cdot \xi} \langle P | 
		T(J^\mu_{a}(\xi) J^\nu_{b}(0)) |P \rangle.
\end{eqnarray}
We have
\begin{equation}	
	T^{\mu \nu}_{ab} (x,Q^2) = 2 \int_{-\infty}^\infty d{q'}^+
		{W^{\mu \nu}_{ab} (x',Q^2) \over {q'}^+ - q^+} ~.
\label{c5disp}
\end{equation} 

Using the BJL expansion \cite{c5jac}, we have
\begin{equation} 
	T^{\mu \nu}_{ab} = -  {1\over q^-}
		 \int d\xi^- d^2 \xi_\bot e^{iq\cdot \xi}
	  \langle P | [J^\mu_{a}(\xi), J^\nu_{b}(0)]_{
		\xi^+=0}| P \rangle \, + ... \label{c5tmn}
\end{equation}
where $ ... $ represents higher order terms in the expansion which we ignore
in the following.
In the limit of large $q^-$, from Eq. (\ref{c5wmn}), we have
\begin{eqnarray}
W^{+-}_{ab} = {1 \over 2 } F_{L(ab)} + (P^\perp)^2 { F_{2(ab)} \over \nu}+ {P^\perp.
q^\perp \over x \nu} F_{2 (ab)}, \label{c5lqwmn}
\end{eqnarray}
with $ x = { -q^2 \over 2 \nu}$ and $ \nu =P.q$. 
On the other hand, from Eq. (\ref{c5tmn}), 
\begin{eqnarray}
Limit_{q^- \rightarrow \infty} ~~ T^{+-}_{ab} = -  {1\over q^-}
		 \int d\xi^- d^2 \xi_\bot e^{iq\cdot \xi}
	  \langle P | [J^+_{a}(\xi), J^-_{b}(0)]_{
		\xi^+=0}| P \rangle \, . \label{c5tpm}
\end{eqnarray}

The components of the 
flavor current $J_a^\mu(x)$
obey the equal- $x^+$ canonical commutation relation (to be specific, we
consider SU(3) of flavors)
\begin{eqnarray} 
\left [ J_a^+(x), J_b^-(y) \right ]_{x^+=y^+}  ~=~ 
 2 i f_{abc} ~{\overline \psi}(x)~ \gamma^-{ \lambda_c \over 2} \psi(x)~
\delta^2(x^\perp - y^\perp) ~ \delta(x^- - y^-) \nonumber \\
 ~~~~~- ~{ 1 \over 2} ~\partial^+_x ~\Bigg [ \epsilon(x^- - y^-)~
  \Big [i f_{abc} ~
{\cal V}_c^-(x \mid y)~ +~ i d_{abc}~ {\overline {\cal V}}_c^- (x \mid y) \Big ]
\delta^2 (x^\perp - y^\perp) \Bigg ] \nonumber \\
~ ~~~~+~ { 1 \over 2}~ i f_{abc} ~\epsilon(x^- - y^-)~ \partial^i_x~ \Bigg [ 
\delta^2(x^\perp - y^\perp)~  \Big [ {\cal V}^i_c(x \mid y)~ - ~\epsilon^{ij}
~{\overline {\cal A}}^j_c (x \mid y) \Big ] \Bigg ] \nonumber \\
~ ~~~~~+~ { 1 \over 2}~ i d_{abc}~ \epsilon(x^- - y^-) ~\partial^i_x~ \Bigg [ 
~\delta^2(x^\perp - y^\perp) ~ \Big [ {\overline {\cal V}}^i_c(x \mid y)
~ +~ \epsilon^{ij}~{\cal A}^j_c (x \mid y) \Big ] \Bigg ]. \label{c5cuco}
\end{eqnarray}
In deriving the above relations, use has been made of the relation
\begin{eqnarray}
\lambda_a \lambda_b = i f_{abc} \lambda_c + d_{abc} \lambda_c.
\end{eqnarray}
Here $a$, $b$, $c$ run from $0$ to $8$.
We have defined the bilocal currents as follows.
\begin{eqnarray}
{\cal V}^\mu_c(x \mid y) &=& { 1 \over 2} \Big [
{\overline \psi}(x)  { \lambda_c \over 2} \gamma^\mu
\psi(y) + {\overline \psi}(y) { \lambda_c \over 2} \gamma^\mu \psi(x) 
\Big ], \nonumber \\
{\overline {\cal V}}^\mu_c(x \mid y ) &=& { 1 \over 2 i} \Big [
{\overline \psi}(x)  { \lambda_c \over 2} \gamma^\mu
\psi(y) - {\overline \psi}(y) { \lambda_c \over 2} \gamma^\mu \psi(x) 
\Big ], \nonumber \\
{\cal A}^\mu_c(x \mid y) & =& { 1 \over 2 } \Big [
{\overline \psi}(x)  { \lambda_c \over 2} \gamma^\mu \gamma^5
\psi(y) + {\overline \psi}(y) { \lambda_c \over 2} \gamma^\mu \gamma^5 \psi(x) 
\Big ], \nonumber \\
{\overline {\cal A}}^\mu_c(x \mid y) &=& { 1 \over 2 i} \Big [
{\overline \psi}(x)  { \lambda_c \over 2} \gamma^\mu \gamma^5
\psi(y) - {\overline \psi}(y) { \lambda_c \over 2} \gamma^\mu \gamma^5 \psi(x) 
\Big ]. \label{c5bi}
\end{eqnarray}  
Further, we introduce the bilocal form factors
\begin{eqnarray}
\langle P \mid {\cal V}^\mu_c(\xi \mid 0) \mid P \rangle &=& P^\mu
V^1_c(\xi^2,
P.\xi) + \xi^\mu V^2_c(\xi^2, P.\xi), \\
\langle P \mid {\overline {\cal V}}^\mu_c(\xi \mid 0) \mid P \rangle &
=& P^\mu {\overline V}^1_c(\xi^2,
P.\xi) + \xi^\mu {\overline V}^2_c(\xi^2, P.\xi).  \label{c5bff}
\end{eqnarray}
From Eqs (\ref{c5tpm}) and  (\ref{c5cuco}),  we get
\begin{eqnarray}
Limit_{q^- \rightarrow \infty} ~~ q^- ~ T^{+-}_{ab}  =
-2 i f_{abc} P^- \Gamma_c ~~~~~~~~~~~~~~~~~~~~~~~\nonumber \\
 + {q^+ \over 2} \int d\xi^- e^{{i \over 2} q^+ \xi^-} \epsilon(\xi^-)
  \Big [ f_{abc} ~
\langle P \mid {\cal V}_c^-(\xi \mid 0)~ \mid P \rangle +~  d_{abc}~
\langle P \mid  {\overline {\cal V}}_c^- (\xi \mid 0) \mid P \rangle \Big ]
\nonumber \\
 - {q^i \over 2} \int d\xi^- e^{{i \over 2} q^+ \xi^-} \epsilon(\xi^-)
  \Big [ f_{abc} ~
\langle P \mid {\cal V}_c^i(\xi \mid 0)~ \mid P \rangle +~  d_{abc}~
\langle P \mid  {\overline {\cal V}}_c^i (\xi \mid 0) \mid P \rangle \Big ].
\label{c5lqtmn}
\end{eqnarray}  
It is to be noted that matrix elements of ${\cal A}^\mu_c(x \mid y)$ do not contribute to
unpolarized scattering. 
Using the dispersion relation given in Eq. (\ref{c5disp}), together with Eqs. 
(\ref{c5lqwmn}) and (\ref{c5lqtmn}) and comparing the coefficient of $q^i$ on
both sides, we get
\begin{eqnarray}
{F_{2(ab)}(x) \over x} = { i \over 4 \pi} \int d \eta e^{-i \eta x} \Big [
f_{abc} V_c^1(\eta) + d_{abc} {\overline V}_c^1(\eta) \Big ].
\end{eqnarray}
Comparing the coefficients of $q^+$ on both sides, we get
\begin{eqnarray}
F_{L(ab)}(x) &=& { 1 \over Q^2} { i \over \pi} 
{ (q^+)^2 \over P^+} \int d \eta e^{ -
i \eta x} \Big [ f_{abc} \langle P \mid
{\cal V}_c^-(\xi \mid 0)\mid P \rangle + 
d_{abc} \langle P \mid {\overline {\cal V}}_c^-
(\xi \mid 0) \mid P \rangle \Big ]  \nonumber \\
& &  - {(P^\perp)^2 \over Q^2} { i \over \pi P^+} x^2
\int d\eta e^{-i \eta x} \Big [ f_{abc} \langle P \mid {\cal V}_c^+(\xi
\mid 0) \mid P \rangle \nonumber\\&&~~~~+ d_{abc} \langle P \mid {\overline
{\cal V}}_c^+(\xi \mid 0) \mid P \rangle\Big ].
\end{eqnarray}
We have introduced $ \eta = { 1 \over 2} P^+ \xi^-$.

Note that our result for $F_L$ differs from the one given in the literature
\cite{c5cjt}. The difference can be traced to the expression for $F_L$ that
one employs. It is customary \cite{c5cjt,c5efp} to ignore target mass $M^2$ in
the expression for $F_L$ (see Appendix B). This leads to an incorrect
expression for $F_L$ which in turn will lead to an incorrect sum rule (see
section 5.5).

The electromagnetic current 
\begin{equation}
J^{\mu}(x) = J^\mu_3(x) + {1 \over \sqrt{3}} J^\mu_8(x).
\end{equation} 
From the flavor structure of electromagnetic current, we observe that, only
$d_{abc}$ contributes to the structure functions in deep-inelastic 
electron-hadron scattering. Explicitly, we have, 
\begin{eqnarray}
{F_2(x) \over x} = { i \over 2 \pi P^+} \int d \eta e^{ - i \eta x} 
  \langle P \mid {\overline {\cal V}}^+( \xi \mid 0) \mid P \rangle.
\end{eqnarray}

The longitudinal structure function is given by 
\begin{eqnarray}
F_{L}(x) &=& { 2 \over Q^2} { i \over \pi} 
{ (q^+)^2 \over P^+} \int d \eta e^{ -
i \eta x} \langle P \mid {\overline {\cal V}}^-(\xi \mid 0) \mid P \rangle  
\nonumber \\
%\nonumber \\ && \qquad \qquad \qquad \qquad \qquad 
%+  y^-   {\overline V}^2 (\eta) \Bigg ] \nonumber \\
&& \qquad \qquad \qquad - 2 {(P^\perp)^2 \over Q^2} { i \over \pi P^+} x^2
\int d\eta e^{-i \eta x}  \langle P \mid {\overline {\cal V}}^+ (\xi \mid 0)
\mid P \rangle  .
\end{eqnarray} 

We have defined the functions

\begin{eqnarray}
{\overline {\cal V}}^\pm (\xi \mid 0)=
\left({2 \over 3}\right)^{3 \over 2}{\overline {\cal V}}^\pm_0(\xi \mid 0) + 
{ 1 \over 3}
{\overline {\cal V}}^\pm_3(\xi \mid 0) +
 { 1 \over 3 \sqrt{3}} 
{\overline {\cal V}}^\pm_8 (\xi \mid 0).
\end{eqnarray}

In arriving at our final results
we have used explicit values of the structure constants 
of $SU(3)$,
\begin{eqnarray}
d_{338} = { 1 \over \sqrt{3}}, ~~ d_{888} = - { 1 \over \sqrt{3}}, ~~
d_{330} = d_{880} = \sqrt{2 \over 3}.
\end{eqnarray}
${\overline {\cal V}}^\mu_0$ is the flavor singlet component of the fermion 
bilocal vector current. 
%%%%%%%%%%%%%%%%%%%%%%%%%%%%%%%%%%%%%%%%%%%%%%%%%%
\section{Flavor Singlet Case}
%%%%%%%%%%%%%%%%%%%%%%%%%%%%%%%%%%%%%%%%%%%%%%%%%
In the flavor singlet channel, for simplicity, we do not explicitly write
the flavor charge dependence. 
The BJL limit, together with light-front current 
algebra \cite{c5jac} in $A^+=0$ gauge, the tools used in the pre-QCD era, lead
to the 
twist four part of the fermionic contribution to the longitudinal 
structure function
\begin{eqnarray}
F^{\tau=4}_{L(f)}(x) = i { 1 \over Q^2} {(xP^+)^2 \over \pi} \int dy^- e^{- {i \over 2}P^+
y^- x}  \langle P \mid \overline{\cal J}^- (y \mid 0) \mid P \rangle -
4 {(P^\perp)^2 \over Q^2} x F_{2(f)}(x) \label{c5flBJL1}
\end{eqnarray}
where $f$ represents a quark ($q$) or anti-quark (${\bar q}$) or both
depending on the target $ \mid P \rangle$, and
where the twist two contribution to the $F_{2(f)}$ structure function 
\begin{eqnarray}
{F_{2(f)} (x) \over x} = i { 1 \over 4 \pi }  \int dy^-  e^{- {i \over 2}P^+
y^- x}  \langle P \mid \overline{\cal J}^+(y \mid 0) \mid P \rangle .
\label{c5f21}  
\end{eqnarray}
%Here $P^- = {(P^\perp)^2 + M^2 \over P^+}$.
The bilocal current operator  
\begin{eqnarray}
\overline{\cal J}^\mu(y \mid 0) = { 1 \over 2 i} \left [ \overline{\psi} (y)
\gamma^\mu \psi(0) - \overline{\psi}(0) \gamma^\mu \psi(y) \right ].
\label{c5bcodef}
\end{eqnarray}
Note that in the case of quark (anti-quark)
contributions, the second (first) term
in the expression for the bilocal current in Eq. (\ref{c5bcodef}) vanishes,
because of a momentum non-conserving delta function.
We have
\begin{eqnarray}
F^{\tau=4}_{L(q)}(x) = {\cal M}_1 + 
{\cal M}_2,  \label{c5fld} 
\end{eqnarray}
with
\begin{eqnarray}
{\cal M}_1~=~{ 1 \over Q^2}~ {x^2 (P^+)^2 \over 2 \pi}~ \int dy^- ~
e^{-{ i \over
2}P^+y^-x}~\langle P \mid \overline{\psi}(y^-) \gamma^- \psi(0) \mid P \rangle,
\label{c5flf1}
\end{eqnarray}
and
\begin{eqnarray}
{\cal M}_2=-{(P^\perp)^2  \over (P^+)^2}
{ 1 \over Q^2} {x^2 (P^+)^2 \over 2 \pi}~ 
\int dy^- ~e^{-{ i \over
2}P^+y^-x}~\langle P \mid \overline{\psi}(y^-) \gamma^+ \psi(0) \mid P \rangle.
\label{c5flf2}
\end{eqnarray}
${\cal M}_1$ contains the `bad' (minus) component of bilocal current
operator. This involves the constrained field ${\psi^-}$ and one has to use
the equation of constraint to eliminate it.

Using 
 $\overline{\psi}(y^-) \gamma^- \psi(0) = 2 {\psi^{-}}^\dagger(y^-)
\psi^-(0)$, where 
\begin{eqnarray}
 \psi^-(z) = { 1 \over 4 i} \int dy^- \epsilon(z^- - y^-) \Big
[ \alpha^\perp . (i \partial^\perp + g A^\perp)+ 
\gamma^0 m \Big ] \psi^+(y^-),
\end{eqnarray}
and
$\epsilon(x^-) = - { i \over \pi} P \int {d \omega \over \omega} e^{{i \over
2 } \omega x^-}$,
we arrive at
\begin{eqnarray}
{\cal M}_1 &=& { 1 \over \pi Q^2 } \int dy^- e^{-{ i \over 2}P^+ y^-x} \langle P
\mid {\psi^{+}}^\dagger(y^-) \nonumber \\
&&\Big [ \alpha^\perp. \big  [ i \partial^\perp + g A^\perp(y)
\big ] + \gamma^0 m  \Big ] \Big [ 
\alpha^\perp . \big [ i \partial^\perp + g A^\perp(0) \big ] + 
\gamma^0 m \Big ] \psi^+(0)
\mid P \rangle, \label{c5fl1} 
\end{eqnarray}
and 
\begin{eqnarray}
{\cal M}_2 &=&  - 2 {(P^\perp)^2  \over Q^2} { 1 \over 2 \pi} x^2 
\int dy^- e^{-
{ i \over 2} P^+y^-x} \langle  P \mid {\psi^{+}}^{\dagger}(y^-) \psi^+(0) \mid P
\rangle . 
%& =& - { 4 \over Q^2}  (P^\perp)^2  x F_{2(q)}(x).
\label{c5fl2}
\end{eqnarray}

We see that the elimination of the constrained field ${\psi^-}$ brings in
interaction ($g$) dependent terms in the operator in ${\cal M}_1$. In fact,
this feature distinguishes higher twist contributions from the leading
twist, since it directly involves quark-gluon interactions.
Thus we have obtained an expression for the twist four part of the fermionic
component of the longitudinal structure function.

After the establishment of QCD as the underlying theory of strong
interactions, 
the twist four part of the quark contributions to 
the longitudinal structure function has been given in the
limits of vanishing target transverse momentum and 
massless quark and in the $A^+=0$ gauge using the QCD 
factorization method \cite{c5efp,c5qiu}
\begin{eqnarray}
 F^{\tau=4}_{L(q)} (x)&=& { 1 \over \pi Q^2 } \int dy^- e^{-{ i \over 2}P^+ y^-x} \langle P
\mid {\psi^{+}}^\dagger(y^-) \nonumber \\
&&~~~~~~~ \big [ i \partial^\perp + g A^\perp(y) \big ].\alpha^\perp
\alpha^\perp . \big [ i \partial^\perp + g A^\perp(0) \big ] \psi^+(0)
\mid P \rangle . \label{c5fact}
\end{eqnarray}

Using Eqs. (\ref{c5fld}), (\ref{c5fl1}), and (\ref{c5fl2}), and taking  
the limit of vanishing target transverse momentum and massless quark,
our result given in Eq. (\ref{c5flBJL1}) reduces to that obtained via the QCD 
factorization method.

Note that the expression for $ F_{L(q)}^{\tau=4}$ given in Eq.
(\ref{c5flBJL1})
appears to violate transverse boost invariance since it involves $P^\perp$. But, we exhibit below
with explicit
calculations in Sec. 5.6 that the  $P^\perp$ dependence cancels between Eqs.
(\ref{c5fl1}) and (\ref{c5fl2}) so that the full $F_{L(q)}^{\tau=4}$ 
is indeed boost invariant.

\vskip .2in
%%%%%%%%%%%%%%%%%%%%%%%%%%%%%%%%%%%%%%%%%%%%%%%%%%%%%%%%%%%%%%%%%%%%%%%%%%
\section{Sum Rule for the Twist Four Longitudinal
Structure Function}
%%%%%%%%%%%%%%%%%%%%%%%%%%%%%%%%%%%%%%%%%%%%%%%%%%%%%%%%%%%%%%%%%%%%%%%%%%
\vskip .2in

From Eq. (\ref{c5f21}) it follows that $ F_{2(f)}(-x) = F_{2(f)}(x)$ and 
from Eq. (\ref{c5flBJL1}) we explicitly find that $
F_{L(f)}^{\tau=4}(-x) = -F^{\tau=4}_{L(f)}(x) $. 
Consider the integral
\begin{eqnarray}
\int_{- \infty}^{+ \infty}dx {F^{\tau=4}_{L(f)}(x) \over x}&=& 
2 \int_{0}^{ \infty}dx
{F^{\tau=4}_{L(f)}(x)
\over x} \nonumber \\
&=& \int_{- \infty}^{+ \infty} {dx \over x}  \Bigg [
 i { 1 \over Q^2} {(xP^+)^2 \over \pi} \int dy^- e^{- {i \over 2}P^+
y^- x}  \nonumber \\
&& \times \Big [ \langle P \mid \overline{\cal J}^- (y \mid 0) \mid P \rangle -
{(P^\perp)^2 \over (P^+)^2} \langle P \mid \overline{\cal J}^+(y \mid 0) 
\mid P \rangle \Big ] \Bigg ].
\end{eqnarray}
The part containing $\overline{\cal J}^-$ can be written as
\be
{(P^+)^2\over {2 \pi Q^2}}\int dx \Big [ \int dy^-{\pp\over {\pp
y^-}}(e^{{i\over 2}P^+y^-})({2\over {iP^+}})\langle P \mid
\overline\psi(0)\gamma^-\psi(y^-)\mid P\rangle
\nonumber\\~~~~~\int dy^-{\pp \over {\pp y^-}} (e^{-{i\over
2}P^+y^-})({-2\over {iP^+}}) \langle P \mid
\overline\psi(0)\gamma^-\psi(y^-)\mid P\rangle,
\e
where in the first term, we have changed $y^-$ to $-y^-$ and used
translational invariance of the matrix element. Interchanging the orders of
$x$ and $y^-$ integrations and carrying out the $x$ integration, we get,
from the above expression,
\be
{2iP^+\over Q^2}\int dy^- \langle P \mid
\overline\psi(0)\gamma^-\pp^+\psi(0)\mid P\rangle \delta({1\over 2}P^+y^-)
\nonumber\\={4i\over Q^2}\langle P \mid
\overline\psi(0)\gamma^-\pp^+\psi(0)\mid P\rangle.
\e
Similar manipulations can be done for the part containing  $\overline{\cal J}^+$
and we arrive at \cite{c5bgj},

\begin{eqnarray}
\int_{- \infty}^{+ \infty} dx  {F_{L(f)}^{\tau=4}(x,Q^2) \over x} &=& 
{ 4 \over Q^2} \Big [
\langle P \mid i \overline{\psi} \gamma^- \partial^+ \psi|_{(0)}
 \mid P \rangle \nonumber\\&&~~~~~~~~~~- {(P^\perp)^2 \over (P^+)^2} 
\langle P \mid i \overline{\psi} \gamma^+ \partial^+  \psi |_{(0)} \mid P 
\rangle \Big ].
\end{eqnarray}
Identifying $ i \overline{\psi} \gamma^- \partial^+ \psi = \theta^{+-}_q$, the
fermionic part of the light-front QCD Hamiltonian density and 
$i \overline{\psi} \gamma^+ \partial^+  \psi = \theta^{++}_q$, the fermionic
part of the light-front QCD longitudinal momentum density, 
(see Eqs. (\ref{c5denp}) and
(\ref{c5denh}) below),
we have:

\begin{eqnarray}
\int_{0}^{1} dx {F_{L(f)}^{\tau=4}(x,Q^2) \over x} = { 2 \over Q^2} \Big [
\langle P \mid \theta_q^{+-}(0) \mid P \rangle - {(P^\perp)^2 \over (P^+)^2} 
\langle P \mid \theta^{++}_q(0) \mid P \rangle \Big ], \label{c5flsr1}
\end{eqnarray}    
where we have used the fact that the physical structure function 
vanishes for $x >1$.

The integral of ${F^{\tau=4}_{L(f)} \over x}$ is therefore 
related to the hadron matrix
element of the (gauge invariant) fermionic part of the light-front 
{\it Hamiltonian density}.
This result
manifests the physical content and the non-perturbative nature of the 
twist-four part of the
longitudinal structure function.

The fermionic operator matrix elements appearing in Eq. (\ref{c5flsr1}) 
change with $Q^2$ as a result of the mixing of quark and gluon operators in
QCD under renormalization.    
Next we analyze the operator mixing and derive a new 
sum rule at the twist four level. 

The symmetric, gauge-invariant energy-momentum tensor in QCD is
given by
\begin{eqnarray}
\theta^{\mu \nu} &=& { 1\over 2} \overline{\psi} i \big [ \gamma^\mu D^\nu +
\gamma^\nu D^\mu \big ] \psi 
-F^{\mu \lambda a} F^{\nu}_{~ \lambda a} + {1 \over 4} g^{\mu \nu} 
(F_{\lambda \sigma a } )^2 \nonumber \\
&& -g^{\mu \nu} \overline{\psi} \left ( i \gamma^\lambda D_\lambda -
m  \right ) \psi.
\label{c5emt}
\end{eqnarray}
The last term vanishes using the equation of motion.
Here $F^{\mu \nu a}=\pp^\mu A^{\nu a}-\pp^\nu A^{\mu a}+ gf_{abc}
A^\mu_bA^\nu_c.$
{\it Formally,} we split the energy momentum tensor into a 
$``$fermionic" part $\theta^{\mu \nu}_{q}$ representing the first term in
Eq. (\ref{c5emt}) and a $``$gauge bosonic" 
part $ \theta^{\mu \nu}_g$ representing the second and third terms in Eq.
(\ref{c5emt}).
%\begin{eqnarray}
%\theta^{\mu \nu}_q = { 1 \over 2} \overline{\psi} i \Big [ \gamma^\mu D^\nu +
%\gamma^\nu D^\mu \Big ] \psi,
%\end{eqnarray}
%and 
%\begin{eqnarray}
%\theta^{\mu \nu}_g =  -F^{\mu \lambda a} F^{\nu}_{ \lambda a } + 
%{1 \over 4} g^{\mu \nu}
%(F_{\lambda \sigma a })^2 .
%\end{eqnarray}
To be consistent with the study of
deep inelastic structure function formulated in the $A^+=0$ gauge,
we shall work in the same gauge. 
We have, for the fermionic part of the longitudinal momentum density,
\begin{eqnarray}
\theta^{++}_{q} = i \overline{\psi} \gamma^+ \partial^+ \psi. \label{c5denp} 
\end{eqnarray}
For the fermionic part of
the Hamiltonian density, we have 
\begin{eqnarray}
\theta^{+-}_q = i {\psi^{+}}^\dagger \partial^- \psi^+ + g {\psi^{+}}^\dagger
A^- \psi^+ + i {\psi^{-}}^\dagger \partial^+ \psi^-.
\end{eqnarray}
%with
%\begin{eqnarray}
%\theta^{+-(1)}_q = i {\psi^{+}}^\dagger \partial^- \psi^+ + 
%g {\psi^{+}}^\dagger
%A^- \psi^+,
%\end{eqnarray}
%and
%\begin{eqnarray}
%\theta^{+-(2)}_q = i {\psi^{-}}^\dagger \partial^+ \psi^-.
%\end{eqnarray}
Using the Dirac equation for the fermion, we find that the sum of the 
first two terms equals the third term in the above equation. Therefore, 
\begin{eqnarray}     
\theta^{+-}_q &=& i \overline{\psi} \gamma^- \partial^+ \psi =
2 i {\psi^{-}}^\dagger \partial^+ \psi^-  \label{c5denh} \\
&=& 2 {\psi^{+}}^\dagger \Big [ \alpha^\perp.(i \partial^\perp + g A^\perp) 
+ \gamma^0 m \Big ]
{ 1 \over i \partial^+} \Big [ \alpha^\perp . (i \partial^\perp + g A^\perp)
+ \gamma^0 m \Big ] \psi^{+} \label{c5thetaqf}.
\end{eqnarray}

The gauge boson part of the Hamiltonian density is given by
\newpage
\begin{eqnarray}
\theta^{+-}_g &=&  - F^{+ \lambda a} F^{-}_{\lambda a}+ { 1 \over 4} 
g^{+-} (F_{\lambda \sigma a})^2 =
{ 1 \over 4} \Big (\partial^+ A^{- a}\Big )^2 + 
{ 1 \over 2} F^{ij a } F^{a}_{ij}
\nonumber \\
&=& (\partial^i A_a^j)^2 + 2gf^{abc}A_a^i A_b^j \partial^i A_c^j
		  + \frac{g^2}{2}
		f^{abc} f^{ade} A_b^i A_c^j A_d^i A_e^j  \nonumber \\
	&& ~~~~~~~~~~ + 2g \partial^i A_a^i \left( \frac{1}{\partial^+}
		\right) (f^{abc} A_b^j \partial^+ A_c^j + 2 (\psi^+)^{\dagger}
		T^a \psi^+ ) \nonumber \\
	&& ~~~~~~~~~~ + g^2 \left( \frac{1}{\partial^+}
		\right) (f^{abc} A_b^i \partial^+ A_c^i + 2 (\psi^+)^{\dagger}
		T^a \psi^+ ) 
	  \left( \frac{1}{\partial^+}\right)
		(f^{ade} A_d^j \partial^+ A_e^j \nonumber\\&&~~~~~~~~~~~~~
             ~~~~~~~~~~~~~~~~  + 2 (\psi^+)^{\dagger} T^a
		\psi^+ )     \nonumber \\
    \label{c5thetagf}
%&=& \left(\partial^i A^i \right)^2 + { 1\over 2} F^{ij}F_{ij} - 
%4 g { 1\over \partial^+}
%\left ( \partial^i A^i \right ) (\psi^+)^\dagger \psi^+ + 4 g^2 \Big ({ 1
%\over \partial^+} (\psi^+)^\dagger \psi^+  \Big )^2 
\end{eqnarray}
where we have used the equation of constraint for the gauge field.     

We define the twist four longitudinal gluon structure function
\begin{eqnarray}
F_{L(g)}^{\tau=4}(x) &=& { 1 \over Q^2} {x P^+ \over 2 \pi} \int dy^- ~
e^{-{i \over 2} P^+ y^- x} \nonumber \\
&& ~~~~~\Big [ \langle P \mid (-) F^{+ \lambda a}(y^-) F^-_{\lambda a}(0) + 
{ 1 \over 4} g^{+-} F^{\lambda \sigma a} (y^-) F_{\lambda \sigma a}(0)+(y^-
\leftrightarrow 0) \mid
P \rangle \nonumber \\
&& ~~~~~~~ - {(P^\perp)^2 \over (P^+)^2} \langle P \mid F^{+ \lambda a}(y^-) 
F^+_{\lambda a}(0) +(y^- \leftrightarrow 0)\mid P \rangle \Big ].
\end{eqnarray}
Then we have
\begin{eqnarray}
\int_0^1 { dx \over x} \Big [ F_{L(q)}^{\tau=4} + F_{L(g)}^{\tau=4} \Big ] =
\int_0^1 { dx \over x} F_{L}^{\tau=4} &=& { 2 \over Q^2} 
\Big [ \langle P \mid \theta^{+-}(0) \mid P \rangle -
\nonumber\\&&~{(P^\perp)^2 \over
(P^+)^2} \langle P \mid \theta^{++}(0) \mid P \rangle \Big ].
\label{c5flsr2}
\end{eqnarray}
Neglect of $M^2$ in the expression of $F_L$ (see Appendix B) will lead to
$(P^\perp)^2+M^2$ instead of $(P^\perp)^2$ in the right hand side of the
above equation and this would spoil the correct sum rule.   

We have
\begin{eqnarray}
\langle P \mid \theta^{+-}(0) \mid P \rangle = 2 P^+ P^- = 2 (M^2 +
(P^\perp)^2)~~ {\rm and} ~~ \langle P \mid \theta^{++}(0) \mid P \rangle = 2
(P^+)^2,
\end{eqnarray}
where $M$ is the invariant mass of the hadron.
Thus we arrive at the new sum rule for the twist four part of the
longitudinal structure function
\begin{eqnarray}
\int_0^1 { dx \over x} F_L^{\tau=4} = 4 {M^2 \over Q^2}. \label{c5flsr}   
\end{eqnarray}
To our knowledge, this is the first sum rule at the twist four level of deep
inelastic scattering or for QCD in general.
\vskip .2in
%%%%%%%%%%%%%%%%%%%%%%%%%%%%%%%%%%%%
\section{Dressed Parton Calculations}
%%%%%%%%%%%%%%%%%%%%%%%%%%%%%%%%%%%%
\subsection{Dressed Quark with Zero Mass}
%%%%%%%%%%%%%%%%%%%%%%%%%%%%%%%%%%%%%%%%%%%%%%%%

Next, we investigate the implications of Eq. (\ref{c5flsr1})
for quadratic divergences in $F_{L(q)}^{\tau=4}$.
For simplicity, we select a dressed quark target 
and evaluate the structure functions 
to order $g^2$. That is, we take the state 
$ \mid P \rangle$ to be a dressed quark
consisting of bare states of a quark and a quark plus 
a gluon:
\begin{eqnarray}
\mid P, \sigma \rangle &=&  \phi_1 b^\dagger(P,\sigma) \mid 0 \rangle
\nonumber \\  
&& + \sum_{\sigma_1,\lambda_2} \int 
{dk_1^+ d^2k_1^\perp \over \sqrt{2 (2 \pi)^3 k_1^+}}  
\int 
{dk_2^+ d^2k_2^\perp \over \sqrt{2 (2 \pi)^3 k_2^+}}  
\sqrt{2 (2 \pi)^3 P^+} \delta^3(P-k_1-k_2) \nonumber \\
&& ~~~~~\phi_2(P,\sigma \mid k_1, \sigma_1; k_2 , \lambda_2) b^\dagger(k_1,
\sigma_1) a^\dagger(k_2, \lambda_2) \mid 0 \rangle. 
\end{eqnarray} 

Here we have truncated the expansion at the quark-gluon level. 
We introduce the Jacobi momenta, $(x_i,\kappa_i^\perp)$, where,
\be
k_i^+=x_iP^+, ~~~~~~~~~~~~~~k_i^\perp=\kappa_i^\perp+x_iP^\perp
\e
so that
\be
\sum x_i=1, ~~~~~~~~~~~~~~~~~~~~~\sum \kappa^\perp_i=0.
\label{c5reljac}
\e
The amplitudes $\phi_1, \phi_2$ are related to the corresponding boost
invariant amplitudes as
\be
\phi_1=\psi_1, ~~~~~~\sqrt {P^+}\phi_2(k_i^+,k_i^\perp)=
\psi_2(x_i,\kappa_i^\perp).
\e
Using the notation, $x=x_1, \kappa^\perp_1=\kappa^\perp$ and Eq.
(\ref{c5reljac}) we have
\be
\lefteqn{\psi_2^{s,\lambda}(x,\kappa^\perp)={1\over
{[M^2-{{m^2+(\kappa^\perp)^2}\over x}- {(\kappa^\perp)^2\over {1-x}}]}}}
\nonumber\\& &{g\over {\sqrt {2(2\pi)^3}}}T^a {1\over {\sqrt {1-x}}} 
[ -2 {\kappa^\perp\over {1-x}}-{{\sigma^\perp \cdot \kappa^\perp-im}\over x}
\sigma^\perp-im \sigma^\perp]\epsilon^{\perp*}_\lambda.\label{c5psi2}
\e
Here $M$ and $m$ are the masses of the dressed and bare quarks respectively.
 The state is
normalized to one, and from the normalization condition, we obtain
\be
{\mid \psi_1 \mid}^2=1-{\alpha_s\over {2 \pi}} C_f
\int_\epsilon^{1-\epsilon} dx {{1+x^2}\over {1-x}}{\rm ln}
{Q^2\over \mu^2}
\label{c5nq}
\e
within order $\alpha_s$.  Here $\epsilon$ is a small cutoff on $x$.

First we work in the massless limits for the
dressed and bare quarks. 
\begin{eqnarray}
{\cal M}_1 &=& { 1 \over \pi Q^2 } \int dy^- e^{-{ i \over 2}P^+ y^-x} \langle P
\mid {\psi^{+}}^\dagger(y^-) \nonumber \\
&&~~~~~~~ \big [ i \partial^\perp + g A^\perp(y) \big ].\alpha^\perp
\alpha^\perp . \big [ i \partial^\perp + g A^\perp(0)  \big ] \psi^+(0)
\mid P \rangle, \label{c5fls1} 
\end{eqnarray}
and 
\begin{eqnarray}
{\cal M}_2 = - 4{ (P^\perp)^2 \over Q^2}   x F_{2(q)}(x).
\label{c5fls2}
\end{eqnarray}
As mentioned before, in this case, it is sufficient to keep only the first
term of the bilocal operator.
Note that the matrix elements appearing in Eqs. (\ref{c5fl1}), (\ref{c5fl2}), 
(\ref{c5thetaqf}), and (\ref{c5thetagf})
involve products of operators. We treat them as normal ordered which is sufficient for our
purposes here.
However, in doing so, the terms we drop are quadratically divergent and they will affect
only the counterterm structure.

%We treat the bilocal operators appearing in Eqs. (\ref{c5fl1}) and (\ref{c5fl2})
%with the normal-ordered prescription. 
We express the operators in terms of two component fermion field, $\xi$, and
use the Fock space expansion of the fields $\xi$ and $A^i$.   
First we evaluate the contribution ${\cal M}_2$ given in 
Eq. (\ref{c5fls2}). The operator in this case involves the `good' (plus)
component of the bilocal current which is independent of interaction
and the matrix element receives only diagonal contribution from the single
particle and two particle Fock space sectors. ${\cal M}_2$ is proportional to
$F_2$. A detailed derivation of the
structure function $F_2$ for a dressed quark state is given in \cite{c5rajen}.
We obtain
\begin{eqnarray}
{\cal M}_2= -4 C_f{(P^\perp)^2 \over Q^2} x^2 \Big [ \delta(1-x) + {\alpha_s \over 2
\pi} {\rm ln} \Lambda^2 \big [ { 1+x^2 \over 1-x} - \delta(1-x) \int dy {1+y^2 \over 1-y}
\big ] \Big ]\label{c5fl2f}
\end{eqnarray}
where we have cutoff the transverse momentum integral at $\Lambda$.
Note that the result in Eq. (\ref{c5fl2f}) violates transverse boost invariance.
In the above expression, we have used Eqs. (\ref{c5nq}) and (\ref{c5psi2}). Here,
$C_f={{N^2-1}\over {2N}}$ for SU(N).  

${\cal M}_1$ involves the `bad' (minus) component of bilocal current and the
operator depends upon interaction. 
We have, from Eq. ({\ref{c5fls1}),
\begin{eqnarray}
{\cal M}_1 
&=& -{ 1 \over \pi Q^2} \int dy^- e^{-{i \over 2} P^+ y^- x} \langle P \mid
{\psi^{+}}^\dagger(y^-) (\partial^\perp)^2 \psi^+(0) \mid P \rangle
\nonumber \\
&& ~~~+ g{ 1 \over \pi Q^2} \int dy^- e^{-{i \over 2} P^+ y^- x} \langle P \mid
{\psi^{+}}^\dagger(y^-) i \partial^\perp . \alpha^\perp \alpha^\perp.
A^\perp(0) \psi^+(0) \mid P \rangle \nonumber \\
&& ~~~+ g{ 1 \over \pi Q^2} \int dy^- e^{-{i \over 2} P^+ y^- x} \langle P \mid
{\psi^{+}}^\dagger(y^-) \alpha^\perp.A^\perp(y) i \partial^\perp .\alpha^\perp
\psi^+(0) \mid P \rangle \nonumber \\
&& ~~~+ g^2{ 1 \over \pi Q^2} \int dy^- e^{-{i \over 2} P^+ y^- x} \langle P \mid
{\psi^{+}}^\dagger(y^-) A^\perp(y).A^\perp(0) \mid P \rangle \\
&&\equiv {\cal M}_1^{a}+{\cal M}_1^{b}+{\cal M}_1^{c}+{\cal M}_1^{d}.
\end{eqnarray} 
Here, the operator in ${\cal M}_1^a$ only does not depend explicitly upon
interaction and will have diagonal contributions. ${\cal M}_1^b$ and ${\cal
M}_1^c$ explicitly depend upon $g$ and the contributions will come due to
the interference between the one particle and two particle sectors. This
type of off-diagonal contributions are absent in the structure function
$F_2$ and they involve quark-gluon dynamics. 
Since the operators in Eq. (\ref{c5fls1}) are taken to be normal ordered, 
the
contribution of ${\cal M}_1^{d}$ vanishes to order $g^2$.
 
Explicit calculation leads to the diagonal Fock basis contributions  
\begin{eqnarray}
({\cal M}_1)_{diag}=
{\cal M}_1^{a} &=& 4 C_f{(P^\perp)^2 \over Q^2} x^2 \Big [ \delta(1-x) + {\alpha_s
\over 2 \pi} {\rm ln} \Lambda^2 \big [ {1+x^2 \over 1-x} - \delta(1-x) \int dy { 1+ y^2
\over 1-y} \big ] \Big ] \nonumber \\
&& ~~~~+  C_f{g^2 \over 2 \pi^2} {\Lambda^2 \over Q^2}{ 1+x^2 \over 1-x} .
\end{eqnarray}
The first term here explicitly cancels the term ${\cal M}_2$ given in
Eq. (\ref{c5fl2f}), thus the $P^\perp$ dependence goes away. Here, we have
used the expression of $\psi_2$ given by Eq. (\ref{c5psi2}) and Eq. (\ref{c5nq}).  

Off-diagonal contributions will come from ${\cal M}_1^b$ and ${\cal M}_1^c$
and these matrix elements will contain $\psi_2^* \psi_1$ and its Hermitian
conjugate. The following results are obtained after substituting $\psi_2$:
\begin{eqnarray}
({\cal M}_1)_{nondiag}={\cal M}_1^{b}+{\cal M}_1^{c} = 
-C_f{g^2 \over \pi^2} {\Lambda^2 \over Q^2}
{1 \over 1-x}.
\end{eqnarray} 

Adding all the contributions, we have 
\begin{eqnarray}
F^{\tau=4}_{L(q)}(x) = - C_f{ g^2 \over 2 \pi^2} 
{\Lambda^2 \over Q^2}(1+x).
\label{c5fltot}
\end{eqnarray}
%Now the question is, do we understand the answer we obtained in the bare
%theory? Do we know how to renormalize the divergences?

%In the cutoff theory we face the following problems:

As anticipated from power counting, we have generated quadratic
divergences for $F_L$ in the bare theory. This quadratic divergence arises
from $\kappa^\perp$ integration. Now we are faced with two
issues:
(a) What is the principle for adding counterterms? (b) What
determines the finite parts of the counterterms?

Further, since $F^{\tau=4}_{L(q)}$ is directly related to the 
physical longitudinal cross
section, we expect $F^{\tau=4}_{L(q)}$ to be positive definite 
(see, for example, Ref.
\cite{c5efp}). 
%Furthermore we expect $F_L$ to be free from end point
%singularities in the longitudinal space. 
From our answers we see that we get a negative answer which is free 
from end point singularities but is quadratically divergent. 
Note that it is easy to show from our expressions in Eqs. (\ref{c5fl1}) and
(\ref{c5fl2})
for a free quark of mass $m$,
$F^{\tau=4}_{L(q)}=4 {m^2 \over Q^2} \delta(1-x) $, a
well-known result. 
%If we include the normal ordering contribution, we formally
%get a positive answer which however suffers from the end point singularity.
%Thus the answers we have generated are both ambiguous and unacceptable.

Now, the question is whether we can invoke the sum rule to get insight into the
answers we obtained in the bare theory and also determine the
counterterm structures to be added for renormalization.
To investigate whether our sum rule can throw some light on this problem
first we have to check the validity of the sum rule to order $g^2$ in
perturbative QCD. Towards this goal 
we have to evaluate
the expectation value of the $+-$ component of the energy-momentum 
tensor which we consider in the following.

Now we show that the sum rule relating the integral of ${F^{\tau=4}_{L(q)}(x)
\over x}$ to the fermionic part of the light-front Hamiltonian density helps
us to understand the results we obtained for $F_L$ for a dressed quark in
the bare theory and indicates how to add counterterms to renormalize 
the theory.

%We again consider the operators to be normal ordered.  
%We have
%\begin{eqnarray}
%F_L^{F-d} = {g^2 \over 2 \pi^2} { 1 + x^2 \over 1-x} {\Lambda^2 \over Q^2}, ~~
%F_L^{F-o-d} =  - {g^2 \over \pi^2 } { 1 \over 1-x} {\Lambda^2 \over Q^2}. 
%\end{eqnarray}
%Using the sum rule, Eq..(\ref{c5flsr}), we get,
A straightforward evaluation leads to
\begin{eqnarray}
\langle P \mid \theta^{+-}_q(0) \mid P \rangle_{nondiag}  = 
-C_f{ 1 \over 2}{g^2
\over \pi^2} \Lambda^2  \int
{dx \over x} { 1 \over 1-x}, \label{c5tfode}
\end{eqnarray}

\begin{eqnarray} 
\langle P \mid \theta^{+-}_q(0) \mid P \rangle_{diag} - 
{(P^\perp)^2 \over (P^+)^2}
\langle P \mid \theta^{++}_{q}(0) \mid P \rangle_{diag} = C_f{ 1 \over 2}{g^2
\over 2 \pi^2} \Lambda^2  \int
{dx \over x} { 1  + x^2 \over 1-x}.\label{c5tfde}
\end{eqnarray}

Adding the diagonal and off-diagonal contributions from the fermionic part
of the Hamiltonian density, we arrive at
\begin{eqnarray}
{2 \over Q^2} \Big [ \langle P \mid \theta^{+-}_q(0) \mid P \rangle 
- {(P^\perp)^2 \over (P^+)^2}\langle P \mid \theta^{++}_q(0) \mid P \rangle  \Big ] =
- C_f { g^2 \over 2 \pi^2} 
{\Lambda^2 \over Q^2}\int { dx \over x} (1+x).
\label{c5thetatot}
\end{eqnarray} 
In all the $x$ integrations we have taken a small $x$ cutoff $\epsilon$. 
Comparison of Eqs. (\ref{c5fltot}) and (\ref{c5thetatot}) immediately shows that
the relation given in Eq. (\ref{c5flsr1}) has been successfully tested to order $g^2$ in perturbative
QCD in the bare theory.

It is now clear that the quadratic divergences we have generated in 
$F^{\tau=4}_{L(q)}$ are
directly related to the fermion mass shift due to the 
fermionic part
of the light-front Hamiltonian density in perturbation theory. Since we want
the renormalized fermion mass to be zero, we need to add counterterms to the 
Hamiltonian density. This precisely dictates the counterterms to be added to
$F^{\tau=4}_{L(q)}$ in order to renormalize via the relation given in Eq.
(\ref{c5flsr1}).
 
To complete the discussion of the fermion mass shift, we now consider the 
contributions from the gluonic part of the Hamiltonian
density. 
%We drop the normal ordering contributions.

Take the off-diagonal contribution arising from 
\begin{eqnarray}
{\cal M}_3= -\langle P \mid 4 g { 1 \over \partial^+} \Big ( \partial^i
A^{ia}
\Big ) {\psi^{+}}^\dagger T^a \psi^+ \mid P \rangle.
\end{eqnarray}
A straightforward evaluation leads to 
\begin{eqnarray}
{\cal M}_3 = -C_f{g^2 \over 2 \pi^2}
\Lambda^2 \int { dx \over x} \Big [ { 2x^2 \over (1-x)^2} + {x \over 1-x}
\Big ]. \label{c5tgode}
\end{eqnarray}
Thus from Eqs. (\ref{c5tfode}) and (\ref{c5tgode}) we have
\begin{eqnarray}
\langle P \mid \Big [ \theta^{+-}_{q} (0) + \theta^{+-}_{g} (0) \Big
] \mid P
\rangle_{nondiag} = - C_f{ g^2 \over 2 \pi^2} \Lambda^2 \int { dx \over x} {1 + x^2
\over (1-x)^2}. \label{c5todtot}
\end{eqnarray}

Next let us consider diagonal contributions.
From the gauge boson part of the Hamiltonian density, we have,
\begin{eqnarray}
\langle P \mid  \theta^{+-}_{g} (0)\mid P \rangle_{diag} 
 - {(P^\perp)^2 \over (P^+)^2} \langle P \mid  \theta^{++}_{g}(0)  
\mid P \rangle_{diag} 
= C_f{ g^2 \over 4 \pi^2} \Lambda^2 \int { dx \over 1-x}{1 + x^2
\over 1-x}. \label{c5tgde}
\end{eqnarray}
Adding the diagonal contributions from the fermion and gauge boson parts,
i.e., Eqs. (\ref{c5tfde}) and (\ref{c5tgde}) we
arrive at
\begin{eqnarray}
\langle P \mid \theta^{+-}(0) \mid P \rangle_{diag} - {(P^\perp)^2 \over
(P^+)^2} \langle P
\mid \theta^{++}(0) \mid P \rangle_{diag} = C_f{ g^2 \over 4 \pi^2} \Lambda^2
\int { dx \over x} {1+x^2 \over (1-x)^2}. \label{c5tdtot}
\end{eqnarray}
Thus the total contribution (quark and gluons) to the expectation
value of the Hamiltonian density is given by
\begin{eqnarray}
\langle P \mid  \theta^{+-} (0) \mid P \rangle - 
{(P^\perp)^2 \over (P^+)^2} \langle P \mid \theta^{++}(0) 
\mid P \rangle  = - C_f{ g^2 \over 4 \pi^2} \Lambda^2 \int {dx \over x} { 1+x^2 \over
(1-x)^2}.
\end{eqnarray}
This result is directly related to the mass shift of the fermion to order
$g^2$ in light-front perturbation theory. In the $x^+$ ordered perturbation
theory, upto order $g^2$, contribution to the mass shift of the quark comes
from three diagrams, 
(a) one gluon exchange interaction, (b) instantaneous fermion interaction,
(c) instantaneous gluon interaction. It can be shown that, the contribution
to the fermion mass shift due to (a) in the limit of zero bare quark 
mass is, (see Eq. (4.10) in Ref. \cite{c5qcd2}),    
\begin{eqnarray}
\delta p_1^- = -{ 1 \over 2 P^+} C_f{g^2 \over 4 \pi^2} \Lambda^2 \int
{dx \over x} {1+x^2 \over (1-x)^2}.
\end{eqnarray}

Note that in the massless limit we encountered only quadratic divergences in the twist
four part of the longitudinal structure function. The case of a massive
quark is discussed in the next section.

%%%%%%%%%%%%%%%%%%%%%%%%%%%%%%%%%%%%%%%%%%%%%%%%%
\subsection{Dressed Quark with Non-zero Mass}
%%%%%%%%%%%%%%%%%%%%%%%%%%%%%%%%%%%%%%%%%%%%%%%%%

We select the target to be a massive dressed quark 
and evaluate the structure functions 
to order $g^2$. That is, we take the state 
$ \mid P \rangle$ to be a dressed quark
consisting of bare states of a quark and a quark plus 
a gluon, as before 
\begin{eqnarray}
\mid P, \sigma \rangle &=& \phi_1 b^\dagger(P,\sigma) \mid 0 \rangle
\nonumber \\  
&& + \sum_{\sigma_1,\lambda_2} \int 
{dk_1^+ d^2k_1^\perp \over \sqrt{2 (2 \pi)^3 k_1^+}}  
\int 
{dk_2^+ d^2k_2^\perp \over \sqrt{2 (2 \pi)^3 k_2^+}}  
\sqrt{2 (2 \pi)^3 P^+} \delta^3(P-k_1-k_2) \nonumber \\
&& ~~~~~\phi_2(P,\sigma \mid k_1, \sigma_1; k_2 , \lambda_2) b^\dagger(k_1,
\sigma_1) a^\dagger(k_2, \lambda_2) \mid 0 \rangle. 
\end{eqnarray} 

In the last section, 
we have shown that the twist four longitudinal structure function  has 
quadratic
divergences in perturbation theory. In this section, we show that for a
massive quark, in addition to quadratic divergences, logarithmic divergences
are generated.
The state is normalized to one. As before, we introduce the Jacobi momenta.
We have
\begin{eqnarray}
F_L = {\cal M}_1 + {\cal M}_2
\end{eqnarray}
where
%The expression for
%$F_L$ is still given in terms of ${\cal M}_1$ and ${\cal M}_2$ with
\begin{eqnarray}
{\cal M}_1 &=& { 1 \over \pi Q^2 } \int dy^- e^{-{ i \over 2}P^+ y^-x} \langle P
\mid {\psi^{+}}^\dagger(y^-) \nonumber \\
&&\Big [ \alpha^\perp. \big  [ i \partial^\perp + g A^\perp(y)
\big ] + \gamma^0 m  \Big ] \Big [ 
\alpha^\perp . \big [ i \partial^\perp + g A^\perp(0) \big ] + 
\gamma^0 m \Big ] \psi^+(0) \nonumber\\&&~~~~~~~~+ h.c.
\mid P \rangle, \label{c5flm} 
\end{eqnarray}
and 
\begin{eqnarray}
{\cal M}_2 = - 4{ (P^\perp)^2 \over Q^2}   x F_{2(q)}(x),
\label{c5fls2m}
\end{eqnarray}
where $\mid P \rangle$ now has a mass $M$ and $m$ is the bare quark mass.

In the case of quark contributions, the second term in the expression for
the bilocal current in Eq. (\ref{c5bi}) vanishes. 
First we evaluate the contribution ${\cal M}_2$ given in 
Eq. (\ref{c5fls2m}).
We obtain
\begin{eqnarray}
{\cal M}_2&=&~ -4 {(P^\perp)^2 \over Q^2} x^2 \Big [ \delta(1-x) 
+ {g^2 \over 8
\pi^3} C_f \Big (\int d^2k_\perp  {{ 1+x^2 \over 1-x} k^2_\perp + (1-x)^3m^2
\over [m^2(1-x)^2 +k^2_\perp]^2}\nonumber\\&&~~~~- \delta(1-x) \int dyd^2k_\perp 
{{ 1+y^2 \over 1-y} k^2_\perp + (1-y)^3m^2
\over [m^2(1-y)^2 + k^2_\perp]^2}
\Big ) \Big ],\label{c5flf}
\end{eqnarray}
where $C_f = {N^2 - 1 \over 2N}$ for $SU(N)$.

Here we have presented the result without working out the transverse
integration to maintain a greater degree of transparency.

It is to be mentioned that, to calculate all the matrix elements involved in
$F_L$, one has to use the expression of $\psi_2$ given in Eq. (\ref{c5psi2}). 
It can be easily seen from Eq. (\ref{c5psi2}) that the term which in linear in 
quark mass $m$ causes helicity flip of the quark. Linear mass terms do not
contribute in $F_L$. However, in the two-particle sector, the diagonal part
of the matrix element involves $\psi_2^* \psi_2$ and contribution comes from
the terms quadratic in $m$.  
 
The contribution from ${\cal M}_1$ is split into four parts 
with additional contributions coming from quark mass terms
and can be written as follows.
\begin{eqnarray}
{\cal M}_1 
& =& { 1 \over \pi Q^2} \int dy^- e^{-{i \over 2} P^+ y^- x} \langle P \mid
{\psi^{+}}^\dagger(y^-)\big ( -(\partial^\perp)^2 + m^2\big)\psi^+(0) \mid P 
\rangle\nonumber \\
&& ~~~+ g{ 1 \over \pi Q^2} \int dy^- e^{-{i \over 2} P^+ y^- x} \langle P \mid
{\psi^{+}}^\dagger(y^-) (i \partial^\perp . \alpha^\perp +\gamma^0m)\alpha^\perp.
A^\perp(0) \psi^+(0) \mid P \rangle \nonumber \\
&& ~~~+ g{ 1 \over \pi Q^2} \int dy^- e^{-{i \over 2} P^+ y^- x} \langle P \mid
{\psi^{+}}^\dagger(y^-) \alpha^\perp.A^\perp(y) (i \partial^\perp .\alpha^\perp
+\gamma^0 m)\psi^+(0) \mid P \rangle \nonumber \\
&& ~~~+ g^2{ 1 \over \pi Q^2} \int dy^- e^{-{i \over 2} P^+ y^- x} \langle P \mid
{\psi^{+}}^\dagger(y^-) A^\perp(y).A^\perp(0) \psi^+(0)\mid P \rangle \\
&&\equiv {\cal M}_1^{a}+{\cal M}_1^{b}+{\cal M}_1^{c}+{\cal M}_1^{d}.
\label{c5mtot}
\end{eqnarray} 
Since the operators in Eq. (\ref{c5flm}) are taken to be normal ordered, 
the
contribution of ${\cal M}_1^{d}$ vanishes to order $g^2$.

In this case also, ${\cal M}_1^a$ gives diagonal contribution and
 ${\cal M}_1^b$ and ${\cal M}_1^c$ give off-diagonal contribution.  
Explicit calculation leads to the diagonal Fock basis contributions  
\begin{eqnarray}
({\cal M}_1)_{diag}=
{\cal M}_1^{a} &=&4 {(P^\perp)^2 \over Q^2} x^2 \Big [ \delta(1-x) 
+ {g^2 \over 8
\pi^3} C_f \Big ( \int d^2k_\perp  {{ 1+x^2 \over 1-x} k^2_\perp + (1-x)^3m^2
\over [m^2(1-x)^2 + k^2_\perp]^2}\nonumber\\&& ~~~
 - \delta(1-x) \int dyd^2k_\perp 
{{ 1+y^2 \over 1-y} k^2_\perp + (1-y)^3m^2
\over [m^2(1-y)^2 + k^2_\perp]^2} \Big )
\Big ]\nonumber\\&&+
{4 m^2 \over Q^2}\delta(1-x)\Big [1- C_f {g^2 \over 8
\pi^3} \int dyd^2k_\perp  {{ 1+y^2 \over 1-y} k^2_\perp + (1-y)^3m^2
\over [m^2(1-y)^2 + k^2_\perp]^2}\Big ]\nonumber\\&& +
{4 C_f \over Q^2} {g^2 \over 8
\pi^3} \int d^2k_\perp (k^2_\perp + m^2) {{ 1+x^2 \over 1-x} k^2_\perp + 
(1-x)^3m^2 \over [m^2(1-x)^2 + k^2_\perp]^2}.
\end{eqnarray}.

The first term here explicitly cancels the term ${\cal M}_2$ given in
Eq. (\ref{c5flf}).

Off-diagonal contributions
\begin{eqnarray}
({\cal M}_1)_{nondiag}={\cal M}_1^{b}+{\cal M}_1^{c} &=& 
{C_f \over Q^2}{g^2 \over \pi^3} \Big [\delta(1-x) \int dyd^2k_\perp 
{m^2(1-y) \over [m^2(1-y)^2
+k^2_\perp]}\nonumber\\&&~~
 - \int d^2k_\perp {k^2_\perp +
m^2(1-x)^2 \over (1-x)[m^2(1-x)^2 + k^2_\perp]}\Big].
\end{eqnarray} 

The linear in $m$
terms in ${\cal M}_1^b$ and ${\cal M}_1^c$ receives contributions from the
mass dependent term in $\psi_2$. As a result, in the final expressions, these are
quadratic in $m$. 

Adding all the contributions, we have 
\begin{eqnarray}
F^{\tau=4}_{L(q)}(x) &=& {4 m^2 \over Q^2} \delta(1-x) 
+ {4 C_f \over Q^2} {g^2 \over 8
\pi^3}\Big [ \int d^2k_\perp (k^2_\perp + m^2) 
{{ 1+x^2 \over 1-x} k^2_\perp + 
(1-x)^3m^2 \over [m^2(1-x)^2 + k^2_\perp]^2}\nonumber\\&&-\delta(1-x) 
m^2 \int dyd^2k_\perp {{ 1+y^2 \over 1-y} k^2_\perp + (1-y)^3m^2
\over [m^2(1-y)^2 + k^2_\perp]^2}\Big ] 
\nonumber\\&&~~~~~~~~~~~~~- {C_f \over Q^2}{g^2 \over \pi^3}\Big [
\int d^2k_\perp {k^2_\perp +
m^2(1-x)^2 \over (1-x)[m^2(1-x)^2 + k^2_\perp]}
\nonumber\\&&~~~~~~~~~~~~~~~~~-\delta(1-x) \int dyd^2k_\perp {m^2(1-y) 
\over [m^2(1-y)^2 + k^2_\perp]}\Big].
\label{c5fltotm}
\end{eqnarray}
Here we have used $M=m$, since the difference that it entails is higher
order in the coupling. Note that we are getting back the free quark answer
once we switch off the interaction. Also, the dressed
mass-less quark answer can be
easily regenerated by putting $M=m=0$. Note that the $k_\perp$-integration
now produces logarithmic divergences with the expected quadratic one.

To check the sum rule explicitly, we evaluate the right hand side of Eq. (\ref{c5flsr1}) next.
A straightforward evaluation leads to
\begin{eqnarray}
\langle P \mid \theta^{+-}_q(0) \mid P \rangle_{nondiag}  = 
-C_f{g^2 \over 2\pi^3} \int dxd^2k_\perp{k^2_\perp +
m^2(1-x)^3 \over x(1-x)}{1 \over [m^2(1-x)^2 + k^2_\perp]}
\end{eqnarray}
\begin{eqnarray} 
\langle P \mid \theta^{+-}_q(0) \mid P \rangle_{diag} - 
{(P^\perp)^2 \over (P^+)^2}
\langle P \mid \theta^{++}_{q}(0) \mid P
\rangle_{diag} =\nonumber\\~~~ 
2m^2 + 
2 C_f {g^2 \over 8
\pi^3} \int dxd^2k_\perp {k^2_\perp +(1-x) m^2 \over x} {{ 1+x^2 \over 1-x} k^2_\perp + 
(1-x)^3m^2 \over [m^2(1-x)^2 + k^2_\perp]^2}.
\end{eqnarray}
Adding the diagonal and off-diagonal contributions from the fermionic part
of the Hamiltonian density and multiplying it by ${2 \over Q^2}$ one obtains
the right hand side of the sum rule. Comparing it with the integral of ${F_L^{\tau=4} 
\over x}$, where $F_L$ is given in Eq. (\ref{c5fltotm}), one easily sees that 
the sum rule is verified. 
%Calculating $\langle P \mid \theta^{+-}_g(0) 
%\mid P \rangle$ and $\langle P \mid \theta^{++}_g(0) 
%\mid P \rangle$ for the massive quark and proceeding exactly through the 
%same steps as in the last
%section, we again see the connection of $F_L$ with the fermion mass shift,
%as we did in the case of mass-less quark.

To see the connection of $F_L$ with the fermionic mass shift, we calculate
the contribution of the gluonic part of the energy momentum 
tensor $\theta^{+-}$ to
the sum rule for the total $F_L$. Explicit calculation gives
\begin{eqnarray}
\langle P \mid \theta^{+-}_g(0) \mid P \rangle_{nondiag}  = 
-C_f{g^2 \over 2\pi^3} \int dxd^2k_\perp{(1+x) k_\perp^2 \over (1-x)^2}{1
\over
[m^2(1-x)^2 + k^2_\perp]}
\end{eqnarray}
\begin{eqnarray} 
\lefteqn{\langle P \mid \theta^{+-}_g(0) \mid P \rangle_{diag} - 
{(P^\perp)^2 \over (P^+)^2}
\langle P \mid \theta^{++}_g(0) \mid P
\rangle_{diag} =}\nonumber\\&&  
2 C_f {g^2 \over 8
\pi^3} \int dxd^2k_\perp {k^2_\perp \over (1-x)} {{ 1+x^2 
\over 1-x} k^2_\perp + 
(1-x)^3m^2 \over  [m^2(1-x)^2 + k^2_\perp]^2}.
\end{eqnarray}
Thus, we get
\begin{eqnarray}
\langle P \mid \theta^{+-}_q(0)+\theta^{+-}_g(0) \mid P \rangle_{nondiag}
= \nonumber \\
-C_f{g^2 \over 2\pi^3} \int dxd^2k_\perp{{(1+x^2) \over 1-x}k_\perp^2 
+(1-x)^3m^2 \over x(1-x)}{1
\over
[m^2(1-x)^2 + k^2_\perp]},
\end{eqnarray}
\begin{eqnarray} 
\langle P \mid \theta^{+-}_q(0)+\theta^{+-}_g(0) \mid P \rangle_{diag} - 
{(P^\perp)^2 \over (P^+)^2}
\langle P \mid \theta^{++}+\theta^{++}_g(0) \mid P
\rangle_{diag} = \nonumber\\  
2 C_f {g^2 \over 8
\pi^3} \int dxd^2k_\perp {{(1+x^2) \over 1-x}k_\perp^2 
+(1-x)^3m^2 \over x(1-x)}{1
\over
[m^2(1-x)^2 + k^2_\perp]}.
\end{eqnarray}
Adding diagonal and off-diagonal contributions, we get,
\begin{eqnarray} 
\langle P \mid \theta^{+-}(0) \mid P \rangle - 
{(P^\perp)^2 \over (P^+)^2}
\langle P \mid \theta^{++}(0) \mid P
\rangle = \nonumber\\~~~  
- C_f {g^2 \over 4
\pi^3} \int dxd^2k_\perp {{(1+x^2) \over 1-x}k_\perp^2 
+(1-x)^3m^2 \over x(1-x)}{1
\over
[m^2(1-x)^2 + k^2_\perp]}.
\end{eqnarray}
 
Note that this result is connected to the full 
fermion mass shift $ \delta p_1^-$ in 
second order perturbation theory. We have (see Eq. (4.10)) in Ref.
\cite{c5qcd2}, in the massive case,
\begin{eqnarray}
\delta p_1^- = - { 1 \over 2 P^+} C_f {g^2 \over 4 \pi^3} \int dx d^2 k_\perp 
{ {(1+x^2) \over 1-x} k_\perp^2 + (1-x)^3 m^2  \over x(1-x)}
{ 1 \over [m^2 (1-x)^2 + k_\perp^2]}.
\end{eqnarray}

%%%%%%%%%%%%%%%%%%%%%%%%%%%%%%%%%%%%%%%%%%%%%%%%%%%%%
\subsection{Dressed Gluon}
%%%%%%%%%%%%%%%%%%%%%%%%%%%%%%%%%%%%%%%%%%%%%%%%%%%%%
In this section we check the sum rule explicitly for a dressed gluon target.
 We consider the gluon to be composed of a bare gluon and a
quark anti-quark pair.
\begin{eqnarray}
\mid P, \sigma \rangle &=& \phi_1 a^\dagger(P,\lambda) \mid 0 \rangle
\nonumber \\  
&& + \sum_{\sigma_1,\sigma_2} \int 
{dk_1^+ d^2k_1^\perp \over \sqrt{2 (2 \pi)^3 k_1^+}}  
\int 
{dk_2^+ d^2k_2^\perp \over \sqrt{2 (2 \pi)^3 k_2^+}}  
\sqrt{2 (2 \pi)^3 P^+} \delta^3(P-k_1-k_2) \nonumber \\
&& ~~~~~\phi_2(P,\sigma \mid k_1, \sigma_1; k_2 , \sigma_2) b^\dagger(k_1,
\sigma_1) d^\dagger(k_2, \sigma_2) \mid 0 \rangle.
\label{c5glu} 
\end{eqnarray}

$\phi_1$ is the probability amplitude of finding one bare gluon and
$\phi_2$ is the  probability amplitude of finding one quark-antiquark pair in
the dressed gluon. We introduce the Jacobi momenta $x, \kappa^\perp$.
The boost invariant amplitude $\psi_2$ is written in terms of $\psi_1$ from
the light-front bound state equation (see chapter 3):
\be
\psi_2(x, \kappa^\perp, 1-x, -\kappa^\perp)= {g\over {2 (2\pi)^3}} {x
(1-x)\over {(\kappa^\perp)^2}} \chi^\dagger_{\sigma'} 
\Big [ -{(\sigma^\perp \cdot \kappa^\perp) \over x} + {\sigma^\perp \cdot
(\sigma^\perp \cdot \kappa^\perp)\over {1-x}}\Big ] \chi_\sigma
\epsilon_\lambda^\perp \psi_1.
\label{c5psiglu}
\e  
From this, we get
\be
\sum_{spins} \int d^2\kappa^\perp {\mid \psi_2 \mid}^2 = {g^2\over (2
\pi)^3} \int {d^2\kappa^\perp \over (\kappa^\perp)^2} [ x^2+(1-x)^2] {\mid
\psi_1 \mid}^2.
\label{c5norglu}
\e
The target gluon and the bare quark and anti-quark masses are taken to be
zero. Note that, to the order $g^2$, there will be a contribution from
the two-gluon Fock sector due to the non-abelian nature of the gauge coupling.
For simplicity, we exclude that contribution. It is easy to
incorporate that contribution by 
trivially extending our calculation presented here.

$F_L$ can be written in terms of ${\cal M}_1$ and ${\cal M}_2$ given 
in Eqs. (\ref{c5flm}-\ref{c5fls2m}),
where $\mid P \rangle$ now stands for the dressed gluon represented by
Eq. (\ref{c5glu}). In this case, contribution comes from both components of
bilocal current. It can be easily seen that $F_L$ for a free gluon is zero. 
Explicit calculation gives
\begin{eqnarray}
{\cal M}_2&=&~ -4 {(P^\perp)^2 \over Q^2} x F_{2(q)}^{dressed-gluon}\nonumber\\
&=&~ - {x^2(P^\perp)^2 \over Q^2} {g^2 \over \pi^2}N_f T_f 
\big[ x^2 + (1-x)^2 \big] {\rm ln} \Lambda^2 .
\end{eqnarray}
Here $T_f={1 \over 2}$ and $N_f$ is the number of flavors.

${\cal M}_1$ is again divided into four parts as in Eq. (\ref{c5mtot}) and explicit
calculation in this case gives the following.
\begin{eqnarray}
{\cal M}_{1(diag)} ={\cal M}_{1(a)}&=&{x^2(P^\perp)^2 \over Q^2} 
{g^2 \over \pi^2}N_f T_f 
\big[ x^2 + (1-x)^2 \big] {\rm ln} \Lambda^2 \nonumber\\&&~~~
+{\Lambda^2 \over Q^2}{g^2\over \pi^2}N_f 
T_f \big[ x^2 + (1-x)^2 \big]\label{c5gl1},
\end{eqnarray}
\begin{eqnarray}
{\cal M}_{1(off-diag)}&=&{\cal M}_{1(b)}+{\cal M}_{1(c)}\nonumber\\
&=&-{\Lambda^2 \over Q^2}{g^2\over \pi^2}N_f T_f 2 (1-x).
\label{c5gl2}
\end{eqnarray}
In the above results, we have used Eq. (\ref{c5psiglu}) and Eq. (\ref{c5norglu}). 
Thus, we get
\begin{eqnarray}
F_L={\Lambda^2 \over Q^2} N_fT_f { g^2 \over \pi^2} \big[x^2 +(1-x)^2
-2(1-x) \big].
\label{c5flgl}
\end{eqnarray}
On the other hand, we get
\begin{eqnarray} 
\lefteqn{\langle P \mid \theta^{+-}_q(0) \mid P \rangle_{diag} - 
{(P^\perp)^2 \over (P^+)^2}
\langle P \mid \theta^{++}_{q}(0) \mid P \rangle_{diag}
=}\nonumber\\&&~~~~~~~~~~~~~~~
\Lambda^2 N_f T_f {g^2 \over 4\pi^2}\int dx \big[ {x^2 +(1-x)^2 \over
x(1-x)}\big ]
\label{c5thd}
\end{eqnarray}
and
\begin{eqnarray}
\langle P \mid \theta^{+-}_q(0) \mid P \rangle_{off-diag} = 
- \Lambda^2 N_f T_f {g^2 \over 2\pi^2}\int dx \big[ {x^2 +(1-x)^2 \over
x(1-x)}\big]
\label{c5thod}.
\end{eqnarray}
Adding diagonal and off-diagonal contributions, we get
\begin{eqnarray} 
\langle P \mid \theta^{+-}_q(0) \mid P \rangle - 
{(P^\perp)^2 \over (P^+)^2}
\langle P \mid \theta^{++}_{q}(0) \mid P \rangle =-
\Lambda^2 N_f T_f {g^2 \over 4\pi^2}\int dx \big[ {x^2 +(1-x)^2 \over
x(1-x)}\big]
\label{c5thgl}.
\end{eqnarray}

Note that this result is connected
to the gluonic mass shift $\delta q_2^-$ due to pair production, since the
contribution from the gluonic part of the energy-momentum tensor
$\theta^{+-}_g$ in this case vanishes. 
In the massless limit, we have (see Eq. (4.40) in Ref. (\cite{c5qcd2}), gluonic
mass shift due to pair production is
\begin{equation}
\delta q_2^- = -{1 \over 2P^+} 
\Lambda^2 N_fT_f{g^2 \over 4\pi^2}\int dx \big[ {x^2 +(1-x)^2 \over 
x(1-x)}\big].
\end{equation}
From Eq. (\ref{c5flgl}) we compute $\int dx {F_L \over x}$. Since
$x$-integration is from $0$ to $1$, it can be written in the following form.
\begin{eqnarray}
\int dx {F_L \over x} =- 
{\Lambda^2 \over Q^2} N_fT_f{g^2 \over 2\pi^2}\int dx \big[ {x^2 +(1-x)^2 \over 
x(1-x)}\big].\label{c5flint}
\end{eqnarray}
Comparing  Eq. (\ref{c5thgl}) and Eq. (\ref{c5flint}), one explicitly 
verifies
the sum rule for a dressed gluon target.

As we have emphasized, in the bare theory, the twist four longitudinal
structure function is afflicted with divergences. We have to add 
counterterms to carry out the renormalization procedure so that we have
physical answers. The sum rule for the bare theory clearly shows that the
quadratic divergences generated are directly related to the gluon mass shift
in second order light-front perturbation theory arising from an intermediate
quark anti-quark pair. As we mentioned earlier, in light-front Hamiltonian
perturbation theory one cannot automatically generate a
massless gluon by dimensional regularization. In order to ensure a massless 
gluon in second order
perturbation theory, we have to add the negative of the shift as a counterterm.
After adding the counterterm, the gluon mass shift in second order perturbation
theory is zero and the twist four longitudinal structure function for a
massless gluon becomes zero. Thus, after renormalization, the sum rule is
satisfied, with a trivial (i.e., zero) gluon longitudinal structure
function.

%%%%%%%%%%%%%%%%%%%%%%%%%%%%%%%%%%%%%%%%%%%%%%%%%%%%%%
\section{Partition  of the Hadron Mass in QCD}
%%%%%%%%%%%%%%%%%%%%%%%%%%%%%%%%%%%%%%%%%%%%%%%%%%%%%%
As is well-known, experiments that measure the twist-two part of the $F_2$
structure function yield information on the fraction of longitudinal momenta
carried by the charged parton  constituents of the
hadron (quarks and anti-quarks). The sum rule 
we have derived yields other useful information about
the hadron structure. Namely, our sum rule shows that experiments to measure
the twist four part of the longitudinal structure function will directly reveal the
fraction of the hadron mass carried by charged parton components of the hadron.
This sum rule is richer in content since it involves dynamics.
The light-front Hamiltonian provides theoretical insight into  this
fraction as follows.

According to our analysis, the twist four part of the longitudinal structure
function is directly related to the fermionic part of the light-front QCD
Hamiltonian density $\theta^{+-}_q$ in the gauge $A^+_a=0$.
explicitly we have 
\begin{eqnarray} 
\theta^{+-}_q = 2 {\psi^{+}}^\dagger \Big [ \alpha^\perp.(i \partial^\perp + g A^\perp) 
+ \gamma^0 m \Big ]
{ 1 \over i \partial^+} \Big [ \alpha^\perp . (i \partial^\perp + g A^\perp)
+ \gamma^0 m \Big ] \psi^{+}.
\end{eqnarray} 
Thus we have the fermion kinetic energy contribution given by
\begin{eqnarray}
\theta^{+-}_{q(free)} = 2 {\psi^+}^\dagger \big [ - (\partial^\perp)^2 + m^2
\big ] \psi^+
\end{eqnarray}
and the interaction dependent part given by
\begin{eqnarray}
\theta^{+-}_{q(int)} &=& 2 g {\psi^+}^\dagger \Big [ \alpha^\perp . A^\perp {
1 \over i \partial^+}( \alpha^\perp . i \partial^\perp + \gamma^0 m ) +
( \alpha^\perp. i \partial^\perp + \gamma^0 m) { 1 \over i \partial^+}
\alpha^\perp . A^\perp \Big ] \psi^+ \nonumber \\
&& ~~~~~~~~+ 2 g^2 {\psi^+}^\dagger \alpha^\perp. A^\perp \alpha^\perp.
A^\perp \psi^+. 
\end{eqnarray}
Note that the fermion kinetic energy constitutes only a
part of the total contribution from fermions. Any theoretical estimate of
the fermionic part of the longitudinal structure function 
 necessarily has to involve off-diagonal contributions from Fock states
differing in the number of gluons by one and two.
      
It is important to emphasize the difference between equal time and light-front 
Hamiltonians in the context of our calculations. The equal-time Hamiltonian
contains the scalar density term (${\bar \psi} \psi$) accompanying the quark
mass $m$. In contrast, the quark mass appears quadratically in the free part of
the light-front Hamiltonian. Recently the question of the partition of hadron
masses in QCD has been addressed by Ji \cite{c5jib} in the context of the
equal-time Hamiltonian and in terms of twist-two and twist-three
observables. In his analysis, the extraction of the fraction of the hadron
mass carried by the fermion constituents is not straightforward  because of the 
presence of the scalar density term. The hadron expectation value of the
strange quark scalar density remains unknown (experimentally). Our analysis,
however, shows that the twist four longitudinal structure function, once
extracted experimentally, directly yields the fraction of the hadron mass
carried by fermionic constituents.   
 
%%%%%%%%%%%%%%%%%%%%%%%%%%%%%%%%%%%%%%%%%%%%%%%%%%%%%%
\section{Discussions}
%%%%%%%%%%%%%%%%%%%%%%%%%%%%%%%%%%%%%%%%%%%%%%%%%%%%%%%

Our results indicate that the experiments to measure the twist four 
longitudinal structure
function reveal the fraction of the hadron mass carried by 
the charged parton components.
Thus these experiments play a complementary role to 
 the longitudinal momentum and helicity
distribution information obtained at the twist two level. It is of interest
to investigate the feasibility of the direct measurement of the twist four
gluon structure function in high energy experiments. Recent work of Qiu,
Sterman and collaborators have shown that semi-inclusive single jet
production in deep inelastic scattering \cite{c5luo} and direct photon
production in hadron nucleus scattering \cite{c5guo} provide direct measurement 
of twist
four gluon matrix elements.

We also note that in the pre-QCD era, there have been discussions about a
possible $\delta (x)$ function contribution to the longitudinal structure function
which may appear to invalidate the sum rule 
derived ignoring such subtleties. In
two-dimensional QCD Burkardt has shown \cite{c5bur} that ${F_L \over x^2}$ has a
delta function contribution which comes from surface terms and he has discussed implications of this for
the sum rule for ${F_L \over x^2}$. Obviously, ${F_L \over x}$ will not be
affected by such a singular contribution and in the next chapter, we show explicitly 
that the sum rule is verified in two-dimensional QCD by virtue of the
't Hooft equation.        

In the next chapter, we present a non-perturbative calculation 
of the longitudinal structure function for a
meson-like bound state in $1+1$ dimensional QCD and for a positronium-like
target in $3+1$ dimensional QED in weak coupling limit.

%%%%%%%%%%%%%%%%%%%%%%%%%%%%%%%%%%%%%%%%%%%%%%%

\chapter{Twist Four $F_L$ for Light-Front Bound
States}
Higher twist or power suppressed contributions to the DIS structure
functions involve non-trivial non-perturbative information on the
structure of hadrons. As we mentioned before, the most important point
of our approach based on light-front Hamiltonian QCD is that both
perturbative and non-perturbative calculations are possible within the
same framework \cite{c6hari1}. We have calculated the twist four part of
the longitudinal structure function $F_L$ in the previous chapter for
a dressed quark and a dressed  gluon state  in perturbation theory.  
When the target
is a bound state in QCD like a nucleon or a meson, one can use the
same technique of Fock space expansion of the target state, but then
has to know the bound state wave function. The analysis of QCD bound
states in $3+1$ dimension in light-front Hamiltonian theory is highly
complicated and it requires the recently developed similarity
renormalization technique \cite{c6sim}. The spectra of the heavy quark 
bound states like
charmonium and bottomonium have been investigated using this technique
\cite{c6mar}. However, the analysis of the structure functions requires
knowledge about the bound state wave functions, the analytic form of
which is not obtained so far.  This is because, the similarity
renormalization technique generates an additional confining
interaction in the effective Hamiltonian in $O(g^2)$ which makes it
impossible to solve the effective Hamiltonian analytically even in the
leading order in bound state Hamiltonian perturbation theory \cite{c6mar}.  

In this chapter, we investigate the twist four part of $F_L$ for a
meson in $1+1$ dimensional QCD. We explicitly show that the sum rule
which we derived in the previous chapter is obeyed in this case by
virtue of t'Hooft equation. In order to understand the calculational
procedure in $3+1$ dimension, we perform a simpler but interesting
analysis.  We calculate the twist-four part of the longitudinal
structure function for a positronium-like bound state in light-front
QED in the weak coupling limit. In this limit, QCD results are not expected
to differ much from the QED results. As a result, this analysis is important since it
tests and illustrates the approach in QCD. The advantage here is that,
in the weak coupling limit, the bound state equation can be solved
analytically and the wave function is known. This allows an analytic
understanding of the problem. We refer to our original works,
\cite{c6prd,c6as}.

The plan of this chapter is as follows. 
In chapter 6.1, we calculate twist four part of $F_L$ for a meson in $1+1$
dimensional QCD and explicitly verify the sum rule. 
In section 6.2, we calculate the twist
four longitudinal structure function for a positronium-like state in
light-front QED in the weak coupling limit and 
verify  sum rule for $F^{\tau=4}_L$ in this case. We also 
show that it reduces to a relation connecting the kinetic and potential
energies to the binding energy of positronium. In section 6.3, we 
analytically calculate $F^{\tau=4}_L$ using the wave function for
positronium.
Discussions and conclusions are given in section 6.4. The bound state equation
for positronium in light front QED in the weak coupling limit is derived in
Appendix C.

%%%%%%%%%%%%%%%%%%%%%%%%%%%%%%%%%%%%%%%%%%%%%%%%%%%%
\section{$1+1$ Dimensional QCD: Explicit Calculations}\label{c611}
%%%%%%%%%%%%%%%%%%%%%%%%%%%%%%%%%%%%%%%%%%%%%%%%%%%%
In the previous chapter, we have derived a sum rule for the twist four part
of the longitudinal structure function $F_L$. We have shown that the sum
rule is obeyed in perturbative QCD for dressed quark and gluon targets. 
In this section, we turn to 
two-dimensional QCD
in order to test the sum rule in a non-perturbative context.
In 1+1 dimensions, in $A^+=0$ gauge, we have,
\begin{eqnarray}
\int_0^1 {dx \over x} F_{L(q)}^{\tau=4}(x) = { 2 \over Q^2} \langle P \mid 
\Big [ \theta^{+-}_q(0) + \theta^{+-}_g(0) \Big ] \mid P \rangle,
\end{eqnarray} 

with $
\theta^{+-}_q = 2 m^2 {\psi^{+}}^\dagger { 1 \over i \partial^+} \psi^+ $
is the fermionic part of the light-front QCD Hamiltonian density in $1+1$
dimension and $ \theta^{+-}_g = - 4 g^2 {\psi^{+}}^\dagger T^a \psi^+ 
{1 \over (\partial^+)^2}
{\psi^{+}}^\dagger T^a \psi^+$ is the gluonic part.
We consider the standard one pair ($q \overline{q}$) approximation to the 
meson ground state. Explicit evaluations show that 
\begin{eqnarray}
{F^{\tau=4}_{L(q)} \over x} = {4 \over Q^2} \psi^*(x) 
{m^2 \over x (1-x)} \psi(x),\\
\int_0^1 dx {F^{\tau=4}_{L(g)} \over x} =  {4 \over Q^2}(-) C_f {
g^2 \over \pi} \int_0^1 dx \int_0^1 dy \psi^*(x) 
{ \psi(y)-\psi(x)\over (x-y)^2}.
\end{eqnarray}

Here $\psi(x)$ is the ground state wave function for the meson. Thus
\cite{c6prd}
%Next we consider the gluonic contribution. The range of $x$ in terms arising
%from normal ordering is necessarily from zero to infinity instead of being
%from zero to one. One can altogether avoid this situation as follows.
%If we take the four fermion operator to be normal ordered, we get a divergent
%contribution in the bare theory, namely
%where $ \psi(x) $ is the ground state wave function for the meson. To carry
%out renormalization, we add a suitable counterterm so that the divergence at
%$x=y$ is removed. In the renormalized theory, we have,
\begin{eqnarray} 
\int_0^1 {dx \over x} F^{\tau=4}_{L}(x) = { 4 \over Q^2} \int_0^1 dx
\psi^*(x) \Big [ {m^2 \over x (1-x)} \psi(x)
- C_f {g^2 \over \pi} \int_0^1 dy { \psi(y) - \psi(x) \over (x-y)^2}
\Big ].
\end{eqnarray} 
The bound state equation obeyed by the
ground state wave function $\psi(x)$ for the meson in the truncated space of
a quark-antiquark pair is the t'Hooft equation: 
\begin{eqnarray}
M^2 \psi(x) = {m^2 \over x (1-x)} \psi(x) - C_f {g^2 \over \pi} \int dy {
\psi(y) - \psi(x) \over (x-y)^2}   
\end{eqnarray}
together with the normalization condition $ \int_0^1 dx \psi^2(x)=1$, 
we easily verify that the twist four longitudinal structure function of the
meson obeys the sum rule
\begin{eqnarray}
\int_0^1 {dx \over x} F_L^{\tau=4} = { 2 \over Q^2} \langle P \mid
\theta^{+-}(0) \mid P \rangle = 4 {M^2 \over Q^2}.
\end{eqnarray}

In the same model, the contribution to the twist two structure function from
the fermionic constituents is given by
\begin{eqnarray}
F_{2(q)}(x) = (x+ 1-x) \psi^*(x) \psi(x)  = \psi^*(x) \psi(x).
\end{eqnarray}
Note that, since there are no partonic gluons or sea in this model, the
longitudinal momentum of the meson is carried entirely by the valence quark
and anti-quark. Thus the momentum sum rule is saturated by the fermionic
part of the longitudinal momentum density. On the other hand, light-front
energy density is shared between fermionic and gauge bosonic parts and as a
consequence the fermions carry only a fraction of the hadron mass.
This seemingly paradoxical situation further illuminates the difference
between the physical content of the $F_2$ and $ F^{\tau=4}_L$ sum rules.

To get a quantitative picture, next, we explicitly calculate the structure
functions $F_{2(q)}(x)$ and ${F_{L(q)}(x) \over x}$ for the ground state
meson in two-dimensional QCD. We have parameterized  the ground state wave
function as 
\be
\psi(x) = {\cal N} x^s (1-x)^s.
\e
 
The factor ${\cal N}$ is determined from the normalization condition
$ \int_0^1 dx \psi^*(x) \psi(x) =1$ and  $s$ is the variational parameter.
We have
\be
M^2=\int dx \psi^*(x){m^2 \over x (1-x)} \psi(x) - C_f {g^2 \over \pi} \int
dx \int dy {
\psi^*(x) \psi(y) - \psi^*(x)\psi(x) \over (x-y)^2}.   
\end{eqnarray}
The first term in the left hand side is the kinetic energy of the fermions
and the other is the interaction. Substituting the above wave function and
also using the normalization condition, we get
\be
M^2=m^2{\beta(2s, 2s)\over \beta(2s+1, 2s+1)}-{g^2\over \pi}C_f {[s(\beta(s,
s+1))^2-\beta(2s, 2s)]\over \beta(2s+1, 2s+1)}
\e
where
$\beta(n, m)= \int_0^1 dx x^{n-1}(1-x)^{m-1}$.

We  determined the value of
$s$ variationally by minimizing $M^2$ for given values of $m^2$ and $g^2$.
The  resulting structure functions are
presented in Fig. 1 for two different values of $m^2$.

\begin{figure}
\centerline{\rotatebox{90}{\epsfxsize=.3\textwidth\epsfbox{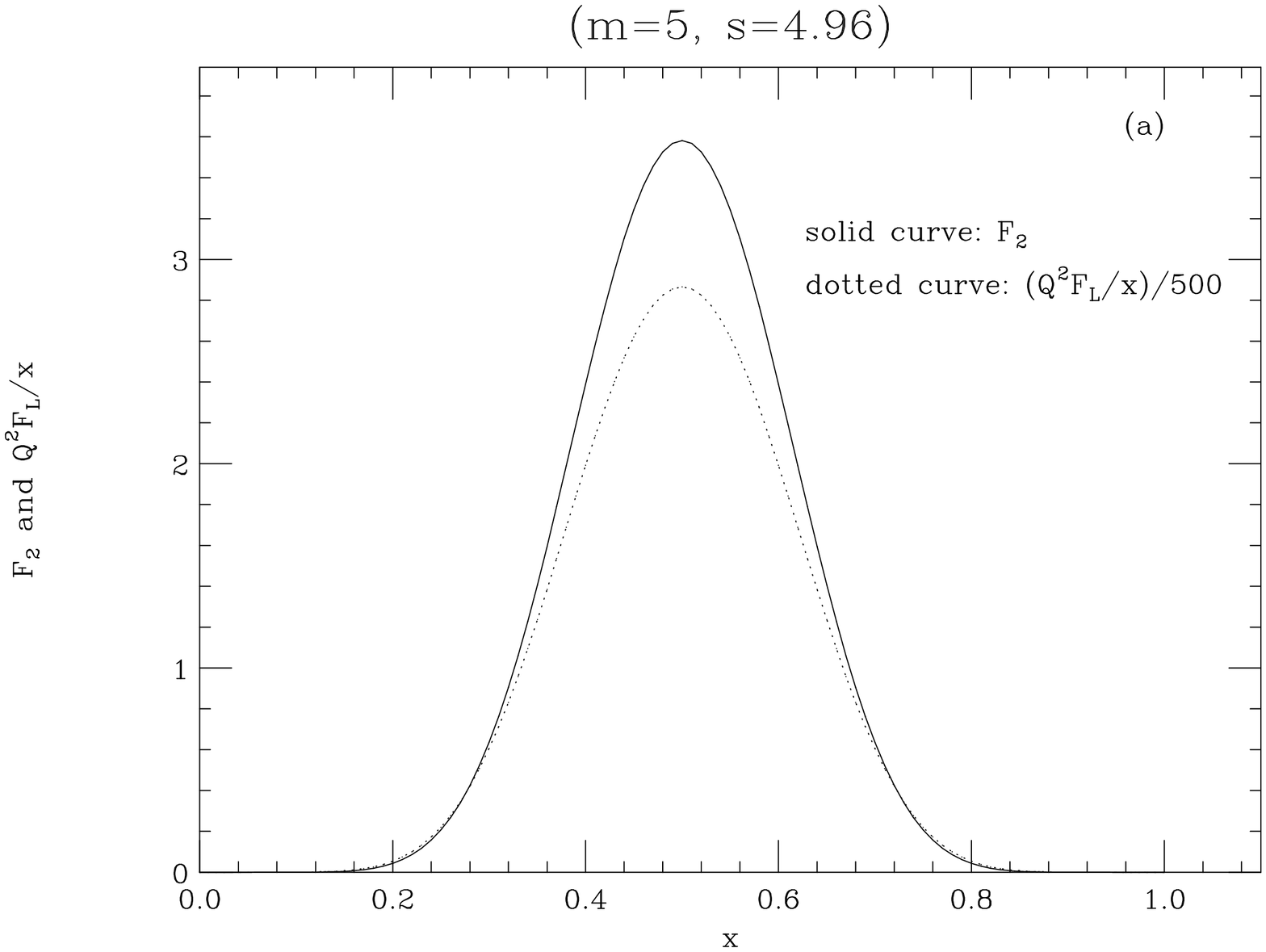}}
\hspace{6mm} 
\rotatebox{90}{\epsfxsize=.3\textwidth\epsfbox{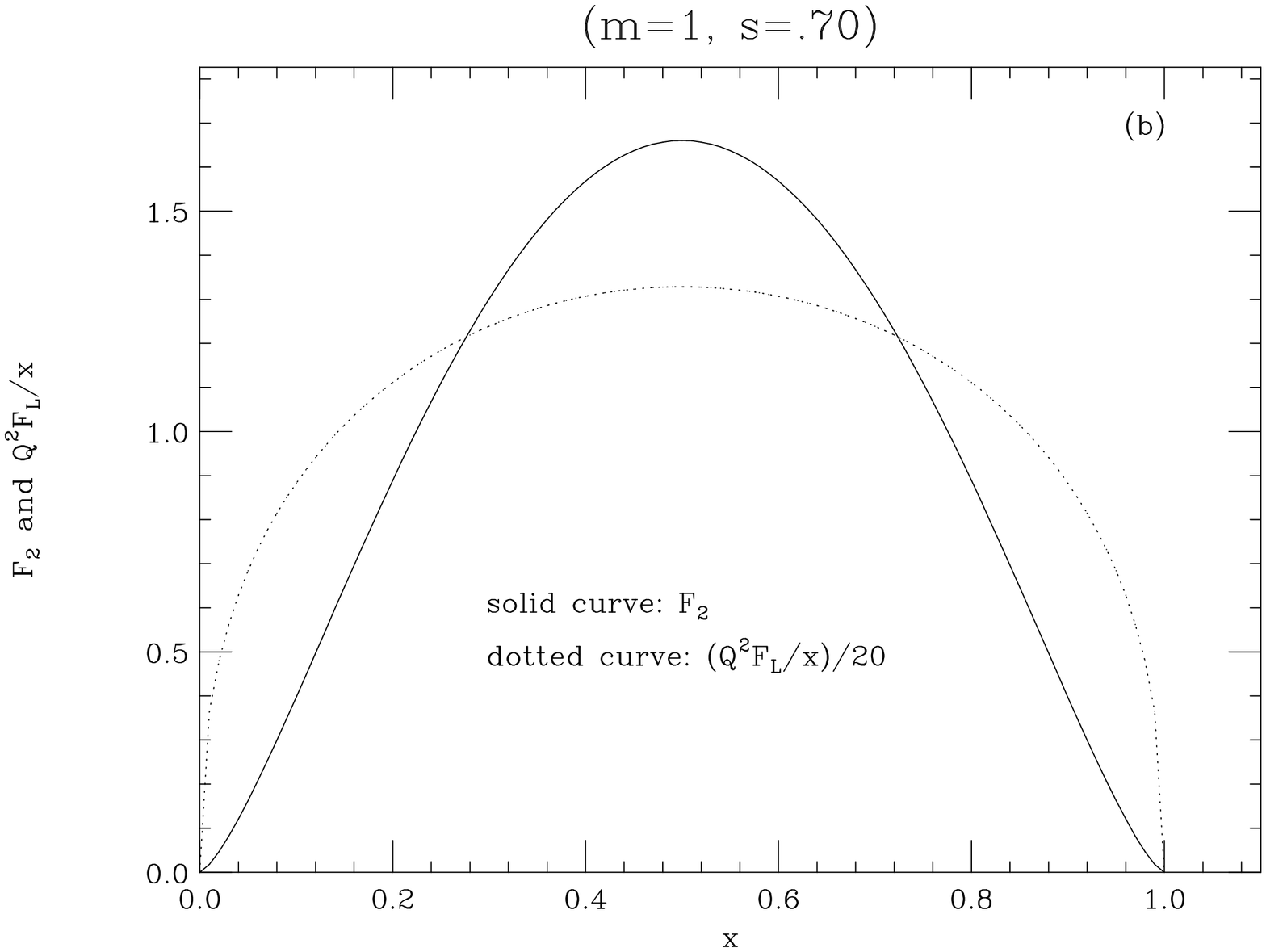}}}
\vskip 5mm
\caption{Fermionic contributions to the structure functions 
$F_2(x)$ and ${F_L^{\tau=4} \over x}$ for the
ground state meson in the 't Hooft model for two different values of $m$,
the quark
mass. (a): $ m =5$, $ s = 4.96$. (b) $ m =1$, $ s = .70$. The parameter 
 $s$ appearing in the wave function is determined by a variational
calculation. We have set $C_f {g^2\over \pi}=1$.}
\end{figure}
\vspace{3mm}

Since both the quark and anti-quark have 
equal mass in the model, both structure
functions are symmetric about $x={1 \over 2}$. When the fermions are heavy
(Fig. 1(a)), the system is essentially non-relativistic and 
the structure functions are significant only near the region $x={1 
\over 2}$. When the fermions become lighter (Fig 1(b)),
contribution to the structure function from the end-point regions become
significant indicating substantial high momentum components in the ground
state wave function.    Note that ${ F_{L(q)} \over x}$ 
measures the fermion kinetic energy (in light-front coordinates). The
exponent $s$ in the wave function is a function of the fermion mass and
$s$ decreases as $ m$ decreases. In the massless limit, $s$ vanishes
\cite{c6berg} so that the wave function for the ground state becomes $ \psi(x) =
\theta(x) \theta(1-x)$. This results in a flat $F_2$ structure function.
However, because of the presence of $m^2$, $F^{\tau=4}_{L(q)}$ vanishes.
Because of an exact cancellation  between the self-energy and gluon exchange
contributions, the gluonic part of the $F^{\tau=4}_{L}$ also vanishes.
Thus the sum rule is satisfied exactly since, in the zero quark mass limit,
the ground state meson is massless in two-dimensional QCD.

In the next section, we investigate the twist four longitudinal structure
function for a positronium in $3+1$ dimension in light-front QED.      

%%%%%%%%%%%%%%%%%%%%%%%%%%%%%%%%%%%%%%%%%%%%%%%%%%%%%%%%%%%%%%%%%%%%%%%
\section{Twist Four Longitudinal Structure Function for a Positronium in
$3+1$ Dimensional QED}\label{c6qed}
%%%%%%%%%%%%%%%%%%%%%%%%%%%%%%%%%%%%%%%%%%%%%%%%%%%%%%%%%%%%%%%%%

We consider a positronium like bound state $\mid P \rangle$ given by
\be
\mid P \rangle &=& \sum_{\sigma_1, \sigma_2} 
\int {dk_1^+ d^2 k_1^\perp \over \sqrt{2 (2 \pi)^3 k_1^+}} 
\int {dk_2^+ d^2 k_2^\perp \over \sqrt{2 (2 \pi)^3 k_2^+}} 
\nonumber \\
&& \phi_2(P \mid k_1, \sigma_1; k_2, \sigma_2) \sqrt{2 ((2 \pi)^3 P^+}
\delta^3(P-k_1-k_2) b^\dagger(k_1, \sigma_1) d^\dagger(k_2,\sigma_2) \mid 0
\rangle \nonumber \\
&&~~~~~~ + \sum_{\sigma_1,\sigma_2,\lambda_3} 
\int {dk_1^+ d^2 k_1^\perp \over \sqrt{2 (2 \pi)^3 k_1^+}} 
\int {dk_2^+ d^2 k_2^\perp \over \sqrt{2 (2 \pi)^3 k_2^+}} 
\int {dk_3^+ d^2 k_3^\perp \over \sqrt{2 (2 \pi)^3 k_3^+}} 
\nonumber \\
&&~~~~~~~~ \phi_3(P \mid k_1, \sigma_1; k_2, \sigma_2; k_3, \lambda_3)
\sqrt{2 (2 \pi)^3 P^+} \delta^3(P-k_1 -k_2 -k_3) 
\nonumber \\
&&~~~~~~~~~~~~~~~~~~b^\dagger(k_1 ,\sigma_1)
d^\dagger (k_2, \sigma_2) a^\dagger(k_3, \lambda_3) \mid 0 \rangle. 
\label{c6e8}
\e

Here $\phi_2$ is the probability amplitude to find an electron and positron in
the positronium, $\phi_3 $ is the probability amplitude to find an electron,
positron and a photon in the positronium.
We restrict ourselves to the three particle sector.

We calculate $F^{\tau=4}_4$ in the weak coupling limit  in light-front QED 
using light front gauge, $A^+=0$. We truncate the Fock state upto the three
particle sector. This approximation includes the
effect of dynamical photon. 
 It can be shown that,
for a weak coupling theory the results are the same as obtained using a non-relativistic
approximation. However, the entire calculation is fully relativistic and
exact in the leading order in $\alpha$ \cite{c6bur}. It can be shown that the recently developed
similarity renormalization scheme for light-front QED gives the same results
in the leading order in bound state perturbation theory as in the weak coupling treatment and particle number
truncation \cite{c6billy}.    

The twist-4 part of the fermionic component of the longitudinal structure
function is given by
\be
F^{\tau=4}_{L(f)}(x) = {M}_1 + {M}_2,
\e
\begin{eqnarray}
{M}_1~=~{ 1 \over Q^2}~ {x^2 (P^+)^2 \over 2 \pi}~ \int dy^- ~
e^{-{ i \over
2}P^+y^-x}~\langle P \mid \overline{\psi}(y^-) \gamma^- \psi(0)-
\overline{\psi}(0) \gamma^- \psi(y^-) \mid P \rangle,
\label{c6e10}
\end{eqnarray}
and
\begin{eqnarray}
{M}_2 &=&-{(P^\perp)^2  \over (P^+)^2}
{ 1 \over Q^2} {x^2 (P^+)^2 \over 2 \pi}~ 
\int dy^- ~e^{-{ i \over
2}P^+y^-x}~\langle P \mid \overline{\psi}(y^-) \gamma^+
\psi(0)\nonumber\\&&~~~~~~~~~~~~~~~~~-
 \overline{\psi}(0) \gamma^+ \psi(y^-)\mid P \rangle.
\label{c6e11}
\end{eqnarray}

We shall take the mass of the state $\mid P \rangle$ to be M and the electron
and positron  mass to be $m$. In the weak coupling (non-relativistic) limit, the helicity
dependence of the wave function factorizes away, so it is sufficient to
consider one helicity sector. Here we shall take the two particle state with
$\s_1$ and $\s_2$ up.

For a positronium state, $\mid P \rangle$ given by Eq. (\ref{c6e8}) we obtain
\be
{F^{\tau=4}_{L(f)}(x)}_{diag} &=&(M_1)_{diag} + (M_2)_{diag}
\nonumber\\&&~~~~= {4\ov Q^2} \int dx_1 d^2\k_1^\p ((\k_1^\p)^2+m^2) {\mid \psi_2
 \mid }^2 \Big [ \delta (x-x_1) + \delta(1-x-x_1) \Big ] \nonumber\\&&~~~~
+{4\ov Q^2}\sum \int dx_1 d^2
\k_1^\p \int dx_2 d^2\k_2^\p {\mid \psi_3 \mid }^2 \Big [
((\k_1^\p)^2+m^2)\delta(x-x_1) + \nonumber\\&&~~~~~~~~~~~~~~~~~~~~
((\k_2^\p)^2+m^2) \delta(x-x_2) \Big ].
\label{c6f1}
\e
The off-diagonal contributions to $ F^{\tau=4}_{L(f)}(x)$ comes from $ M_1 $
alone.
\be
(M_1)_{off-diag} = -{4\ov Q^2} {e^2\ov {2(2\pi)^3}}\int dx_1d^2\k_1^\p \int
dyd^2\k^\p \Big [ M_1^a + M_1^b + M_1^c + M_1^d \Big ]
\e
where
\be
M_1^a = {1\ov {E(x_1-y)^2}}\Big [ {2(\k_1^\p)^2 y \ov x_1}\delta(x-x_1) -
{2(\k^\p)^2 x_1\ov y} \delta (x-y) \Big ]
{\mid \psi_2^{\s_1\s_2} (x_1, \k_1^\p)\mid }^2,   
\label{c6u1}
\e
\be
M_1^b = {1\ov {E(x-y)^2}}\Big [ -(\k_1^\p)^2\delta(x-x_1) + (\k^\p)^2
\delta(x-y) \Big ]  \nonumber\\ ~~~~~~~~~~~~~\Big (
 \psi_2^{*\s_1\s_2} (x_1, \k_1^\p) \psi_2^{\s_1\s_2} ( y,\k^\p) + h.c. \Big
),
\label{c6u2}
\e
\be
M_1^c = {1\ov {E'(y-x_1)^2}} \Big [ {2(\k_1^\p)^2y\ov
(1-x_1)}\delta(1-x-x_1) - {2(\k^\p)^2 (1-x_1)\ov y} \delta (x-1+y) \Big ]
\nonumber\\~~~~~~~~~~~~~~~~~~~~~~~~~~~~~~~~~~~~{\mid \psi_2^{\s_1\s_2} 
(x_1, \k_1^\p)\mid}^2,
\label{c6u3}
\e  
\be
M_1^d = {1\ov {E'(y-x_1)}}\Big [ (\k^\p)^2 \delta (x+y-1) - (\k_1^\p)^2
\delta (1-x-x_1) \Big ]\nonumber\\~~~~~~~~~~~~\Big ( \psi_2^{*\s_1\s_2} (x_1, \k_1^\p)
\psi_2^{\s_1\s_2}( y,\k^\p) + h. c. \Big ).
\label{c6u4}
\e
In this calculation we have taken all operators to be normal ordered.
Also, in Eqs. (\ref{c6u1})- (\ref{c6u4})  we have neglected
all mass terms in the vertex, since as described in Appendix C, these terms are
suppressed in the weak coupling limit. The energy denominators are given
by
\be
E=M^2-{{(\k^\perp)^2+m^2}\over y}-{{(\kappa_1^\perp)^2+m^2}\over
1-x_1}-{(\kappa_1^\perp-\k^\perp)^2\over x_1-y}\nonumber\\
E'=M^2-{{(\kappa_1^\perp)^2+m^2}\over x_1}-{{(\k^\perp)^2+m^2}\over
y}-{(\kappa_1^\perp-\k^\perp)^2\over (y-x_1)}.
\label{c6u5}
\e 

We define the twist four longitudinal photon structure function as in the
previous chapter,
\be
F_{L(g)}^{\tau=4}(x) &=& { 1 \over Q^2} {x P^+ \over 2 \pi} \int dy^- ~
e^{-{i \over 2} P^+ y^- x} \nonumber \\
&& ~~~~~\Big [ \langle P \mid (-) F^{+ \lambda}(y^-) F^-_{\lambda}(0) + 
{ 1 \over 4} g^{+-} F^{\lambda \sigma} (y^-) F_{\lambda \sigma}(0) \mid
P \rangle \nonumber \\
&& ~~~~~~~ - {(P^\perp)^2 \over (P^+)^2} \langle P \mid F^{+ \lambda}(y^-) 
F^+_{\lambda}(0) \mid P \rangle + ( y^- - 0)\Big ].
\e

Now, from the definition of $ F_{L(f)}^{\tau=4}(x)$ and  $F_{L(g)}^{\tau=4}(x) $ 
it can be verified explicitly that
\be
 F_{L(f)}^{\tau=4}(-x) = -  F_{L(f)}^{\tau=4}(x),\\
 F_{L(g)}^{\tau=4}(-x) =-  F_{L(g)}^{\tau=4}(x).
\e 
$ F_{L(g)}^{\tau=4}(x)$ has both diagonal and off-diagonal parts. As before,
we take all operators to be normal ordered and we get 
\be
{{ F_{L(g)}^{\tau=4}(x)}\ov x}_{diag}&=& {4\ov Q^2}  \int dx_1d^2\k_1^\p \int dy
d^2\k^\p
{\mid \psi_3 \mid }^2 {(-\k_1^\p-\k^\p)^2 \ov (1-x_1-y)} \delta ( 1-x_1-x-y) 
\nonumber\\&&-{4\ov Q^2} {4e^2\ov
{2(2\pi)^3}}\int dx_1d^2\k_1^\p \int dyd^2\k^\p \psi_2^*(x_1,
\k_1^\p)\nonumber\\&&~~~~~~~~~~~~~~~~~~~~~~~~~~~~~~~~~~
 \psi_2(y,\k^\p){1\ov {(x_1 - y)^2}}\delta(x_1-x-y).
\e
The second term in the right hand side is the contribution of the instantaneous
interaction.
The off-diagonal contribution is
\be
{ F_{L(g)}^{\tau=4}(x)}_{off-diag} = G_1 + G_2,
\e
where
\be
G_1 &=& -{4\ov Q^2} {e^2\ov {2(2\pi)^3}}  \int dx_1 d^2\k_1^\p \int dy
d^2\k^\p  {x\ov E(x_1-y)^2} \nonumber\\&&~\Big [ \Big ( -4{ (\k_1^\p - \k^\p)^2\ov
(x_1-y)} + {2(\k_1^\p)^2\ov x_1} - {2(\k^\p)^2\ov y} \Big ) 
{\mid \psi_2^{\s_1\s_2} (x_1, \k_1^\p)\mid
}^2\nonumber\\&&~~~~~~~~~
+\Big ( {2(\k_1^\p-\k^\p)^2\ov {x_1-y)}} + {(\k_1^\p)^2\ov (1-x_1)} -
{(\k^\p)^2\ov (1-y)} \Big ) 
\nonumber\\&&~~~~~~~~~~~~~~~\Big ( \psi_2^{*\s_1\s_2} (x_1, \k_1^\p)
 \psi_2^{\s_1\s_2} (y,\k^\p) + h.c. \Big )\Big ]   
\delta (x-x_1 +y), 
\e
\be
G_2 &=& -{4\ov Q^2} {e^2\ov {2(2\pi)^3}} \int dx_1 d^2\k_1^\p \int dy
d^2\k^\p  {x\ov {E'(y-x_1)^2}} \nonumber\\&&\Big [\Big (-{4(\k_1^\p -\k^\p)^2\ov
(y-x_1)} - {2(\k^\p)^2\ov (1-y)} + {2(\k_1^\p)^2\ov (1-x_1)} \Big )
{\mid \psi_2^{\s_1\s_2} (x_1, \k_1^\p)\mid
}^2\nonumber\\&&~~~~~~~~~~~~~
+\Big (-{2(\k_1^\p -\k^\p)^2\ov
(y-x_1)} - {(\k^\p)^2\ov y} + {(\k_1^\p)^2\ov x_1} \Big ) 
\nonumber\\ &&~~~~~~~~~~~~~~~~~~~\Big ( \psi_2^{*\s_1\s_2} (x_1, \k_1^\p)
 \psi_2^{\s_1\s_2} ( y,\k^\p) + h.c. \Big )\Big ]
\delta (x_1+x-y)
\e
where $E$ and $E'$ are given by Eq. (\ref{c6u5}).

From these expressions, we calculate
\be
\lefteqn{\int_0^1 {{ F_{L(q)}^{\tau=4}(x) + F_{L(g)}^{\tau=4}(x)}\ov x} dx
= {4\ov Q^2}  \int dxd^2\k^\p {\psi_2}^* \psi_2 \Big [ {(\k^\p)^2\ov
x} + { (\k^\p)^2\ov {1-x}} \Big ]}  \nonumber\\&&
+{4\ov Q^2} \sum \int dxd^2\k^\p 
\int dy d^2q^\p {\psi_3}^* \psi_3 \Big [ {(\k^\p)^2\ov
{x}} + { (q^\p)^2 \ov y}+ {(-\k^\p - q^\p)^2\ov (1-x-y)} \Big ]
\nonumber\\&&~~~~~~~+{4\ov Q^2}{e^2\ov {2(2\pi)^3}}\int dxd^2\k^\p 
\int dy d^2q^\p [ A_1 + A_2 + B_1 + B_2] \nonumber\\&&~~~~~~~~
- {4\ov Q^2} { 4e^2\ov {2(2\pi)^3}}\int dxd^2\k^\p \int dy d^2q^\p {1\ov
(x-y)^2} \nonumber\\&&~~~~~~~~~~~~~~~~~~~ \psi_2^*(x,\k^\p)\psi_2( y, q^\p) 
\label{c6c4}
\e
where
\be
A_1 = {1\ov E}{1\ov (x-y)}\Big [V_1
{\mid \psi_2^{\s_1 \s_2} (x,\k^\p) \mid }^2 + V_2
\psi_2^{*\s_1 \s_2} (x,\k^\p)\psi_2^{\sigma_1\sigma_2}(y,q^\p)\Big ] 
\label{c6a1}
\e
and
\be
A_2 &=&{1\ov E'}{1\ov (y-x)}\Big [{V'}_2
 \psi_2^{*\sigma_1\sigma_2}
(x,\kappa^\perp)\psi_2^{\sigma_1\sigma_2}(y,q^\p) 
\nonumber\\&&~~~~~~~~~~~~~~~~~~~~~~
+ {V'}_1
{\mid \psi_2^{\s_1 \s_2} (x,\k^\p) \mid }^2\Big ], 
\label{c6a2}
\e 
\be
B_1&=& {1\ov E}{1\ov (x-y)} \Big [ V_1
{\mid \psi_2^{\s_1 \s_2} (x,\k^\p) \mid }^2  
\nonumber\\&&~~~~~~~~~~~~~~~~+V_2
\psi_2^{\s_1 \s_2} (x,\k^\p)\psi_2^{*\sigma_1\sigma_2}
(y,q^\p)\Big ], 
\label{c6c1}
\e
\be
B_2&=& {1\ov E'}{1\ov (y-x)} \Big [{V'}_2
\psi_2^{\sigma_1\sigma_2}
(x,\kappa^\perp)\psi_2^{*\sigma_1\sigma_2}
(y,q^\p) \Big ] 
\nonumber\\&&~~~~~~~~~~~~~~~~~~~~~
{V'}_1{\mid \psi_2^{\s_1 \s_2} (x,\k^\p) \mid }^2\Big ].  
\label{c6c2}
\e
Here $V_1, V_2, {V'}_1$ and ${V'}_2$ are given by
\be
V_1 = \Big [{2(\k^\p-q^\p)^2\ov (x-y)^2} + {(x+y)\ov (x-y)}\Big ( {(\k^\p)^2 
\ov x^2} +
{{(q^\p)^2}\ov y^2}\Big )\Big ],
\e
\be
V_2 = {V'}_2 = \Big [{(\k^\p)^2(1-2x)\ov {x(1-x)(x-y)}}+
{(q^\p)^2(2y-1) \ov {y(1-y)(x-y)}} -{2(\k^\p-q^\p)^2\ov (x-y)^2} \Big ],
\e
\be
{V'}_1 = \Big [{2(\k^\p-q^\p)^2\ov (x-y)^2} + {(2-x-y)\ov (y-x)
}
\Big ( {(\k^\p)^2 \ov {(1-x)^2}} +
{{(q^\p)^2}\ov {(1-y)^2}}\Big )\Big ].
\e 

 In these
expressions we have kept only those terms in the vertex which survive in the
non-relativistic limit. The helicities $\s_1$ and $\s_2$ are both up and 
 we have neglected all mass terms in the vertex since
in this limit they are suppressed. Also in the weak coupling (non-relativistic) limit, 
we consider only photon exchange interactions and the
off-diagonal terms proportional to ${\mid \psi_2 \mid
}^2 $ originating from self energy effects can be neglected. 

Now, from the expression of $\psi_3$ given in Appendix C, we get
\be
\int dxd^2\k^\p \int dy d^2q^\p {\mid \psi_3 \mid }^2 = { e^2\ov
{2(2\pi)^3}} \int dx d^2\k^\p \int dyd^2q^\p \Big [ {1\ov E} B_1 + 
{1\ov E'} B_2 \Big ]
\label{c6c3}
\e
where $B_1$ and $B_2$ are given earlier.
Using this, one can write the second term in the right hand side of Eq. (\ref{c6c4}) as
\be
\lefteqn{{4\ov Q^2} {e^2\ov {2(2\pi)^3}} \int dxd^2\k^\p \int dyd^2q^\p 
\Big [ {1\ov E} B_1 + {1\ov E'} B_2 \Big ]} \nonumber\\&&~~~~~~~~~~
\Big [ {{(\k^\p)^2 + m^2}\ov x} + {{(q^\p)^2
+ m^2} \ov y} + {(-\k^\p-q^\p)^2\ov (1-x-y)} \Big ].
\e
Considering the fact that the total energy is conserved, one can write this
as
\be
-{4\ov Q^2} {e^2\ov {2(2\pi)^3}}\int dxd^2\k^\p \int dyd^2q^\p
( B_1 + B_2).
\e
So we get
\be
\int_0^1 {{ F_{L(q)}^{\tau=4}(x) + F_{L(g)}^{\tau=4}(x)}\ov x} dx
&=& {4\ov Q^2} \int dxd^2\k^\p {\psi_2}^* \psi_2 [ {(\k^\p)^2\ov
x} + { (\k^\p)^2\ov {1-x}} ]\nonumber\\&& 
+{4\ov Q^2}{e^2\ov {2(2\pi)^3}}\int dxd^2\k^\p 
\int dy d^2q^\p [ A_1 + A_2 ] \nonumber\\
&&~~~~~~~~~
- {4\ov Q^2} { 4e^2\ov {2(2\pi)^3}}\int dxd^2\k^\p \int dy d^2q^\p {1\ov
(x-y)^2} \nonumber\\&&~~~~~~~~~~~~~
\psi_2^*(x,\k^\p, 1-x, -\k^\p)\psi_2( y, q^\p). 
\e
Also, if we denote ${(\k^\p)^2 + m^2} \ov {x(1-x)}$ by $M^2_o$, then in the
non-relativistic limit, it can be shown that
\be
M^2 - M^2_0 \simeq O(e^4).
\e
So we neglect this difference in the energy denominators 
and replace the bound state mass in $E$ and $E'$ by
 $ M^2 = {{(\k^\p)^2 + m^2}\over x(1-x)}$. The  energy
denominators then become
\be
E = {{(\k^\p)^2 + m^2 } \ov x} - {{ (q^\p)^2 + m^2 }\ov y} - {{ (\k^\p -
q^\p)^2} \ov {x-y}} \nonumber\\
= -{1\ov (x-y)} [ ({m\ov x})^2 (x-y)^2 + (\k^\p- q^\p)^2]
\e
and\\
\be
E' =  {{(\k^\p)^2 + m^2 } \ov {1-x}} - {{ (q^\p)^2 + m^2 }\ov {1-y}} 
+ {{ (\k^\p -
q^\p)^2 }\ov {x-y}} \nonumber\\
= {1\ov (x-y)} [ ({m\ov {1-x}})^2 (x-y)^2 + (\k^\p-q^\p)^2].
\e
We get, in this limit,
\be
\int_0^1 {{ F_L^{\tau=4}(x)}\ov x} dx&=& {4\ov Q^2} \int dx d^2\k^\p {\mid
\psi_2 \mid }^2 {{(\k^\p)^2 + m^2}\ov {x(1-x)}} \nonumber\\&&
-{4\ov Q^2}{2e^2\ov {2(2\pi)^3}} \int dyd^2q^\p\psi^*_2( x, \k^\p, 1-x,
-\k^\p) \psi_2(y,q^\p,
1-y,-q^\p) \nonumber\\&&~~~~
\Big [ \Big ({m\ov x}\Big )^2 {1\ov { (\k^\p-q^\p)^2 +({m\ov
x})^2(x-y)^2}}\nonumber\\&&~~~~~~ + 
      \Big ({m\ov (1-x)}\Big )^2 {1\ov { (\k^\p-q^\p)^2 + ({m\ov (1-x)})^2(x-y)^2}}\Big
].
\label{c6c5}
\e
Here in the weak coupling limit we have omitted the spin indices. We can see that the interaction part of $\int_0^1 {{ F_L^{\tau=4}(x)}\ov x}
dx$ has exactly the same form as in non-relativistic bound state equation
(see Appendix C) and hence in the leading order
it can be related to the expectation value of the Coulomb interaction. Since
$\k^\p$ and $q^\p$ are small in the non-relativistic limit, all $(\k^\p)^2$
and 
$(q^\p)^2$ dependence in the numerator of the interaction terms are neglected
compared to the $m^2$ dependent terms. However, the term proportional to
${(\k^\p - q^\p)^2\ov (x-y)^2}$ cannot be neglected because both $x$ and $y$
are almost equal and  this term cancels the contribution from the 
instantaneous interaction in the non-relativistic limit. Both of
these terms originate from $F_{L(g)}$ and one can see that only the gauge
bosonic part of the longitudinal structure function is important for the
Coulomb interaction in the weak coupling limit. The $m^2$ terms in the
energy denominators combine with the other terms to give the non-vanishing
contribution.  The fermionic part of the longitudinal structure function 
 gives contribution to the kinetic
energy of the fermions. Reminding oneself that we are working in the
light-front gauge and not in the Coulomb gauge, this is a manifestation of
the gauge invariance of the separation of the Hamiltonian density into a
fermionic and a gauge bosonic part.

The Fermionic part of the Hamiltonian density is given by
\begin{eqnarray}
\theta^{+-}_f = i \overline{\psi}\gamma^- \partial^+ \psi 
  = 2 {\psi^{+}}^\dagger \Big [ \alpha^\perp.(i \partial^\perp + gA^\perp) +
\gamma^0m \Big ] {1\over {i \partial^+}} \Big [ \alpha^\perp . (i
\partial^\perp + gA^\perp) + \gamma^0m \Big ] \psi^+.
\end{eqnarray}
The gauge bosonic part of the Hamiltonian density is given by\\
\begin{eqnarray}
\theta^{+-}_g  =  - F^{+ \lambda} F^{-}_{\lambda}+ { 1 \over 4} 
g^{+-} (F_{\lambda \sigma})^2 =
{ 1 \over 4} \Big (\partial^+ A^{- }\Big )^2 + 
{ 1 \over 2} F^{ij} F_{ij}
\nonumber \\
 = (\partial^i A^j)^2  
	 + 2e \partial^i A^i \left( \frac{1}{\partial^+}
		\right)  2 (\psi^+)^{\dagger}
		\psi^+ \nonumber \\~~~  + e^2 \left( \frac{1}{\partial^+}
		\right)  2 (\psi^+)^{\dagger}
		 \psi^+  
	  \left( \frac{1}{\partial^+}\right)
		 2 (\psi^+)^{\dagger} 
		\psi^+.   
\end{eqnarray}
The fermionic part of the longitudinal momentum density is given by
\begin{eqnarray}
\theta^{++}_{f} = i \overline{\psi} \gamma^+ \partial^+ \psi.
\end{eqnarray}

The gauge bosonic part of the longitudinal momentum density
\begin{eqnarray}
\theta^{++}_{g} = - F^{+ \lambda } F^-_{\lambda }.
\end{eqnarray}
For a positronium like bound state, we calculate the matrix element of
$\theta_f^{+-}$ and $ \theta_g^{+-}$. The matrix elements have both diagonal
and off diagonal contribution. The diagonal contribution to the matrix element 
from the fermionic and the gauge bosonic part is given by 
\begin{eqnarray}
\lefteqn{{\Big[\langle P \mid\theta^{+-}\mid P \rangle - 
{(P^\perp)^2 \over (P^+)^2}
\langle P \mid \theta^{++} \mid P \rangle \Big
]}_{diag}=}\nonumber\\&&~~~~~~~~2\int
dx_1d^2\kappa_1^\perp \psi^*_2\psi_2 \Big [{(\kappa_1^\perp)^2 \over {x_1}} +
{(\kappa_2^\perp)^2 \over (1- x_1)}\Big ]+ \nonumber\\& &{2\int dx_1d^2\kappa_1^\perp
\int dx_2d^2\kappa_2^\perp \psi_3^*\psi_3 \Big [{(\kappa_1^\perp)^2 \over {x_1}} +
{(\kappa_2^\perp)\over {x_2}} + {(-\kappa_1^\perp -\kappa_2^\perp)^2 
\over (1-x_1 -x_2)}}\Big ]\nonumber\\& &~~~~
-{8e^2\ov {2(2\pi)^3}} \int dx_1 d^2\k_1^\p \int dy d^2\k^\p
\psi_2^*(x_1,\k_1^\p)
 \psi_2(y, \k^\p) {1\ov
{(x-y)^2}} 
\end{eqnarray}
where $\theta^{+-}=\theta^{+-}_f + \theta^{+-}_g$.

The off-diagonal part can be written as
\be
{ \Big [ \langle P \mid \theta^{+-} \mid P \rangle - {(P^\perp)^2 \over (P^+)^2}
\langle P \mid \theta^{++} \mid P\rangle \Big ]}_{off-diag}
={\cal V}_1 +{\cal V}_2
\e
where,
\be
{\cal V}_1= {2e^2\over {2(2\pi)^3}}   \int dx d^2\kappa^\perp 
\int dy d^2q^\p \Big [ {1\ov E}{1\ov (x-y)} \Big [2V_1
{\mid \psi_2^{\s_1 \s_2} (x,\k^\p) \mid }^2  
\nonumber\\+V_2
(\psi_2^{*\s_1 \s_2} (x,\k^\p)\psi_2^{\sigma_1\sigma_2}(y,q^\p)+ h. c.) \Big
], 
\e

\be
{\cal V}_2= {2e^2\over {2(2\pi)^3}} \int dx d^2\kappa^\perp \int dy
d^2q^\perp
{1\ov E'}{1\ov (y-x)} \Big [
2{V'}_1
{\mid \psi_2^{\s_1 \s_2} (x,\k^\p) \mid }^2 \nonumber\\+{V'}_2
 (\psi_2^{\sigma_1\sigma_2}
(x,\kappa^\perp,1-x,-\kappa^\perp)\psi_2^{\sigma_1\sigma_2}(y,q^\p)+ 
h. c.) \Big ]. 
\e
Here the expressions for $V_1, V_2, {V'}_1, {V'}_2, E$ and $E'$ are
 given earlier. As before, we have taken the two particle state with both
$\s_1$ ,$\s_2$ up.

Considering only the photon exchange interactions and putting $M^2=M_0^2$ in
the energy denominators as before, one obtains in the non-relativistic limit
for a weak coupling theory 
\be
\lefteqn{ \Big [ \langle P \mid \theta^{+-} \mid P \rangle - {(P^\perp)^2 \over (P^+)^2}
\langle P \mid \theta^{++} \mid P\rangle \Big ] = 
2 \int dx d^2\k^\p {\mid
\psi_2 \mid }^2 {{(\k^\p)^2 + m^2}\ov {x(1-x)}}} \nonumber\\~~&&
-2{2e^2\ov {2(2\pi)^3}} \int dyd^2q^\p\psi^*_2( x, \k^\p, 1-x,
-\k^\p) \psi_2(y,q^\p,
1-y,-q^\p) \nonumber\\&&~~~~~~~~~~~~~~~~~~
\Big [ ({m\ov x})^2 {1\ov { (\k^\p-q^\p)^2 + ({m\ov
x})^2(x-y)^2}} \nonumber\\&&~~~~~~~~~~~~~~~~~~~~~~~~+ 
      ({m\ov (1-x)})^2 {1\ov { (\k^\p-q^\p)^2
+ ({m\ov (1-x)})^2(x-y)^2}}\Big ].
\label{c6d3}
\e
Introducing the three vector $\vec{p}$ (see Appendix C), this can be written
as
\be
 \Big [ \langle P \mid \theta^{+-} \mid P \rangle - {(P^\perp)^2 \over (P^+)^2}
\langle P \mid \theta^{++} \mid P\rangle \Big ]  = 2 \int d^3\vec{p} {\mid
\phi(\vec{p})\mid }^2 4 [(\vec{p})^2 + m^2] \nonumber\\~~~~ -{{4e^2}\ov
{2(2\pi)^3}} \int d^3\vec{p}\int d^3\vec{p'} \phi^*(\vec{p}) \phi(\vec{p'})
{4m\ov ( \vec{p} - \vec{p'})^2 }.
\e 
Multiplying the bound state equation (see Appendix C) by $\phi^*(\vec{p})$ and
integrating we get
\be
M^2 =  \int d^3\vec{p} {\mid
\phi(\vec{p})\mid }^2 4 [(\vec{p})^2 + m^2]  -{{2e^2}\ov
{2(2\pi)^3}} \int d^3\vec{p}\int d^3\vec{p'} \phi^*(\vec{p}) \phi(\vec{p'})
{4m\ov ( \vec{p} - \vec{p'})^2 }.
\label{c6d4}
\e 
Hence from Eqs. (\ref{c6c5}), (\ref{c6d3}) and (\ref{c6d4}) it can be seen that the
sum rule is satisfied in the lowest order in weak coupling limit 
for a positronium target in light-front QED and it can be written as
\be
\int_0^1 {{ F_L^{\tau=4}(x)}\ov x} dx  = {2\ov Q^2}\Big [ \langle P \mid \theta^{+-} \mid P \rangle - {(P^\perp)^2 \over (P^+)^2}
\langle P \mid \theta^{++} \mid P\rangle \Big ]
= 4{M^2\ov Q^2}.
\e
In the non-relativistic limit for a weak coupling theory,
\be
M^2 = 4m^2 + 4mB_e,
\e
where $B_e$ is the binding energy of positronium. 

From Eq. (\ref{c6d4}) we obtain
\be
B_e =  \int d^3\vec{p} {\mid
\phi(\vec{p})\mid }^2 {(\vec{p})^2\ov m} -{{2e^2}\ov
{2(2\pi)^3}} \int d^3\vec{p}\int d^3\vec{p'} \phi^*(\vec{p}) \phi(\vec{p'})
{1\ov ( \vec{p} - \vec{p'})^2}.
\e
The first term in the right hand side is the kinetic energy with ${m\over 2}$ being the reduced
mass of the two body system and the second term is the expectation value of
the Coulomb interaction. So we see that in the weak coupling limit, the sum
rule reduces to a relation connecting the kinetic and potential energies to
the binding energy.

%%%%%%%%%%%%%%%%%%%%%%%%%%%%%%%%%%%%%%%%%%%%%%%%%%%%%%%%%%%%%%%%%%%
\section{$F_L^{\tau=4}$ for the Ground State of Positronium}\label{c6pos}
%%%%%%%%%%%%%%%%%%%%%%%%%%%%%%%%%%%%%%%%%%%%%%%%%%%%%%%%%%%%%%%%%%%

The bound state equation (see Appendix C) can be analytically solved for QED,
which is the primary motivation for studying QED. The ground state wave
function of positronium is given by
\be
\phi_{\nu,s_e,s_{e'}}( \vec{p},s,s') = \phi_{\nu}(\vec{p}) \delta_{s_e,s}
\delta_{s_{e'},s'}
\e
where $s_e$ and $s_{e'}$ label the spin quantum numbers of the electron and
positron respectively and $\nu$ denotes all the other quantum numbers, $\nu
= n,l,m$ correspond with the standard non-relativistic quantum numbers of
hydrogen. 
The spin part factorizes out and the normalization condition is given in
Appendix C.
The wave function is given by
\be
\phi_{\nu}(\vec{p}) = {4(e_n)^{5\ov 2}\ov ((e_n)^2 +
(\vec{p})^2)^2}Y_{\nu}(\Omega_p),
\e

where
\be
e_n = {m\alpha\ov {2n}},
\e
and
\be
Y_\nu(\Omega_p) = Y_{n,l,m}(\Omega)
\e
are Hyperspherical harmonics. Here $0 \leq \mid m \mid \leq l \leq n-1$.

For 1s state of positronium, we have
\be
Y_{1,0,0} = {1\ov {\sqrt {2{\pi}^2}}}.
\e
In terms of $x$ and $\k^\p$, the 1s state wave function can be written as
\be
\phi_2(x,\k^\p) = \sqrt{m\ov {\pi}^2} {4(e_1)^{5\ov 2}\ov {\Big [ (e_1)^2 - m^2 +
{1\ov 4} { {(\k^\p)^2 + m^2} \ov {x(1-x)}} \Big ]}^2}
\e
which agrees with \cite{c6brod} for non-relativistic $x \simeq {1\ov 2}$.

The leading  order contributions to the structure functions $F_2(x)$ and
$F_L^{\tau=4}(x)$ can be directly evaluated using this wave function.
\be
F_2(x)&=& \int d^2\k^\p {\mid \psi_2(x,\k^\p)\mid }^2
\nonumber\\&=& \int d^2\k^\p {Ax^4(1-x)^4 \ov {\Big [ m^2[(1-2x)^2 +
{\alpha}^2x(1-x)] + (\k^\p)^2\Big ]}^4} 
\e
where $A=4.78 \times {10}^{-12} (Mev)^6$.
The integral is convergent and can be evaluated analytically 
introducing a cutoff
$\Lambda$ and taking the limit $\Lambda \rightarrow \infty$ in the end. We get
\be
F_2(x) = 28.15 \times {10}^{-11} {x^4(1-x)^4\ov {\Big [ (1-2x)^2 +
{\alpha}^2 x(1-x) \Big ]}^3}.
\e
Near $x={1\ov 2}$, $F_2(x) \simeq 1861.22 x(1-x)$.
$F_2(x)$ is very sharply peaked at $x={1\ov 2}$. 

The twist four longitudinal structure function is given by
\be
{F_L^{\tau=4}\ov x}&=& {4\ov Q^2} \int d^2\k^\p {(\k^\p)^2 \ov {x(1-x)}}
{\mid \psi_2 \mid }^2
\nonumber\\&&
={4\ov Q^2} \int d^2\k^\p {{A(\k^\p)^2 x^3 (1-x)^3 }\ov
{\Big [ m^2 [(1-2x)^2 + {\alpha}^2 x(1-x)] + (\k^\p)^2 \Big ]}^4}.
\e
Here we have considered only the $(\k^\p)^2$ dependent part, the integral of
which is directly connected to the kinetic energy of the electron-positron
pair.
Evaluating the integral analytically, we get
\be
{F_L(x)\ov x} = 14.7 \times {10}^{-11} {1\ov Q^2}{x^3(1-x)^3\ov {\Big [
(1-2x)^2 + {\alpha}^2 x(1-x) \Big ]}^2}.
\e 
$Q^2 {F_L(x)\ov x}$ is  sharply peaked at $x={1\over 2}$. Near the maximum 
\be
{{F_L(x)\ov x}Q^2 \simeq 14.7 \times {10}^{-11}
{1\ov {{\alpha}^4 x(1-x)}}}.
\e
%%%%%%%%%%%%%%%%%%%%%%%%%%%%%%%%%%%%%%%%%%%%%%%%%%%%
\section{Discussions}\label{c6dis}
%%%%%%%%%%%%%%%%%%%%%%%%%%%%%%%%%%%%%%%%%%%%%%%%%%%%
To summarize, in this chapter, we have investigated the twist-four longitudinal structure
function for a meson in $1+1$ dimensional QCD and also for a positronium like bound state in light-front QED in the weak
coupling limit. 

In the previous chapter, we have derived a new sum rule for the twist four
part of the longitudinal structure function. 
The validity of the sum rule has been explicitly checked in two dimensional
QCD ('t Hooft model). To get a qualitative picture of the twist four
structure function we have computed numerically both $F_2$ and $F_L$
structure functions in the 't Hooft model for the
ground state wave function calculated using a variational ansatz.

In the weak coupling limit in light-front QED, we get expressions that look similar to the
familiar non-relativistic expressions, but the entire calculation is fully
relativistic in the leading order in bound state perturbation theory. We have explicitly verified a sum rule for $F^{\tau=4}_L$ that we
previously proposed.  We have shown that in the weak coupling
limit, the sum rule reduces to a relation connecting the kinetic and the
potential energies to the binding energy of positronium. We have also shown
that, in this limit, the fermionic part of  $F^{\tau=4}_L$ contributes only
to the kinetic energy of the fermions and not to the interactions.      

The twist four part of the longitudinal structure function is important
since it is the leading non-perturbative contribution to $F_L$. The leading
twist contribution to  $F_L$ is perturbative, in contrast to the case of
$F_2$. This analysis for a bound state in weak-coupling light-front QED is
quite interesting since it gives an idea of what goes in such a calculation
in light-front QCD.
The situation in light-front QCD is somewhat different
because similarity renormalization group technique generates a confining
interaction even in $O(g^2)$ in the effective Hamiltonian which
makes it impossible to solve the bound state equation analytically. 
However, similarity renormalization upto $O(e^2)$ does not produce any
additional interaction in the effective QED Hamiltonian and in the weak
coupling limit, one can work with the canonical Hamiltonian. Similarity
renormalization results agree with the results using particle number
truncation in the weak coupling limit for light-front QED. Inspite of
these differences, the overall computational framework in QCD is the same
and this analysis in light-front QED allows an analytic understanding of the
problem.
%%%%%%%%%%%%%%%%%%%%%%%%%%%%%%%%%%%%%%%%%%%%%%%%%%%%%%%%%%%%%%%%%%%%%%%%%%%%

%%%%%%%%%%%%%%%%%%%%%%%%%%%%%%%%%%%%%%%%%%%%%%%%%%%%%%%%%%

%\end{document}

\chapter{Transverse Spin in QCD and Transverse Polarized DIS}

The spin structure of the proton is one of the most challenging problems in
present day particle physics. A large number of theoretical and experimental
investigations are going on in this subject. The famous EMC result
gave rise to the so called `spin crisis' by 
indicating that only a small fraction of
the proton helicity is carried by the quarks and antiqurks, and so
Ellis-Jaffe sum rule is violated. A much more debated and interesting issue  
  is the transverse polarized structure function $g_T$, which can be
measured in transverse polarized DIS. The contribution to the cross section
in this case is proportional to $g_T$ and this is suppressed by a factor
${1\over Q}$ compared to the leading contribution of $g_1$ in the
longitudinally polarized case. So, the structure function $g_T$ is measured
less accurately than $g_1$. Also, $g_T$ cannot be expressed as an incoherent
sum over on-mass-shell partons like $g_1$. The partons must be interacting
in order to contribute to transverse polarized scattering. Due to these
reasons, $g_T$ is called a higher twist effect.

The literature on transverse polarized structure function $g_T$ or $g_2$
($g_T=g_1+g_2$) is vast and even contradictory and confusing \cite{c7jnp,c7efre}.
Early works on $g_2$ include \cite{c7hei}.
The discussion of $g_2$ at large $Q^2$ using OPE has been done in \cite{c7sh}. 
 Quark mass plays an important role in
transversely polarized scattering  (see chapter 4 for
references). The recent polarized DIS data has opened up new avenues to
explore $g_T$.

We analyze the transverse polarized structure function $g_T$ in light-front
Hamiltonian QCD. We show that $g_T$ is related to the interaction
dependent transverse spin operator in light-front QCD which is the reason
for its complexity. In this chapter, we
present  our analysis of the transverse spin operator and transverse
polarized structure function.       

The plan of this chapter is as follows. In Sec. 7.1, first, we briefly
review the complexities associated with the description of the spin of a
composite system in a moving frame in the conventional equal time
quantization. Then in Sec. 7.2 we give
the canonical
structure of light-front Poincare algebra. In Sec. 7.3 and 7.4 we introduce
the light-front spin operators for massive and massless cases respectively
 of arbitrary transverse momentum.
The
explicit form of transverse rotation operators in light-front QCD is derived
in Sec. 7.5. The connection of the transverse spin operators with $g_T$ is
given in Sec. 7.6. 
Summary and conclusions are presented in Sec. 7.7. 
For the sake of completeness and clarity,
the explicit form of the kinematical operators and the Hamiltonian in
light-front QCD starting from the gauge invariant, symmetric, interaction
dependent,  energy momentum
tensor is derived in Appendix D. A complete discussion of 
transverse spin operators in free fermion
field theory and free massless, spin one boson field theory 
is presented in detail in Appendices E and F.     

 %%%%%%%%%%%%%%%%%%%%%%%%%%%%%%%%%%%%%%%%%%%%%%%%%%%%%%%%%%%%%%%%%
\section{The Problem of Spin in Relativistic Quantum Field Theory}
%%%%%%%%%%%%%%%%%%%%%%%%%%%%%%%%%%%%%%%%%%%%%%%%%%%%%%%%%%%%%%%%%%%

From the early days of quantum field theory, it has been recognized that the
issues associated with the spin of a composite system in an arbitrary
reference frame are
highly complex and non-trivial \cite{c7alfaro}.
The familiar Pauli-Lubanski operators readily qualify for spin operators 
{\it only} in the rest frame of
the particle. For a single particle in a moving frame it is known \cite{c7gur} 
how to construct the appropriate spin operators starting from the 
Pauli-Lubanski operators. 
How to construct the spin operators for a composite system in an arbitrary
reference frame is a nontrivial problem. In equal-time quantization,  
the complexities arise
from the facts that for a moving composite object, {\it Pauli-Lubanski
operators are necessarily interaction dependent} and, further, it is quite
difficult \cite{c7os} to separate the center of mass and internal variables which is
mandatory in the calculation of spin. Due to these difficulties there has been
rarely any attempt to study the spin of a moving composite system in the
conventional equal time formulation of even simple field theoretic models,
let alone Quantum Chromo Dynamics (QCD).

In order to resolve the above mentioned problems and puzzles, we have
undertaken an  investigation of the spin of a composite
system in an arbitrary reference frame in QCD. We have compared and
contrasted both the instant form and front form formulations. We emphasize
that the interaction
dependence of the spin of a composite system in an arbitrary reference frame
is not a peculiarity of light-front dynamics, it is a general feature in any
formulation of quantum field theory. What is peculiar to light-front
dynamics is that one can at most go only to the transverse rest frame of the
particle. No frame exists in which $P^+=0$ and one is so to speak $``$always in
a moving frame". As a consequence, spin measured in any direction other than
that of $P^+$ cannot be separated into orbital and intrinsic parts. This is
to be contrasted with the light-front helicity ${\cal J}^3$ which is independent
of interactions and further can be separated in to orbital and intrinsic
parts. The situation is quite analogous to that of a light-like particle. In
this case it is well known that since there is no rest frame, one can
uniquely identify the spin of the particle only along the direction of motion
since only along this direction one can disentangle rotation from translation
for a massless particle. Also, in any direction other than the direction of
motion, one cannot separate the angular momentum into orbital and intrinsic 
parts.

%%%%%%%%%%%%%%%%%%%%%%%%%%%%%%%%%%%%%%%%%%%%%%%%%%%%%%%%%%%%%%%%%%%%
\subsection{Intrinsic Spin in Non-Relativistic Quantum Mechanics}
%%%%%%%%%%%%%%%%%%%%%%%%%%%%%%%%%%%%%%%%%%%%%%%%%%%%%%%%%%%%%%%%%%%%
In the non-relativistic case, the transformation from one inertial frame into
another is obtained by a Galelian transformation, which is generated by ten
generators, namely, the Hamiltonian $H$, which generates time translation,
three momenta ${\bf P}$ which produce translation in space, three
angular momenta ${\bf J}$ which produce rotation and the Galelian boost
operators ${\bf N}$. Among these, only the Hamiltonian is dynamical, all the
remaining generators are kinematical. These generators obey the following 
commutation relations among themselves:
\be
\lefteqn{[J_k, J_l]=i\ep_{klm}J_m, ~~~~[P_k, J_l]=i\ep_{klm}P_m, ~~~[N_k, J_l]=
i\ep_{klm}N_m}\nonumber\\& &~~~[P_k, P_l]=0, ~~~~~[N_k, N_l]=0,~~~~~~
[P_k, N_l]=i\delta_{kl} M \nonumber\\& & [P_k,H]=0,~~~~~~~~~[J_k,H]=0,~~~~~~~~~
[N_k,H]=-iP_k,
\e
where we have used summation convention over repeated indices. Here $M$ is
the mass operator. 
From the above commutation relations, it can be seen that the angular momenta
are not translationally invariant. The angular momenta can be written as
\be
{\bf J}={1\over M}({\bf N} \times {\bf P})+{\bf I}
\e
where ${1\over M}$ is the inverse of the mass operator and $I$ is called the
intrinsic spin \cite{c7jsc}. The intrinsic spin operators are
translationally invariant and obey $SU(2)$ commutation relations among
themselves. The non-relativistic spin operators are simple due to the fact 
that the Galelian boost generators are kinematical and
commute within themselves.
%%%%%%%%%%%%%%%%%%%%%%%%%%%%%%%%%%%%%%%%%%%%%%%%%%%%%%%%%%%%%%%%%%%%%%
\subsection{Relativistic Spin Operators in Arbitrary Frame}
%%%%%%%%%%%%%%%%%%%%%%%%%%%%%%%%%%%%%%%%%%%%%%%%%%%%%%%%%%%%%%%%%%%%%%
A relativistic dynamical system in equal-time formulation is transformed
from one frame to another by Poincare transformation, the generators of
which are, the relativistic Hamiltonian $H$,
three momenta ${\bf P}$, three
angular momenta ${\bf J}$  and the boost operators ${\bf K}$.  The
commutation relations obeyed by these generators are given below:
\be
\lefteqn{[J_k, J_l]=i\ep_{klm}J_m, ~~~~[P_k, J_l]=i\ep_{klm}P_m, ~~~[K_k, J_l]=
i\ep_{klm}K_m}\nonumber\\& &[P_k, P_l]=0, ~~~~~[K_k, K_l]=-i\ep_{klm}J_m,~~ ~~~~
[P_k, K_l]=i\delta_{kl}H \nonumber\\& &[P_k,H]=0,~~~~~~~~~[J_k,H]=0,~~~~~~~~~
[K_k,H]=-iP_k.
\e 
The crucial difference from the non-relativistic case is that the
relativistic boost generators are dynamical and they do not commute with
each other. The angular momentum $J$ is not translationally
invariant and therefore does not qualify as spin operators. 
The Casimir invariants of the Lie algebra are 
\be
M^2=P^\mu P_\mu, ~~~~~~~~~W^2=W^\mu W_\mu,
\e
where
\be
W^\mu=-{1\over 2} \ep^{\mu \nu \rho \sigma}J_{\nu \rho} P_\sigma
\e
is the Pauli-Lubanski operator. It is translationally invariant and its
components obey the commutation relations:
\be
[W^\mu, W^\nu]=i\ep^{\mu \nu \lambda \rho}W_\lambda P_\rho.
\e
For a particle at rest, the 4-momentum is, $P^\mu=(M, 0, 0, 0)$. The
commutation relation then becomes
\be
[W^i, W^j]=i\ep^{ijk} W^k M
\e
so that if we define three operators, $S^i={W^i\over M}$, then $S^i$ obey
$SU(2)$ algebra. Thus, we can define spin in terms of Pauli-Lubanski
operators. However, in an arbitrary frame, it is much more complex, because
$W^\mu$ are interaction dependent.

We consider two classes of representations which are of physical importance:
\begin{itemize}
{\item Positive time-like representations: $M^{2}>0~~H>0$}
{\item Positive light-like representations: $M^{2}=0~~H>0$}.
\end{itemize}
In either cases we do not demand that the representations be irreducible (this
allows us to deal with elementary and composite systems simultaneously).
%%%%%%%%%%%%%%%%%%%%%%%%%%%%%%%%%%%%%%%%%%%%%%%%%%%%%
\subsubsection{A. Positive Time-like Representations}
%%%%%%%%%%%%%%%%%%%%%%%%%%%%%%%%%%%%%%%%%%%%%%%%%%%%%
Starting from the Pauli-Lubanski operator, 
one can
construct an operator ${\bf S}$ such that it is translationally invariant, transforms
as a three vector under pure rotations and its components  obey $SU(2)$ commutation
relations among themselves.
\be
[S^j,P^\mu]=0, ~~[J^j,S^k]=i \epsilon^{jkl}S^l, ~~[S^j,S^k]=i
\epsilon^{jkl}S^l.
\e
A suitable solution to the above requirements is provided by
\be
{\bf S}&=&{1\over M}\left[{\bf W}-{{\bf P}W^{0}\over {M+H}}\right]\nonumber\\
  &=&{\bf J}~{P^0 \over M} - {\bf K} \times  {{\bf P}\over M} -
{({\bf J} \cdot {\bf P})\over {M+P^0}}{{\bf P}\over M}    
\e
where ${\bf W}$ are the space components of the Pauli-Lubanski
operator. $ H $, ${\bf P}$ are equal time Hamiltonian and momentum
operators respectively obtained by integrating the energy
momentum tensor over a spacelike surface and $ {\bf J}$ and $ {\bf K}$ are
the equal time rotation and boost generators respectively, which are obtained by
integrating the angular momentum density over a spacelike surface.  
Since boost ${\bf K}$ is dynamical, {\it all the three components of ${\bf S}$
are interaction dependent} in the equal time quantization. 
Nevertheless, the component of {\bf S} along {\bf P} remains kinematical.
We shall see later in this section that, this is to be
compared with light-front quantization where {\it the third component of the
light-front spin operator ${\cal J}^3$ is kinematical}.
This arises from the facts that boost operators are kinematical on the light
front, the interaction dependence of light-front spin
operators ${\cal J}^i$ arises solely from the rotation operators, and the
third component of the rotation operator $J^3$ is kinematical on the light-
front.

The operators ${\bf S}$ cease to be defined when $M$ tends to zero.
The commutation relations among ${\bf P}, {\bf S}$ and $M$ are
given by  
\be
 [ P^j,S^k ]=0,~~
[S^j,S^k]=i \epsilon^{jkl}S^l,~~[S^j,M]=0.\label{c7rel1}
\e
Since ${\bf P}$ and $M$ stand for the momentum and invariant mass of the system,
the above relations make clear that ${\bf S}$  should represent
`intrinsic spin' of the system. 

The invariant $W^{2}$ can be completely expressed in terms of $M$ and ${\bf S}$ as
\be
W^{2}=-M^{2}{\bf S}^{2}.
\e
\vskip .2in
%%%%%%%%%%%%%%%%%%%%%%%%%%%%%%%%%%%%%%%%%%%%%%%%%%%%%%
{\subsubsection{B. Positive Light-like Representations}
%%%%%%%%%%%%%%%%%%%%%%%%%%%%%%%%%%%%%%%%%%%%%%%%%%%%%%%%
In the massless case, the spin operators do not obey $SU(2)$ algebra. 
Beginning from the basic generators ${\bf P}$, $\bf J$ and $\bf K$ (here
$H=|{\bf P}|$) one has to construct operators $S$, ${\cal T}^1$ and
${\cal T}^2$ such that they commute with four momentum $P^{\mu}$ and amongst 
themselves satisfy $E(2)$ commutation relations:
\begin{equation}
\begin{array}{lll}
[S,{\cal T}^1]=i{\cal T}^2, & [S,{\cal T}^2]=-i{\cal T}^1, & 
[{\cal T}^1,{\cal T}^2]=0.
\end{array} \end{equation}
A suitable solution consistent with the above requirements is:
\be
S={W^{0}\over \mid{\bf P}\mid }, \nonumber\\
{\cal T}^1=W^{1}-P^{1}{(W^{3}+W^{0})\over (\mid {\bf P}\mid+P^{3})},
\nonumber\\
{\cal T}^2=W^{2}-P^{2}{(W^{3}+W^{0})\over (\mid {\bf P}\mid +P^{3})}.
\e
Note that here
$S$ is the component of angular momentum in the direction of motion. 
Another point is, $S$ is a scalar
under pure spatial rotation, while shows complicated behavior under
pure boosts. 
%%%%%%%%%%%%%%%%%%%%%%%%%%%%%%%%%%%%%%%%%%%%%%%%
\subsubsection{C. Comments}
%%%%%%%%%%%%%%%%%%%%%%%%%%%%%%%%%%%%%%%%%%%%%%
The spin operators for a single particle in an arbitrary frame can be
defined easily in the above way.  
The generators for a multi-particle relativistic 
system have been analyzed by several authors \cite{c7os} in the equal-time
formalism. The expressions
obtained are too complicated to be used in any practical calculations.
In order
to describe the intrinsic spin of a composite system, one should be able to
separate the center of mass motion from the internal motion. Even in free field
theory, this turns out to be quite involved (See Ref. \cite{c7os} and references
therein). The generators cannot
be easily separated into the center of mass and internal variables.
Moreover, the derivations have been done neglecting the field theoretical
effects such as pair creation and crossing and so are expected to be valid
in the relatively low energy region where an expansion in ${v\over c}$ is
permissible. Interactions are to be incorporatated by introducing an
effective potential which vanish sufficiently rapidly for large distance.
On the other hand, in light-front theory, since transverse boosts
are simply Galilean boosts, separation of center of mass motion and internal
motion is as simple as in non-relativistic theory. (See Appendix G).     

Next, we discuss the light-front case.

%%%%%%%%%%%%%%%%%%%%%%%%%%%%%%%%%%%%%%%%%%%%%%%%%%%%%%%%%%%%%%
\section{Light-Front Poincare Generators}
%%%%%%%%%%%%%%%%%%%%%%%%%%%%%%%%%%%%%%%%%%%%%%%%%%%%%%%%%%%%%%
The light-front Poincare generators have been introduced in chapter 3. 
In terms of the gauge invariant, symmetric energy momentum tensor
$\Theta^{\mu \nu}$, the four-vector $P^\mu$ and the 
tensor $M^{\mu \nu}$ are given by
\begin{eqnarray}
P^\mu &=& {1 \over 2} \int dx^- d^2 x^\perp \Theta^{+ \mu}, \\
M^{\mu \nu} &=& {1 \over 2} \int dx^- d^2 x^\perp \left [ x^\mu \Theta^{+
\nu} - x^\nu \Theta^{+ \mu} \right ]. \label{c7def1}
\end{eqnarray}
The boost operators are $ M^{+-} = 2 K^3$ and $M^{+i}=E^i$. The rotation
operators are $ M^{12}=J^3$ and $ M^{-i} = F^i$. The Hamiltonian $P^-$ and
the transverse rotation operators $F^i$ are dynamical (depend on the
interaction) while other seven operators are kinematical. The 
rotation operators obey the $E(2)$-like algebra of two dimensional Euclidean 
group, namely,
\begin{eqnarray}
[F^1,F^2]=0, ~ [J^3,F^i] = i \epsilon^{ij} F^j
\end{eqnarray}
where $\epsilon^{ij}$ is the two-dimensional antisymmetric tensor. 
The remaining non-zero commutators are:
\be
\lefteqn{[P^+,K^3]=-iP^+, ~~~~[P^+, F^i]=2iP^i, ~~~[P^i, E^j]=
i\delta^{ij}P^+} \nonumber\\& &~~~[P^i, J^3]=i \ep^{ij}P^j, ~~~~~[P^i, F^j]=
i \delta^{ij} P^-,~~ ~~~~
[K^3, E^i]=iE^i \nonumber\\& &~~~~[K^3, F^i]=-iF^i,~~~~~~~~~[K^3, P^-]=-iP^-,~~~~~~~~~
[E^i, J^3]=i\ep^{ij}E^j\nonumber\\& &~~~~[E^i,
F^j]=2iK^3\delta^{ij}-2iJ^3\ep^{ij},~~~[E^i, P^j]=-2iP^i.
\e 
Another important point is that, the transverse boost generators $E^i$
commute with each other like the Galelian boosts.

From the early days of light-front field theory, the complications
associated with
transverse rotation operators $F^i$ 
have been recognized. They are interaction
dependent just like the Hamiltonian. Furthermore, together with the third
component of the rotation operator $J^3$, which is kinematical, $F^i$  do 
not obey the angular momentum algebra. Instead they obey the algebra of 
two dimensional Euclidean group which is appropriate only for massless
particles. For massive particles, one can define transverse spin operators
\cite{c7ls78} which together with the third component (helicity) obey the angular
momentum algebra. However, they cannot be separated into orbital and spin
parts unlike the helicity operator \cite{c7hk}. 
Most of the studies of the transverse spin operators in light-front field
theory, so far, are
restricted to free field theory \cite{c7except}. 
Even in this case the operators have a
complicated structure. However, one can 
write these operators as a
sum of orbital and spin parts, which can be achieved via a unitary
transformation, the famous Melosh transformation \cite{c7melosh}. 
In interacting
theory, presumably this can be achieved order by order \cite{c7bp} in a suitable
expansion parameter 
which is justifiable only in a weakly  coupled theory.

Knowledge about 
transverse rotation operators and transverse spin operators is important  
for addressing issues concerning Lorentz invariance in
light-front theory. Unfortunately,
very little is
known \cite{c7review} regarding the field theoretic aspects of the
interaction dependent spin operators,  {\it We emphasize that in a moving 
frame, the spin
operators are interaction dependent irrespective of whether one considers
equal-time field theory or light-front field theory}. 
To the best of our knowledge, in gauge field theory, the canonical 
structure of spin
operators of a composite system in a moving frame  has 
never been studied.
In this chapter, we present a systematic investigation of the spin of a
composite system in a moving frame
in QCD. We refer to our papers and preprints, \cite{c7lett,c7tran1}.
We show that, in spite of the complexities,
light-front field theory
offers a unique opportunity to address the issue of relativistic spin
operators in an arbitrary reference frame since boost is kinematical in this
formulation.

%%%%%%%%%%%%%%%%%%%%%%%%%%%%%%%%%%%%%%%%%%%%%%%%%%%%%%%%%%%%%%%%%%%%%%
\section{Transverse Spin Operators}
%%%%%%%%%%%%%%%%%%%%%%%%%%%%%%%%%%%%%%%%%%%%%%%%%%%%%%%%%%%%%%%%%%%%%%
\subsection{Massive particle}
%%%%%%%%%%%%%%%%%%%%%%%%%%%%%%%%
The Pauli-Lubanski spin operator 
\begin{eqnarray}
W^\mu = - { 1 \over 2} \epsilon^{\mu \nu \rho \sigma} M_{\nu \rho}
P_\sigma
\end{eqnarray}
with $ \epsilon^{+-12} = -2$.
For a massive particle, the transverse spin operators \cite{c7ls78} ${\cal J}^i$ in 
light-front theory are given in terms of Poincare generators by
\begin{eqnarray}
M{\cal J}^1 &=& W^1 - P^1 {\cal J}^3 = { 1 \over 2} F^2 P^+ + K^3 P^2   - { 1
\over 2} E^2 P^- - P^1 {\cal J}^3, \label{c7j1}
\e
\be  
M{\cal J}^2 &=& W^2 - P^2 {\cal J}^3 = - { 1 \over 2 } F^1 P^+ -K^3 P^1  + { 1 \over 2} E^1 P^- -
P^2 {\cal J}^3. \label{c7j2}
\end{eqnarray}
The first term in Eqs. (\ref{c7j1}) and (\ref{c7j2}) contains both center of
mass motion and internal motion and the next three terms in these equations
serve to remove the center of mass motion.  
 
The helicity operator is given by
\begin{eqnarray}
{\cal J}^3 &=& {W^+ \over P^+} = J^3 + { 1 \over P^+}(E^1P^2 - E^2 P^1).
\end{eqnarray} 
Here, $J^3$ contain both center of mass motion and internal motion and the
other two terms serve to remove the center of mass motion. 
The operators ${\cal J}^i$  obey the angular momentum commutation relations 
\begin{eqnarray}
\left [ {\cal J}^i, {\cal J}^j \right ] = i \epsilon^{ijk} {\cal J}^k
\label{c7j3} .
\end{eqnarray}
They commute with $P^\mu$. 
%%%%%%%%%%%%%%%%%%%%%%%%%%%%%%%%%%%%%%%%%%%%%%%%%%%%%%%%%%%%%
\subsection{Massless Case}
%%%%%%%%%%%%%%%%%%%%%%%%%%%%%%%%%%%%%%%%%%%%%%%%%%%%%%%%%%%%%
Again, we start from the Pauli-Lubanski spin operator
\begin{eqnarray}
W^\mu = - { 1 \over 2} \epsilon^{\mu \nu \rho \sigma} M_{\nu \rho } p_\sigma
.
\end{eqnarray}
For the light-like vector $ p^\mu$, usually the collinear choice is
made \cite{c7Tung,c7Weinberg}, namely, $p^+ \neq 0$, $
p^\perp=0$. Then we get, $ W^-=0$, $ W^+ = J^3 p^+$, $W^1 = { 1 \over 2} F^2
p^+$, $ W^2 = - { 1 \over 2} F^1 p^+$. 

In free field theory, we have explicitly
constructed the Poincare generators for a massless spin one particle in
$A^+=0$ gauge in Appendix F. 
Consider the single particle state $ \mid p \lambda \rangle$ with $
p^\perp=0$. 
From the explicit form of the operators, we find that 
\begin{eqnarray}
J^3 \mid p \lambda \rangle &=& \lambda \mid p \lambda \rangle, \nonumber \\
F^i \mid p \lambda \rangle && = 0, ~ i=1,2 
\end{eqnarray}
since $ p^\perp=0$.

For calculations with composite states (dressed partons, for example) we
have to consider light-like particles with arbitrary
transverse momenta. Let us try a light like momentum $ P^\mu$ with $ P^\perp
\neq 0$, but $ P^- = {(P^\perp)^2 \over P^+}$ so that $P^2 = 0$. 
Then we get, as in the case of massive particle,
\begin{eqnarray}
W^+ &=& J^3 P^+ + E^1 P^2 - E^2 P^1 ,\nonumber \\
W^1 &=& { 1 \over 2} F^2 P^+ +K^3 P^2     - { 1 \over 2} E^2 P^-,\nonumber \\
W^2 &=& - { 1 \over 2} F^1 P^+ - K^3 P^1  + { 1 \over 2} E^1 P^-, \nonumber
\\
W^- &=& F^2 P^1 - F^1 P^2 - J^3 P^-.    
\end{eqnarray}
Thus even though $W^1$ and $W^2$ do not annihilate the state, we do get
$W^\mu W_\mu(={1\over 2}W^+W^- + {1\over 2}W^-W^+ - (W^1)^2 - (W^2)^2)
  \mid k \lambda \rangle =0$ 
as it should be for a massless particle.

Just as in the case of massive particle, we have the helicity operator for
the massless particle,
\begin{eqnarray}
{\cal J}^3 &=& {W^+ \over P^+} = J^3 + { 1 \over P^+}(E^1P^2 - E^2 P^1).
\end{eqnarray}
In analogy with 
the transverse spin for massive particles, we define the transverse spin 
operators for massless particles as
\begin{eqnarray}
{\cal J}^i = W^i - P^i {\cal J}^3.
\end{eqnarray}
They do satisfy
\begin{eqnarray}
{\cal J}^i \mid k \lambda \rangle &=&0, \nonumber \\
{\cal J}^3 \mid k \lambda \rangle && = \lambda \mid k \lambda \rangle, 
\end{eqnarray}
where $k$ is an arbitrary momentum. 
The operators ${\cal J}^i$ and ${\cal J}^3$ obey the $E(2)$-like algebra
\begin{eqnarray}
\left [{\cal J}^1, {\cal J}^2 \right ] =0, ~
\left [{\cal J}^3, {\cal J}^1 \right ] = i {\cal J}^2,~
\left [{\cal J}^3, {\cal J}^2 \right ] = -i {\cal J}^1.  
\end{eqnarray}
%%%%%%%%%%%%%%%%%%%%%%%%%%%%%%%%%%%%%%%%%%%%%%%%%%%%%%%%%%%
\subsection{Comments}
%%%%%%%%%%%%%%%%%%%%%%%%%%%%%%%%%%%%%%%%%%%%%%%%%%%%%%%%%%
In order to calculate the transverse spin operators, first we need to
construct the Poincare generators $P^+$, $P^i$, $P^-$, 
$K^3$, $E^i$, $J^3$ and
$F^i$ in light-front QCD. The explicit form of the operator $J^3$ is given
Ref. \cite{c7hk}. The construction of $F^i$ which is algebraically 
quite involved is carried out in the next section. The construction of the
rest of the kinematical operators is given in Appendix D. In this appendix we also present
the Hamiltonian in a manifestly Hermitian form.    

In order to have a physical picture of the complicated situation at hand it
is instructive to calculate the spin operator in free 
field
theory. The case of free massive fermion is carried out in Appendix E. 
In free field theory one can
indeed show that (see Appendix E) ${\cal J}^i \mid k \lambda \rangle = 
{ 1 \over 2} \sum_{\lambda'} \sigma^i_{\lambda' \lambda} \mid k \lambda'
\rangle$.   The case of free massless spin one
particle is carried out in Appendix F.
%%%%%%%%%%%%%%%%%%%%%%%%%%%%%%%%%%%%%%%%%%%%%%%%%%%%%%%%%%%%%%%%%%%%%%%%
\section{The Transverse Rotation Operator in QCD}
%%%%%%%%%%%%%%%%%%%%%%%%%%%%%%%%%%%%%%%%%%%%%%%%%%%%%%%%%%%%%%%%%%%%%%%%
In this section we derive the expressions for interaction dependent
transverse rotation operators in light-front QCD starting from the
manifestly gauge invariant energy momentum tensor. It is extremely
interesting to compare and contrast the situation in the equal time and
light-front case. 
The angular momentum density
\begin{eqnarray}
{\cal M}^{\alpha \mu \nu} = x^\mu \Theta^{\alpha \nu} - x^\nu
\Theta^{\alpha \mu}. 
\end{eqnarray}
In equal time theory, generalized angular momentum
\begin{eqnarray}
M^{\mu \nu} = \int d^3x {\cal M}^{0 \mu \nu}.
\end{eqnarray}
The rotation operators are $ J^i = \epsilon^{ijk} M^{jk}$. Thus in a
non-gauge theory, all the three components of the rotation operators are
manifestly interaction independent. However, as we have seen in the last
section, the spin operators $S^i$ for a 
composite system in
a moving frame involves, in addition to  $J^i$, the  boost operators
$K^i = M^{0i}$ which are interaction dependent. {\it Thus all the three
components of $S^i$ become interaction dependent.}

A gauge invariant separation of the nucleon angular momentum is performed in
Ref. \cite{c7ji}. However, as far the spin operator in an arbitrary reference
frame is concerned, 
the analysis of this reference is valid only in the rest frame where spin
coincides with total angular momentum operator and in an arbitrary 
reference frame the need to project out the
center of mass motion, which is quite complicated in equal time theory is
not emphasized. 
Moreover, the distinction between the longitudinal and transverse components
of the spin is never made.
It is crucial to make this distinction since physically
the longitudinal and transverse components of the spin carry quite distinct
information (as is clear, for example, from the spin of a massless particle). 
Moreover, even for the third component of the spin of a composite system
in a moving frame, there is crucial difference between equal time and light
front cases. ${\cal J}^3$
(helicity) is interaction independent whereas $S^3$ is interaction
dependent in general except when measured along the direction of {\bf P}.

In light-front theory, generalized angular momentum 
\begin{eqnarray}
M^{\mu \nu} =
{ 1 \over 2} \int dx^- d^2 x^\perp {\cal M}^{+ \mu \nu}.
\end{eqnarray}
$J^3$ which is related to the helicity is given by  
\begin{eqnarray}
J^3 = M^{12} = { 1
\over 2} \int dx^1 d^2 x^\perp [x^1
\Theta^{+2} - x^2 \Theta^{+1} ] 
\end{eqnarray}
and is interaction independent.
On the other hand, the transverse rotation operators which are related
to the transverse spin are given by
$$ F^i =M^{-i}= { 1 \over 2} \int dx^- d^2 x^\perp [ x^- \Theta^{+i} - x^i
\Theta^{+-} ] . $$
They are interaction dependent {even in a non-gauge theory} since
$\Theta^{+-}$ is the Hamiltonian density.

In light-front theory we set the gauge $A^+=0$
and eliminate the dependent variables $\psi^-$ and $A^-$ using the equations
of constraint. In this paper we restrict to the topologically trivial sector
of the theory and set the boundary condition $A^i(x^-, x^i) \rightarrow 0 $
as $ x^{-,i} \rightarrow \infty$. This completely fixes the gauge and put
all surface terms to zero.

The transverse rotation operator 
\begin{eqnarray}
F^i = {1 \over 2} \int dx^- d^2 x^\perp \Big [ x^- \Theta^{+i} - x^i
\Theta^{+-} \Big ].
\end{eqnarray}
The symmetric, gauge invariant energy momentum tensor 
\begin{eqnarray}
\Theta^{\mu \nu} &=& { 1 \over 2} {\overline \psi} \Big [ 
  \gamma^\mu i D^\nu + \gamma^\nu i D^\mu \Big ] \psi - F^{\mu \lambda a}
F^{\nu a}_{\, \, \lambda} 
 \nonumber\\&&~~~~~~~~~~~~~~~- g^{\mu \nu} \Big [ - { 1 \over 4}
 (F_{\lambda \sigma a})^2 +
{\overline \psi} ( \gamma^\lambda i D_\lambda - m) \psi \Big ],
\end{eqnarray}
where 
\begin{eqnarray}
i D^\mu &=& {1 \over 2} \st{\lh}{i\pp^\mu} + g A^\mu, \nonumber \\
F^{\mu \lambda a} &=& \partial^\mu A^{\lambda a} - \partial^\lambda A^{\mu
a} + g f^{abc} A^{\mu b} A^{\lambda c}, \nonumber \\
F^{\nu a}_{\, \, \lambda} &=& \partial^\nu A_{\lambda}^a - \partial_\lambda
A^{\nu a} + g f^{abc} A^{\nu b} A_\lambda^c.
\end{eqnarray}
First consider the fermionic part of $ \Theta^{\mu \nu}$:
\begin{eqnarray}
\Theta^{\mu \nu}_F = { 1 \over 2} {\overline \psi} \Big [ \gamma^\mu i D^\nu
+ \gamma^\nu i D^\mu \Big ]\psi - g^{\mu \nu } {\overline \psi} (\gamma^\lambda
i D_\lambda - m)\psi.
\end{eqnarray}
The coefficient of $g^{\mu \nu}$ vanishes because of the equation of motion. 

Explicitly, the contribution to $F^2$ from the fermionic part of
$\Theta^{\mu \nu}$ is given by
\begin{eqnarray}
F^2_F && = { 1 \over 2} \int dx^- d^2 x^\perp \left [ x^- { 1 \over 2}
{\overline \psi} (\gamma^+ i D^2 + \gamma^2 i D^+) \psi 
 - x^2 { 1 \over 2}{\overline \psi} (\gamma^+ i D^- + \gamma^- i D^+) \psi
\right ],  \nonumber \\
&& = F^2_{F(I)} + F^2_{F(II)},
\end{eqnarray} 
where 
\begin{eqnarray}
F^2_{F(I)}= { 1 \over 2} \int dx^- d^2 x^\perp x^- \Big [ 
{\psi^+}^\dagger {1 \over 2} \st{\lh}{i\pp^2} \psi^+ + {\psi^+}^\dagger g A^2 \psi^+ + 
{ 1 \over 4} {\overline
\psi} \gamma^i  \st{\lh}{i\pp^+}
\psi \Big ],
\end{eqnarray}
\begin{eqnarray}
F^2_{F(II)}= -{ 1 \over 2} \int dx^- d^2 x^\perp x^2 \Big [ 
{\psi^+}^\dagger \Big ({1 \over 2} \st{\lh}{i\pp^-} + gA^- \Big ) \psi^+ 
 + 
{ 1 \over 4}
{\psi^-}^\dagger \gamma^i \st{\lh}{i\pp^+}
\psi^- \Big ].
\end{eqnarray} 
We have the equation of constraint
\begin{eqnarray}
i \partial^+ \psi^- = \big [ \alpha^\perp \cdot ( i \partial^\perp + g
A^\perp) + \gamma^0 m \big ] \psi^+ \label{c7eoc}
\end{eqnarray}
and the equation of motion
\begin{eqnarray}
i \partial^- \psi^+ = -g A^- \psi^+ + \big [ \alpha^\perp \cdot  (i
\partial^\perp + g A^\perp) + \gamma^0 m \big]{ 1 \over i \partial^+}
\big [ \alpha^\perp \cdot  (i
\partial^\perp + g A^\perp) + \gamma^0 m \big]   \psi^+. \label{c7eom}
\end{eqnarray}
Using the Eqs. (\ref{c7eoc}) and (\ref{c7eom}) we arrive at free ($g$
independent) and
interaction ($g$ dependent) parts of $F^2_F$.
The free part of $F^2_F$ is given by
\begin{eqnarray}
F^2_{F(free)} &=& { 1 \over 2} \int dx^- d^2 x^\perp \Bigg \{x^- \Bigg [
\xi^\dagger \Big [i \partial^2 \xi\Big] - 
\Big [i \partial^2 \xi^\dagger\Big ] \xi \Bigg ]  \nonumber \\
&&~~~~~~ - x^2 \Bigg [ \xi^\dagger \Big [{ - (\partial^\perp)^2 +m^2 \over i
\partial^+} \xi \Big ] -  \Big [{ - (\partial^\perp)^2 +m^2 \over i
\partial^+} \xi^\dagger\Big ] \xi \Bigg ] \nonumber \\
&& ~~~~~~+ \Bigg [ \xi^\dagger \Big [ \sigma^3 \partial^1 + i \partial^2
\Big]{ 1 \over
\partial^+} \xi + \Big [ { 1 \over \partial^+} (\partial^1 \xi^\dagger \sigma^3 -
i \partial^2 \xi^\dagger) \Big ] \xi \Bigg ] \nonumber \\ 
&&~~~~~~ + m \Bigg [ \xi^\dagger \Big [{ \sigma^1 \over i \partial^+} 
\xi\Big ] -
\Big [{ 1 \over i \partial^+} \xi^\dagger\sigma^1\Big ] \xi \Bigg ]
\Bigg \}.
\end{eqnarray}
We have introduced the two-component field $\xi$, 
\begin{eqnarray} \psi^+ = 
\left [ \begin{array}{c} \xi \\
                       0 \end{array} \right ].
\end{eqnarray} 
The interaction dependent part of $F^2_{F(I)}$ is 
\begin{eqnarray}
F^2_{F(I)int} &=& g \int dx^- d^2 x^\perp x^- \xi^\dagger A^2 \xi \nonumber
\\
&& ~~~~+ { 1 \over 4} g \int dx^- d^2 x^\perp \Big [ \xi^\dagger { 1 \over
\partial^+}[(-i \sigma^3 A^1 + A^2)\xi] \nonumber\\&&~~~~~~~~~~~~~~
+ { 1 \over \partial^+}
[ \xi^\dagger (i \sigma^3 A^1 + A^2)]\xi \Big ].
\end{eqnarray} 

The interaction dependent part of $F^2_{F(II)}$ is 
\begin{eqnarray}
F^2_{F(II)int} =  { 1 \over 4} g \int dx^- d^2 x^\perp \Big 
[ \xi^\dagger { 1 \over
\partial^+}[(-i \sigma^3 A^1 + A^2)\xi] + { 1 \over \partial^+}
[ \xi^\dagger (i \sigma^3 A^1 + A^2)]\xi \Big ] \nonumber \\
- { 1 \over 2} g \int dx^- d^2 x^\perp x^2 \Bigg [
{\pp^\p\over {\pp^+}} [ \xi^\dagger ({\tilde \s}^\p \cdot A^\p) ] {\tilde \s}^\p
\xi + \xi^\dagger ({\tilde \s}^\p \cdot A^\p) {1\over {\pp^+}} ({\tilde
\s}^\p \cdot \pp^\p) \xi \nonumber\\ ~~~~~ + ({ \pp^\p \over {\pp^+}}
\xi^\dagger) {\tilde \s}^\p ( { \tilde \s}^\p \cdot A^\p) \xi + \xi^\dagger
{1\over {\pp^+}} ({\tilde \s}^\p \cdot  \pp^\p) ( {\tilde \s}^\p \cdot
A^\p) \xi \nonumber\\ ~~~~~~ -m {1\over {\pp^+}} [ \xi^\dagger ({\tilde
\s}^\p \cdot A^\p) ] \xi + m \xi^\dagger ( {\tilde \s}^\p \cdot A^\p){1\over
{\pp^+}} \xi
\nonumber\\~~~~~~~~~~ + m ( {1\over {\pp^+}} \xi^\dagger) ({\tilde \s}^\p
\cdot A^\p) \xi - m \xi^\dagger {1\over {\pp^+}} [( {\tilde \s}^\p \cdot
A^\p) \xi] \Bigg ] \nonumber\\  
~~~~- {1 \over 2 } g^2 \int dx^- d^2 x^\perp x^2 \Bigg [ 
\xi^\dagger {\tilde \sigma}^\perp \cdot A^\perp  { 1 \over i \partial^+} 
{\tilde \sigma}^\perp \cdot (A^\perp \xi)
- { 1 \over i \partial^+} (\xi^\dagger {\tilde \sigma}^\perp \cdot A^\perp) 
{\tilde \sigma}^\perp \cdot A^\perp \xi \Bigg ].
\end{eqnarray} 
We have introduced $ {\tilde \sigma}^1 =  \sigma^2$ and $ {\tilde
\sigma}^2 = - \sigma^1$.

Next consider the gluonic part of the operator $F^2$:
\begin{eqnarray}
F^2_g = { 1 \over 2} \int dx^- d^2 x^\perp \Big [ x^- \Theta^{+2}_g - x^2
\Theta^{+-}_g \Big ],
\end{eqnarray} 
where
\begin{eqnarray} 
\Theta^{+2}_g &=& - F^{+ \lambda a} F^{2a}_{\, \, \lambda}, \nonumber \\
\Theta^{+-}_g &=& - F^{+\lambda a} F^{- a }_{\, \, \lambda} + { 1 \over 4}
g^{+-}(F_{\lambda \sigma a})^2.
\end{eqnarray}
Using the constraint equation
\begin{eqnarray}
{ 1 \over 2} \partial^+ A^{-a} = \partial^i A^{ia} + g f^{abc} { 1 \over
\partial^+}(A^{ib} \partial^+A^{ic}) + 2 g { 1 \over \partial^+} \Big (
\xi^\dagger T^a \xi \Big ),
\end{eqnarray}
we arrive at
\begin{eqnarray}
F^2_g = F^2_{g(free)} + F^2_{g(int)}
\end{eqnarray}
where
\begin{eqnarray}
F^2_{g(free)} &=& { 1 \over 2} \int dx^- d^2 x^\perp \Bigg \{ x^- \Big (
A^{ja}\partial^+\partial^j A^{2a} - A^{2a}\partial^+ \partial^j A^{ja}+
A^{ja}\partial^+\partial^2 A^{ja}\Big ) \nonumber \\
&& ~~~~~~~~~~~ -x^2  \Big ( A^{ka}(\partial^j)^2 A^{ka} \Big ) \Bigg
\}
\nonumber \\
&& ~~~~~~~~~~~~~~~~~~~- 2\int dx^- d^2 x^\perp A^{2a} \partial^1  A^{1a}.
\end{eqnarray}
The interaction part
\begin{eqnarray}
F^2_{g(int)} &=& { 1 \over 2} \int dx^- d^2 x^\perp x^- \Bigg \{ 
gf^{abc} \partial^+ A^{ia} A^{2b} A^{ic} \nonumber \\
&&~~~~ + g\Big (  f^{abc} { 1 \over \partial^+}(A^{ib} 
\partial^+ A^{ic}) + 2{
1 \over \partial^+} (\xi^\dagger T^a \xi) \Big ) \partial^+ A^{2a} \Bigg \}
\nonumber \\
&&~~ - {1 \over 2} \int dx^- d^2 x^\perp  x^2 \Bigg \{
2g f^{abc} \partial^i A^{ja} A^{ib} A^{jc} + {g^2 \over 2} f^{abc} f^{ade}
A^{ib} A^{jc} A^{id} A^{je} \nonumber \\
&& ~~~~ + 2g \partial^i A^{ia} { 1 \over \partial^+}
\Big ( f^{abc}  A^{jb} \partial^+ A^{jc} + 2 \xi^\dagger T^a \xi \Big )
\nonumber \\
&& ~~~~ + g^2 \Big ( f^{abc} { 1\over \partial^+} (A^{ib}\partial^{+} A^{ic})
+2 { 1 \over \partial^+} \xi^\dagger T^a \xi \Big
)\nonumber\\&&~~~~~~~~~~~~~~~~~~~~~~~
\Big ( f^{ade} { 1\over \partial^+} (A^{jd}\partial^{+} A^{je})
+ 2{ 1 \over \partial^+} \xi^\dagger T^a \xi \Big ) \Bigg \}.  
\end{eqnarray}
So the full transverse rotation operator in QCD can be written as
\begin{eqnarray}
F^2  = F^2_{I} + F^2_{II} + F^2_{III},
\end{eqnarray}
where
\begin{eqnarray}
F^2_{I}&=& {1\over 2} \int dx^- d^2x^\p [ x^- {\cal P}^2_0 - x^2 ({\cal H}_0 +
{\cal V}) ], \\
F^2_{II} &=& 
{1\over 2} \int dx^- d^2x^\p \Bigg [\xi^\dagger \Big [ \sigma^3 \partial^1 + i \partial^2
\Big]{ 1 \over
\partial^+} \xi + \Big [ { 1 \over \partial^+} (\partial^1 \xi^\dagger \sigma^3 -
i \partial^2 \xi^\dagger) \Big ] \xi \Bigg ] \nonumber \\ 
&&~~~~~~ + {1\over 2} \int dx^- d^2x^\p m \Bigg [ \xi^\dagger \Big [{ \sigma^1 \over i \partial^+} 
\xi\Big ] -
\Big [{ 1 \over i \partial^+} \xi^\dagger\sigma^1\Big ] \xi \Bigg ]
\nonumber \\
&& ~~+ {1\over 2} \int dx^- d^2x^\p  g \Bigg [ \xi^\dagger { 1 \over
\partial^+}[(-i \sigma^3 A^1 + A^2)\xi] + { 1 \over \partial^+}
[ \xi^\dagger (i \sigma^3 A^1 + A^2)]\xi \Bigg ], \\
F^2_{III}&=& 
- \int dx^- d^2 x^\perp  2(\partial^1 A^{1})A^2 \nonumber \\
&&~~-{1\over 2} \int dx^- d^2x^\p g {4\over {\pp^+}} (\xi^\dagger T^a
\xi) A^{2a} \nonumber\\&&~~~~~~~~~~~~~~~~~- {1\over 2} \int dx^- d^2x^\p g f^{abc} {2\over
{\pp^+}} (
A^{ib} \pp^+ A^{ic} ) A^{2a}
\end{eqnarray}
where $ {\cal P}^i_0$ is the free momentum density, $ {\cal H}_o$ is the
free Hamiltonian density and ${\cal V}$ are the interaction terms in the
Hamiltonian in manifestly Hermitian form. The operators $F^2_{II}$ and $F^2_{III}$ 
whose integrands do not
explicitly depend upon coordinates arise from the fermionic and bosonic
parts respectively of the gauge invariant, symmetric, energy momentum tensor
in QCD.  
From Eq. (\ref{c7j1}) in Sec. 7.3 it follows that the transverse spin
operators ${\cal J}^i$, ($i=1,2$) can also be written as the sum of three
parts, ${\cal J}^i_{I}$ whose integrand 
has explicit coordinate dependence, ${\cal
J}^i_{II}$ which arises from the fermionic part, and  ${\cal J}^i_{III}$ which
arises from the bosonic part of the energy momentum tensor.   
%%%%%%%%%%%%%%%%%%%%%%%%%%%%%%%%%%%%%%%%%%%%%%%%%%%%%%%%%%%%%%%%%%%
\section{Connection with Transverse Polarized Scattering}
%%%%%%%%%%%%%%%%%%%%%%%%%%%%%%%%%%%%%%%%%%%%%%%%%%%%%%%%%%%%%%%%%%%%

From the phenomenological point of view, the issue of transverse spin has 
become very important in high energy physics thanks to recent 
experimental advances \cite{c7expt}. Since transverse spin for a free massless
gluon is identically zero, transverse spin measurements for gluonic
observables directly probe the long distance, nonperturbative features of
QCD. Analogous to longitudinally polarized scattering, where 
quark helicity carries roughly only 25 \% of the proton helicity, 
one may ask what is the
situation in transversely polarized scattering. 
In particular can one relate the operators appearing in the transverse spin
to the integrals of structure functions appearing in transverse polarized
scattering? 

In the previous section, we have shown that, though the transverse spin
operators cannot be separated into an orbital and a spin part, one can still
define a decomposition of it into three different parts.
In this section, we establish the physical relevance of this decomposition by   
exploring the connection between hadron expectation values of the
transverse spin operators and the 
quark and gluon distribution functions that
appear in transversely polarized deep inelastic scattering.

It is known that the transverse polarized distribution function in deep 
inelastic scattering is
given by (we have taken transverse polarization along the $x$-axis)
\begin{eqnarray}
g_T(x) &=& {1\over 8\pi M} \int d\eta 
		e^{-i\eta x}\times 
  \langle PS^1|\overline{
		\psi}(\eta) \Big(\gamma^1 -{P^1\over P^+}
		\gamma^+ \Big)\gamma_5 \psi(0) +~ h.c. |PS^1 \rangle 
		\, ,
\end{eqnarray}
where $P^\mu$ and 
$S^\mu$ are the four momentum and the polarization vector of the target. 
Using the constraint equation for $\psi^-$, we arrive at
\begin{eqnarray}
\int_{-\infty}^{+ \infty} dx g_T(x) &=& 
\int_{-\infty}^{+ \infty} dx (g_{T(I)}(x) + g_{T(II)}(x)) 
\end{eqnarray}
\begin{eqnarray}
\int_{-\infty}^{+ \infty} dx g_{T(I)}(x) & =& 
{ 1 \over 2 M} \langle P S^1 \mid
\Bigg [ \xi^\dagger \Big [ \sigma^3 \partial^1 + i \partial^2 \Big ] { 1 \over
\partial^+} \xi +
{\partial^1 \over \partial^+}(\xi^\dagger) \sigma^3 \xi - i
{\partial^2 \over \partial^+} (\xi^\dagger) \xi 
\nonumber\\&& + m \xi^\dagger \sigma^1 { 1 \over i \partial^+}(\xi) - m { 1 \over
i \partial^+} (\xi^\dagger) \sigma^1 \xi  
 +g
\Big 
[ \xi^\dagger { 1 \over
\partial^+}[(-i \sigma^3 A^1 + A^2)\xi]\nonumber\\&&~~~~~~~~~~~~~  
+ { 1 \over \partial^+}
[ \xi^\dagger (i \sigma^3 A^1 + A^2)]\xi \Big ]
 \Bigg] \mid P S^1 \rangle .
\end{eqnarray}
Thus the integral of $g_{T(I)}(x)$ is directly proportional to the nucleon
expectation value of $F^2_{II}$. Both $g_{T(I)}$ and $F^2_{II}$ depend on
the center of mass motion whereas both $g_T$ and ${\cal J}^i$ are
independent of the center of mass motion. The removal of the center of mass
motion from $g_{T(I)}$ is achieved by $g_{T(II)}$. We have 
\begin{eqnarray}
\int_{-\infty}^{+ \infty} dx g_{T(II)}(x)=   
{ 1 \over M} { P^1 \over P^+}\langle P S^1 \mid\xi^\dagger 
\sigma^3 \xi \mid P S^1 \rangle. 
\end{eqnarray} 
The integral of $g_{T(II)}(x)$ is directly proportional to the nucleon
expectation value of the quark intrinsic helicity operator
\begin{eqnarray}
J^3_{q(i)} = { 1 \over 2} \int dx^- d^2 x^\perp \xi^\dagger \sigma^3 \xi.
 \end{eqnarray}

Consider the polarized gluon distribution function that appears in
transversely polarized scattering (see Ref. \cite{c7Ji}) 
\begin{eqnarray}
G_T(x) = { 1 \over 8 \pi x (P^+)^2} { 1 \over ({S^i}^2)} i \epsilon^{\mu \nu
\alpha \beta} S_\alpha P_\beta 
\int d \eta e^{- i \eta x  }
\langle P S^\perp \mid F^{+a}_{~~ \mu} (\eta) F^{+a}_{~~\nu}(0) \mid
P S^\perp \rangle.  
\end{eqnarray}
Here $S^\mu$ is the polarization vector,
\be
S^\mu={1\over 2}\overline u \gamma^\mu \gamma^5 u
\e
where $u$ is the Dirac spinor. For longitudinal polarization, the normalized 
spinors are \cite{c7ped}
\be
u_{\uparrow} (P) = {1\over \sqrt {2P^+}}\left (\begin{array}{c}P^++M\\P^1+iP^2\\
P^+-M\\P^1+iP^2\end{array}\right),~~~~~~~~~~~~~~~~
u_{\downarrow} (P) = {1\over \sqrt {2P^+}} 
\left (\begin{array}{c}-P^1+iP^2\\P^++M\\
P^1-iP^2\\-P^++M\end{array}\right).
\e
These give $S^+=P^+$, $S^\perp=0$.
For transverse polarization (along $x$ direction), the normalized spinors
are
\be
u_{\uparrow} (P) = {1\over \sqrt {2P^+}}\left (\begin{array}{c}M+P^+-iP^2\\
-P^1\\
P^1\\-M+P^++iP^2\end{array}\right),~
u_{\downarrow} (P) = {1\over \sqrt {2P^+}} 
\left (\begin{array}{c}-M+P^+-iP^2\\-P^1\\
P^1\\M+P^++iP^2\end{array}\right).
\e
Using these, we get $S^+=0$, 
 $S^2=0$, $S^1=M$, $ S^-=2 {P^1 \over P^+}M$.

For a transversely polarized nucleon,  $F^+_{~~ -}=0$.
Since for $ \alpha, \mu , \nu =- $, the contribution is automatically zero, 
$ \beta = -$. 
Further, let us pick, without loss of generality, the transverse
polarization along the $ x $ axis. 
Thus $\alpha$ is
forced to be $1$ or $+$. 
Then 
\begin{eqnarray}
G_T(x) = G_{T(I)}(x) + G_{T(II)}(x) 
\end{eqnarray}
where
\begin{eqnarray}
G_{T(I)}(x)  = { i \over 8 \pi x MP^+}  
\int d \eta e^{-i \eta x} 
\langle P S^1 \mid  F^{+a}_{~~ 2} (\eta) F^{+a}_{~~ +}(0) - 
F^{+a}_{~~
+}(\eta) F^{+a}_{~~ 2}(0)  \mid P S^1 \rangle,  
\end{eqnarray}
and
\begin{eqnarray}
G_{T(II)}(x)  &=& - 
{ i \over 16 \pi x P^+M^2} S^-  \int d \eta 
e^{-i \eta x} 
\langle P S^1 \mid F^{+a}_{~~1}(\eta) F^{+a}_{~~2}(0)
\nonumber\\&&~~~~~~~~~~~~~~~~~~-
F^{+a}_{~~2}(\eta) F^{+a}_{~~1}(0) \mid P S^1 \rangle.  
\end{eqnarray}
We get
\begin{eqnarray}
\int_{ - \infty }^{+ \infty} dx G_{T(I)}(x)= 
 { 1 \over 2 M} \langle P S^1
\mid
A^{2a}(0) { 1 \over 2} \partial^+ A^{-a}(0) \mid P S^1 \rangle. 
\end{eqnarray} 
From the constraint equation,
we explicitly see that the operator structure of the integral of
$G_{T(I)}$ is similar to $F^2_{III}$.

We also have
\begin{eqnarray}
\int_{- \infty}^{\infty} dx G_{T(II)}(x) =
 { 1 \over 4 M}{P^1 \over P^+} 
\langle P S^1 \mid \left 
(A^a_1 \partial^+ A_2^a - A^a_2 \partial^+ A_1^a \right )
\mid P S^1 \rangle . 
\end{eqnarray}
Thus the integral of $G_{T(II)}(x)$ is proportional to the 
nucleon expectation value of the
gluon intrinsic helicity operator 
\begin{eqnarray}
J^3_{g(i)} = { 1 \over 2} \int dx^- d^2 x^\perp \left [ 
A^a_1 \partial^+ A^a_2 - A^a_2 \partial^+ A^a_1 \right ]. 
\end{eqnarray}
Thus, provided the interchange of the order of
integrations is legal, we have shown that a direct relation exists 
between the
coordinate independent part of ${\cal J}^i$ which arises from the gauge
invariant fermionic and gluonic parts of the symmetric 
energy momentum tensor and the
integrals of the quark and gluon distribution functions $g_T$ and $G_T$ that
appear in polarized scattering.

%%%%%%%%%%%%%%%%%%%%%%%%%%%%%%%%%%%%%%%%%%%%%%%%%%%%%%%%%%%%%%%%%%
\section{Summary}
%%%%%%%%%%%%%%%%%%%%%%%%%%%%%%%%%%%%%%%%%%%%%%%%%%%%%%%%%%%%%%%%%%

In this chapter,  we have investigated the transverse spin operator in QCD
and showed its connection with the transverse polarized structure function
$g_T$. In equal
time quantization, one encounters two major difficulties in the description
of the spin of a composite system in an arbitrary reference frame. They are
1) the complicated interaction dependence arising from dynamical boost
operators and 2) the
difficulty in the separation of center of mass motion from the internal
motion. Due to these severe difficulties, there have been hardly any attempt
to study spin operators of a moving composite system in the conventional
equal time formulation of quantum field theory. 

In light-front theory, on the other hand, the longitudinal spin
operator (light-front helicity) is interaction independent and the
interaction dependence of transverse spin operators arises solely from that
of transverse rotation operators. Moreover, in this case the separation of
center of mass motion from internal motion is trivial since light-front
transverse boosts are simple Galilean boosts.    

We have investigated the case of transverse spin operators for both 
massive and massless particles. We have introduced 
the transverse spin operators for massless particles with arbitrary 
transverse momentum. This is done for the
first time in light-front field theory. 
To provide physical intuition for transverse spin 
operators which have a complicated structure in interaction theory, 
we have provided the explicit form of these operators in Fock space 
basis for both free fermion field theory and free massless spin 
one field theory in the appendix. 

In QCD, our starting point is the formula for transverse rotation operators
expressed as the integral of generalized angular
momentum density given in terms of gauge invariant, symmetric, energy
momentum tensor. We have emphasized  the differences between spin operators
in field theory in equal time and light-front quantization schemes.  

Appropriate to light-front quantization, we choose the
light-front gauge. 
We use the constraint equations for $\psi^-$ and $A^-$ to eliminate them in
favor of dynamical degrees of freedom.  
In this initial study, we restrict to topologically
trivial sector of QCD and set the requirement that the transverse gauge
fields vanish as $x^{-,i} \rightarrow \infty$. This eliminates the
surface terms and completely fixes the gauge.
In the gauge fixed theory we found that the transverse rotation operators
can be decomposed as the sum of three distinct terms: $F^i_{I}$ which has
explicit coordinate dependence in its integrand, and $F^i_{II}$ and $F^i_{III}$ which have no
explicit coordinate dependence in their integrand. 
Further, $F^i_{II}$ and $F^i_{III}$ arise
from the fermionic and bosonic parts of the energy momentum tensor. 
Since transverse spin is responsible for the helicity flip of the nucleon in
light-front theory, we now have identified the complete set of
operators responsible for the helicity flip of the nucleon.

Our construction and decomposition of the transverse spin operators in QCD also
have important phenomenological consequences. We have 
shown \cite{c7lett} that nucleon expectation values of $F^i_{II}$ and
$F^i_{III}$ are directly related to the integrals of quark and gluon
distribution functions that appear in transversely polarized deep inelastic
scattering.
After the experimental discovery of the so-called spin crisis, the question
of the sharing of nucleon helicity among its constituents has become an
active research area. On the theoretical side, the first step involves the
identification of the complete set of operators contributing to nucleon
helicity. In this work, we have made this identification in 
the case of transverse spin. We have explicitly shown that the operators
involved in the case of the helicity and transverse spin are very
different. Because of their interaction dependence, operators contributing
to transverse spin are more interesting from the theoretical point of view
since they provide valuable information on the non-perturbative 
structure of the hadron.

It is extremely interesting to contrast the 
cases of longitudinal and transverse spin operators
in light-front field theory. In the case of longitudinal spin operator
(light-front helicity), in the gauge fixed theory, the operator is
interaction independent and can be separated into orbital and spin parts for
quarks and gluons. It is known for a long time that the transverse spin 
operators in
light-front field theory cannot be separated into orbital and spin parts
except in the trivial case of free field theory.
We have shown that, in spite of the complexities, 
a physically interesting separation is indeed
possible for the transverse spin operators which is quite different from
the separation into orbital and spin parts in the rest frame familiar in the
equal time picture.    

In short, in this chapter we have explored in detail the theoretical aspects of spin 
operators in quantum field theory in the context of QCD and their
consequences.
 It is
interesting to establish a transverse spin sum rule in analogy to the
helicity sum rule and explore its phenomenological consequences. 
Since  transverse rotational symmetry is not manifest in light-front theory 
a study of these operators is essential for questions regarding Lorentz invariance
in the theory \cite{c7glazek}. 
An important issue in the case of transverse spin operators concerns
renormalization. Since they are interaction dependent, they will acquire
divergences in perturbation theory just like the Hamiltonian. It is of
interest to find the physical meaning of these divergences and their
renormalization. We address these issues in the next chapter
by computing the
expectation value of the transverse spin operators in a dressed quark state.

%%%%%%%%%%%%%%%%%%%%%%%%%%%%%%%%%%%%%%%%%%%%%%%%%%%%%%%%%%%%%%%%%%%%%%%%%%%

\chapter{Transverse Spin in Light-Front QCD: Radiative Corrections}
In this chapter, we continue our investigation of the transverse spin
operators in QCD. 
We have shown in chapter 7 that there exists a physically
interesting separation of the transverse spin operators into a coordinate
dependent part and two coordinate independent parts. 
We have also shown that the
transverse spin operators are  manifestly interaction dependent. They will
acquire divergences in perturbation theory just like the Hamiltonian and it
is important to find the physical meaning of the divergences and their
renormalization. 

In light-front QCD
Hamiltonian, quark mass appears as $m^2$ and $m$ terms, $m^2$  in the
free helicity non-flip part of the Hamiltonian and $m$  in the
interaction dependent helicity flip part of the Hamiltonian. It is known
that $m^2$ and $m$ renormalize differently. $m^2$ and $m$ also appear
in ${\cal J}^i$. 
Do these terms  require new counterterms in addition to those
necessary to renormalize the Hamiltonian? 
  
Besides renormalization, there is another important issue regarding the
transverse spin operators.
Recently it was shown that \cite{c8hk}, starting from the manifestly gauge 
invariant
symmetric energy momentum tensor, in light-front QCD (the gauge $A^+=0$ and
light-front variables), after the elimination of constrained variables,
${\cal J}^3$ becomes explicitly interaction independent and can be separated
into quark and gluon orbital and spin operators. Thus one can write down a
helicity sum rule which has a clear physical meaning. The orbital and
intrinsic parts of the light-front helicity operator have also been analyzed
recently in \cite{c8ang}. Even though ${\cal
J}^i$ cannot be separated into orbital and spin parts and they are
interaction dependent, one can still ask whether one can identify distinct 
operator structures in ${\cal J}^i$ and whether one can propose a physically
interesting sum rule. In chapter 7, we have seen that, in the gauge
$A^+=0$, by eliminating the constrained variables, one can decompose ${\cal
J}$ into three parts, 
${\cal J}^i= {\cal J}^i_I + {\cal J}^i_{II} + {\cal J}^i_{III}$ where only 
${\cal J}^i_{I}$ has explicit coordinate ($x^-, x^i$) dependence in its 
integrand. The operators ${\cal J}^i_{II}$ and ${\cal J}^i_{III}$ arise from 
the fermionic and bosonic parts respectively of the gauge invariant energy 
momentum tensor. We have also shown its connection with transversely
polarized scattering. One has to investigate whether this decomposition 
is protected by radiative corrections. 

In this chapter, we explore the theoretical consequences of the decomposition of 
${\cal J}^i$. We compare and contrast the
consequences of this decomposition  and the corresponding decomposition of
the helicity operator into orbital and spin parts.
Next we address the issue of radiative corrections  by
carrying out the calculation of the transverse spin of a dressed quark in
pQCD in the old-fashioned Hamiltonian formalism. To the best of our
knowledge, this is for the first time that such a calculation has been
performed in quantum field theory. This calculation is
facilitated by the fact that boost is kinematical in the light-front
formalism. Thus we are able to isolate the internal motion which is only
physically relevant  from the spurious center of mass motion. We carry out the
calculations in a reference frame with arbitrary transverse momentum
$P^\perp$ and explicitly verify the frame independence of our results.
 We find that
because of cancellation between various interaction independent and dependent 
operator matrix elements, only one counterterm is needed. We establish the
fact the mass counterterm for the renormalization of ${\cal J}^i$ is the
same mass counterterm  required for the linear mass term appearing in
the interaction dependent helicity flip vertex in QCD. It is important to
mention that the divergence structure and renormalization in light-front
theory is entirely different from the usual equal-time theory.
If one uses constituent momentum cutoff, one violates boost invariance and also 
encounters non-analytic behavior in the structure of counterterms
\cite{c8wilson}.
We have done one loop renormalization of the transverse spin
operators by imposing cutoff on the relative transverse momenta and on the
longitudinal momentum fraction.
 Upto one loop, we find that all infrared divergences (in the
longitudinal momentum fraction) get canceled in the result.

The plan of the chapter is as follows. 
In section 8.1, we discuss the physical relevance of the decomposition of
the transverse spin operator and also compare and contrast it with the
helicity operator. In 
section 8.2, we present the calculation of the transverse spin for a dressed
quark state upto $O(\alpha_s)$ in perturbation theory. 
Comparison of this calculation with the perturbative calculation of helicity
is given in section 8.3.
Discussion and summary are given in section 8.4. 
The calculation of transverse spin for a system of two free fermions is
given in Appendix G. The full evaluation of
the transverse spin operator for a dressed quark in an arbitrary reference
frame is given in appendix H. There we also show the manifest cancellation
of all the center of mass momentum dependent terms. Some details of the
calculation are provided in appendix I. We also refer to our original
papers, \cite{c8spin,c8hei}.

%%%%%%%%%%%%%%%%%%%%%%%%%%%%%%%%%%%%%%%%%%%%%%%%%%%%%%%%%%%%%%%%%%%%
\section{Decomposition of Transverse Spin and Comparison with Helicity Sum
Rule}
%%%%%%%%%%%%%%%%%%%%%%%%%%%%%%%%%%%%%%%%%%%%%%%%%%%%%%%%%%%%%%%%%%%%
The helicity ${\cal J}^3$ and the transverse spin operators ${\cal J}^i$ in 
light-front theory for a
massive particle is given in terms of Poincare generators in chapter 7.  
In \cite{c8hk} it has been shown explicitly 
that the helicity operator ${\cal J}^3$ in the light-front gauge, 
in terms of the dynamical fields in the topologically
trivial sector of QCD can be written as
\be
{\cal J}^3 = {\cal J}_{fi}^3 + {\cal J}_{fo}^3 +{\cal J}_{gi}^3 +{\cal J}_{
go}^3
\e
where  ${\cal J}_{fi}^3$ is the fermion intrinsic part, ${\cal J}_{fo}^3$ 
is the fermion orbital part, ${\cal J}_{gi}^3$ is the gluon intrinsic part 
and ${\cal J}_{go}^3$ is the gluon orbital part. 
The helicity sum rule is
given by, for a longitudinally polarized fermion state
\be
{1\over {\cal N}} \langle P S^\parallel \mid {\cal J}_{fi}^3 + {\cal J}_{fo}^3 +{\cal J}_{go}^3 +{\cal J}_{
gi}^3\mid P S^\parallel \rangle = \pm {1\over 2}.
\e
In the transverse rest frame ($P^\perp=0$), ${\cal J}^3$ coincides with
$J^3$ and the helicity sum rule takes the
form
\be
{1\over {\cal N}} \langle P S^\parallel \mid J_{fi}^3 + J_{fo}^3 +
J_{go}^3 +J_{
gi}^3\mid P S^\parallel \rangle = \pm {1\over 2}.
\e

For a boson state, right hand side of the above equation should be replaced with the
corresponding helicity. Here, ${\cal N}$ is the normalization constant of
the state.

Unlike the helicity operator, which can be
separated into orbital and spin parts, the transverse spin operators cannot
be written as a sum of orbital and spin contributions. Only in the free theory, 
one can write them as a sum of orbital and spin parts by a unitary
transformation called Melosh transformation. In chapter 7, we have
shown that the transverse rotation operators can be separated into three 
distinct components 
\be
F^2  = F^2_{I} + F^2_{II} + F^2_{III},
\end{eqnarray}
where
\begin{eqnarray}
F^2_{I}&=& {1\over 2} \int dx^- d^2x^\p [ x^- {\cal P}^2_0 - x^2 ({\cal H}_0 +
{\cal V}) ], \\
F^2_{II} &=& 
{1\over 2} \int dx^- d^2x^\p \Bigg [\xi^\dagger \Big [ \sigma^3 \partial^1 + i \partial^2
\Big]{ 1 \over
\partial^+} \xi + \Big [ { 1 \over \partial^+} (\partial^1 \xi^\dagger \sigma^3 -
i \partial^2 \xi^\dagger) \Big ] \xi \Bigg ] \nonumber \\ 
&&~~~~~~ + {1\over 2} \int dx^- d^2x^\p m \Bigg [ \xi^\dagger \Big [{ \sigma^1 \over i \partial^+} 
\xi\Big ] -
\Big [{ 1 \over i \partial^+} \xi^\dagger\sigma^1\Big ] \xi \Bigg ]
\nonumber \\
&& ~~+ {1\over 2} \int dx^- d^2x^\p  g \Bigg [ \xi^\dagger { 1 \over
\partial^+}[(-i \sigma^3 A^1 + A^2)\xi] + { 1 \over \partial^+}
[ \xi^\dagger (i \sigma^3 A^1 + A^2)]\xi \Bigg ], \\
F^2_{III}&=& 
- \int dx^- d^2 x^\perp  2(\partial^1 A^{1})A^2 \nonumber \\
&&-{1\over 2} \int dx^- d^2x^\p g {4\over {\pp^+}} (\xi^\dagger T^a
\xi) A^{2a} - {1\over 2} \int dx^- d^2x^\p g f^{abc} {2\over {\pp^+}} (
A^{ib} \pp^+ A^{ic} ) A^{2a}
\end{eqnarray}
where $ {\cal P}^i_0$ is the free momentum density, $ {\cal H}_o$ is the
free Hamiltonian density and ${\cal V}$ are the interaction terms in the
Hamiltonian in full Hermitian form. It follows that 
the transverse spin
operators ${\cal J}^i$, ($i=1,2$) can also be written as the sum of three
parts, ${\cal J}^i_{I}$ whose integrand 
has explicit coordinate dependence, ${\cal
J}^i_{II}$ which arises from the fermionic part, and  ${\cal J}^i_{III}$ which
arises from the bosonic part of the energy momentum tensor. Thus although
the transverse spin operators cannot be separated into an orbital and
intrinsic parts, we can still talk of a decomposition into an orbital-like
part and two intrinsic-like parts.

 At this
point, we would also like to contrast our work with \cite{c8ji}, 
where a gauge
invariant decomposition of nucleon spin has been done. The analysis in
\cite{c8ji} has been performed in the rest frame of the hadron and no distinction
is made between  helicity and transverse spin, whereas, we have worked in the
gauge fixed theory in an arbitrary reference frame. 

In analogy with the helicity sum rule, we propose a decomposition of the
transverse spin, which can be written as 
\be
{1\over {\cal N}} \langle P S^\perp \mid {\cal J}^i_I + {\cal J}^i_{II}+ {\cal
J}^i_{III} \mid P S^\perp \rangle = \pm {1\over 2}
\label{c8tsum}
\e
for a fermion state polarized in the transverse direction. For a bosonic state, right hand side will be replaced with the
corresponding transverse component of spin.

The question that now naturally arises is, 
does the above decomposition have any physical relevance ? 
In the case of the helicity operator, we know that the hadron expectation
value of the fermion intrinsic part ${\cal J}^3_{fi}$ is related to the
first moment of the quark helicity distribution function $g_1$ measured in
longitudinally polarized DIS. Also ${\cal J}^3_{gi}$ is related to the first
moment of the intrinsic gluon helicity distribution. In light-front QCD,
helicity is kinematical and independent of interaction, as a result, the
helicity sum rule has parton interpretation, that is, the different terms 
 can be expressed as parton helicity distributions. But the transverse spin
is dynamical and the corresponding sum rule has no parton interpretation.
However,  we have shown in chapter 7 that the above decomposition has
relevance in the context of transversely polarized DIS.  
There exists 
a direct connection between
the hadron expectation value of the fermionic intrinsic-like part of the transverse spin
operator ${\cal J}^i_{II}$ and the integral of the quark distribution 
function $g_T$ that
appear in transversely polarized deep inelastic scattering. Also we can
identify \cite{c8let} the operators that are present in the hadron expectation value
of  
${\cal J}^i_{III}$ with the operator structures that are present in 
the integral of the gluon distribution function that appear in transverse
polarized hard scattering. 
The physical relevance of the
decomposition is made clear from the identification. Our results show 
 the intimate
connection between transverse spin in light-front QCD and transverse
polarized deep inelastic scattering. As far as we know, such connections are
not established so far  in instant form of field theory and this is the
first time that the first moment of $g_T$ is related to a conserved
quantity. Another
important point is that in perturbation theory, 
the helicity flip interactions which are proportional
to mass play a crucial role both in $g_T$ and in the transverse spin
operator whereas they are not important in the case of the helicity
operator. In fact, one cannot get a transversely polarized state in the
absence of mass.

Because the transverse spin operators are interaction dependent, they
acquire divergences in perturbation theory. One has to regularize them by
imposing momentum cutoffs and
in the regularized theory the Poincare algebra as well as the commutation 
relation obeyed by the spin operators are violated \cite{c8glazek}. One has to
introduce appropriate counterterms to restore the algebra. 
In the next section, we perform the renormalization of the full
transverse spin operator upto $O(\alpha_s)$ in light-front Hamiltonian
perturbation theory by evaluating the matrix element for a quark state
dressed with one gluon. This calculation also verifies the relation
(\ref{c8tsum})  
 upto $O(\alpha_s)$ in perturbation theory.

%%%%%%%%%%%%%%%%%%%%%%%%%%%%%%%%%%%%%%%%%%%%%%%%%%%%%%%%%%%%%%%%%%%%%
\section{Transverse Spin of a Dressed Quark in Perturbation Theory}
%%%%%%%%%%%%%%%%%%%%%%%%%%%%%%%%%%%%%%%%%%%%%%%%%%%%%%%%%%%%%%%%%%%%%

In this section, we evaluate the expectation value of the transverse spin
operator in perturbative QCD for a dressed quark state.

The dressed quark state with fixed helicity $\sigma$ can be expanded in Fock 
space as
\be
\mid P, \sigma \rangle &=& \phi^\lambda_1 b^\dagger(P,\sigma) \mid 0 \rangle
\nonumber \\  
&& + \sum_{\sigma_1,\lambda_2} \int 
{dk_1^+ d^2k_1^\perp \over \sqrt{2 (2 \pi)^3 k_1^+}}  
\int 
{dk_2^+ d^2k_2^\perp \over \sqrt{2 (2 \pi)^3 k_2^+}}  
\sqrt{2 (2 \pi)^3 P^+} \delta^3(P-k_1-k_2) \nonumber \\
&& ~~~~~\phi^\sigma_{\sigma_1, \lambda_2}(P,\mid k_1,; k_2 ) b^\dagger(k_1,
\sigma_1) a^\dagger(k_2, \lambda_2) \mid 0 \rangle. 
\label{c8dr0}
\e 
We are considering  dressing with one gluon since we shall evaluate
the expectation value upto $O(g^2)$. The normalization of the state is given by
\be
\langle k',\lambda' \mid k,\lambda \rangle = 2(2\pi)^3 k^+ \delta_{\lambda
\lambda'} \delta(k^+ -k'^+) \delta ( k^\perp - k'^\perp).
\label{c8nor}
\e  
The quark target transversely polarized in the $x$ direction can be expressed
in terms of helicity up and down states by
\be
\mid k^+, k^\perp, s^1 \rangle = {1\over {\sqrt 2}}(\mid k^+, k^\perp,
\uparrow \rangle \pm \mid k^+, k^\perp, \downarrow \rangle)
\label{c8dr1}
\e 
with $s^1 = \pm m_R$, where $m_R$ is the renormalized mass of the quark.

We introduce the boost invariant amplitudes $\Phi_1^\lambda$ and
$\Phi^\lambda_{\sigma_1 \lambda_2}(x, q^\perp)$ respectively by
$\phi^\lambda (k) = \Phi_1^\lambda$ and $\phi^\lambda_{\lambda_1 \lambda_2}
(k; k_1, k_2)= {1\over {\sqrt k^+}}\Phi^\lambda_{\lambda_1 \lambda_2}(x,
q^\perp)$, where $x={k_1^+ \over P^+}$ and $q^\perp = k_1^\perp -
x P^\perp$ . From the light-front QCD Hamiltonian, to lowest order in
 perturbative QCD, we have
\be
	&& \Phi_{\sigma_1, \sigma_2}^\lambda (x,q^\perp) 
		= -{ x(1-x) \over { (q^\perp)^2 + m^2(1-x)^2}}{1\over
{\sqrt {1-x}}} \nonumber \\
	&&~~~~~~~~~~\times   { g \over {\sqrt {2 (2 \pi)^3}}} T^a 
		\chi^\dagger_{\sigma_1} \Big[  2 {q^\perp \over 1-x} + 
		{{\tilde \sigma^\perp.q^\perp}\over x} \tilde \sigma^\perp - 
		\tilde \sigma^\perp i m {{1-x}\over x} \Big ] \chi_{\lambda} .{(\epsilon^\perp_{
		\sigma_2})}^*\Phi^\lambda_1. \label{c8dr2} 
\label{c8wf}
\e 
Here $m$ is the quark mass and $x$ is the longitudinal momentum carried by
the quark. Also, $\tilde \sigma^1 = \sigma^2$ and $ \tilde \sigma^2 = -
\sigma^1$. It is to
be noted that the $m$ dependence in the above wave function arises from
the helicity flip part of the light-front QCD Hamiltonian. This term plays a
very important role in the case of transversely polarized target states.
 Also, we have seen in chapter 7 that a transversely polarized
dressed quark state cannot be obtained when quark mass is zero.

For simplicity, in this section, we calculate the matrix element of the transverse spin
operator for a dressed quark state in a frame where the transverse momentum of
the quark is zero. It can be seen from the expressions of ${\cal J}^i$ in
chapter 7 that the sole contribution
in this case comes from the first term in the right hand side, namely the
transverse rotation operator. A detailed calculation of the matrix elements
of the transverse spin operator in an arbitrary reference frame is given in
appendix H where we have explicitly shown that all the terms depending on
$P^\perp$ get canceled.

The matrix elements presented below have been evaluated between 
states of different helicities, namely $\sigma$ and
$\sigma'$. Since the transversely polarized  state can be
expressed in terms of the longitudinally polarized (helicity) states by
Eq. (\ref{c8dr1}), the matrix elements of these operators between transversely
polarized states can be easily obtained from these expressions.   

Here, we have used the manifest Hermitian form of all the operators. It is
necessary to keep manifest Hermiticity at each intermediate step to cancel
terms containing derivative of delta function.
 
The operator ${1\over 2}F^2P^+$ can be separated into three parts,
\be
{1\over 2}F^2P^+ = {1\over 2}F^2_IP^+ + {1\over 2}F^2_{II}P^+ +
 {1\over 2}F^2_{III}P^+
\e
where $F^2_I$, $F^2_{II}$ and $F^2_{III}$ have been defined earlier.
The matrix elements of the different parts of these for a dressed quark state
are given below. The evaluation of the matrix element of ${1\over 2}F^2_IP^+$
 is quite complicated since it involves derivatives of delta functions. A
part of this calculation has been given in some detail in appendix I.
We first consider the operator
\be
{1\over 2}F^2_IP^+ = {1\over 2}F^2_I(1)P^+ - {1\over 2}F^2_I(2)P^+ -
 {1\over 2}F^2_I(3)P^+.
\label{c8f2i}
\e
The first term contains the momentum density, the second and the third terms
contain the free and the interaction parts of the Hamiltonian density respectively.
The matrix elements are given by
\be
\langle P,\sigma \mid {1\over 2}F^2_I(1)P^+ \mid P, \sigma' \rangle  &=&
\langle P,\sigma \mid {1\over 2} \int dx d^2q^\perp x^- P^2_0 {1\over
2}P^+ \mid P, \sigma' \rangle \nonumber\\&=& -{i\over 2}\sum_{spin} 
\int dx d^2q^\perp q^2 
\Phi^{*\sigma}_{\sigma_1\lambda}{\pp\Phi^{*\sigma'}_{\sigma'_1\lambda'} \over
{\pp x}} + h. c. 
\e
\be
\lefteqn{\langle P,\sigma \mid {1\over 2}F^2_I(2)P^+\mid P, \sigma' \rangle = 
\langle P,\sigma \mid {1\over 2} \int dx d^2q^\perp x^2 P^-_0 {1\over
2}P^+\mid P, \sigma' \rangle }\nonumber\\&=&{i\over 4}\sum_{spin}\int dxd^2q^\perp 
\Phi^{*\sigma}_{\sigma_1\lambda}{\pp \Phi^{\sigma'}_{\sigma'_1\lambda'}
\over {\pp q^2}} (q^\perp)^2 \left( {{1-x}\over x} - {x\over {1-x}} \right ) 
\nonumber\\&&~~~~~~~~~ +{i\over 4}\sum_{spin} \int dx d^2q^\perp m^2 {{1-x}\over x}
\Phi^{*\sigma}_{\sigma_1
\lambda}{\pp \Phi^{\sigma'}_{\sigma'_1\lambda'}\over {\pp q^2}}+ h. c. 
\e
In the above two equations, both the single particle and two particle
diagonal matrix elements contribute. Here, {\it h.c.} is the Hermitian conjugate, $\sum_{spin}$ is the summation over $\sigma_1, {\sigma'}_1, \lambda_1,
{\lambda'}_1$. $P^-_0$ is the free part of the Hamiltonian density. The
interaction part:
\be
\lefteqn{\langle P, \sigma \mid {1\over 2}F^2_I(3)P^+\mid P, \sigma' \rangle =
\langle P, \sigma \mid  {1\over 2} \int dx d^2q^\perp x^2 P^-_{int} {1\over
2}P^+\mid P, \sigma' \rangle} \nonumber\\&=& {g\over \sqrt {2(2\pi)^3}}
\sum_{spin} \int dxd^2q^\perp {1\over {\sqrt {1-x}}} \Big (
-{i\over 4} 
\Phi_1^{* \sigma} \chi^\dagger_\sigma [ {\tilde \sigma}^2 ( {\tilde
\sigma}^\perp \cdot \ep^\perp ) \nonumber\\&&~~~~~~~~~~~~~~~~~~~+ {({\tilde 
\sigma}^\perp \cdot \ep^\perp)
{\tilde \sigma}^2 \over x}]\chi_{\sigma_1} \Phi^{\sigma'}_{\sigma_1
\lambda}+ h. c.\Big ).
\e
$P^-_{int}$ is the interaction part of the light-front QCD Hamiltonian
density. The manifestly Hermitian form of it is given in Appendix D. 
The interaction part of the Hamiltonian contains $qqg$, $ggg$, $qqgg$,
$qqqq$ and $gggg$ terms. Only the $qqg$ part of it contributes to the dressed quark matrix
element.

The operator ${1\over 2}F^2_{II}P^+$ which originates from the fermionic
part of the energy momentum tensor, can be separated into three parts,

\be
{1\over 2}F^2_{II}P^+ = {1\over 2}F^2_{mII}P^+ + {1\over 2}F^2_{q^\perp II}P^++ 
{1\over 2}F^2_{gII}P^+
\e
where ${1\over 2}F^2_{mII}P^+$ is the  explicit mass
dependent part of the operator, ${1\over 2}F^2_{q^\perp II}P^+$ is the part
containing derivatives with respect to $x^\perp$ and ${1\over 2}F^2_{g II}P^+$ is
the interaction part. The matrix elements are given by 
\be
\langle P, \sigma \mid {1\over 2}F^2_{mII}P^+ \mid P, \sigma'
\rangle = {m\over 2}\Phi^{*\sigma}_1
\Phi^{\sigma'}_1 + {m\over 2}\sum_{spin} \int dx d^2q^\perp  \Phi^{*\sigma}_{\sigma_1
\lambda} \chi^\dagger_{\sigma_1} \sigma^1 \chi_{\sigma'_1}
\Phi^{\sigma'}_{\sigma'_1
\lambda'} {1\over x}, \label{c8m2}
\e
\be
\langle P, \sigma \mid {1\over 2}F^2_{q^\perp II} P^+ \mid P, \sigma'
 \rangle &=& 
{1\over 2}\sum_{spin}\int dx d^2q^\perp \Phi^{*\sigma}_{\sigma_1
\lambda} \chi^\dagger_{\sigma_1} \sigma^3 q^1 \chi_{\sigma'_1}
\Phi^{\sigma'}_{\sigma'_1 \lambda'} {1\over x},
\label{c8q2}
\e
\be
\langle P, \sigma \mid {1\over 2}F^2_{g II} P^+ \mid P, \sigma'
 \rangle &=& {1\over 4} {g\over {\sqrt {2(2 \pi)^3}}}\sum_{spin} \int dx d^2q^\perp
{1\over {\sqrt {1-x}}} \Big ( i\Phi^{*\sigma}_1 \nonumber\\&&\Big [ 
\chi^\dagger_\sigma (
-i\sigma^3 \ep^1_\lambda + \ep^2_\lambda ) \chi_{\sigma_1} - {1\over x}
\chi^\dagger_\sigma (i\sigma^3 \ep^1_\lambda + \ep^2_\lambda ) \chi_{\sigma_1}
 \Big ] \Phi^{\sigma'}_{\sigma_1 \lambda}
\nonumber\\&&~~~~~~~~~~~~~~~~~~~~~~ + h. c.\Big ) . \label{c8g2}
\e
In Eqs. (\ref{c8m2}) and (\ref{c8q2}), contributions come from only diagonal
matrix elements whereas Eq. (\ref{c8g2}) contain only off-diagonal matrix
elements.
The matrix element of ${1\over 2}F^2_{III} P^+$, which comes from the
gluonic part, is given by
\be
\lefteqn{\langle P, \sigma \mid {1\over 2}F^2_{III} P^+ \mid P, \sigma'
 \rangle = -{g \over {\sqrt {2(2\pi)^3}}}
\sum_{spin} \int dx d^2q^\perp {1\over
{\sqrt {1-x}}}}\nonumber\\&&~~~ \Big (  \Phi^{*\sigma}_1 \ep^2_\lambda 
\Phi^{\sigma'}_{\sigma_1 \lambda}{1\over {i(1-x)}} + h. c. \Big )
 - \int dx d^2q^\perp {q^1\over (1-x)} \sum_{\lambda,
\sigma_1, \sigma'_1} \lambda
\Phi^{*\sigma}_{\sigma_1 \lambda}\Phi^{\sigma'}_{\sigma'_1
\lambda} .
\e
The first term in the right hand side is the off-diagonal contribution which comes from
the interaction dependent part of the operator. The second term is 
 the diagonal contribution coming from the free part.

The expectation value of the transverse spin operator between transversely
polarized states is given by
\be
\langle P, S^1 \mid M{\cal J}^1 \mid P, S^1
 \rangle = \langle P,S^1 \mid {1\over 2} F^2 P^+ + K^3P^2 - {1\over 2}
E^2 P^- - P^1 {\cal J}^3 \mid P, S^1  \rangle.
\e
 Since we are in the reference frame with zero $P^\perp$, only the first term
in the right hand side, i.e. the ${1\over 2} F^2 P^+$ term will contribute, as mentioned
earlier. We substitute
for $\Phi^\sigma_{\sigma_1 \lambda}$ using Eq. (\ref{c8wf}). The final forms of
the matrix elements are given by 
\be
\langle P, S^1 \mid M {\cal J}^1_I(1) \mid P, S^1 \rangle = -{m
\alpha_s\over {4 \pi}}C_f {\rm ln} {Q^2\over \mu^2} \int_\ep^{1-\ep} dx (1+x),
\e
\be
\langle P, S^1 \mid M {\cal J}^1_I(2) \mid P, S^1 \rangle = {m
\alpha_s\over {4 \pi}}C_f {\rm ln} {Q^2\over \mu^2} \int_\ep^{1-\ep} dx (1-2x) ,
\e
\be
\langle P, S^1 \mid M {\cal J}^1_I(3) \mid P, S^1 \rangle = -{m
\alpha_s\over {4 \pi}}C_f{\rm ln} {Q^2\over \mu^2} \int_\ep^{1-\ep} dx (1-x)
\e
where $M{\cal J}^1_I(1), M{\cal J}^1_I(2)$ and $ M{\cal J}^1_I(3)$ are
 related respectively to $F^2_I(1), F^2_I(2)$ and $ F^2_I(3)$ defined
earlier. 
 $\mu $ is the hadronic factorization scale for separating the `hard'
and `soft' dynamics of QCD, i. e. we have set a hadronic scale such that
${\mid q^\perp \mid }^2 >> \mu^2 >>m^2$. $\ep$ is a small cutoff on the
longitudinal momentum fraction which can be safely taken to zero at the end
of the calculation.

So we obtain, from the above three expressions, using
Eq. (\ref{c8f2i}),
\be
\langle P, S^1 \mid M {\cal J}^1_I \mid P, S^1 \rangle = -{m
\alpha_s\over {4 \pi}}C_f {\rm ln} {Q^2\over \mu^2} .
\label{c8dr4}
\e
The contribution to the matrix element of $M{\cal J}^1_{II}$ entirely comes from
$F^2_{II}$. The various parts of this matrix element are given by
\be
\langle P, S^1 \mid M {\cal J}^1_{mII} \mid P, S^1 \rangle = {1\over 2} m
{\mid \Phi^\sigma_1 \mid }^2 +
{m\alpha_s \over {2 \pi}} C_f {\rm ln}{Q^2\over \mu^2}\int_\ep^{1-\ep} dx{1\over {1-x}},
\label{c8dr6}
\e
\be
\langle P, S^1 \mid M {\cal J}^1_{q^\perp II} \mid P, S^1 \rangle = 
-{m\alpha_s \over {4\pi}} C_f {\rm ln}{Q^2\over \mu^2}\int_\ep^{1-\ep} dx(1-x),
\e
\be
\langle P, S^1 \mid M {\cal J}^1_{g II} \mid P, S^1 \rangle = 
{m\alpha_s \over {4\pi}} C_f {\rm ln}{Q^2\over \mu^2}{1\over 2},
\e
where $M{\cal J}^1_{m II}, M{\cal J}^1_{q^\perp II}$ and $ M{\cal J}^1_{g
II}$ are related respectively to $F^2_{mII}, F^2_{q^\perp II}$ and $ F^2_{g
II}$. In Eq. (\ref{c8dr6}) we have to use the normalization condition
\be
{\mid \Phi^\sigma_1 \mid }^2 = 1- {\alpha_s\over {2\pi}} C_f {\rm ln}{Q^2 \over
\mu^2} \int_\ep^{1-\ep}dx {1+x^2\over {1-x}}.
\e 
Upto $O(\alpha_s)$, the normalization condition will contribute only in the
first term of Eq. (\ref{c8dr6}). 
We get, from Eq. (\ref{c8dr6}),
\be
\langle P, S^1 \mid M {\cal J}^1_{mII} \mid P, S^1 \rangle = {1\over 2} m
+{m\alpha_s \over {4 \pi}} C_f {\rm ln}{Q^2\over \mu^2}\int_\ep^{1-\ep} dx
\left (
{2\over {1-x}}-{{1+x^2}\over {1-x}} \right ).
\e
It is clear that the singularity at $x=1$ is canceled due to
the contribution from the normalization condition.
The overall contribution coming from $M{\cal J}^1_{II}$ is given by
\be
\langle P, S^1 \mid M {\cal J}^1_{II} \mid P, S^1 \rangle = {m\over 2} \left (1+
{3\alpha_s \over {4\pi}} C_f {\rm ln}{Q^2\over \mu^2}\right ),
\e
which does not involve any $x$ divergence.
 The matrix element of $M{\cal J}^1_{III}$ is given by
\be
\langle P, S^1 \mid M {\cal J}^1_{III} \mid P, S^1 \rangle = 
{2m\alpha_s \over {4\pi}} C_f {\rm ln}{Q^2\over \mu^2}\int_\ep^{1-\ep} ( 1-x) dx.
\label{c8dr5}
\e 
It is to be noted that all the contributing matrix elements are proportional
to the quark mass. Among the different parts of the operator, only ${\cal
J}^i_{mII}$ and a part of the interaction terms in ${\cal J}^i_I$ 
are proportional to the quark mass $m$. These mass dependent
terms flip the quark helicity. It is also to be noted that the terms
proportional to $m^2$ do not flip the helicity. In all the other terms, though the operators
do not depend on $m$ explicitly, the contributions to the matrix elements
arise from the interference of the $m$ terms in the wave function of
Eq. (\ref{c8wf}), with the non-$m$ dependent terms through the different parts
of the transverse spin operator. Since in light-front formulation, helicity
and chirality are the same, these linear in $m$ terms are explicit 
chiral symmetry
breaking terms.
From Eq. (\ref{c8dr4}) and Eq. (\ref{c8dr5}) we find that
\be
\langle P, S^1 \mid M {\cal J}^1_{I}+ M {\cal J}^1_{III} \mid P, S^1 \rangle =  
{m \alpha_s \over {4 \pi}} C_f {\rm ln} {Q^2 \over \mu^2} \int_\ep^{1-\ep} (1-2x) dx
= 0
\e
which means that the entire contribution to the matrix element of the
transverse spin operator is given by
\be
\langle P, S^1 \mid M {\cal J}^1 \mid P, S^1 \rangle =
{m\over 2} \left ( 1 + {3 \alpha_s \over {4\pi}} C_f {\rm ln}{Q^2\over \mu^2}
\right).
\e
This contribution entirely comes from $M{\cal J}^1_{II}$. Contribution from the
orbital-like part $(M{\cal J}^1_I)$ exactly cancels the contribution from the
gluon intrinsic-like part $(M{\cal J}^1_{III})$. 

The renormalized mass $m_R$ of the quark is given in terms of the bare
mass upto order $\alpha_s$ in light-front Hamiltonian perturbation theory
by \cite{c8hari3},
\be
m_R = m \left( 1 + { 3 \alpha_s \over {4 \pi}}C_f{\rm ln} {Q^2 \over \mu^2} \right
).
\e
In the light-front formulation of QCD, there are two mass terms in the
Hamiltonian, one is quadratic in $m$ which is present in the free part and
does not break chiral symmetry, the other is linear in $m$ which we discuss
here and which explicitly cause chiral symmetry breaking. An important
feature of light-front QCD is that, these two mass scales are renormalized
differently even in the perturbative region. The renormalization of $m^2$ is
different from the result stated above. 

Adding all the parts, for a dressed quark in perturbation theory upto $O(g^2)$, the
expectation value of the transverse spin operator is given by 
\be
\langle P, S^1 \mid M {\cal J}^1 \mid P, S^1 \rangle &=& 
\langle P, S^1 \mid M {\cal J}^1_{I}+M {\cal J}^1_{II}+M {\cal J}^1_{III} 
 \mid P, S^1\rangle \nonumber\\&=& {m_R\over 2} .
\e    
It is important to mention that here we are calculating the expectation
value of the operator $M{\cal J}^i$. In order to extract the eigenvalue of 
${\cal J}^i$ 
one has to know the eigenvalue of $M$. Both $M{\cal J}^i$ and $M$ are
dynamical operators. However, in this case, 
the mass $M$ in the left hand side in the renormalized theory is nothing but the
renormalized mass of the quark, which therefore gets canceled from the
above equation, and we get
\be
\langle P, S^1 \mid  {\cal J}^1 \mid P, S^1 \rangle &=& 
\langle P, S^1 \mid  {\cal J}^1_{I}+ {\cal J}^1_{II}+ {\cal J}^1_{III} 
 \mid P, S^1 \rangle \nonumber\\&=& {1\over 2}.
\e

The identification of ${\cal J}$ with spin, therefore, requires knowledge of
the mass eigenvalue, independently of the boost invariance properties of the
light-front dynamics.

We can explicitly verify the relation between the integral of $g_T$ and the
expectation value of the fermion intrinsic-like part of the transverse spin
operator to order $\alpha_s$ in perturbative QCD.
The transverse polarized structure function for a dressed quark is 
given \cite{c8hari1} by,
\be
g_T(x, Q^2) &=&{e^2_q\over 2}{m\over S^1} \Big \{ \delta (1-x) +
 {\alpha_s \over {2\pi}}  C_f {\rm ln}{Q^2\over \mu^2} \Big [ {1+2x -x^2\over
 {1-x}} - \delta(1-x) \int_0^1 dx' {1+{x'}^2\over {1-x'}}
\nonumber\\&&~~~~~~~~~~~~~~~~~~~~~~~~~+{1\over 2}
\delta (1-x) \Big ] \Big \} , 
\e
so we get
\be
\int_0^1 g_T (x) dx = {e_q^2\over {2S^1}}\langle P, S^1 \mid M{\cal J}^1_{II} \mid P, S^1\rangle
\e
which explicitly shows the connection between the integral of the transverse
polarized structure function and the matrix element of the fermion
intrinsic-like part of
the transverse spin operator. It can be seen that the operator structure of $M{\cal
J}^1_{mII}$ as well as the corresponding part of $g_T$ is related to the 
quark transversity distribution $h_1$ \cite{c8dip}. 

%%%%%%%%%%%%%%%%%%%%%%%%%%%%%%%%%%%%%%%%%%%%%%%%%%%%%%%%%%%%%%%%%%%%%%%
\section{Comparison with the Perturbative Calculation of Helicity ${\cal
J}^3$}  
%%%%%%%%%%%%%%%%%%%%%%%%%%%%%%%%%%%%%%%%%%%%%%%%%%%%%%%%%%%%%%%%%%%%%%%%
It is quite instructive to compare our calculation of the transverse spin of
the dressed quark with the helicity of the dressed quark \cite{c8hk} in
perturbative QCD. All the operators contributing to helicity are kinematical
(interaction independent) and hence all of them give rise to only diagonal
contributions. Further, in this calculation mass of the quark can be
completely ignored since they give rise to only power-suppressed
contribution. In the massless limit, helicity is conserved at the quark
gluon vertex. This means that the quark in the quark-gluon state has the
same helicity as the parent quark. 
Since the transverse gluon carry helicity $\pm 1$, we get a non-vanishing
contribution from the gluon intrinsic helicity operator. However, both the
quark and the gluon in the quark-gluon state have non-vanishing orbital
angular momentum due to transverse motion. Conservation of total helicity
implies that orbital contribution has to cancel gluon intrinsic helicity
contribution. This is precisely what happens \cite{c8hk} and
we find that the total quark plus gluon
orbital part exactly canceled the intrinsic gluon contribution and the
overall contribution to the helicity is $\pm{1\over 2}$, which entirely
comes from the intrinsic part of the fermionic helicity operator.

In contrast, in the case of transverse spin operator, it 
has both interaction independent and interaction dependent parts. The latter
gives rise to off-diagonal matrix elements and they play a very important
role. Of special interest is the gluon intrinsic-like transverse spin
operator. This operator gives vanishing matrix elements for a free gluon.
However, since gluon in the quark-gluon state has intrinsic transverse
momentum, both diagonal and off-diagonal terms give rise to non-vanishing
contributions and we get a net non-vanishing matrix element for the gluon
intrinsic-like transverse spin operator. However, we find that 
contribution from this matrix element is completely canceled by that from the matrix elements of
orbital-like transverse spin operators. This is analogous to what
happens in the helicity case.  
\vskip .2in
In this section, the calculation of the matrix elements has been done 
in the frame with
$P^\perp = 0$. The complete calculation of the matrix element of the
transverse spin operator in an arbitrary reference frame is given in 
appendix H. 
It is clear from the expressions there that all the terms explicitly
dependent on $P^\perp$ get canceled in the expectation value of $M{\cal
J}^1$. The parts that remain after the cancellation of
the $P^\perp$ dependent terms are those given above.
In the above expressions, we have used the manifest Hermitian form of the operators. 
We again stress the fact that this manifest cancellation of contributions
from center of mass motion is
typical in light-front  field theory because the transverse boost
operators are kinematical. The situation in the equal time relativistic case
is completely different and there one cannot separate out the center of mass
motion from the internal motion in a straightforward way even in the 
free theory case \cite{c8os}
because of the complicated boost generators. Due to the manifest cancellation
of the center of mass momenta, ${\cal J}^i$ can truly be identified as the
transverse spin operator.

%%%%%%%%%%%%%%%%%%%%%%%%%%%%%%%%%%%%%%%%%%%%%%%%%%%%%%%%%%%%%%%%%%
\section{Summary and Discussions}
%%%%%%%%%%%%%%%%%%%%%%%%%%%%%%%%%%%%%%%%%%%%%%%%%%%%%%%%%%%%%%%%%%

In chapter 7, we initiated the investigation of the transverse spin
operators in QCD. In this chapter, we have continued the analysis and
discussed two very important aspects of it. 
In analogy with the helicity sum rule, we have proposed a 
decomposition for the transverse spin. Earlier we have shown 
the relationship between 
nucleon matrix elements of ${\cal J}^i_{II}$  and ${\cal J}^i_{III}$
and  
the first moments of quark and gluon structure functions respectively,
appearing in
transverse polarized hard scattering. This is the first time that
the integral of $g_T$ is related to a conserved quantity, namely the
transverse spin operator. It is important to mention here that the proposed
decomposition of the transverse spin operator will not be affected if one
adds a total derivative term to the angular momentum density. Such a term can
at most produce a surface term which we are neglecting since we have
restricted ourselves to the topologically trivial sector of the theory.
We have started with the angular momentum density defined in terms of the
symmetric gauge invariant stress-energy tensor, which is obtained from the
Noether's stress-energy tensor by  adding a total derivative term. 
Even though the angular momentum density differs from the Noether angular
momentum density by a total
derivative term, both give rise to the same generators.
Another point worth mentioning is that we have worked in the gauge fixed
theory. In the light-front gauge, $A^+=0$, the transverse spin operator can
be separated into three parts, and ${\cal J}_{II}^i$ is related to
the first moment of $g_T$ measured in transverse polarized scattering, which
is a gauge invariant object. This is similar to the helicity case, where
only in the light-front gauge and using light-front quantization, the intrinsic fermionic
helicity is related to  the gauge invariant first moment of $g_1$ measured in
longitudinally polarized scattering. The corresponding gluon intrinsic
helicity cannot be
measured directly in polarized deep inelastic lepton-nucleon scattering but in some other process
like polarized hadron-hadron scattering. A similar situation holds in the
case of transverse spin.

We have also discussed another important issue, namely, the renormalization
of the transverse spin operators.
In light-front theory, in addition to the
Hamiltonian, transverse spin operators  also contain interactions  and have a 
complicated structure. They are dynamical operators.  
Since  transverse rotational symmetry is not manifest in light-front theory, 
a study of these operators is essential for questions regarding Lorentz invariance
in the theory \cite{c8glazek}. 
Since they are interaction dependent, they will acquire
divergences in perturbation theory just like the Hamiltonian. It is of
interest to find the physical meaning of these divergences and their
renormalization. 
The renormalization of only the intrinsic-like fermion 
part of the 
transverse spin operator has been discussed in the literature so far. 
In this chapter, we have carried out the renormalization of the
full transverse spin operator for the first time upto $O(\alpha_s)$ in 
light-front Hamiltonian perturbation theory by evaluating the matrix
elements for a dressed quark target.  We have shown that the entire
contribution to the matrix element comes from the fermion intrinsic-like 
part of the
transverse spin operator and is equal to ${1\over 2}$. The 
contributions from ${\cal J}^i_{I}$ and 
${\cal J}^i_{III}$  exactly get canceled. Also, the mass of the quark is
very crucial in this case, since the helicity flip interactions which are
proportional to the quark mass play a very important role.
However, the terms proportional to $m^2$ do not flip the helicity and do not
contribute. Since helicity flip is involved, we do not encounter any
quadratic divergence unlike the case of renormalization of the
light-front Hamiltonian.
Further, we have compared and contrasted the calculations of transverse spin
and helicity of a dressed quark in perturbation theory.       

In this chapter, we have also verified the frame independence of our results. We have
explicitly shown that, in an arbitrary reference frame, all the terms
depending on the center of mass momenta manifestly get canceled in the
matrix element. The cancellation is as simple as in non-relativistic theory
since boost is kinematical on the light-front. For future studies, it is an
interesting problem to evaluate non-pertubatively \cite{c8bk} the matrix
element of the transverse spin operator in light-front QCD. Also, in this work,
we have used cutoff on the relative transverse momenta and the small $x$
divergence gets canceled in the one loop result. It is interesting to study
the renormalization of
the transverse spin operators using similarity renormalization technique
 \cite{c8wilson}.

%%%%%%%%%%%%%%%%%%%%%%%%%%%%%%%%%%%%%%%%%%%%%%%%%%%%%%%%%%%%%%%%%%%%%%%%%%%%%%%%

%\documentstyle[aps,preprint]{revtex}
%\tighten
%\setcounter{page}{118}
%\oddsidemargin .6cm
%\textwidth 16cm
%\begin{document}

%\baselineskip=20 true pt
%\setcounter{section}{0}
%\renewcommand{\thesection}{3.\arabic{section}}
%\setcounter{equation}{0}
%\renewcommand{\theequation}{3.\arabic{equation}}

\chapter{Summary and Conclusions}
In this thesis, we have investigated the higher twist structure functions,
namely the twist four longitudinal structure function $F_L$ and the transverse
polarized structure function $g_T$ in the recently developed method based
on light-front Hamiltonian QCD. Deep inelastic scattering is a light cone
dominated process and the general features of DIS are most suitably realized
in terms of parton distributions as provided by Feynman's parton model. The
field theoretic realization of the parton model is possible in light-front
Hamiltonian QCD approach. It is known that the constituent quark model (CQM)
where the hadrons are thought of as composites of only valence quarks and
antiquarks, is very successful in describing the non-relativistic properties
of hadrons. In QCD, this description is not possible because of the
complicated structure of the vacuum. The simplicity of light-front vacuum
gives us hope of reconciling CQM with QCD.
Our approach is more intuitive than the conventional approach based on OPE.
Furthermore, here we deal with the structure functions themselves, instead
of the moments, which naturally arise in the OPE method. It is to be
remembered that it is the structure
functions  which are actually measured in DIS experiments. 

Because our
formulation is entirely different, we have obtained various new and
interesting results. Also, it is possible to address the issues which are not
easy to  address in the conventional approach. Both the structure functions 
twist four $F_L$ and $g_T$   contain non-trivial interaction dependence in
the operator structure and therefore involve  quark-gluon
dynamics. In chapter 5, we have derived a sum rule which connects the twist
four part of $F_L$ to the mass square of the target hadron. This is a new
sum rule and it has important consequences regarding the understanding of
how the mass of the hadron is distributed among its quark and gluon
constituents. A very important aspect of our approach is that we can analyze
both the perturbative and non-perturbative contents of the structure
functions. The sum rule has both perturbative and non-perturbative
consequences. The Fock space expansion of the target state and particle number
truncation allow us to express the structure functions in terms of
light-front multiparton wave functions. The structure functions can be
calculated once these wave functions are known. Major investigations are
under way to calculate these wave functions by solving the light-front bound state
equation. In chapter 6, we have
investigated the twist four longitudinal structure function 
for bound states like a meson in $1+1$ dimensional QCD 
and also for positronium in $3+1$ dimensional QED in weak coupling limit. These calculations give
analytic understanding of the underlying non-perturbative dynamics.
We have also verified the sum rule non-perturbatively in both these cases.

Another most interesting issue that we have addressed is the transverse
polarized stucture function $g_T$. The operator here also involves
non-trivial quark-gluon dynamics. In chapter 7 we have shown that $g_T$ is
related to a part of the light-front transverse spin operator. The problem
of relativistic spin operators for a composite system in an arbitrary
reference frame is non-trivial in equal time theory because of the
interaction dependence of the Pauli-Lubansky operators. One cannot separate
the center of mass motion easily because equal time boost operators
are dynamical. We have shown in chapters 7 and 8 that the light-front
formulation is a much more straightforward way to address the issue of spin
since light-front boost operators are kinematical and one can separate the
center of mass motion from the relative  motion. In chapters 7 and 8 we have
shown that though the light-front transverse spin operators cannot be
separated into  orbital and spin parts, there exists a physically
interesting decomposition of them into three parts, one orbital-like and two
intrinsic-like. We have worked in the light-front gauge. One of the intrinsic-like part is related to the fermionic
part of the gauge invariant symmetric QCD energy momentum tensor, and this
in turn is connected with $g_T$. We have compared and contrasted this
decomposition of the transverse spin with the corresponding decomposition of
helicity for the nucleon. 

In perturbative calculations using field theory one encounters divergences.
The actual results are, of course, observables and are finite. Therefore one
has to renormalize the divergent results. Renormalization in light-front
Hamiltonian theory is entirely different from the usual equal time case 
because it is manifestly non-covariant and also a gauge 
fixed theory. In chapters 5 and 8, we have renormalized the twist four
longitudinal structure function and the full transverse spin operator
respectively in light-front Hamiltonian perturbation theory. Replacing the
target by a dressed quark, we have shown that the counterterm required to
renormalize the twist four part of $F_L$ has the same structure as the mass
counterterm and that needed for the renormalization of
transverse spin operator is related to the linear mass counterterm in
light-front Hamiltonian perturbation theory.  It is to be remembered that
$m^2$ and $m$ are renormalized  differently in light-front theory. Thus, our
results show that the regularization and renormalization of the higher twist
structure functions $F_L$ and $g_T$ are dictated by the renormalization of
the light-front QCD Hamiltonian.   
     
Our alternative approach gives a new insight to DIS. We have shown
that the twist four longitudinal structure function is related to the
light-front QCD Hamiltonian density and the transverse polarized structure
function is related to the transverse spin operator. Earlier, it has been
shown that $F_2$ is related to the longitudinal momentum density and $g_1$
is related to the helicity. Thus it gives a unified picture of DIS and
connects the electroproduction structure functions to the Poincare generators
in light-front theory. It is important to mention that, these relations are
non-perturbative in general. 

There are various new and interesting problems along this line that needs to
be investigated in future. One of the most challenging problems is
to calculate the structure functions for a bound state like a nucleon or a
meson in $3+1$ dimensional QCD using the recently developed similarity
renormalization technique \cite{ssim}. In our work, we have used large ultraviolet
transverse momentum cutoff and a small infrared longitudinal momentum
cutoff to regularize the theory. It is interesting to address the issue of
renormalization of the transverse spin operator using similarity
renormalization. Another possible way of a non-perturbative calculation of
the hadronic matrix elements is the transverse lattice formalism
\cite{sdalley}.   
%%%%%%%%%%%%%%%%%%%%%%%%%%%%%%%%%%%%%%%%%%%%%%%%%%%%%%%%%%%%%%%%%%%%%%%

% If you have appendices in your dissertation, you will need the
% following, else keep it commented. The following appendices are in
% files called ``app1.tex'',  ``app2.tex'', ... and they
% look just like any chapter.
%

\appendix
%\documentstyle[aps,preprint,eqsecnum]{revtex}

%\tighten
%\setcounter{page}{122}
%\oddsidemargin .6cm
%\textwidth 16cm
%\begin{document}

%\baselineskip=20 true pt
%\setcounter{section}{0}
%\renewcommand{\thesection}{\arabic{section}}
%\setcounter{equation}{0}
%\renewcommand{\theequation}{\arabic{equation}}
%\appendix

%{\flushleft\huge\bf {Appendix}}
%\vskip .2in

%%%%%%%%%%%%%%%%%%%%%%%%%%%%%%%%%%%%%%%%%%%%%%
\chapter{Notations and Conventions}
%%%%%%%%%%%%%%%%%%%%%%%%%%%%%%%%%%%%%%%%%%%%%%
Light front variables are defined as,
\be
x^+=x^0+x^3, ~~~~~~~~~~~~~~~x^-=x^0-x^3.
\e
We denote the four vector $x^\mu$ by, 
\be
x^\mu =(x^+,x^-,x^\perp).
\e
Scalar product of two four-vectors,
\be
x \cdot y = {1\over 2}x^+y^-+ {1\over 2}x^-y^+-x^\perp \cdot y^\perp.
\e
The metric tensor is,
\be
g_{\mu \nu} = \left (\begin{array}{cccc}0&{1\over 2}&0&0\\{1\over 2}&0&0&0\\
0&0&{-1}&0\\0&0&0&{-1}\end{array}\right),~~~~~~~~~~~
g^{\mu \nu} = \left (\begin{array}{cccc}0&2&0&0\\2&0&0&0\\
0&0&{-1}&0\\0&0&0&{-1}\end{array}\right).
\e
Partial derivatives are given by,
\be
\pp^+=2 \pp_-=2{\pp\over {\pp x^-}},
\e
\be
\pp^-=2 \pp_+=2{\pp\over {\pp x^+}}.
\e
Four dimensional volume element: $d^4x= {1\over 2} dx^+dx^-d^2x^\perp$.
Lorentz invariant volume element in momentum space
\be
[d^3k]={{dk^+d^3k^\perp}\over {2(2\pi)^3k^+}}.
\e
We define the integral operator,
\be
{1\over \pp^+}f(x^-)={1\over 4}\int dy^- \epsilon(x^--y^-)f(y^-).
\e
\be
{1\over (\pp^+)^2}f(x^-)={1\over 8}\int dy^- \mid x^--y^- \mid f(y^-).
\e
Unless otherwise specified, we choose the following convention for the gamma matrices:
\be
\gamma^\pm=\gamma^0 \pm \gamma^3.
\e
\be
\gamma^+=\left (\begin{array}{cc}0&0\\2i&0\end{array}\right),~~
\gamma^-=\left (\begin{array}{cc}0&-2i\\0&0\end{array}\right).
\e
\be
\gamma^i=\left (\begin{array}{cc}-i\sigma^i&0\\
0&i\sigma^i\end{array}\right)
\e
where $\sigma$ are the Pauli spin matrices. 
\be
\gamma_5=i\gamma_0\gamma_1\gamma_2 \gamma_3, ~~~~~~\vec{\alpha}=\gamma^0
\vec{\gamma}.
\e
Projection operators
\be
\Lambda^\pm={1\over 4} \gamma^\mp \gamma^\pm.
\e
%%%%%%%%%%%%%%%%%%%%%%%%%%%%%%%%%%%%%%%%%%%%%%%%
\chapter{$F_L$: Comparison with Other Conventions}
%%%%%%%%%%%%%%%%%%%%%%%%%%%%%%%%%%%%%%%%%%%%%%%
The hadron tensor relevant to unpolarized electron-hadron deep inelastic scattering
is given by 
\begin{eqnarray}
	W^{\mu \nu} &=&\Big(-g^{\mu \nu} + {q^\mu q^\nu 
		\over q^2} \Big) W_1(x,Q^2) + \Big(P^\mu - {P.q 
		\over q^2} q^\mu\Big)\Big(P^\nu -{P.q \over q^2} 
		q^\nu\Big)W_2(x,Q^2). \label{awmn}
\end{eqnarray}
This is valid for any $Q^2$ and here we have not made any approximations.
Here $W_1$ is dimensionless and $W_2$ has dimension ${1\over M^2}$. 
The dimensionless functions
\begin{equation}
	F_L(x,Q^2)= 2 \Big [-W_1 + \big [ M^2 -{(P.q)^2 \over q^2}\big ] W_2
\Big ], \label{afpara}
\end{equation}
and 
\begin{equation}
 F_2(x,Q^2) = \nu W_2(x,Q^2)
\end{equation}
are the unpolarized structure functions.

Our notations are 
different from the definition of Close \cite{aclose} where both $W_1$ and
$W_2$ are dimensionless. The longitudinal photo absorption
cross section is given by,
\be
\sigma_L = {{4 \pi^2 \alpha }\over K} W_L
\e
where $K$ is the incident flux of photons and $W_L = {F_L\over 2} =
-W_1-{\nu^2\over q^2} W_2 + M^2 W_2$.

In the limit $M=0$, we get $F_L= (-W_1-{\nu^2\over q^2} W_2)$. Our
definition of $F_L$ is also different from that of Reya \cite{areya}.
\be
P^\mu P^\nu W_{\mu \nu}={Q^2\over {4 x^2}} {1\over 2x} F_L^{Reya}.
\e
In terms of $W_1$ and $W_2$, 
\be
F_L^{Reya}= 2x [-W_1-{\nu^2\over q^2}W_2 + M^2 W_2] -M^2[W_1-M^2 W_2].
\e
In the limit $M=0$, we get, $F_L^{Reya}=x F_L$. Thus, $P^\mu P^\nu$ projects
out $F_L$ from $W^{\mu \nu}$ only in the massless limit.

Next, we compare our conventions with those of Ellis, Furmanski and
Petronzio  \cite{asimple}. We write $W^{\mu \nu}$ in terms
of  $F_L$ and $F_T$ where $F_T= {F_2\over x}$. 
\be
W^{\mu \nu}&=&{1\over 2}(g^{\mu \nu}-{q^\mu q^\nu\over q^2})F_L\nonumber\\
&&~~~~~~+[x{P^\mu P^\nu\over {P.q}}+{1\over
{2P.q}}(P^\mu q^\nu+P^\nu q^\mu)-{1\over 2}g^{\mu
\nu}]F_T\nonumber\\&&~~~~~~~~~~~~~~~~~+M^2W_2(-g^{\mu \nu} + {q^\mu
q^\nu\over q^2}).
\e
EFP has used the same expression but $M=0$. Towards the end, they have
considered the mass effects separately. In our work, we have taken both the
quark and the target to be massive from the beginning.     

%{\it discuss the difference between our definition and the definition in
%DJT, Reya, EFP. refer  to following section }.

%%%%%%%%%%%%%%%%%%%%%%%%%%%%%%%%%%%%%%%%%%%%%%%%%%%%%%%%%%%%%%%
\chapter{Bound State Equation for Positronium in the Weak Coupling  Limit}
%%%%%%%%%%%%%%%%%%%%%%%%%%%%%%%%%%%%%%%%%%%%%%%%%%%%%%%%%%%%%%%%%%%%%

The Fock space expansion of the
positronium state is given in chapter 6.
The relativistic LF version of the Schroedinger equation is,
\begin{eqnarray}
P^- \mid P \rangle = { M^2+ (P^\perp)^2\over P^+} \mid P \rangle
\label{ae9}
\end{eqnarray}

where $P^-$ is the LFQED Hamiltonian. Substituting the Fock space expansion
of the state 
 in this expression and taking
 projection with the three particle state,\\ $b^\dagger ( k_1,
\sigma_1) d^\dagger ( k_2, \sigma_2 ) a^\dagger ( k_3, \lambda_3) \mid 0
\rangle$ we get

\begin{eqnarray}
\psi_3^{\sigma_1 \sigma_2 \lambda_3}(x, \kappa_1; x_2, \kappa_2;
1-x-x_2, \kappa_3) = {\cal M}_1 + {\cal M}_2,
\label{ab2}
\end{eqnarray}
where the amplitudes are given by

\begin{eqnarray}
	{\cal M}_1 &=& { 1 \over E} (-) { e \over \sqrt{2 (2 \pi)^3}} 
		{ 1 \over \sqrt{1 - x - x_2}} ~W_1~ \psi_2^{\sigma_1' 
		\sigma_2}(1-x_2, -\kappa_2^\perp; x_2,\kappa_2^\perp) 
\end{eqnarray}
and
\begin{eqnarray}
	{\cal M}_2 &=& { 1 \over E} { e \over \sqrt{2 (2 \pi)^3}} 
		{ 1 \over \sqrt{1 - x - x_2}} ~W_2~ \psi_2^{\sigma_1 
		\sigma_2'}(x,\kappa_1^\perp;1-x,-\kappa_1^\perp) 
\end{eqnarray}
with the energy denominator
\begin{eqnarray}
	E= \big[ M^2  - {m^2 + (\kappa_1^\perp)^2 \over x} -
		{m^2 + (\kappa_2^\perp)^2 \over x_2} - {(\kappa_3^\perp)^2 
		\over 1 - x -x_2} \big ].
\end{eqnarray}  
The vertices  are,   
\begin{eqnarray}
	W_1=\chi_{\sigma_1}^\dagger \sum_{\sigma_1'}\big [ { 2 
	\kappa_3^\perp \over 1 - x -x_2} - { (\sigma^\perp. \kappa_1^\perp
	- i m) \over x} \sigma^\perp + \sigma^\perp {(\sigma^\perp. 
	\kappa_2^\perp -im) \over 1-x_2} \big] \chi_{\sigma_1'}. 
		(\epsilon^\perp_{\lambda_1})^*
\end{eqnarray}
and
\begin{eqnarray}
	W_2=\chi_{-\sigma_2}^\dagger \sum_{\sigma_2'}
		\big [ { 2 \kappa_3^\perp \over 1 - x -x_2} - \sigma^\perp
		{ (\sigma^\perp. \kappa_2^\perp
		- i m) \over x_2}  +  {(\sigma^\perp. \kappa_1^\perp -
		im) \over 1-x} \sigma^\perp 
		\big] \chi_{-\sigma_2'}. (\epsilon^\perp_{\lambda_1})^*.
\end{eqnarray}
Here,
\be
\k_3^\p = -\k_1^\p -\k_2^\p.
\end{eqnarray}

Also, taking projection with a two particle state $ b^\dagger ( k_1, \s_1)
d^\dagger(k_2, \s_2)\mid 0 \rangle$ we get
\begin{eqnarray}
\Big [M^2-{{(\kappa_1^\perp)^2+m^2}\over x_1}-{{(\kappa_1^\perp)^2+m^2}\over 1-x_1}
\Big
]\psi_2^{\sigma_1\sigma_2}(x_1,\kappa_1^\perp,1-x_1,-\kappa_1^\perp)
= {\cal N}_1 + {\cal N}_2 + I. 
\label{ab3}
\e
Here
\be
\lefteqn{{\cal N}_1 = {e\over
\sqrt{2(2\pi)^3}}\sum_{\sigma_1'\sigma_2'}\int_\epsilon^{x_1-\epsilon} 
dy\int d^2\kappa^\perp{1\over
\sqrt{x_1-y}}}\nonumber\\&&
U_1\psi_3^{\sigma_1'\sigma_2\lambda}(y,\kappa^\perp,1-x_1,
-\kappa_1^\perp,x_1-y,\kappa_1^\perp-\kappa^\perp)
\e
\be
\lefteqn{{\cal N}_2 = {e\over
\sqrt{2(2\pi)^3}}\sum_{\sigma_1'\sigma_2'}\int_\epsilon^{1-x_1-\epsilon}
 dy\int d^2
\kappa^\perp{1\over
\sqrt{1-x_1-y}}}\nonumber\\&&
U_2\psi_3^{\sigma_1\sigma_2'
\lambda}(x_1,\kappa_1^\perp,y,
\kappa^\perp,1-x_1-y,-\kappa_1^\perp-\kappa^\perp).
\e

The vertices are
\be
U_1 = \chi^\dagger_{\sigma_1}\Big
[{-2((\kappa_1^\perp-\kappa^\perp).\epsilon^\perp)\over
{x_1-y}}+{(\sigma^\perp.\epsilon^\perp)(\sigma^\perp.\kappa_1^\perp)\over y}
+{(\sigma^\perp.\kappa_1^\perp)(\sigma^\perp.\epsilon^\perp)\over
x_1}\chi_{\sigma_1'}\Big ]
\e
and
\be
U_2 = \chi^\dagger_{-\sigma_2'}\Big
[{2((-\kappa^\perp-\kappa_1^\perp).\epsilon^\perp)\over
1-x_1-y}+{(\sigma^\perp.\epsilon^\perp)(\sigma^\perp.\kappa_1^\perp)\over
1-x_1}-{(\sigma^\perp.\kappa^\perp)(\sigma^\perp.\epsilon^\perp)\over
y}\Big]\chi_{-\sigma_2}\Big ].
\e
$I$ is the instantaneous interaction given by
\be
I = -{4e^2\over {2(2\pi)^3}} \int dy d^2\k^\p
\psi_2(y,\k^\p,1-y,-\k^\p){1\over
(x-y)^2}.
\e

Substituting the expressions for $\psi_3$ in terms of $\psi_2$ in Eq.
(\ref{ab3})
we get the bound state equation for $\psi_2$.

The bound state equation is
\begin{eqnarray}
\lefteqn{\Big [ M^2-{{(\kappa^\perp)^2+m^2}\over x}-{{(\kappa^\perp)^2+m^2}\over 1-x}\Big
]\psi_2(x,\kappa^\perp,1-x,-\kappa^\perp)
={e^2\over{2(2\pi)^3}}\int dy\int d^2q^\perp}\nonumber\\&& \Big ({1\over
E}{1\over (x-y)}\Big [ V_1\psi_2(x,\kappa^\perp,1-x,-\kappa
^\perp)+ V_2
 \psi_2(y,q^\perp,1-y,-q^\perp)\Big ] 
\nonumber\\&&+{1\ov E'}{1\ov (y-x)} \Big [{V'}_1 \psi_2
(x,\kappa^\perp,1-x,-\kappa^\perp)+ {V'}_2
 \psi_2(y,q^\perp,1-y,-q^\perp)\Big ]\Big )
\nonumber\\&&~~~~~~~~~~~
-{4e^2\ov {2(2\pi)^3}} \int dy d^2q^\p \psi_2 (y,q^\p, 1-y,
-q^\p) { 1\ov (x-y)^2}. 
\label{ab4}
\end{eqnarray}
Since we are interested in the non-relativistic limit of the bound state
equation, we have kept only those terms of the vertex which survive in this
limit. Also, it has been shown that in this limit the helicity dependence of
the wave function goes away, so it is sufficient to look into one helicity
sector. We are looking into the helicity up-up sector and the
linear in $m$ terms in the vertex, where $m$ is the electron mass, that cause helicity flip are
not considered. However, both the linear $m$ terms and the terms quadratic
in $m$ in the bound state equation
coming from the vertex are suppressed in the non-relativistic limit.  The last term
 in the right hand side of Eq.(\ref{ab4})) is the instantaneous interaction where
we have neglected annihilation effects in this limit.
The energy denominators are
\be
E=M^2-{{(q^\perp)^2+m^2}\over y}-{{(\kappa^\perp)^2+m^2}\over
1-x}-{(\kappa^\perp-q^\perp)^2\over x-y}\nonumber\\
E'=M^2-{{(\kappa^\perp)^2+m^2}\over x}-{{(q^\perp)^2+m^2}\over
y}-{(\kappa^\perp-q^\perp)^2\over (y-x)}.
\label{ab5}
\e
It is to be noted that the entire mass dependence comes from these energy
denominators in the non-relativistic limit.

Substituting these expressions in the bound state equation and considering
only the photon exchange interactions in the non-relativistic limit,
we obtain the
non-relativistic bound state equation
\be
\lefteqn{[M^2 -{{(\k^\p)^2 + m^2 } \ov {x(1-x)}}] \psi_2 ( x, \k^\p , 1-x, -\k^\p) 
=}\nonumber\\&&-{4e^2\ov {2(2\pi)^3}} {1\ov 2} \int dyd^2q^\p \psi_2(y,q^\p,
1-y,-q^\p) \Big [ \left ({m\ov x}\right )^2 {1\ov { (\k^\p-q^\p)^2 + ({m\ov
x})^2(x-y)^2}}\nonumber\\&&~~~~~~~~~~~~~~~~~~~~~~~~~~ + 
      \left ({m\ov (1-x)}\right )^2 {1\ov { (\k^\p-q^\p)^2
+ ({m\ov (1-x)})^2(x-y)^2}}\Big ].
\label{ab9}
\e
We introduce a new three vector 
\be
\vec{p}= (\k, \k_z)
\e
where $\k_z$ is defined through a coordinate transformation from $x \in
[0,1]$ to $\k_z \in [ -\infty, \infty]$ by
\be
x \equiv {1\ov 2} + {\k_z\ov { 2\sqrt { \k^{\p2} + \k_z^2 +m^2}}}.
\e
In the non-relativistic limit, 
\be
x \simeq {1\ov 2m} (m + \k_z),
\e
\be
x-y \simeq {(\k_z - \k_z')\ov 2m},
\e
\be
{{(\k^\p)^2 + m^2}\ov { x(1-x)}} = 4( (\vec{p})^2 + m^2 ).
\e
Then
\be
{\left ({m\ov x}\right )^2 {1\ov { (\k^\p-q^\p)^2 + ({m\ov
x})^2(x-y)^2}} \simeq \left ({m\ov (1-x)}\right )^2 {1\ov { (\k^\p-q^\p)^2 + ({m\ov
(1-x)})^2(x-y)^2}}}\nonumber\\~~~~~~~~~~~~~~~~~~~~~~~~~~~~~~~~~~~~~
 \simeq {4m^2 \ov { ( \vec{p} - \vec{p'} )^2}}.
\e
We introduce the bound state wave function $ \tilde \phi(\vec{p})$ which is
normalized as
\be
\int dxd^2\k^\p {\mid \psi_2(x,\k^\p) \mid }^2 = \int d^3 \vec{p} J(\vec{p})
{\tilde \phi}^*(\vec{p})\tilde \phi(\vec{p}) = 1
\e
where in the non-relativistic limit we have suppressed the spin which
factorizes out.
The Jacobian of the coordinate change can be written as
\be
J(\vec{p})= {dx\ov {d\k_z}} \simeq {1\ov 2m}.
\e 
The tilde on the wave function is removed by suitable normalization,
\be
\int d^3 \vec{p} J(\vec{p})
{\tilde \phi}^*(\vec{p})\tilde \phi(\vec{p}) = \int d^3\vec{p}
\phi^*(\vec{p})
\phi(\vec{p})= 1.
\label{an1}
\e
The bound state equation then becomes
\be
\Big [ M^2 - 4( (\vec{p})^2 + m^2)\Big ]\phi(\vec{p}) = - {2e^2 \ov
{2(2\pi)^3}} \int d^3 \vec{p} \sqrt{ J(\vec{p}) J(\vec{p'})}\phi(\vec{p'}) 
{8m^2 \ov {( \vec{p} - \vec{p'})^2}}. 
\e
This can be written as
\be
 \Big [ M^2 - 4( (\vec{p})^2 + m^2)\Big ]\phi(\vec{p}) = - {2e^2 \ov
{2(2\pi)^3}} \int d^3 \vec{p} \phi(\vec{p'}) 
{4m \ov {( \vec{p} - \vec{p'})^2}}. 
\label{ad1}
\e
The instantaneous interaction and the photon exchange interactions combine to
give the Coulomb interaction in the non-relativistic limit.

%%%%%%%%%%%%%%%%%%%%%%%%%%%%%%%%%%%%%%%%%%%%%%%%%%%%%
\chapter{Poincare Generators in Light-Front QCD}
%%%%%%%%%%%%%%%%%%%%%%%%%%%%%%%%%%%%%%%%%%%%%%%%%%%%%
In this appendix we derive the manifestly hermitian kinematical Poincare 
generators (except
$J^3$) and the Hamiltonian in light-front QCD
starting from the gauge invariant symmetric energy momentum tensor
$\Theta^{\mu \nu}$. To begin with, $\Theta^{\mu \nu}$ is interaction
dependent. In the {\it gauge fixed} theory we find that the seven kinematical
generators are manifestly independent of the interaction.

We shall work in the gauge $A^+=0$ and ignore all surface terms. Thus we are
working in the completely gauge fixed sector of the theory \cite{ahk}. The
explicit form of the operator $J^3$ in this case is given in 
Ref. \cite{ahk} which is manifestly free of interaction at the operator level.
The rotation operators are given in chapter 7.

At $x^+=0$, the operators $K^3$ and $E^i$ depend only on the density
$\Theta^{++}$. A straightforward calculation leads to 
\begin{eqnarray}
\Theta^{++} =  {\psi^+}^\dagger \st{\lh}{i\pp^+}\psi^+ + 
\partial^+ A^i \partial^+ A^i.
\end{eqnarray}
Then, longitudinal momentum operator,
\begin{eqnarray}
P^+ &=& { 1 \over 2} \int dx^- d^2 x^\perp \Theta^{++} \nonumber \\
&=& { 1 \over 2} \int dx^- d^2 x^\perp \left [ {\psi^+}^\dagger 
\st{\lh}{i\pp^+} \psi^+ + \partial^+ A^j \partial^+ A^j \right ].
\end{eqnarray}
Generator of longitudinal scaling
\begin{eqnarray}
K^3 &=& - { 1 \over 4} \int dx^- d^2 x^\perp x ^- \Theta^{++}, \nonumber \\
&=&  - { 1 \over 4} \int dx^- d^2 x^\perp x ^- \left [
 {\psi^+}^\dagger \st{\lh}{i\pp^+}\psi^+ + \partial^+ A^j \partial^+ A^j
\right ].
\end{eqnarray}
Transverse boost generators
\begin{eqnarray}
E^i &=&  - { 1 \over 2} \int dx^- d^2 x^\perp x^i \Theta^{++}, \nonumber \\
&=& - { 1 \over 2} \int dx^- d^2 x^\perp x^i \left [  {\psi^+}^\dagger 
\st{\lh}{i\pp^+} \psi^+ + \partial^+ A^j \partial^+ A^j \right ].
\end{eqnarray}
The transverse momentum operator 
\begin{eqnarray}
P^i = { 1 \over 2} \int dx^- d^2 x^\perp \Theta^{+i}
\end{eqnarray}
which appears to have explicit interaction dependence.
Using the constraint equations for $\psi^-$ and $A^-$, we  have
\begin{eqnarray}
\Theta^{+i} &=& \Theta^{+i}_F + \Theta^{+i}_G, \nonumber \\
\Theta^{+i}_F &=& 2 {\psi^+}^\dagger i \partial^i \psi^+ + 2 g
{\psi^+}^\dagger A^i \psi^+, \\
\Theta^{+i}_G &=& \partial^+ A^j \partial^i A^j - \partial^+ A^j \partial^j A^i
+ \partial^+ A^i \partial^j A^j - 2 g {\psi^+}^\dagger A^i \psi^+.
\end{eqnarray}
Thus 
\begin{eqnarray} 
P^i = { 1 \over 2} \int dx^- d^2 x^\perp \left [ 
 {\psi^+}^\dagger \st{\lh}{i\pp^i} \psi^+ +
  A^j \partial^+\partial^j A^i - A^i \partial^+\partial^j A^j 
- A^j\partial^+ \partial^i A^j  \right ].
\end{eqnarray}
Thus we indeed verify that all the kinematical operators are explicitly
independent of interactions. 

Lastly, the Hamiltonian operator can be written in the manifestly Hermitian
form as
\begin{eqnarray}
P^- = {1 \over 2} \int dx^- d^2 x^\perp \Theta^{+-}
= {1\over 2}\int dx^-d^2x^\perp ( {\cal H}_0 + {\cal H}_{int}) 
\end{eqnarray}

where ${\cal H}_0$ is the free part given by
\be
{\cal H}_0 =- A^j_a {(\pp^i)}^2 A^j_a + \xi^\dagger \Big [ {{-(\pp^\p)^2 +
m^2}\over {i\pp^+}}\Big ] \xi- \Big [ {{-(\pp^\p)^2 +
m^2}\over {i\pp^+}}\xi^\dagger\Big ] \xi .
\e   
The interaction terms are given by
\be
{\cal H}_{int} = {\cal H}_{qqg} + {\cal H}_{ggg} + {\cal H}_{qqgg}+ 
{\cal H}_{qqqq} + {\cal H}_{gggg} ,
\e
where
\be
 {\cal H}_{qqg} = -4g \xi^\dagger {1\over \pp^+} (\pp^\p.A^\p)\xi
 +  g{\pp^\p\over {\pp^+}} [ \xi^\dagger ({\tilde \s}^\p \cdot A^\p) ] {\tilde \s}^\p
\xi + g\xi^\dagger ({\tilde \s}^\p \cdot A^\p) {1\over {\pp^+}} ({\tilde
\s}^\p \cdot \pp^\p) \xi \nonumber\\ ~~~~~ + g({ \pp^\p \over {\pp^+}}
\xi^\dagger) {\tilde \s}^\p ( { \tilde \s}^\p \cdot A^\p) \xi + g\xi^\dagger
{1\over {\pp^+}} ({\tilde \s}^\p \cdot  \pp^\p) ( {\tilde \s}^\p \cdot
A^\p) \xi \nonumber\\ ~~~~~~ -mg {1\over {\pp^+}} [ \xi^\dagger ({\tilde
\s}^\p \cdot A^\p) ] \xi + m g\xi^\dagger ( {\tilde \s}^\p \cdot A^\p){1\over
{\pp^+}} \xi
\nonumber\\~~~~~~~~~~ + mg ( {1\over {\pp^+}} \xi^\dagger) ({\tilde \s}^\p
\cdot A^\p) \xi - mg \xi^\dagger {1\over {\pp^+}} [( {\tilde \s}^\p \cdot
A^\p) \xi] ,
\e
\be
{\cal H}_{ggg} = 2gf^{abc} \Big [ \pp^iA^j_aA^i_bA^j_c + (\pp^i
A^i_a){1\over {\pp^+}}(A^j_b\pp^+A^j_c)\Big ] ,
\e
\be
{\cal H}_{qqgg}&=& g^2 \Big [ \xi^\dagger ({\tilde \s}^\p.A^\p){1\over
{i\pp^+}} ({\tilde \s}^\p.A^\p) \xi - {1\over {i\pp^+}} (\xi^\dagger {\tilde
\s}^\p.A^\p) {\tilde \s}^\p.A^\p \xi
\nonumber\\&&~~~+ 4{1\over \pp^+} (f^{abc} A^i_b\pp^+A^i_c){1\over \pp^+}
(\xi^\dagger T^a \xi)\Big ] ,
\e 
\be
{\cal H}_{qqqq} = 4g^2 {1\over \pp^+} (\xi^\dagger T^a \xi){1\over \pp^+}
(\xi^\dagger T^a \xi) ,
\e
\be
{\cal H}_{gggg} &=&{g^2\over 2} f^{abc}f^{ade} \Big [ A^i_bA^j_cA^i_dA^j_e
\nonumber\\&&~~~~~~~~~+ 2{1\over \pp^+} ( A^i_b\pp^+A^i_c){1\over
\pp^+}(A^j_d \pp^+A^j_e)\Big ] .
\e
%%%%%%%%%%%%%%%%%%%%%%%%%%%%%%%%%%%%%%%%%%%%%%%%%%%%%%
\chapter{Transverse Spin in Free Fermion Field Theory}
%%%%%%%%%%%%%%%%%%%%%%%%%%%%%%%%%%%%%%%%%%%%%%%%%%%%
%\subsection{Free fermion field}
%%%%%%%%%%%%%%%%%%%%%%%%%%%%%%%%%%%%%%%%%%%%%%%%%%%%%%%%%%%%%%
\section{Poincare Generators: Operator Forms}
%%%%%%%%%%%%%%%%%%%%%%%%%%%%%%%%%%%%%%%%%%%%%%%%%%%%%%%%%%%%%%
The symmetric energy momentum tensor 
\begin{eqnarray} \Theta^{\mu \nu} =   \Big [{\overline \psi}
  \gamma^{\mu} {1 \over 4}
\st{\lh}{i\pp^\nu} \psi   +
 {\overline \psi} \gamma^{\nu}  {1 \over 4}\st{\lh}{i\pp^\mu} \psi \Big ] . \end{eqnarray}
\noindent The momentum operators are given by 
\begin{eqnarray}  P^{+}  &=& 
 { 1 \over 2} \int dx^- d^2 x^\perp  {\overline \psi} \gamma^{+} 
{1 \over 2}\st{\lh}{i\pp^+}
  \psi \nonumber \\
&=& {1\over 2} \int dx^- d^2 x^\perp [\xi^{\dagger} i\partial^+  
 -(i\partial^+ \xi^{\dagger})]  \xi.
\end{eqnarray}
\begin{eqnarray} P^{i}  &=& 
 { 1 \over 2} \int dx^- d^2 x^\perp
 \Bigg  [ {\overline \psi} \Big \{ \gamma^{+}{1 \over 4}
\st{\lh}{i\pp^i}  + 
 \gamma^{i} {1 \over 4}
\st{\lh}{i\pp^+}\Big \}  \psi \Bigg ]  \nonumber \\
&=& {1\over 2} \int dx^- d^2 x^\perp  [\xi^{\dagger} i \partial^{i}-(
i\partial^i \xi^\dagger )] \xi.
 \end{eqnarray}
 The Hamiltonian operator is  
\begin{eqnarray}  P^- &=& 
{ 1 \over 2} \int  dx^- d^2 x^\perp   
{\overline \psi} \Big [ \gamma^-
  {1 \over 4} 
\st{\lh}{i\pp^+}
+ \gamma^+ { 1 \over 4}
\st{\lh}{i\pp^-}
\Big ] \psi \nonumber \\
&=& {1\over 2}\int dx^- d^2 x^\perp \Big [ \xi^{\dagger} {1 \over i \partial^{+}}[ m_F^2 -
(\partial^{\perp})^{2} ] - ({1 \over i \partial^{+}}[ m_F^2 -
(\partial^{\perp})^{2}]\xi^\dagger)\Big ]\xi
 . \end{eqnarray}
The longitudinal scaling operator (at $x^+=0$) is 
\begin{eqnarray} K^{3} &=& - { 1 \over 2} \int dx^- d^2 x^\perp
 x^{-}  
\Bigg[{\overline\psi} \gamma^{+}{1 \over 4} \st{\lh}{i\pp^+} \psi \Bigg]  \cr
&=& - {i \over 4}  \int dx^- d^2 x^\perp x^{-} \big [\xi^{\dagger} \partial^{+}
\xi-(\partial^+ \xi^\dagger ) \xi \big ].  \end{eqnarray}
The transverse boost operators are
\begin{eqnarray} E^{i} &=&  -{ 1 \over 4} \int dx^- d^2 x^\perp  x^{i}
 \Bigg [{\overline \psi} \gamma^+{1 \over 4} \st{\lh}{i\pp^+}\psi \Bigg ] \nonumber \\
&=& -  { 1 \over 4} \int dx^- d^2 x^\perp x^i\big [\xi^{\dagger}
i \partial^+ -(i\partial^+ \xi^\dagger)\big ]\xi
  . \end{eqnarray}
The generators of rotations are
\begin{eqnarray} J^{3} &=&{ 1 \over 2 } \int dx^- d^2
x^\perp \Bigg \{
x^{1} \Bigg [ {\overline \psi} \Big \{ \gamma^{+}{ 1 \over 4} 
\st{\lh}{i\pp^2}+
 \gamma^{2}{ 1 \over 4} \st{\lh}{i\pp^+}\Big \} \psi \Bigg ] \nonumber \\
&&~~~~- x^{2} \Bigg [ {\overline \psi} \Big \{ \gamma^{+}{ 1 \over 4}
\st{\lh}{i\pp^1} 
 + 
\gamma^{1}{ 1 \over 4} \st{\lh}{i\pp^+}  \Big \}\psi \Bigg ] 
\Bigg \} \nonumber \\
&=& \int dx^- d^2 x^\perp \Big [\xi^{\dagger} 
\big [ {i\over 2} (x^1 {\overrightarrow \partial^2}- x^2 {\overrightarrow
\partial^1})\xi -\big [ {i\over 2} (x^1 {\overrightarrow \partial^2}- x^2 {\overrightarrow
\partial^1})\xi^\dagger\big ]\xi\nonumber\\&&~~~~~~~~~~~~~~~~~~~~~ 
 + \xi^\dagger{\sigma_{3} \over 2}\xi \big ].  \end{eqnarray}
 and 
\begin{eqnarray} F^{i}  &=&  { 1 \over 2} \int dx^- d^2 x^\perp 
 \Bigg \{ 
x^{-} \Bigg [ {\overline \psi} \Big \{  \gamma^{+}{1\over 4} \st{\lh}{i\pp^i}
 + \gamma^{i}{1 \over 4} 
\st{\lh}{i\pp^+}
  \Big \} \psi \Bigg ] \nonumber \\
&&~~~- x^{i} \Bigg [{\overline \psi}\Big \{\gamma^{+}{1 \over 4} 
\st{\lh}{i\pp^-} 
+{1 \over 4} \gamma^{-} 
\st{\lh}{i\pp^+}
\Big \} \psi \Bigg ]
 \Bigg \} \nonumber \\
&=&   {i\over 2}  \int dx^- d^2 x^\perp \xi^{\dagger} \Bigg [ 
x^i (m^2 -(\partial^{\perp})^{2}){1 \over \partial^{+}} - x^{-}
{\partial \over \partial x^{i}} \nonumber \\
&& ~~~+ {1 \over \partial^{+}}
\Big \{ - { \partial \over \partial x^i} - i \epsilon^{ij} \sigma^3 {
\partial \over \partial x^j} + \epsilon^{ij} m\sigma^j \Big \} \Bigg ] \xi
\nonumber\\&&~~~- {i\over 2}  \int dx^- d^2 x^\perp  \Bigg [ 
\big [x^i (m^2 -(\partial^{\perp})^{2}){1 \over \partial^{+}} - x^{-}
{\partial \over \partial x^{i}} \nonumber \\
&& ~~~+ {1 \over \partial^{+}}
\Big \{ - { \partial \over \partial x^i} + i \epsilon^{ij} \sigma^3 {
\partial \over \partial x^j} + \epsilon^{ij} m\sigma^j \Big \}\xi^\dagger
\big ] \Bigg ] \xi
 . \end{eqnarray}
%%%%%%%%%%%%%%%%%%%%%%%%%%%%%%%%%%%%%%%%%%%%%%%%%%%%%%%%%%%%%%%%%%%%%
\section{Fock Representation}
%%%%%%%%%%%%%%%%%%%%%%%%%%%%%%%%%%%%%%%%%%%%%%%%%%%%%%%%%%%%%%%%%%%%%
Free spin-half field operator is
\begin{eqnarray} \xi(x) = \sum_{\lambda}\chi_\lambda 
 \int {dk^+ d^2 k^\perp \over 2 (2 \pi)^3 \sqrt {k^+}}  
[ b(k,\lambda)  e^{-ik.x}
+ d^{\dagger}(k,-\lambda)  e ^{ik.x}]  . \end{eqnarray}
In terms of Fock space operators
\begin{eqnarray} P^{+} = \int {dk^+ d^2 k^\perp \over 2 (2 \pi)^3 k^+} k^{+} \sum_{\lambda} \big 
[ b^{\dagger}(k, \lambda) b(k,\lambda)  +
d^{\dagger}(k,-\lambda) d(k,-\lambda) \big ].  \end{eqnarray}
\begin{eqnarray}
 P^{i} = \int {dk^+ d^2 k^\perp \over 2 (2 \pi)^3 k^+} k^{i} \sum_{\lambda} \big 
[ b^{\dagger}(k,\lambda) b(k,\lambda) + 
d^{\dagger}(k,-\lambda) d(k,-\lambda) \big ] . \end{eqnarray}
\begin{eqnarray}
 P^{-} = \int {dk^+ d^2 k^\perp \over 2 (2 \pi)^3 k^+} {m_F^2 + (k^{\perp})^2 \over k^{+}} \sum_{\lambda}
\big [ b^{\dagger}(k,\lambda) b(k,\lambda) + d^{\dagger}(k,-\lambda)
 d(k,-\lambda) \big ].  \end{eqnarray}
\begin{eqnarray} K^{3} &=&  - {i\over 2} \int {dk^+ d^2 k^\perp \over 2 (2 \pi)^3 k^+} k^{+} \sum_{\lambda}
\Big (\big [  {\partial b^{\dagger}(k,\lambda) \over 
\partial k^{+}} b(k,\lambda) +   {\partial
d^{\dagger}(k,-\lambda) \over \partial k^{+}}d(k,-\lambda) 
\big ]\nonumber\\&&~~~~~-\big [ b^{\dagger}(k,\lambda) {\partial b(k,\lambda)
\over {\partial k^+}} + 
d^{\dagger}(k,-\lambda){\partial d(k,-\lambda)\over {\partial k^+}} 
\big ] \Big ) . \end{eqnarray}
\begin{eqnarray} E^{i} &=& { i\over 2} \int {dk^+ d^2 k^\perp \over 2 (2 \pi)^3 k^+} \sum_{\lambda}
k^{+}\Big (\big [   {\partial b^{\dagger}(k,\lambda)
\over \partial k^{i}}b(k,\lambda)  +  
{\partial d^{\dagger}(k,-\lambda) \over \partial k^{i}} d(k,-\lambda)
 \big ]\nonumber\\&&~~~~~-\big [ b^{\dagger}(k,\lambda){\partial b(k,\lambda)
\over {\partial k^i}}  +  
d^{\dagger}(k,-\lambda){\partial d(k,-\lambda)\over {\partial k^i}}
 \big ] \Big ). 
\end{eqnarray}
\begin{eqnarray} J^{3} &=& {i\over 2} \int {dk^+ d^2 k^\perp \over 2 (2 \pi)^3 k^+} \sum_{\lambda}
\big [ \Big( [ k^{1} {\partial \over \partial k^{2}}
- k^{2} {\partial \over \partial k^{1}} ] b^{\dagger}(k,\lambda)  \Big )
b(k,\lambda)-b^\dagger(k,\lambda) \nonumber\\&&[ k^{1} {\partial \over \partial k^{2}}
- k^{2} {\partial \over \partial k^{1}} ] 
b(k,\lambda) + \Big ( \big [ k^{1} {\partial \over \partial k^{2}}
- k^{2} {\partial \over \partial k^{1}} ] d^{\dagger}(k,-\lambda)  \big ]
\Big )
 d(k,-\lambda)\nonumber\\&&~~~~~~~~~~~~~- d^\dagger(k,-\lambda)\big [ k^{1}
 {\partial \over \partial k^{2}}
- k^{2} {\partial \over \partial k^{1}} ]d(k,-\lambda) \big ]
 \nonumber \\
&&~~~~~~~~~~~~~~ + { 1 \over 2} \int {dk^+ d^2 k^\perp \over 2 (2 \pi)^3 k^+}
 \sum_{\lambda} \lambda \big [ b^{\dagger}(k,\lambda) b(k,\lambda) + 
d^{\dagger}(k,\lambda) d(k,\lambda) \big ] \end{eqnarray}
with $\lambda = \pm 1$. 
\begin{eqnarray} F^{i} &=&   i \int {dk^+ d^2 k^\perp \over 2 (2 \pi)^3 k^+} k^i \sum_{\lambda}
\Big (\Big[ {\partial b^{\dagger}(k,\lambda)
\over \partial k^{+}} b(k,\lambda) +
{\partial d^{\dagger}(k,-\lambda)
\over \partial k^{+}} d(k,-\lambda) \Big ]\nonumber\\&&~~~~~~~~~~~~~-
\Big[ b^\dagger(k,\lambda){\partial b(k,\lambda)\over \partial k^{+}} +
d^\dagger(k, -\lambda){\partial d(k,-\lambda)
\over \partial k^{+}} \Big ]\Big )  \nonumber \\
&& + {i\over 2} \int {dk^+ d^2 k^\perp \over 2 (2 \pi)^3 k^+} 
 {m_F^2 + (k^{\perp})^2 \over k^{+}}\sum_{\lambda}
\Big ( \Big[ {\partial b^{\dagger}(k,\lambda)
\over \partial k^i} b(k,\lambda) +{\partial d^{\dagger}(k,-\lambda)
\over \partial k^i} d(k,-\lambda) \Big ]\nonumber\\&&~~~~~~~~~~~~~~~~
-\Big [ b^\dagger(k, \lambda){\partial b(k,\lambda)
\over \partial k^i} +
d^\dagger (k, -\lambda){\partial d(k, -\lambda) \over \partial k^i}
 \Big] \Big ) \nonumber \\&& - \int {dk^+ d^2 k^\perp 
\over 2 (2 \pi)^3 k^+}  { \epsilon^{ij} \over k^+}
 k^j \sum_{\lambda \lambda'}\sigma^{3}_{\lambda \lambda'}
\Big[ b^{\dagger}(k,\lambda) b(k,\lambda')
- d^{\dagger}(k,-\lambda') d(k,-\lambda)\Big ] \nonumber \\
&& - \int {dk^+ d^2 k^\perp \over 2 (2 \pi)^3 k^+}  
{ \epsilon^{ij} \over k^+}
 m_F \sum_{\lambda \lambda'}\sigma^{j}_{\lambda \lambda'}
\Big[ b^{\dagger}(k,\lambda) b(k,\lambda')
+ d^{\dagger}(k,-\lambda) d(k,-\lambda')\Big ]. 
\end{eqnarray}
%%%%%%%%%%%%%%%%%%%%%%%%%%%%%%%%%%%%%%%%%%%%%%%%%%%%%%%%%%%%%%
\section{Transverse Spin of a Single Fermion}
%%%%%%%%%%%%%%%%%%%%%%%%%%%%%%%%%%%%%%%%%%%%%%%%%%%%%%%%%%%%%%
For a single fermion of mass $m$ and momenta ($k^+,k^\perp$), we have
\begin{eqnarray}
P^+ \mid k \lambda \rangle &=&k^+ \mid k \lambda \rangle,
~P^1\mid k \lambda \rangle=k^1\mid k \lambda \rangle,
 ~ P^2\mid k \lambda \rangle=k^2\mid k \lambda \rangle, \nonumber \\
 P^- \mid k \lambda \rangle &=& {(k^\perp)^2 + m^2 \over k^+}\mid k \lambda \rangle,
~{\cal J}^3\mid k \lambda \rangle={1 \over 2} \lambda \mid k \lambda \rangle, 
\nonumber \\
~K^3 \mid k \lambda \rangle &=& -i k^+ { \partial \over \partial k^+}
\mid k \lambda \rangle, ~ E^1
\mid k \lambda \rangle = i k^+ { \partial \over
\partial k^1}
\mid k \lambda \rangle, ~E^2 
\mid k \lambda \rangle= i k^+ { \partial \over \partial k^2}
\mid k \lambda \rangle, \nonumber \\
F^1 \mid k \lambda \rangle &=& \left (2 i k^1 {\partial \over \partial k^+} + i {(k^\perp)^2 + m^2 \over k^+}
{\partial \over \partial k^1} - {k^2 \over k^+} \lambda \right ) \mid k
\lambda \rangle  - { m
\over k^+} \sum_{\lambda'}\sigma^2_{\lambda' \lambda} 
\mid k \lambda' \rangle, \nonumber \\
F^2 \mid k \lambda \rangle &=& \left (2 i k^2 {\partial \over \partial k^+} + i {(k^\perp)^2 + m^2 \over k^+}
{\partial \over \partial k^2} + {k^1 \over k^+} \lambda \right )
\mid k \lambda \rangle
  + { m \over k^+} \sum_{\lambda'} \sigma^1_{\lambda' \lambda} 
\mid k \lambda' \rangle.
\end{eqnarray}
We arrive at 
\begin{eqnarray}
m{\cal J}^1
\mid k \lambda \rangle &=& \left ({ 1 \over 2} F^2 P^+ +K^3P^2 
 - { 1 \over 2} E^2 P^- - P^1
{\cal J}^3 \right ) 
\mid k \lambda \rangle\nonumber \\
&=& m \sum_{\lambda'}{\sigma^1_{\lambda' \lambda} \over 2}
\mid k \lambda' \rangle,
\end{eqnarray}
\begin{eqnarray}
m {\cal J}^2 
\mid k \lambda \rangle&=& \left (- { 1 \over 2} F^1 P^+- K^3 P^1  + { 1 \over 2} E^1 P^- -
P^2 {\cal J}^3 \right )
\mid k \lambda \rangle\nonumber \\
&=& m \sum_{\lambda'}{\sigma^2_{\lambda' \lambda} \over 2}
\mid k \lambda' \rangle.
\end{eqnarray} 

%%%%%%%%%%%%%%%%%%%%%%%%%%%%%%%%%%%%%%%%%%%%%%%%%%%%%%%%%%%%%%%%
\chapter{Transverse Spin in Free Massless Spin One Field Theory}
%%%%%%%%%%%%%%%%%%%%%%%%%%%%%%%%%%%%%%%%%%%%%%%%%%%%%%%%%%%%%%%%
\section{Poincare Generators: Operator Forms}
%%%%%%%%%%%%%%%%%%%%%%%%%%%%%%%%%%%%%%%%%%%%%%%%%%%%%%%%%%%%%%%%
The symmetric gauge invariant energy momentum tensor
\begin{eqnarray} \Theta^{\mu \nu} = 
F^{\lambda \mu} F^{\nu}_{\lambda} - g^{\mu \nu} {\cal L} . \end{eqnarray}
where the Lagrangian density  
\begin{eqnarray} {\cal L} = -{1 \over 4} F^{\mu \nu} F_{\mu \nu}
\end{eqnarray}
with
\begin{eqnarray} F^{\mu \nu} = \partial^{\nu} A^{\mu} - \partial^{\mu} A^{\nu}
. \end{eqnarray}

We choose $A^{+}=0$ gauge. Only the transverse fields $A^{i}$ are dynamical
variables.  
The momentum operators are given by
\begin{eqnarray} P^{+} = { 1 \over 2}\int dx^- d^2 x^\perp  \partial^{+} A^{j}
\partial^{+} A^{j}   , \end{eqnarray}
\begin{eqnarray} P^i = { 1 \over 2} \int dx^- d^2 x^\perp \Big (
A^{j}\partial^+\partial^j A^{i} - A^{i}\partial^+ \partial^j A^{j}
-A^{j}\partial^+\partial^i A^{j}\Big ). \end{eqnarray}
The Hamiltonian operator is 
\begin{eqnarray} P^- &=& 
{1 \over 2} \int dx^- d^2 x^\perp \Big 
[ {1 \over 4} (\partial^+ A^-)^2 + { 1 \over 2} F^{ij} F_{ij} \Big ]
\nonumber \\
&=& {1 \over 2}\int dx^- d^2 x^\perp  \partial^{i} A^{j} \partial^{i} A^{j}  
  = -{1\over 2}\int dx^- d^2x^\perp A^j {(\partial^i)}^2 A^j.
\end{eqnarray}
The longitudinal scale generator (at $x^+=0$) is  
\begin{eqnarray}  K^3 = -{1 \over 2} \int dx^- d^2 x^\perp x^{-}  \partial^{+} A^{j} \partial^{+} A^{j} 
 . \end{eqnarray}
The transverse boost generators are
\begin{eqnarray} E^{i} = - {1 \over 2}\int dx^- d^2 x^\perp x^{i}  \partial^{+} A^{j} \partial^{+} A^{j} 
. \end{eqnarray}
The generators 
 of rotations are
\begin{eqnarray} J^3 &=& {1 \over 2}\int dx^- d^2 x^\perp \Big ( 
x^1 [ \partial^+ A^2  \partial^i A^i + \partial^+ A^1
( \partial^2 A^1 - \partial^1 A^2)] \nonumber \\
&& - x^2 [ \partial^+ A^1  \partial^i A^i + \partial^+ A^2
(- \partial^2 A^1 + \partial^1 A^2)] \Big)\nonumber \\
&=& { 1 \over 2} \int dx^- d^2 x^\perp \left (x^1 
[ \partial^+ A^1 \partial^2 A^1 + \partial^+ A^2 \partial^2 A^2]
-x^2 [ \partial^+ A^1 \partial^1 A^1 + \partial^+ A^2 \partial^1 A^2] \right )
\nonumber \\
&& ~~~ + { 1 \over 2} \int dx^- d^2 x^\perp 
[ A^1 \partial^+ A^2 - A^2 \partial^+ A^1 ]. 
   \end{eqnarray}
and 
\begin{eqnarray} F^i &=& {1 \over 2}\int dx^- d^2 x^\perp \Big (
x^- \Big (
A^{ja}\partial^+\partial^j A^{i} - A^{i}\partial^+  \partial^j A^{j}
-A^{j}\partial^+\partial^i A^{j}\Big ) \nonumber \\
&& - x^i [  A^k (\partial^j)^2 A^k] \Big ) 
- 2\int dx^- d^2 x^\perp A^i \eta^{ij} \partial^j A^j, ~~{\rm
no~summation~over}~i,j ~
, \end{eqnarray}
with $ \eta^{12}=\eta^{21}=1$, $\eta^{11}=\eta^{22}=0$.
%%%%%%%%%%%%%%%%%%%%%%%%%%%%%%%%%%%%%%%%%%%%%%%%%%%%%%%%%%%%%%%%%%%%%%%%
\section{Fock Representation}
%%%%%%%%%%%%%%%%%%%%%%%%%%%%%%%%%%%%%%%%%%%%%%%%%%%%%%%%%%%%%%%%%%%%%%%%
The dynamical components of the free massless spin field operator 
in $A^{+}=0$ gauge are
\begin{eqnarray}  A^{i}(x) = \sum_{\lambda=1}^{2} \int {dk^+ d^2 k^\perp \over 2 (2 \pi)^3 k^+} \delta^{i \lambda}
[ a(k,\lambda)e^{-ik.x} + a^{\dagger}(k,\lambda) e^{ik.x} ]. \end{eqnarray}
In terms of Fock space operators, we have
\begin{eqnarray} P^{+} =  \int {dk^+ d^2 k^\perp \over 2 (2 \pi)^3 k^+} k^{+} \sum_{\lambda}  
a^{\dagger}(k,\lambda) a (k,\lambda).  \end{eqnarray}
\begin{eqnarray} P^{i} =  \int {dk^+ d^2 k^\perp \over 2 (2 \pi)^3 k^+} k^{i} \sum_{\lambda} 
a^{\dagger}(k,\lambda) a(k,\lambda)  . \end{eqnarray}

 \begin{eqnarray} H =  \int {dk^+ d^2 k^\perp \over 2 (2 \pi)^3 k^+} 
 {{k^{\perp}}^{2} \over k^{+}}\sum_{\lambda} 
a^{\dagger}(k,\lambda) a(k,\lambda)  . \end{eqnarray}
\begin{eqnarray} K^3 = - { i\over 2}\int {dk^+ d^2 k^\perp \over 2 (2 \pi)^3 k^+} 
 k^{+}\sum_{\lambda} \Big [
\big ( {\partial a^{\dagger}(k,\lambda) \over \partial
k^{+}}\big )  a(k,\lambda)- a^\dagger ( k, \lambda){\partial a(k,\lambda)
 \over \partial k^{+}}\big )\Big ]  . \end{eqnarray}
\begin{eqnarray} E^{i} =  {i\over 2}  \int {dk^+ d^2 k^\perp \over 2 (2 \pi)^3 k^+} 
k^{+}\sum_{\lambda} \big [
 \big ({\partial a^{\dagger}(k,\lambda) \over \partial k^{i}} \big )
a(k,\lambda)- a^\dagger ( k, \lambda){\partial a(k,\lambda)
 \over \partial k^{i}}\big ) \big ] 
 . \end{eqnarray}
\begin{eqnarray} J^3 &=& 
{i\over 2}  \int {dk^+ d^2 k^\perp \over 2 (2 \pi)^3 k^+} \sum_{\lambda}\big
[ \Big ( 
(k^1 {\partial \over \partial k^2} - k^2 {\partial \over \partial k^1}
) a^{\dagger}(k,\lambda) \Big ) a(k,\lambda)\nonumber\\&&~~~~~~~~~~~~~~~~~~~
-a^\dagger (k, \lambda)(k^1 {
\partial \over \partial k^2} - k^2 {\partial \over \partial k^1}
) a(k,\lambda)\big ] \nonumber \\
&&~~~~~~~~~~~~~~~~~~~~~~~ + i \int {dk^+ d^2 k^\perp
 \over 2 (2 \pi)^3 k^+}  \Big (
a^{\dagger}(k,2) a(k,1) - a^{\dagger}(k,1) a(k,2) \Big )  . \end{eqnarray} 
Introduce creation and annihilation operators 
\begin{eqnarray} a(k, \uparrow) = { - 1 \over \sqrt{2}} [ a(k,1) - i a(k,2)] ,
 a(k, \downarrow) = {  1 \over \sqrt{2}} [ a(k,1) + i a(k,2)]
.\end{eqnarray}
Then
\begin{eqnarray} J^3 &=& {i\over 2}
 \int {dk^+ d^2 k^\perp \over 2 (2 \pi)^3 k^+} \sum_{\lambda}\big [ \Big ( 
(k^1 {\partial \over \partial k^2} - k^2 {\partial \over \partial k^1}
) a^{\dagger}(k,\lambda) \Big ) a(k,
\lambda)\nonumber\\&&~~~~~~~~~~~~~~~~~~~~- a^\dagger (k, \lambda)(k^1 {
\partial \over \partial k^2} - k^2 {\partial \over \partial k^1}
) a(k,\lambda)\big ] \nonumber \\
&&~~~~~~~~~~~~~~~~~~~~ +  \int {dk^+ d^2 k^\perp \over 2 (2 \pi)^3 k^+}
 \sum_{\lambda} \lambda a^{\dagger}(k,\lambda) a(k,\lambda)   . \end{eqnarray} 

where $\lambda$ now denotes circular polarization.
\begin{eqnarray}  F^i &=&
 i  \int {dk^+ d^2 k^\perp \over 2 (2 \pi)^3 k^+} k^i
\sum_{\lambda} \big ({\partial a^{\dagger}(k,\lambda) \over
\partial k^{+}} a(k,\lambda) - a^\dagger ( k, \lambda){\partial a(k,\lambda)
 \over \partial k^{+}}\big ) \nonumber \\ 
&& + {i\over 2}  \int {dk^+ d^2 k^\perp \over 2 (2 \pi)^3 k^+}
 {(k^{\perp})^2 \over k^+}\sum_{\lambda} 
\big ( {\partial a^{\dagger}(k,\lambda) \over \partial k^i}a( k, \lambda) -a^\dagger 
( k, \lambda){\partial a(k,\lambda) \over \partial k^{i}}\big )
 \nonumber \\
&& - 2  \epsilon^{ij} \int {dk^+ d^2 k^\perp \over 2 (2 \pi)^3 k^+}
{k^j \over k^+} \sum_\lambda \lambda a^\dagger(k,\lambda) a(k,\lambda).  
\end{eqnarray}
%%%%%%%%%%%%%%%%%%%%%%%%%%%%%%%%%%%%%%%%%%%%%%%%%%%%%%%%%%%%%%%%%%%%%%%
\section{Transverse Spin}
%%%%%%%%%%%%%%%%%%%%%%%%%%%%%%%%%%%%%%%%%%%%%%%%%%%%%%%%%%%%%%%%%%%%%%%
Using the explicit form of the operators, we get for a state of
momentum $k(k^+, k^\perp)$ and helicity $\lambda$, 
\begin{eqnarray}
{\cal J}^3 \mid k \lambda \rangle ={W^+ \over P^+} \mid k \lambda \rangle &=& \lambda \mid k \lambda \rangle,
\nonumber \\
W^1 \mid k \lambda \rangle &=& k^1 \lambda \mid k \lambda \rangle, \nonumber \\
W^2 \mid k \lambda \rangle &=& k^2 \lambda \mid k \lambda \rangle, \nonumber \\
W^- \mid k \lambda \rangle &=& {(k^\perp)^2 \over k^+} \lambda \mid k
\lambda \rangle.
\end{eqnarray}
\begin{eqnarray}
{\cal J}^i \mid k \lambda \rangle =0.
\end{eqnarray}

%%%%%%%%%%%%%%%%%%%%%%%%%%%%%%%%%%%%%%%%%%%%%%%%%%%%%%%%%%%%%%%%%%%%%

\chapter{Transverse Spin for a System of Two Non-interacting Fermions}
%%%%%%%%%%%%%%%%%%%%%%%%%%%%%%%%%%%%%%%%%%%%%%%%%%%%%%%%%%%%%%%%%%%%%%

In order to show the non-triviality of the transverse spin operators even in the free
theory and the manifest cancellation of the center of mass motion in this
case, here we
 evaluate the transverse spin for a composite system of two free
fermions. The manifest cancellation of the center of mass motion for the
interacting theory is much more complicated and is given in
appendix H.  

Let the mass of each fermion be $m$ and momenta $(k^+_i, k^\p_i)$, $ i=1,2$.
We take the state to be $\mid P \rangle = b^\dagger(k_1,s_1)b^\dagger(k_2,
s_2)\mid 0 \rangle $, where $s_1$ and $s_2$ are the helicities. 
\be
M{\cal J}^1
\mid P \rangle &=& \left ({ 1 \over 2} F^2 P^+ +K^3P^2 
 - { 1 \over 2} E^2 P^- - P^1
{\cal J}^3 \right ) 
\mid P \rangle .
\e

We introduce Jacobi momenta, $(x_i,q^\p)$ defined as
\be
k_1^\p = q^\p + x_1P^\p, ~~k_2^\p = -q^\p + x_2P^\p~~~~~~~~~ k_i^+ = x_iP^+
\e
with $
\sum x_i = 1$.

Here $M$ is the mass of the composite system and $(P^+, P^\p)$ are the
momenta of the center of mass.

The partial derivatives with respect to the particle momenta can be
expressed in terms of these variables as
\be
{\pp \over \pp k_1^i} = x_2{\pp \over \pp q^i} + {\pp \over \pp P^i},
~~~~~~~~~~~~~~~~~~{\pp \over \pp k_2^i} =  -x_1{\pp \over \pp q^i}
 + {\pp \over \pp P^i},
\e
and
\be
{\pp \over \pp k_1^+} = {x_2\over P^+}{\pp \over \pp x_1}+ {\pp \over \pp P^+}
- x_2{P^\p \over P^+}\cdot{\pp \over \pp q^\p}~, 
\e
\be
{\pp \over \pp k_2^+} = {x_1\over P^+}{\pp \over \pp x_2}+ {\pp \over \pp P^+}
+ x_1{P^\p \over P^+}\cdot{\pp \over \pp q^\p} ~.
\e  
Then we have
\be
K^3P^2\mid P \rangle = [ -iP^2 x_1x_2({\pp \over \pp x_1} + {\pp \over \pp
x_2} ) -i P^2P^+ {\pp \over \pp P^+} ]\mid P \rangle ~,
\e 
\be
-{1\over 2}E^2P^-\mid P \rangle = \Big [-{i\over 2}( (P^\p)^2 + M^2 ) {\pp \over
\pp P^2} +iP^2 \Big ]\mid P \rangle~,
\e
\be
P^1{\cal J}^3 \mid P \rangle = [-i P^1(q^2{\pp \over \pp q^1} - 
q^1{\pp \over \pp q^2}) + P^1{s_1\over 2}+ P^1{s_2\over 2}]\mid P \rangle ~.
\e
\be
{1\over 2}F^2P^+ \mid P \rangle &=& \Big [iq^2 ( x_2{\pp \over \pp x_1} -
x_1{ \pp
\over \pp x_2})+ {s_1\over 2}{q^1\over x_1}
- {s_2\over 2}{q^1\over x_2}+ {i\over 2} ( m^2 + (q^\p)^2) ({x_2\over x_1} 
- {x_1\over x_2}) {\pp \over
\pp q^2}\nonumber\\&&~~~~+ {m\over 2} \sum_\lambda ( {\s^1_{\lambda s_1}\over
x_1} + {\s^1_{\lambda s_2}\over x_2}) 
 - iq^2P^\p \cdot{\pp \over \pp q^\p} +iP^2  x_1x_2({\pp \over \pp x_1}
+ {\pp \over \pp x_2}) \nonumber\\&&~~~~~~~+iP^+ P^2 {\pp \over \pp P^+} +
+{i\over 2}( m^2 + (q^\p)^2) {1\over {x_1 x_2}}
 {\pp \over \pp P^2} + {i\over 2} (P^\p)^2 {\pp \over \pp P^2}
\nonumber\\&&~~~~~~~~~+ i (
q^\p \cdot P^\p){\pp \over \pp q^2} 
~~  +{P^1\over 2}(s_1 + s_2)-iP^2 \Big ] \mid P \rangle ~ .
\e
Substituting
\be
M^2 = {( m^2 + (q^\p)^2)\over x_1x_2}~, 
\e
we get
\be
M{\cal J}^1\mid P \rangle &=&\Big [iq^2 (x_2{\pp \over \pp x_1} - x_1{\pp \over \pp x_2}) + {i\over 2}(
m^2 + (q^\p)^2) ({x_2\over x_1} - {x_1\over x_2}) {\pp \over \pp q^2}
\nonumber\\&&~~~~ +{q^1\over 2}({s_1\over x_1} - {s_2\over x_2}) + {m\over
2}\sum_\lambda ({\s^1_{\lambda s_1}\over x_1}+{\s^1_{\lambda s_2}
\over x_2})\Big ] \mid P \rangle ~ .
\e 
Explicitly we see that $M{\cal J}^1$ does not depend on the center of mass 
momenta.

%%%%%%%%%%%%%%%%%%%%%%%%%%%%%%%%%%%%%%%%%%%%%%%%%%%%%%%%%%%%%%%%%%%%%%%%%%%%%
\chapter{Transverse Spin of a Dressed Quark in an Arbitrary Reference Frame}
%%%%%%%%%%%%%%%%%%%%%%%%%%%%%%%%%%%%%%%%%%%%%%%%%%%%%%%%%%%%%%%%%%%%%%%%%%%%%

We introduce a wave packet state
\be
\mid \psi_\sigma \rangle = {1\over 2} \int dP^+ d^2P^\perp f(P) \mid P,
\sigma \rangle  \label{awp}
\e
which is normalized as
\be
\langle \psi_\sigma \mid \psi_{\sigma'} \rangle = \delta_{\sigma \sigma'}.
\e
Here $f(P)$ is a function of $P$, the exact form of which is not important.
Using the normalization condition of the state we get
\be
{1\over 2}\int dP^+ d^2P^\perp f^*(P) f(P)(2\pi)^3 P^+ = 1.
\label{anorm}
\e 
The expectation values of the various operators involved in the definition
of $M{\cal J}^i$ are given below.
It is to be noted that we have done the calculation in an
arbitrary reference frame, in order to show that the dependence on the total
center of mass momenta $(P^+, P^\perp)$ actually gets canceled in the
expectation value of $M{\cal J}^i$.

The matrix elements presented below have been evaluated between wave packet
states of different helicities, namely $\sigma$ and
$\sigma'$. Since the transversely polarized  state can be
expressed in terms of the longitudinally polarized (helicity) states
(see chapter 8), the matrix elements of these operators between transversely
polarized states can be easily obtained from these expressions.   
We introduce
\be
\psi^\sigma_1 = f(P)\Phi^\sigma_1, ~~~~~~~~~~~~~~\psi^\sigma_{\sigma_1
\lambda} = f(P)\Phi^\sigma_{\sigma_1\lambda}.
\e
The matrix elements are given by 
\be
\langle \psi_\sigma \mid K^3 P^2 \mid \psi_{\sigma' }\rangle&=&
{1\over 2} \int dP^+d^2P^\perp (2\pi)^3 P^+\Big ({i\over 2}\psi^{*\sigma}_1 
{\pp \psi^{\sigma'}_1 \over
{\pp P^+}} P^+ P^2  \nonumber\\&&~~~~
 + {i\over 2}\sum_{spin} \int dxd^2q^\perp P^2 
\psi^{\sigma*}_{\sigma_1 \lambda} {\pp \psi^{\sigma'}_{\sigma_1'
\lambda'}\over {\pp P^+}}P^+ + h. c.\Big ).
\e   
\be
\langle \psi_\sigma \mid {1\over 2}E^2P_{free}^- \mid \psi_{\sigma'} \rangle&=& 
{1\over 2} \int dP^+d^2P^\perp (2\pi)^3 P^+\Big (-{i\over
4}\psi^{*\sigma}_1 {\pp \psi^{\sigma'}_1\over {\pp P^2}}P^+{ (P^\perp)^2 +
m^2\over P^+}\nonumber\\&&~~~~~~~~~~~~~-{i\over 4}\sum_{spin}
 \int dxd^2q^\perp \psi^{*\sigma}_{\sigma_1 \lambda} {\pp \psi^{\sigma'}_{
\sigma'_1 \lambda'} \over {\pp P^2}}
 \Big [ {m^2 + (q^\perp+xP^\perp)^2\over x} \nonumber\\&&~~~~~~~~~~~~~~~~~~~~~~~~~~~~
+ {(-q^\perp + (1-x)P^\perp )^2 \over {1-x}}\Big ]+ h.c.\Big ).
\e
\be
\langle \psi_\sigma \mid {1\over 2}E^2P_{int}^- \mid \psi_{\sigma'} \rangle&& 
= -{g\over \sqrt {2(2\pi)^3}}{1\over 2}\int dP^+ d^2P^\perp(2\pi)^3 P^+
\sum_{spin} \int dx d^2q^\perp {1\over \sqrt {1-x}}\nonumber\\&&~~~~~
\Big (i\psi^{\sigma *}_1 {\pp \psi^{\sigma'}_{\sigma_1 \lambda}\over {\pp
P^2}}\chi^\dagger_{\sigma} ~\Big [ 
-{(q^\perp \cdot \epsilon^\perp) \over {1-x}} -{1\over 2} { (\tilde
\sigma^\perp \cdot \epsilon^\perp)(\tilde \sigma^\perp \cdot q^\perp)\over
x}\nonumber\\&&~~~~~~~~~~~~~~~~ -{1\over 2} im {(1-x)\over x}(\tilde 
\sigma^\perp \cdot \epsilon^\perp) \Big ] \chi_{\sigma_1} + h. c. \Big ).
\e
Here $h. c.$ is the Hermitian conjugate, $\sum_{spin}$ is summation over $\sigma_1, \sigma'_1, \lambda,
\lambda'$. $P^-_{free}$ is the free part and $P^-_{int}$ is the interaction part
of the light-front QCD Hamiltonian density.
\be
\langle \psi_\sigma \mid P^1 {\cal J}^3 \mid \psi{\sigma'} \rangle&=&
{1\over 2}\int dP^+ d^2P^\perp (2\pi)^3 P^+ \Big [\sum_{spin}
\int dxd^2q^\perp \nonumber\\&&~~~~~P^1 \Big ( {i\over 2}\psi^{*\sigma}_{\sigma_1 \lambda}
 ( q^2 {\pp \over{\pp q^1}} - q^1 {\pp \over {\pp q^2}} ) \psi^{\sigma'}_{
\sigma'_1 \lambda'}+ h.c. \Big )\nonumber\\&& +{1\over 2} \int dxd^2q^\perp P^1 \sum_{
\lambda,\sigma_2,\sigma'_2}  \lambda \psi^{*\sigma}_{\lambda \sigma_2} 
\psi^{\sigma'}_{\lambda \sigma'_2}
+ \int dxd^2q^\perp P^1\nonumber\\&&~~~~~~~~~~~~~~~ \sum_{\lambda,
\sigma_1, \sigma'_1}  \lambda \psi^{*\sigma}_{\sigma_1\lambda} 
\psi^{\sigma'}_{\sigma'_1 \lambda}\Big ] .
\e 
The first term in the above expression is the quark-gluon orbital part, the
second and the third terms are the intrinsic helicities of the quark and
gluon respectively.
Finally, the operator ${1\over 2}F^2P^+$ can be separated into three parts,
\be
{1\over 2}F^2P^+ = {1\over 2}F^2_IP^+ + {1\over 2}F^2_{II}P^+ +
 {1\over 2}F^2_{III}P^+
\e
where $F^2_I$,$F^2_{II}$ and $F^2_{III}$ have been defined earlier.
The matrix elements of the different parts of these operators for a dressed quark state
in an arbitrary reference frame are given below. A
part of this calculation has been given in some detail in appendix I.
We have
\be
{1\over 2}F^2_IP^+ = {1\over 2}F^2_I(1)P^+ - {1\over 2}F^2_I(2)P^+ -
 {1\over 2}F^2_I(3)P^+.
\e
The matrix elements of these three parts are
\be
\langle \psi_\sigma \mid {1\over 2}F^2_I(1)P^+ \mid \psi_{\sigma'} \rangle
&=&
\langle \psi_\sigma \mid {1\over 2} \int dx d^2q^\perp x^- P^2_0 {1\over
2}P^+ \mid \psi_{\sigma'} \rangle \nonumber\\&& = 
{1\over 2}\int dP^+ d^2P^\perp (2\pi)^3 P^+\Big [ -{i\over
2}\psi^{*\sigma}_1{\pp\psi^{\sigma'}_1\over {\pp P^+}}P^+
P^2\nonumber\\&&~~~~+{i\over 2}\sum_{spin}\int dxd^2q^\perp q^2 
\psi^{*\sigma}_{\sigma_1\lambda} P^\perp  {\pp 
\psi^{\sigma'}_{\sigma'_1\lambda'}\over {\pp q^\perp}}
\nonumber\\&&~~~~~~~~ -{i\over 2}\sum_{spin} \int dx d^2q^\perp  \psi^{*\sigma}_{
\sigma_1\lambda} P^2  {\pp \psi^{\sigma'}_{\sigma'_1\lambda'}\over {\pp
P^+}}P^+\nonumber\\&&~~~~~~~~~~ -{i\over 2}\sum_{spin} \int dx d^2
q^\perp q^2 \psi^{*\sigma}_{\sigma_1\lambda}{\pp\psi^{*\sigma'}_{\sigma'_1
\lambda'} \over {\pp x}} + h.c. \Big ].
\e
\be
\lefteqn{\langle \psi_\sigma \mid {1\over 2}F^2_I(2)P^+\mid \psi_{\sigma'} \rangle = 
\langle \psi_\sigma \mid {1\over 2} \int dx d^2q^\perp x^2 P^-_0 {1\over
2}P^+\mid \psi_{\sigma'} \rangle} \nonumber\\&=& 
{1\over 2} \int dP^+ d^2P^\perp (2\pi)^3 P^+ \Big [{i\over
4}\psi^{*\sigma}_1 {\pp \psi^{\sigma'}_1\over {\pp P^2}}{ (P^\perp)^2 + m^2
\over P^+}P^+ \nonumber\\&&~~~~~+{i\over 4}\sum_{spin}
 \int dxd^2q^\perp \psi^{*\sigma}_{\sigma_1 \lambda} {\pp \psi^{\sigma'}_{
\sigma'_1 \lambda'} \over {\pp P^2}}
 \Big [ {m^2 + (q^\perp+xP^\perp)^2\over x} 
+ {(-q^\perp + (1-x)P^\perp )^2 \over {1-x}}\Big ]\nonumber\\&&~~~~~~~~+ 
{i\over 2}\sum_{spin} \int dx d^2q^\perp  \psi^{*\sigma}_{\sigma_1\lambda}{\pp \psi^{\sigma'}_{
\sigma'_1\lambda'}\over {\pp q^2}} (q^\perp \cdot P^\perp)
\nonumber\\&&~~~~~~~~~~~~~~~+{i\over 4}\sum_{spin}\int dxd^2q^\perp 
\psi^{*\sigma}_{\sigma_1\lambda}{\pp \psi^{\sigma'}_{\sigma'_1\lambda'}
\over {\pp q^2}} (q^\perp)^2 ( {{1-x}\over x} - {x\over {1-x}} ) 
\nonumber\\&&~~~~~~~~~~~~~~~~~~~~ +{i\over 4}\sum_{spin} \int dx d^2q^\perp m^2 {{1-x}\over x} \psi^{*\sigma}_{\sigma_1
\lambda}{\pp \psi^{\sigma'}_{\sigma'_1\lambda'}\over {\pp q^2}} 
 + h.c. \Big ].
\e
In the above two equations, both the single particle and two particle
diagonal matrix elements contribute.
\be
\lefteqn{\langle \psi_\sigma \mid {1\over 2}F^2_I(3)P^+\mid \psi_{\sigma'} \rangle =
\langle \psi_\sigma \mid  {1\over 2} \int dx d^2q^\perp x^2 P^-_{int} {1\over
2}P^+\mid \psi_{\sigma'} \rangle} \nonumber\\&=& {g\over \sqrt {2(2\pi)^3}}
\sum_{spin}{1\over 2}\int dP^+ d^2P^\perp (2\pi)^3P^+ \int dxd^2q^\perp {1\over {\sqrt {1-x}}} \Big ( i\psi^{\sigma *}_1 {\pp \psi^{\sigma'}_{\sigma_1 \lambda}\over {\pp
P^2}}\nonumber\\&&~~\chi^\dagger_{\sigma} \Big [ 
-{(q^\perp \cdot \epsilon^\perp) \over {1-x}} -{1\over 2} { (\tilde
\sigma^\perp \cdot \epsilon^\perp)(\tilde \sigma^\perp \cdot q^\perp)\over
x} -{1\over 2} im {(1-x)\over x}(\tilde 
\sigma^\perp \cdot \epsilon^\perp)
 \Big ]\chi_{\sigma_1}\nonumber\\&&~~~~~~ -{i\over 4} 
\psi^{* \sigma} \chi^\dagger_\sigma [ {\tilde \sigma}^2 ( {\tilde
\sigma}^\perp \cdot \ep^\perp )+ {({\tilde 
\sigma}^\perp \cdot \ep^\perp)
{\tilde \sigma}^2 \over x}]\chi_{\sigma_1} \psi^{\sigma'}_{\sigma_1
\lambda}+ h. c.\Big ).
\e
Only the off-diagonal matrix elements contribute in the above equation.
The matrix elements of the three different parts of ${1\over 2}F^2_{II}P^+$ are given by
      
\be
\langle \psi_\sigma \mid {1\over 2}F^2_{mII}P^+ \mid \psi_{\sigma'}
\rangle&=&    
{1\over 2}\int dP^+ d^2P^\perp (2\pi)^3 P^+\Big [{m\over 2}\psi^{*\sigma}_1
\psi^{\sigma'}_1 \nonumber\\&&~~~~~~~~~+ {m\over 2}\sum_{spin} \int dx d^2q^\perp  \psi^{*\sigma}_{\sigma_1
\lambda} \chi^\dagger_{\sigma_1} \sigma^1 \chi_{\sigma'_1}
\psi^{\sigma'}_{\sigma'_1
\lambda'} {1\over x}\Big ],\label{ab14}
\e
\be
\langle \psi_\sigma \mid {1\over 2}F^2_{q^\perp II} P^+ \mid \psi_{\sigma'}
 \rangle &=& {1\over 2}\int dP^+ d^2P^\perp (2\pi)^3 P^+\Big [
{1\over 2}\sum_{spin}\int dx d^2q^\perp \psi^{*\sigma}_{\sigma_1
\lambda} \chi^\dagger_{\sigma_1} \sigma^3 q^1 \chi_{\sigma'_1}
\psi^{\sigma'}_{\sigma'_1 \lambda'} {1\over x} \nonumber\\&&~~~~~~~~~~~~~
+ {1\over 2} \int dxd^2q^\perp \sum_{\lambda, \sigma_2, \sigma_2'} \lambda 
P^1\psi^{*\sigma}_{\lambda
\sigma_2} \psi^{\sigma'}_{\lambda \sigma_2'}\Big ],\label{ab15}
\e
\be
\langle \psi_\sigma \mid {1\over 2}F^2_{g II} P^+ \mid \psi_{\sigma'}
 \rangle &=& {1\over 4} {g\over {\sqrt {2(2 \pi)^3}}}\sum_{spin} {1\over 2}
\int dP^+ d^2P^\perp (2\pi)^3 P^+\int dx d^2q^\perp\nonumber\\&&~
{1\over {\sqrt {1-x}}} \Big ( i\psi^{*\sigma}_1 \Big [ 
\chi^\dagger_\sigma (
-i\sigma^3 \ep^1_\lambda + \ep^2_\lambda ) \chi_{\sigma_1} - {1\over x}
\chi^\dagger_\sigma (i\sigma^3 \ep^1_\lambda + \ep^2_\lambda ) \chi_{\sigma_1}
 \Big ] \psi^{\sigma'}_{\sigma_1 \lambda}
\nonumber\\&&~~~~~~~~~~~~~~~~~~~~~~~~~ + h. c.\Big ).\label{ab16}
\e
In Eqs. (\ref{ab14}) and (\ref{ab15}), contributions come from only diagonal
matrix elements whereas Eq. (\ref{ab16}) contain only off-diagonal matrix
elements.
The matrix element of ${1\over 2}F^2_{III} P^+$ is given by
\be
\lefteqn{\langle \psi_\sigma \mid {1\over 2}F^2_{III} P^+ \mid \psi_{\sigma'}
 \rangle = -{g \over {\sqrt {2(2\pi)^3}}}{1\over 2}
\int dP^+ d^2P^\perp (2\pi)^3 P^+\Big [\sum_{spin} \int dx d^2q^\perp {1\over
{\sqrt {1-x}}}}\nonumber\\&&~~~ \Big (  \psi^{*\sigma}_1 \ep^2_\lambda 
\psi^{\sigma'}_{\sigma_1 \lambda}{1\over {i(1-x)}} + h. c. \Big )
 - \int dx d^2q^\perp {q^1\over (1-x)} \sum_{\lambda,
\sigma_1, \sigma'_1} \lambda
\psi^{*\sigma}_{\sigma_1 \lambda}\psi^{\sigma'}_{\sigma'_1
\lambda}\nonumber\\&&~~~~~~~~~~~~~~~~~~~~~~~~~~~~~~~~+ 
\int dx d^2q^\perp P^1 \sum_{\lambda, \sigma_1, {\sigma'}_1} \lambda 
\psi^{*\sigma}_{\sigma_1 \lambda} 
\psi^{\sigma'}_{\sigma'_1 \lambda}\Big ] . 
\e

Finally, the expectation value of the transverse spin operator is given by
\be
\langle \psi_{s^1} \mid M{\cal J}^1 \mid \psi_{s^1}
 \rangle = \langle \psi_{s^1} \mid {1\over 2} F^2 P^+ + K^3P^2 - {1\over 2}
E^2 P^- - P^1 {\cal J}^3 \mid \psi_{s^1} \rangle .
\e
From the above expressions it is clear that all the explicit $P^\perp$
dependent terms get canceled in the final expression. 
To be specific, it can be easily seen that all the terms in the expectation
value of $K^3P^2-{1\over 2}E^2P^-_{free} - P^1{\cal J}^3_{orbital}$ are
$P^\perp$ dependent and they exactly cancel the $P^\perp$ dependent terms in
${1\over 2} F^2_{I(free)}P^+$; the two $P^1$ dependent terms in the
intrinsic part of $P^1{\cal J}^3$ exactly cancel the two similar terms in
the expectation value of ${1\over 2}F^2_{II}P^+ + {1\over 2}F^2_{III}P^+$
and the expectation value of ${1\over 2} E^2 P^-_{int}$ completely cancel all
the $P^\perp$ dependent terms in the expectation value of ${1\over
2}F^2_{I(int)}P^+$.  
%%%%%%%%%%%%%%%%%%%%%%%%%%%%%%%%%%%%%%%%%%%%%%%%%%%%%%%%%%%%%%%%%%%%%%%%%%%%
\chapter{ Details of the Calculation}
%%%%%%%%%%%%%%%%%%%%%%%%%%%%%%%%%%%%%%%%%%%%%%%%%%%%%%%%%%%%%%%%%%%%%%%%%%%%
Here, we explicitly show the evaluation of one of the matrix elements of the
interaction part of $F^2_I$.
Consider the operator,
\be
O_g = {1\over 2} \int dx^- d^2x^\perp x^2 g \Big [{( \tilde \sigma^\perp 
\cdot \pp^\perp + m)\over {\pp^+}}\xi^\dagger \Big ] ( \tilde \sigma \cdot
A^\perp) \xi {P^+\over 2}.
\e 
This can be written in Fock space as
\be
O_g&=& {g\over 2}\sum_{s_1, s_2, \lambda}\int (dk_1) \int (dk_2) \int [dk_3] \Big ( b^\dagger ( k_1, s_1 )
a ( k_3, \lambda) b( k_2, s_2 )\nonumber\\&&~ \chi^\dagger_{s_1} {( \tilde \sigma^\perp 
\cdot k_1^\perp - im)\over k_1^+}( \tilde \sigma^\perp
\cdot \ep^\perp_\lambda)  \chi_{s_2} i {\pp\over {\pp k_1^2}} 2(2\pi)^3 \delta^3(
k_1 - k_2 - k_3 ) \nonumber\\&&~~~~~~~~~~~~+ b^\dagger ( k_1, s_1 )
a^\dagger ( k_3, \lambda) b( k_2, s_2 ) \chi^\dagger_{s_1} 
{( \tilde \sigma^\perp 
\cdot k_1^\perp - im)\over k_1^+}( \tilde \sigma^\perp \cdot 
\ep^{*\perp}_\lambda)  \chi_{s_2}\nonumber\\&&~~~~~~~~~~~~~~~~~~~~~~~~~~~~~
i {\pp\over {\pp k_1^2}} 2(2\pi)^3 \delta^3(k_1 - k_2 + k_3 ) \Big ){P^+ 
\over 2}
\e
where $ (dk) = {dk^+ d^2k^\perp\over {2(2\pi)^3}\sqrt k^+}$ and $[dk] = {dk^+ d^2
k^\perp\over { 2(2\pi)^3 k^+}}$.
We evaluate the expectation value of this operator for the dressed quark
state given by Eq.(\ref{awp}). Only the off-diagonal parts of the matrix element will
give non-zero contribution. The matrix element is given by
\be
\langle \psi_\sigma \mid O_g \mid \psi_{\sigma'} \rangle &=&
{g\over 2} \sum_{\sigma_1, \lambda}\int (dP)'\int \{ dp_1 \} \int \{ dp_2 \} \sqrt {2(2\pi)^3}
P^+\delta^3(P-p_1 - p_2)\nonumber\\&&~~~~~\{ \phi_1^{*\sigma}\sqrt {p^+_1}
 \chi^\dagger_\sigma {{(\tilde \sigma^\perp \cdot P^\perp -
im) }\over P^+}( \tilde \sigma^\perp \cdot \ep_\lambda^\perp) \chi_{\sigma_1}
\phi^{\sigma'}_{ \sigma_1 \lambda'} i{\pp\over \pp P^2}\nonumber\\&&~~~~~~~~~~~~~~~~~~~~~~
 2(2\pi)^3 \delta^3 (P-p_1-p_2) {1\over 2} P^+ + h. c. \}
\e  
\be
\lefteqn{=-{ig\over 4}\int (dP)' \int \{ dp_1 \} \int \{ dp_2 \}\sum_{\sigma_1, \lambda} \sqrt {2(2\pi)^3} P^+ 
\Big [\phi_1^{*\sigma}[{\pp \over {\pp P^2}}\phi^{\sigma'}_{ 
\sigma_1 \lambda}\delta^3(
P-p_1 - p_2)\sqrt {p^+_1}}\nonumber\\&&~~~~~~ \chi^\dagger_\sigma 
{{( \tilde \sigma^\perp \cdot P^\perp -im) }\over P^+}( \tilde 
\sigma^\perp \cdot \ep_\lambda^\perp)\chi_{\sigma'}\Big ]
  2(2\pi)^3 \delta^3 (P-p_1-p_2) P^++ h. c. \Big ]\nonumber
\e  
\be
=-{ig\over 4}{1\over {\sqrt {2(2\pi)^3}}}&&\sum_{\sigma_1, \lambda}\int (dP)' \int dx d^2 q^\perp
{1\over {\sqrt {1-x}}}\Big [ \psi_1^{*\sigma} {\pp \psi^{\sigma'}_{\sigma_1 
\lambda}\over {\pp
P^2}}\chi^\dagger_\sigma ( \tilde \sigma^\perp \cdot P^\perp - im)(\tilde 
\sigma^\perp \cdot \ep_\lambda^\perp)\chi_{\sigma'}\nonumber\\&&~~~~~
 + h. c. \Big ] -{ig\over 4}{1\over {\sqrt {2(2\pi)^3}}}\sum_{
\sigma_1, \lambda}\int (dP)' \int dx d^2 q^\perp{1\over {\sqrt
{1-x}}}\Big [ \psi_1^{*\sigma} \psi^{\sigma'}_{\sigma_1 \lambda}
\nonumber\\&&~~~~~~~~~~~~~~~~~~\chi^\dagger_{\sigma} {\tilde \sigma}^2 ( \tilde \sigma^\perp \cdot
\epsilon^\perp_\lambda ) \chi_{\sigma'} + h. c. \Big ].   
\e     
where $\{dp \} = {dp^+ d^2p^\perp \over {\sqrt {2 (2\pi)^3 p^+}}}$ and $ (dP)'
= {1\over 2}dP^+ d^2p^\perp 2(2\pi)^3P^+$. 
The other terms can also be evaluated in a similar method.
%%%%%%%%%%%%%%%%%%%%%%%%%%%%%%%%%%%%%%

%\end{document}

%\include{app2}
\begin{plist}
%\begin{center}
%{\bf LIST OF PUBLICATIONS}
%\end{center}
\noindent Following 
is the list of publications and preprints based on our 
collaborative work. Main results of this thesis are included  in
1-6 of the list. {\it In these papers, my abbreviated surname `Mukherjee'
has been used, athough my official surname is `Mukhopadhyay'}. 
\vskip 1cm
\begin{enumerate}
\item {SUM RULE FOR THE TWIST FOUR LONGITUDINAL STRUCTURE FUNCTION.\\
By A. Harindranath, Rajen Kundu, Asmita Mukherjee and  
James P. Vary;\\
Published in Phys. Lett. {\bf B417}, 361 (1998);\\
e-Print Archive: hep-ph/9711298.}

\item {TWIST FOUR LONGITUDINAL STRUCTURE FUNCTION IN LIGHT FRONT QCD.\\
By A. Harindranath, Rajen Kundu, Asmita Mukherjee and  James P.
Vary;\\
Published in Phys. Rev. {\bf D58}, 114022 (1998);\\
e-Print Archive: hep-ph/9808231.}  

\item {TRANSVERSE SPIN IN QCD AND TRANSVERSE POLARIZED DEEP INELASTIC
SCATTERING.\\
By A. Harindranath, Asmita Mukherjee and Raghunath Ratabole;\\
Published in Phys. Lett. {\bf B476}, 471 (2000);\\
e-Print Archive: hep-ph/9908424.}

\item {TRANSVERSE SPIN IN QCD. I. CANONICAL STRUCTURE.\\
By A. Harindranath, Asmita Mukherjee and Raghunath Ratabole;\\
e-Print Archive: hep-th/0004192.} 

\item {TRANSVERSE SPIN IN QCD : RADIATIVE CORRECTIONS.\\
By A. Harindranath, Asmita Mukherjee and Raghunath Ratabole;\\
Published in Phys. Rev. {\bf D63}, 045006 (2001);\\
e-Print Archive: hep-th/0004193.}

\item {TWIST FOUR LONGITUDINAL STRUCTURE FUNCTION FOR A POSITRONIUM-LIKE
BOUND STATE IN WEAK COUPLING LIGHT-FRONT QED.\\
By Asmita Mukherjee;\\
e-Print Archive: hep-ph/0104175.}

\item {ELECTROMAGNETIC DUALITY ON THE LIGHT-FRONT IN THE PRESENCE OF
EXTERNAL SOURCES.\\
By Asmita Mukherjee and Somdatta Bhattacharya;\\
Published in Int. Jour. of Mod. Phys. {\bf A15}, No. 30, 4739 (2000);\\
e-Print Archive: hep-th/9811105.}

\item {A NUMERICAL EXPERIMENT IN DLCQ: MICROCAUSALITY, CONTINUUM LIMIT AND
ALL THAT.\\
By Dipankar Chakrabarti, Asmita Mukherjee, Rajen Kundu and A.
Harindranath;\\
Published in Phys. Lett. {\bf B480}, 409 (2000);\\
e-Print Archive: hep-th/9910108.}

\item {QUARK TRANSVERSITY DISTRIBUTION IN PERTURBATIVE QCD: LIGHT-FRONT
HAMILTONIAN APPROACH.\\
By Asmita Mukherjee and Dipankar Chakrabarti;\\
Published in Phys. Lett. {\bf B506}, 283 (2001);\\
e-Print Archive: hep-ph/0102003.}

\item {ON THE SIZE OF HADRONS.\\
By H. C. Pauli and Asmita Mukherjee;\\
e-Print Archive: hep-ph/0103150.}

\end{enumerate}
\vskip .5cm 
In addition, following is the list of publications in various conference
proceedings:

\begin{enumerate}
\item {WHEPP-6 QCD WORKING GROUP REPORT.\\
By S. Gupta, D. Indumathi, S. Banerjee, R. Basu, M. Dittmer, R. V. Gavai, F.
   Gelis, D. Ghosh, A. Mukherjee;\\
Published in Pramana, Indian Journal of Physics, Vol. {\bf 55}, 327 (2000);\\
e-Print Archive: hep-ph/0004009.}

\item {A FIELD THEORETIC INVESTIGATION OF SPIN IN QCD.\\
By Asmita Mukherjee;\\
Published in Nucl. Phys. Proc. Suppl. {\bf B90}, 31 (2000);\\
Proceedings of the Xth International Light-Cone Meeting on Non-perturbative
QCD and Hadron Phenomenology, June 12-17, 2000, Heidelberg, Germany;\\
e-Print Archive: hep-ph/0008055.}

\end{enumerate}
\end{plist}


\begin{thebibliography}{99}

\bibitem{c2cheng} {\it Gauge Theory of Elementary Particle Physics}, T. P.
Cheng and L.F. Li, Oxford University Press, New York, 1996.

\bibitem{c2wein} S. Weinberg, {\it The Quantum Theory of Fields}, Vol. I,
Cambridge University Press, 1998.

\bibitem{c2reya} E. Reya, Phys. Rep. {\bf 69}, 196 (1981).

\bibitem{c2wil} See for example, T. Muta, {\it Foundations of Quantum
Chromodynamics} (World Scientific, Singapore, 1987).


\bibitem{c2natch} O. Nachtmann, Nucl. Phys. {\bf B63}, 237 (1973).


\bibitem{c2gorgi} H. Georgi and H. D. Politzer, Phys. Rev. {\bf D14}, 1829
(1976).

\bibitem{c2soldate} R. L. Jaffe and M. Soldate, Phys. Rev. {\bf D26}, 49
(1982). 

\bibitem{c2EFP} R. K. Ellis, W. Furmanski and R. Petronzio, Nucl. Phys. {\bf
B212}, 29 (1983).


\bibitem{c2shur} E. V. Shuryak and A. L. Vainshtein, Nucl. Phys. {\bf B199},
951 (1982); {\bf B201}, 141 (1982).

\bibitem{c2pol} H. D. Politzer, Nucl. Phys. {\bf B172}, 349 (1980).

\end{thebibliography}

\begin{thebibliography}{99}


\bibitem{c3dir} P. A. M. Dirac, Rev. of Mod. Phys. {\bf 21}, 392 (1949).
\bibitem{c3chi} W. M. Zhang, Chi. Jour. of Phys. {\bf 32}, 717 (1994).

\bibitem{c3ped} See A. Harindranath, An Introduction to the Light Front
Dynamics for Pedestrians in {\it Light-Front Quantization and
Non-perturbative QCD}, J. P. Vary and F. Wolz (ed), published by
International Institute of Theoretical and Applied Physics, Ames, Iowa, USA  
(1997), and references therein.  

\bibitem{c3suss} S. Weinberg, Phys. Rev. {\bf 150}, 1313 (1966); L. Susskind,
Phys. Rev. {\bf 165}, 1535 (1968).

\bibitem{c3wil} K. G. Wilson, T. S. Walhout, A. Harindranath, W. M. Zhang 
, R. J. Perry, S. D. Glazek, Phys. Rev. {\bf D 49}, 6270 (1994).

\bibitem{c3ji} C. R. Ji, Bound State and Scattering Problems on the Light Front
in {\it Light-Front Quantization and Non-perturbative QCD}, J. P. Vary and
F. Wolz (ed), published by
International Institute of Theoretical and Applied Physics, Ames, Iowa, USA 
(1997).



\bibitem{c3brod} G. P. Leapage, S. J. Brodsky, T. Huang, P. B. Mackenzie in
Particles and Fields 2, edited by A. Z. Capri and A. N. Kamal (Plenum, New
York, 1983).

\bibitem{c3per} R. J. Perry and A. Harindranath, Phys. Rev. {\bf D43}, 4051
(1991) and references therein.

\bibitem{c3sop} J. B. Kogut and D. E. Soper, Phys. Rev. {\bf D1},2901, (1970).  

\bibitem{c3tom} E. Tomboulis, Phys. Rev. {\bf D8}, 2736 (1973).

\bibitem{c3cas} A. Casher, Phys. Rev. {\bf D14}, 452 (1976).

\bibitem{c3bar} W. A. Bardeen and R. B. Pearson, Phys. Rev. {\bf D13}, 547
(1976).

\bibitem{c3thorn} C. B. Thorn, Phys. Rev. {\bf D19}, 369 (1979); {\bf D20},
1435 (1979); {\bf D20}, 1934 (1979).

\bibitem{c3leap} G. P. Leapage and S. J. Brodsky, Phys. Rev. {\bf D22}, 2157
(1980). 

\bibitem{c3light} W. M. Zhang and A.
Harindranath, Phys. Rev.{\bf D48}, 4868 (1993) and references therein. 

\bibitem{c3similarity} M. Brisudova, S. Szpigel, R. J. Perry, Phys. Lett. {\bf
B421}, 334, (1998); B. H. Allen, R. J. Perry, Phys. Rev. {\bf D58}, 125017
(1998); {\bf D62}, 025005 (2000); R. D. Kylin, B. H. Allen, R. J. Perry, 
Phys. Rev. {\bf D60}, 067704
(1999); T. S. Walhout, Phys. Rev. {\bf D59}, 065009 (1999).


\bibitem{c3two} W. M. Zhang and A.
Harindranath, Phys. Rev.{\bf D48}, 4881. 

\bibitem{c3sti} P. A. M. Dirac, {\it Lectures on Quantum Mechanics} (Yeshiva
University, New York, 1964).

\bibitem{c3jac} R. Jackiw, in Spinger Tracts in Modern Physics, Vol. 62, G.
Hohler (ed), (Springer Verlag, 1972).

\bibitem{c3gla} S. D. Glazek and K. G. Wilson, Phys. Rev. {\bf D48}, 5863
(1993); {\bf D49}, 4214 (1994).

\end{thebibliography}

\begin{thebibliography}{99}
%%%%%%%%%%%%%%%%%%%%%%%%%%%%%%%%%%%%%%%%%%%

\bibitem{c4nonpert} A. Harindranath, R. Kundu and W. M. Zhang, Phys. Rev. {\bf
D59}, 094012, (1999).

\bibitem{c4par} R. P. Feynman, {\it Photon Hadron Interactions} (Benjamin, New
York, 1972).

\bibitem{c4jack} R. Jackiw, 
    {\it Diverse Topics in Theoretical and 
    Mathematical Physics}, (World Scientific, Singapore, 1995), p.309.
    Also see R. Jackiw, Spri. Tracts in Mod. Phys. Vol. 
    {\bf 62} (Springer, Berlin, 1971), p.1. 

\bibitem{c4EPF} R. K. Ellis, W. Furmanski and R. Petronzio, Nucl. Phys. {\bf
B212}, 29 (1983).


\bibitem{c4jaji} R. L. Jaffe and X. Ji, Phys. Rev. {\bf D43}, 724 (1991).

\bibitem{c4efre} A. V. Efremov and O. V. Teryaev, Sov. J. Nucl. Phys. {\bf
39}, 962 (1984).

\bibitem{c4jaffe} R. L. Jaffe and X. Ji, Nucl. Phys. {\bf B375}, 527 (1992).

\bibitem{c4matrix} A. Harindranath and W. M. Zhang, Phys. Lett. {\bf B390},
359 (1997). 

\bibitem{c4fac} J. Collins, D. Soper and G. Sterman, in {\it Perturbative
Quantum Chromodynamics}, edited by A. Mueller (World Scientific, Singapore,
1989), p.1.

\bibitem{c4rad} A. Harindranath, R. Kundu, W. M. Zhang, Phys. Rev. {\bf
D59}, 094013, (1999).

%\bibitem{c4emc} Ashman {\it et al}, Phys. Lett. {\bf B206}, 364 (1988).

\bibitem{c4ano} R. D. Carlitz, J. C. Collins and A. H. Mueller, Phys. Lett.
{\bf B214}, 229 (1988); G. Altarelli and G. G. Ross, Phys. Lett. {\bf B212},
391 (1988). 

\bibitem{c4or} A. Harindranath and R. Kundu, Phys. Rev. {\bf D59}, 116013
(1999).

\bibitem{c4wand} S. Wandzura and F. Wilczek, Phys. Lett. {\bf B72}, 195
(1977).

\bibitem{c4alt} G. Altarelli and S. Muzzetto, Preprint LPTENS 79/27 (1979); J.
Kodaira et al, Nucl. Phys. {\bf B159}, 99 (1979), {\bf B165}, 129 (1979). 

\bibitem{c4shur} E. V. Shuryak and A. I. Vainshtein, Nucl. Phys. {\bf B201},
141 (1982).

\bibitem{c4wan} A. Harindranath and W. M. Zhang, Phys. Lett. {\bf B408}, 347
(1997).
\end{thebibliography}

\begin{thebibliography}{99}

    \bibitem{c5dasu} 
     S. D. Dasu {\it et al.}, Phys. Rev. Lett. {\bf 61}, 1061 (1988);
     L. W. Whitlow, S. Rock, A. Bodek, S. Dasu, and E. M.
     Riordan, Phys. Lett. {\bf B 250}, 193  (1990).   


\bibitem{c5amanda} A. M. Cooper-Sarkar, R. C. E. Devenish, A. De Roeck, Int.
Jour. Mod. Phys. {\bf A 13}, 3385 (1998).
  
\bibitem{c5js} R. L. Jaffe and M. Soldate, Phys. Lett. {\bf 105B}, 467
     (1981); Phys. Rev. {\bf D 26}, 49 (1982).  

    \bibitem{c5efp} R. K. Ellis, W. Furmanski, and R. Petronzio, Nucl. Phys.
       {\bf B  212}, 29 (1983).

     \bibitem{c5qiu} J. Qiu, Phys. Rev. {\bf D 42}, 30 (1990).   

     \bibitem{c5fl1} A. Harindranath, R. Kundu, A. Mukherjee, and
     J. P. Vary, Phys. Lett. {\bf B417}, 361 (1998).


    \bibitem{c5fl2} A. Harindranath, R. Kundu, A. Mukherjee and J. P. Vary,
Phys. Rev. {\bf D58}, 114022 (1998). 

\bibitem{c5jac} R. Jackiw, in Springer Tracts in Modern Physics, Vol.
62, edited by G. Hohler (Springer-Verlag, New York (1972));
  R. Jackiw, {\it Diverse Topics in Theoretical and 
	Mathematical Physics}, (World Scientific, Singapore, 1995), p.309.

\bibitem{c5cjt} D. A. Dicus, R. Jackiw, and V. L. Teplitz, Phys. Rev. {\bf D 4}, 
               1733 (1971).          

%   \bibitem{c5polits} H. D. Politzer, Nucl. 
%      Phys. {\bf B 172}, 349 (1980).
    

\bibitem{c5bgj} In the pre-QCD era this result, for the case of
vanishing target transverse momentum, was obtained by D. J.
Broadhurst, J. F. Gunion, and R. L. Jaffe, Phys. Rev. {\bf D 8}, 566
(1973).

\bibitem{c5rajen}  {\it `Selected Topics in Light-Front Field Theory and
Applications to the High Energy Phenomena'}, R. Kundu, Ph. D. thesis, University of
Calcutta. 

  % \bibitem{c5mira} J. L. Miramontes, M. A. Miramontes, and J. Sanchez Guillen, 
 %    Phys. Rev. {\bf D 40}, 2184 (1989). 
 
%  \bibitem{c5ji} X. Ji, Nucl. Phys. {\bf B 448},  30 (1995).    
  

\bibitem{c5qcd2} Wei-Min Zhang and A. Harindranath, Phys. Rev. {\bf D 48}
  (1993) 4881.

\bibitem{c5jib} X. Ji, Phys. Rev. Lett. {\bf 74}, 1071 (1995); X. Ji, Phys.
Rev. {\bf D 52}, 271 (1995). 

\bibitem{c5luo} M. Luo, J. Qiu, and G. Sterman, Phys. Rev. {\bf D 50}, 1951
(1994).

\bibitem{c5guo} X. Guo, and J. Qiu, Phys. Rev. {\bf D 53},  6144 (1996).

\bibitem{c5bur} M. Burkardt, Nucl. Phys. {\bf B373}, 613 (1992).

%\bibitem{c5bsp} K.G. Wilson, T.S. Walhout, A. Harindranath, W.-M. Zhang, R.J.
%Perry, and St. D. G{\l}azek, Phys. Rev. {\bf D 49},  6720 (1994); 
%Wei-Min Zhang, Phys. Rev. {\bf D 56}, 1528 (1997); M. M.
%Brisudova, R.J. Perry, and K.G. Wilson, Phys. Rev. Lett. {\bf 78} 1227
%(1997). 
%Also see, S.J. Brodsky, H.-C. Pauli, and S.S. Pinsky, {\it Quantum
%Chromodynamics and other field theories on the light cone}, SLAC-PUB-7484, 
%hep-ph/9705477 and references therein.

% \bibitem{c5hari1} A. Harindranath and Wei-Min Zhang, Phys. Lett. {\bf B
%        390}, 359 (1997).

%    \bibitem{c5wh} Wei-Min Zhang and A. Harindranath, preprint (1996)
%        hep-ph/9606347.

%    \bibitem{c5hari2} A. Harindranath and Wei-Min Zhang, 
%   Phys. Lett. {\bf B 408}, 347 (1997); hep-ph/9706419.


%    \bibitem{c5pqcd} J. Collins, D. Soper and G. Sterman, in {\it Perturbative 
%	Quantum Chromo Dynamics}, ed. by A. Mueller 
%    (World Scientific, Singapore 1989) p.1.


%\bibitem{c5paper2} A. Harindranath, Rajen Kundu, and Wei Min Zhang, {\it Deep
%Inelastic Structure Functions in Light-Front QCD: Radiative Corrections},
%preprint, April 1998.



\end{thebibliography}

\begin{thebibliography}{99}


\bibitem{c6hari1} A. Harindranath, R. Kundu, W. M. Zhang, Phys. Rev. {\bf D
59}, 094012, (1999).



\bibitem{c6sim} S. D. Glazek and K. G. Wilson, Phys. Rev. {\bf D 48}, 5863,
(1993); S. D. Glazek and K. G. Wilson, Phys. Rev. {\bf D 49}, 4214,
(1993); K. G. Wilson, T. S. Walhout, A. Harindranath, W. M. Zhang, R. J.
Perry and S. D. Glazek, Phys. Rev {\bf D 49}, 6720 (1994).  

\bibitem{c6mar} M. Brisudova and R. J. Perry, Phys. Rev. {\bf D 54}, 1831,
(1996); M. Brisudova, R. J. Perry and K. G. Wilson, Phys. Rev. Lett. {\bf
78}, 1227 (1997); W. M. Zhang, Phys. Rev. {\bf D56}, 1528 (1997).

\bibitem{c6prd} A. Harindranath, R. Kundu, A. Mukherjee, J. P. Vary, Phys. Lett
{\bf B 417}, 361, (1998); A. Harindranath, R. Kundu, A. Mukherjee, J. P. Vary,
Phys. Rev. {\bf D 58}, 114022, (1998).

\bibitem{c6as} A. Mukherjee, hep-ph/0104175.

\bibitem{c6berg} H. Bergknoff, Nucl. Phys. {\bf B122}, 215 (1977).

\bibitem{c6bur} M. Burkardt, Nucl. Phys. {\bf B 373}, 371, (1992).

\bibitem{c6billy} {\it `Light-Front Hamiltonian Approach to the Bound State
Problem in Quantum Electrodynamics'}, B. D. Jones, Ph. D. thesis, The Ohio
State University, 1997.


\bibitem{c6brod} S. J. Brodsky and G. P. Leapage, Phys. Rev. {\bf D22}, 2157,
(1980). 

\end{thebibliography}

\begin{thebibliography}{99}

\bibitem{c7jnp} R. L. Jaffe, Comments. Nucl. Part. Phys. Vol 19, No. 5, 239
(1900) and references therein.

\bibitem{c7efre} M. Anselmino, A. Efremov, E. Leader, Phys. Rept. {\bf 261}, 1
(1995).

\bibitem{c7hei}R. L. Heiman, Nucl. Phys. {\bf B64}, 429 (1973).
 
\bibitem{c7sh} S. Wandzura, Nucl. Phys. {\bf B122}, 412 (1977); E. V. Shuryak and A. I Vainshtein, Nucl. Phys. {\bf B201}, 141
(1982); R. L. Jaffe and M. Soldate, Phys. Lett. {\bf B105}, 467
(1981); Phys. Rev. {\bf D26}, 49 (1982).

\bibitem{c7alfaro} V. de Alfaro, S. Fubini, G. Furlan, and G. Rossetti, {\it
Currents in Hadron Physics} (North-Holland, Amsterdam, (1973)).


\bibitem{c7gur} See,  for example, F. Gursey in {\it High Energy Physics}, 
C. DeWitt \& R. Omnes (eds.)
(Gordon \& Breach Science Publishers, (1965)).  

\bibitem{c7os} See, for example, B. Bakamjian and
L. H. Thomas, Phys. Rev. {\bf 92} 1300 (1955); 
H. Osborn, Phys. Rev. {\bf 176} 1514 (1968). 

\bibitem{c7jsc} J. S. Schwinger, {\it Particles, Sources and Fields, Vol 1},
(Addison-Wesley Publishing Company, Inc., California (1989)),
chapter 1; E.C.G. Sudarshan and N. Mukunda {\it Classical Dynamics: A Modern
Perspective} (John Wiley \& Sons, New York (1974)); J. S. Lomont and H.
E. Moses, Jour. of Math. Phys {\bf 3} 405 (1962).


\bibitem{c7ls78} 
K. Bardakci and M. B. Halpern, Phys. Rev. {\bf 176}, 1686 (1968);
D. E. Soper, {\it Field Theories
in the Infinite Momentum Frame}, Ph. D. Thesis, Stanford University, (1971),
SLAC-137; 
H. Leutwyler and J. Stern, Ann. Phy. {\bf 112}, 94 (1978). 

\bibitem{c7hk} A. Harindranath and R. Kundu, Phys. Rev. D {\bf 59},
             116013 (1999).


\bibitem{c7except} For attempts to incorporate interactions, see for example,
E. Eichten, F. Feinberg, and J.F. Willemsen, Phys. Rev. D {\bf 8}, 1204
(1973); R. Carlitz and W.K. Tung, Phys.
     Rev. D {\bf 13}, 3446 (1976).


\bibitem{c7melosh} H.J. Melosh, Phys. Rev. D {\bf 9}, 1095 (1974).

\bibitem{c7bp} M. Brisudova and R.J. Perry, Phys. Rev. D {\bf 54}, 6453
(1996).

\bibitem{c7review} S. J. Brodsky, H.C. Pauli, and S. S. Pinsky, Phys. Rep. 
{\bf 301}, 299 (1998).

\bibitem{c7lett} A. Harindranath, A. Mukherjee and R. Ratabole,
 Phys. Lett. {\bf B476}, 471 (2000).

\bibitem{c7tran1} A. Harindranath, A. Mukherjee and R. Ratabole,
hep-th/0004192.
         
\bibitem{c7Tung} Wu Ki Tung, {\it Group Theory in Physics} (World
Scientific, Singapore (1985)).


\bibitem{c7Weinberg} S. Weinberg, {\it Quantum Theory of Fields, Vol. I,}
(Cambridge University Press, Cambridge (1995)).
 
\bibitem{c7ji} X. Ji, Phys. Rev. Lett. {\bf 78} 610 (1997).


\bibitem{c7expt} D. Adams {\it et al.}, Phys. Lett. {\bf B336}, 125
(1994); K. Abe {\it et al.}, Phys. Rev. Lett. {\bf 76}, 587
(1996); P.L. Anthony {\it et al.}, SLAC-PUB-7983, January 1999,
        hep-ex/9901006.

\bibitem{c7Ji} X. Ji, Phys. Lett. {\bf B289} 137 (1992).

\bibitem{c7ped} See A. Harindranath, {\it An Introduction to Light-Front
Dynamics for Pedestrians} in Light Front Quantization and Non-perturbative
QCD, J. P. Vary and F. Woltz (ed.); published by International Institute of
Theoretical and Applied Physics, Ames, Iowa, USA (1997).

\bibitem{c7glazek} T. Maslowski and  S. D. Glazek, 
{\it Construction of the Effective Poincare Algebra}, hep-th/9906140. 

\end{thebibliography}

\begin{thebibliography}{99}
%%%%%%%%%%%%%%%%%%%%%%%%%%%%%%%%%%%%%%%%%%%%%%%%%%%%%%%%%%%%%%%%%%%%%%%%%%%%%%%

\bibitem{c8hk} A. Harindranath and R. Kundu, Phys. Rev. D {\bf 59},
             116013 (1999).

\bibitem{c8ang} S. J. Brodsky, D. S. Hwang, B. Q. Ma and I. Schmidt,
hep-th/0003082.

\bibitem{c8wilson} K. G. Wilson, T. S. Walhout, A. Harindranath, W. M. Zhang,
R. J. Perry, S. D. Glazek, Phys. Rev. {\bf D49}, 6720, (1994).

\bibitem{c8spin} A. Harindranath, A. Mukherjee and R. Ratabole, Phys. Rev.
{\bf D63}, 045006 (2001).

\bibitem{c8hei} A. Mukherjee, Nucl. Phys. Proc. Suppl. {\bf B90}, 31 (2000).

\bibitem{c8ji} X. Ji, Phys. Rev. Lett. {\bf 78}, 610(1997).
 
\bibitem{c8let} A. Harindranath, A. Mukherjee and R. Ratabole,
 Phys. Lett. {\bf B476}, 471 (2000).

\bibitem{c8glazek} T. Maslowski and S. Glazek, hep-th/9906140.

\bibitem{c8hari3} R.J. Perry, Phys. Lett. {\bf B 300}, 8 (1993); 
A. Harindranath and W.-M. Zhang, Phys. Rev. {\bf D
48}, 4903 (1993).

\bibitem{c8hari1} A. Harindranath and W.-M. Zhang, Phys. Lett. {\bf B 408}, 347
(1997).
 
\bibitem{c8dip} A. Mukherjee and D. Chakrabarti, hep-ph/0102003, to appear in
Phys. Lett. B.

\bibitem{c8os} B. Bakamjian and L. H. Thomas, Phys. Rev. {\bf 92}, 1300
(1950);
H. Osborn, Phys. Rev. {\bf 176}, 1514 (1968).

\bibitem{c8bk} See for example, M. Burkardt and H. El-Khozondar,
Phys. Rev. D {\bf 60}, 054504 (1999). 

 
\end{thebibliography}

\begin{thebibliography}{99}

\bibitem{ssim}S. D. Glazek and K. G. Wilson, Phys. Rev. {\bf D 48}, 5863,
(1993); S. D. Glazek and K. G. Wilson, Phys. Rev. {\bf D 49}, 4214,
(1993); K. G. Wilson, T. S. Walhout, A. Harindranath, W. M. Zhang, R. J.
Perry and S. D. Glazek, Phys. Rev {\bf D 49}, 6720 (1994).  

\bibitem{sdalley} S. Dalley and  B. Van de Sande, Phys. Rev. {\bf D56}, 7917
(1997); M. Burkardt and B. Klindworth, Phys. Rev. {\bf D55}, 1001 (1997); S.
Dalley and B. Van de Sande, Phys. Rev. Lett. {\bf 82}, 1088 (1999); S. Dalley, 
hep-ph/0101318; M. Burkardt and S. K. Seal, hep-ph/0102245.

\end{thebibliography}

\begin{thebibliography}{99}
%%%%%%%%%%%%%%%%%%%%%%%%%%%%%%%%%%%%%%%%
\bibitem{aclose} F. E. Close, {\it An Introduction to Quarks and Partons},
Academic Press, 1979.

\bibitem{areya} E. Reya, Phys. Rep. {\bf 69}, 195 (1981).

\bibitem{asimple} R. K. Ellis, W. Furmanski, and R. Petronzio, Nucl. Phys.
{\bf B 207}, 1 (1982). 

\bibitem{ahk} A. Harindranath and R. Kundu, Phys. Rev. D {\bf 59},
             116013 (1999).

\end{thebibliography}
\end{document}